\documentclass[a4paper, fontsize=12pt, twoside=true, numbers=noenddot]{kaobook}

\ifxetexorluatex
	\usepackage{polyglossia}
	\setmainlanguage{english}	
\else
	\usepackage[english]{babel} 
\fi
\usepackage[english=british]{csquotes}	
\usepackage{amsmath,nicefrac,bbm,kaobiblio,xcolor,kaorefs,physics,xfrac}
\usepackage{caption}
\usepackage{enumitem}
\usepackage{adjustbox}
\usepackage{wrapfig}
\usepackage{graphicx}
\usepackage{pdfpages}
\usepackage{nicematrix}
\usepackage{tabularx, multirow, makecell}
\usepackage{tikz}

\usetikzlibrary{arrows}
\usetikzlibrary{positioning}
\usetikzlibrary{calc}
\DeclareUnicodeCharacter{0301}{'}
\DeclareUnicodeCharacter{0302}{\^{}}
\DeclareUnicodeCharacter{2217}{*}

\definecolor{definitioncolor}{RGB}{240,240,255}
\definecolor{theoremcolor}{RGB}{200, 218, 191}
\definecolor{lemmacolor}{RGB}{233, 190, 190}
\definecolor{observationcolor}{RGB}{255, 232, 205}
\definecolor{propositioncolor}{RGB}{217, 213, 209}
\definecolor{corollarycolor}{RGB}{194, 185, 211}
\definecolor{kaoboxcolor}{RGB}{238,237,238}
\definecolor{equationcolor}{RGB}{222,94,100}
\definecolor{citationcolor}{RGB}{94,102,228}

\definecolor{boxcolor}{RGB}{215,215,253}
\definecolor{block2c}{RGB}{215,185,212}
\definecolor{block3c}{RGB}{216,155,172}
\definecolor{block4c}{RGB}{216,125,132}

\hypersetup{
    linkcolor=equationcolor,
    citecolor=equationcolor,
    filecolor=black,
    urlcolor=equationcolor
}

\newcommand{\mA}{\mathcal{A}}
\newcommand{\mB}{\mathcal{B}}
\newcommand{\mC}{\mathcal{C}}
\newcommand{\mD}{\mathcal{D}}
\newcommand{\mE}{\mathcal{E}}
\newcommand{\mF}{\mathcal{F}}

\newcommand{\mH}{\mathcal{H}}

\newcommand{\mK}{\mathcal{K}}
\newcommand{\mL}{\mathcal{L}}
\newcommand{\mM}{\mathcal{M}}

\newcommand{\mO}{\mathcal{O}}

\newcommand{\mR}{\mathcal{R}}
\newcommand{\mS}{\mathcal{S}}
\newcommand{\mT}{\mathcal{T}}
\newcommand{\mU}{\mathcal{U}}
\newcommand{\mV}{\mathcal{V}}

\newcommand{\mbD}{\mathbb{D}}

\newcommand{\mbM}{\mathbb{M}}
\newcommand{\mbR}{\mathbb{R}}
\newcommand{\mbP}{\mathbb{P}}
\newcommand{\mbPrev}{\scalebox{-1}[1]{$\mbP$}}
\newcommand{\revrelent}[1]{\scalebox{-1}[1]{R}\pqty{#1}}

\newcommand{\pstate}{\mathbf{p}}
\newcommand{\qstate}{\mathbf{q}}
\newcommand{\rstate}{\mathbf{r}}

\newcommand{\estate}{\mathbf{e}}
\newcommand{\slope}{\mathbf{g}}
\newcommand{\catstate}{\mathbf{c}}
\newcommand{\probspace}{\mV}
\newcommand{\dsucc}{\succ_{\beta}}

\DeclareMathAlphabet{\mathmybb}{U}{bbold}{m}{n}
\newcommand{\1}{\mathmybb{1}}

\newcommand{\psivec}[1]{\ket{\psi_{#1}}}
\newcommand{\Upsirec}[1]{\hat{U}^{[#1]}}
\newcommand{\Upsirecch}[1]{\hat{\mathbf{U}}^{[#1]}}
\newcommand{\Upsirecapp}[1]{\hat{\mathcal{U}}^{[#1]}}

\newcommand{\Eopch}[2]{\hat{\mathbf{E}}^{[#2]}_{#1}}
\newcommand{\hE}[2]{\hat{\mE}^{[#2]}_{#1}}

\newcommand{\frs}[1]{\mS_{#1}}
\newcommand{\fro}[2]{\mO_{#1 \to #2}}
\newcommand{\RNG}[2]{\mO_{#1 \to #2}^{\mathrm{RNG}}}
\newcommand{\CRNG}[2]{\mO_{#1 \to #2}^{\mathrm{CRNG}}}
\newcommand{\mincomp}[2]{\frs{#1}\!\otimes_{\text{min}}\!\frs{#2}}
\newcommand{\maxcomp}[2]{\frs{#1}\!\otimes_{\text{max}}\!\frs{#2}}
\newcommand{\sepcomp}[2]{\frs{#1}\!\otimes_{\text{sep}}\!\frs{#2}}

\newcommand{\GP}[2]{\mO_{#1 \to #2}^{\mathrm{GP}}}
\newcommand{\COV}[2]{\mO_{#1 \to #2}^{\mathrm{COV}}}
\newcommand{\GPC}[2]{\mO_{#1 \to #2}^{\mathrm{GPC}}}
\newcommand{\TO}[2]{\mO_{#1 \to #2}^{\mathrm{TO}}}
\newcommand{\eTO}[2]{\mO_{#1 \to #2}^{\mathrm{TO}_{2}}}
\newcommand{\ETO}[2]{\mO_{#1 \to #2}^{\mathrm{ETO}}}
\newcommand{\MTO}[2]{\mO_{#1 \to #2}^{\mathrm{MTO}}}
\newcommand{\SYM}[1]{\mD_{#1}^{\mathrm{SYM}}}
\newcommand{\setXTO}{\mT_{\mathrm{X}}}
\newcommand{\setETO}{\mT_{\mathrm{ETO}}}
\newcommand{\setMTO}{\mT_{\mathrm{MTO}}}
\newcommand{\setTO}{\mT_{\mathrm{TO}}}
\newcommand{\setCETO}{\mT_{\mathrm{CETO}}}
\newcommand{\setCETOqb}{\mT_{\mathrm{CETO}}^{(2)}}
\newcommand{\setCETOqd}[1]{\mT_{\mathrm{CETO}}^{(#1)}}
\newcommand{\setCTO}{\mT_{\mathrm{CTO}}}
\newcommand{\setCMTO}{\mT_{\mathrm{CMTO}}}
\newcommand{\setGCETO}{\mT_{\mathrm{GC-ETO}}}

\newcommand{\setGCXTO}{\mT_{\mathrm{GC-X}}}

\newcommand{\setGCMTO}{\mT_{\mathrm{GC-MTO}}}

\newcommand{\toeto}{\xrightarrow{\mathrm{ETO}}}
\newcommand{\toto}{\xrightarrow{\mathrm{TO}}}

\DeclareMathOperator{\extr}{extr}

\DeclareMathOperator{\id}{id}

\DeclareMathOperator{\sgn}{sgn}

\DeclareMathOperator{\SEP}{SEP}
\newcommand{\CPTP}[2]{\mathrm{CPTP}_{#1 \to #2}}
\DeclareMathOperator{\conv}{conv}
\DeclareMathOperator{\spec}{spec}

\DeclareMathOperator{\supp}{supp}

\DeclareMathOperator{\SWAP}{SWAP}
\newcommand{\KL}{\mathrm{KL}}
\newcommand{\gbs}{\gamma^{\beta}}

\newcommand{\scto}{\xrightarrow{\mathrm{sc}}}
\newcommand{\ccto}{\xrightarrow{\mathrm{cc}}}

\renewcommand{\ket}[1]{|{#1}\rangle}
\renewcommand{\bra}[1]{\langle{#1}|}
\renewcommand{\braket}[2]{\langle{#1}|{#2}\rangle}
\renewcommand{\ketbra}[2]{|{#1}\rangle\!\langle{#2}|}
\newcommand{\dm}[1]{\ketbra{#1}{#1}}
\renewcommand{\ev}[2]{\bra{#2}#1\ket{#2}}

\usepackage[framed=true]{kaotheorems}

\begin{document}

\frontmatter 


%
\includepdf[pages=-]{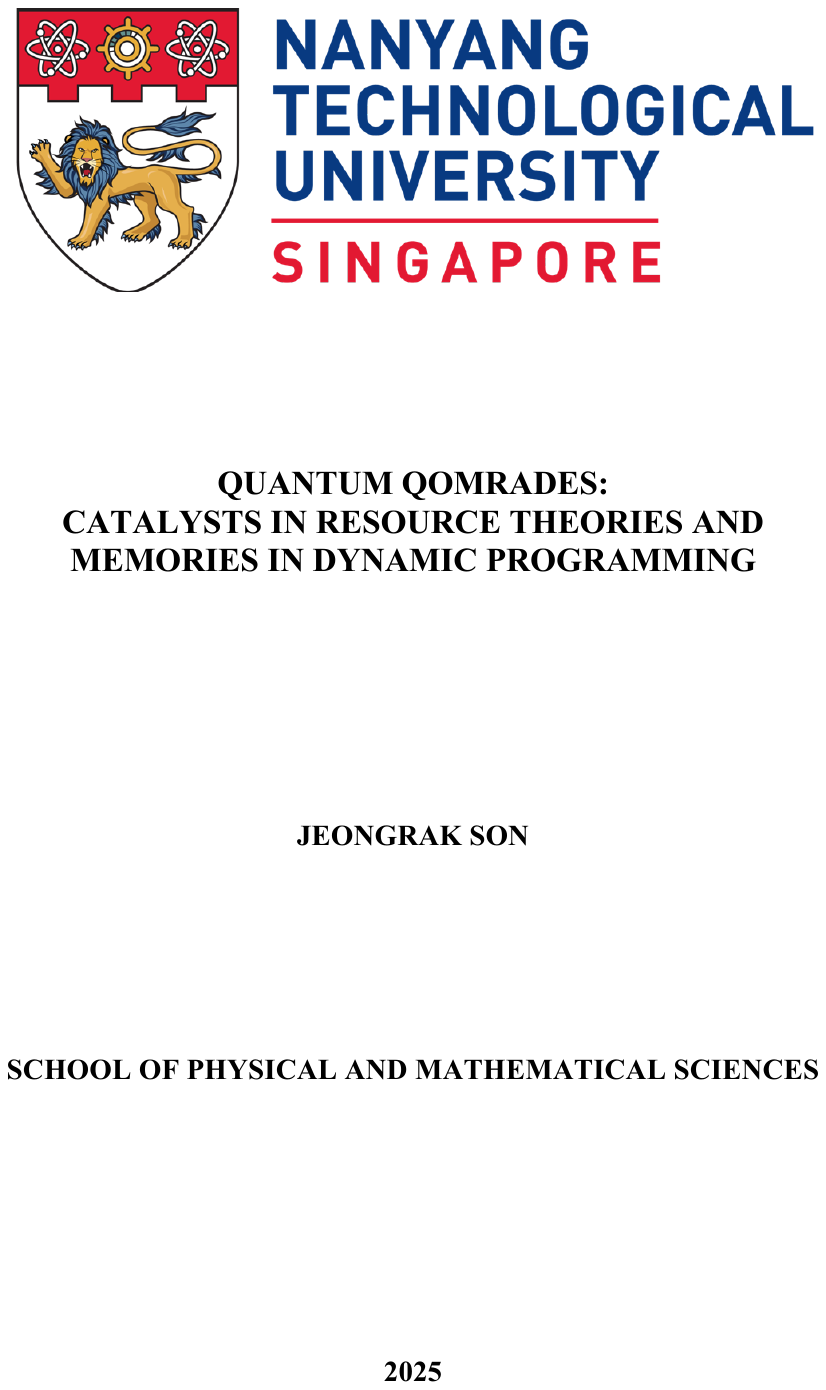}
%
%
%
\begin{center}
	\large \textbf{Authorship Attribution Statement}
\end{center}

This thesis contains material from five papers published in the following peer-reviewed journal(s) / posted as preprints in which I am listed as an author. 

\begin{itemize}
	\item[\cite{Son2024_CETO}] \textbf{Catalysis in action via elementary thermal operations}, Jeongrak Son and Nelly H. Y. Ng,  {\href{https://doi.org/10.1088/1367-2630/ad2413}{\color{teal}\textit{New J. Phys.} \textbf{26}, 033029 (2024).}}
	
	\item[\cite{Son2024hierarchy}] \textbf{A hierarchy of thermal processes collapses under catalysis}, Jeongrak Son and Nelly H. Y. Ng,  {\href{https://doi.org/10.1088/2058-9565/ad7ef5}{\color{teal}\textit{Quantum Sci. Technol.} \textbf{10}, 015011 (2024).}}
	
	\item[\cite{SonQDP}] \textbf{Quantum Dynamic Programming}, Jeongrak Son, Marek Gluza, Ryuji Takagi, and Nelly H. Y. Ng, {\href{https://doi.org/10.1103/PhysRevLett.134.180602}{\color{teal}\textit{Phys. Rev. Lett.} \textbf{134}, 180602 (2025).}}
	
	\item[\cite{SonDBQITE}] \textbf{Double-bracket quantum algorithms for quantum imaginary-time evolution}, Marek Gluza, Jeongrak Son, Bi Hong Tiang, René Zander, Raphael Seidel, Yudai Suzuki, Zoë Holmes, and Nelly H. Y. Ng, {\href{https://doi.org/10.48550/arXiv.2412.04554}{\color{teal} arXiv:2412.04554 (2025).}}
	
	\item[\cite{SonRCRB}] \textbf{Robust Catalysis and Resource Broadcasting: The Possible and the Impossible}, Jeongrak Son, Ray Ganardi, Shintaro Minagawa, Francesco Buscemi, Seok Hyung Lie, and Nelly H.Y. Ng, {\href{https://doi.org/10.48550/arXiv.2412.06900}{\color{teal} arXiv:2412.06900 (2024).}}
	
\end{itemize}

Chapter~\ref{chapter: CETO} and Appendix~\ref{chapter: ETO cone characterisation} contain material from Ref.~\cite{Son2024_CETO}.
	
Chapter~\ref{chapter: CETO} and Appendix~\ref{chapter: Lie groups} contain material from Ref.~\cite{Son2024hierarchy}.
	
Chapter~\ref{chapter: resource broadcasting} contains material from Ref.~\cite{SonRCRB}.
	
Chapter~\ref{chapter: quantum recursions} contains material from Ref.~\cite{SonDBQITE}.
	
Chapter~\ref{chapter: QDP} contains material from Ref.~\cite{SonQDP}.

In all five papers listed, all authors contributed to deriving main results and writing the manuscript.
I acted as the lead author and wrote the first draft for Refs.~\cite{Son2024_CETO, Son2024hierarchy, SonQDP, SonRCRB}.

During my PhD studies, I have also completed the following publications and preprints. 
These works are not included in this thesis to maintain a singular narrative focus.

\begin{itemize}
	\item[\cite{Son2021Monitoring}] \textbf{Monitoring Quantum Otto Engines}, Jeongrak Son, Peter Talkner, and Juzar Thingna,  {\href{https://doi.org/10.1103/PRXQuantum.2.040328}{\color{teal}\textit{PRX Quantum} \textbf{2}, 040328 (2021).}}
	
	\item[\cite{Son2022Battery}] \textbf{Charging quantum batteries via Otto machines: Influence of monitoring}, Jeongrak Son, Peter Talkner, and Juzar Thingna,  {\href{https://doi.org/10.1103/PhysRevA.106.052202}{\color{teal}\textit{Phys. Rev. A} \textbf{106}, 052202 (2022).}}
	
	\item[\cite{SonEntGen}] \textbf{Entanglement generation from athermality}, A. de Oliveira Junior*, Jeongrak Son*, Jakub Czartowski, and Nelly H. Y. Ng,  {\href{https://doi.org/10.1103/PhysRevResearch.6.033236}{\color{teal}\textit{Phys. Rev. Research} \textbf{6}, 033236 (2024).}}
	
	\item[\cite{SonDBINumerics}] \textbf{Double-bracket quantum algorithms for high-fidelity ground state preparation}, Matteo Robbiati*, Edoardo Pedicillo*, Andrea Pasquale*, Xiaoyue Li, Andrew Wright, Renato M. S. Farias, Khanh Uyen Giang, Jeongrak Son, Johannes Knörzer, Siong Thye Goh, Jun Yong Khoo, Nelly H.Y. Ng, Zoë Holmes, Stefano Carrazza, and Marek Gluza, {\href{https://doi.org/10.48550/arXiv.2408.03987}{\color{teal} arXiv:2408.03987 (2024).}}
	
	\item[\cite{SonDBQSP}] \textbf{Double-bracket algorithm for quantum signal processing without post-selection}, Yudai Suzuki, Bi Hong Tiang, Jeongrak Son, Nelly H. Y. Ng, Zoë Holmes, and Marek Gluza, {\href{https://doi.org/10.48550/arXiv.2504.01077}{\color{teal} arXiv:2504.01077 (2025).}}
	
	\item[*:] co-first authors
\end{itemize}

\noindent\rule{6cm}{0.4pt} \hfill \rule{6cm}{0.4pt} \\ 
Date: 16 May 2025 \hfill Jeongrak Son

\index{list-of-publication}

%
\newpage

\begin{center}
\large \textbf{Abstract}    
\end{center}

Quantum information theory seeks to delineate the ultimate limits and efficient pathways for manipulating quantum states. 
One common strategy for enhancing various information processing tasks is to use auxiliary systems, which often act as remarkably effective aids. 
Yet, a systematic and unified explanation for why these systems are so powerful, and how their capabilities can be exploited, has remained incomplete. 
This thesis addresses this gap by investigating the underlying sources of strength in auxiliary systems. 
We then demonstrate how these insights can be applied in practical quantum information protocols, with the aim of building a more thorough and intuitive understanding of how to best harness these effects.

In particular, our research operates within two frameworks: resource theories and recursive algorithms in quantum computation.
Within resource theories, which represent an idealisation of physical settings with limited capabilities, our focus is on the catalytic role of auxiliary systems.
This choice stems from the stringent resource accounting inherent in such theories; catalysts, by definition, must remain unchanged in their resource content, thus integrating seamlessly into this framework.

We identify three sources of catalytic advantage. 
First, catalysts can provide a memory effect. 
Within resource theories imposing Markovian restrictions, even states typically considered 'free' (useless) can be catalytically employed to circumvent these constraints. 
We demonstrate, specifically, that this form of catalytic advantage renders multiple models of quantum thermodynamics equivalent. 
Second, we establish that the capacity to fine-tune catalyst states relative to the system's initial state is critical for most catalytic advantages. 
Introducing the concept of state-agnostic catalysis, we prove that for a broad class of resource theories, it enables no non-trivial operations. 
However, in scenarios where state-agnostic catalysis does confer advantages, its impact is substantial. 
We prove an equivalence between state-agnostic catalysis and a process wherein catalysts act as `seeds', disseminating resources to other systems whilst retaining their own state. 
We also provide an explicit construction of such processes for a generic class of resource theories.
This role as a seed for the resource distribution constitutes the third identified source of catalytic power.

In the second part of this thesis, auxiliary systems are examined within the computational context.
They function similarly to catalysts: as memories operate in a state-agnostic manner. 
We introduce the notion of quantum recursion and illustrate its versatility through an example algorithm simulating non-linear, non-unitary quantum evolution. 
The primary bottleneck in such algorithms---exponential growth of circuit depth with each recursion step---is addressed by employing auxiliary states as memory registers that instruct the computation. 
Crucially, these memory states need not be learnt or measured due to the state-agnostic nature of the framework. 
This strategy achieves an exponential reduction in circuit depth, albeit at the cost of an exponential increase in circuit width. 
This depth-width trade-off represents a quantum analogue of classical dynamic programming, where a large problem is solved by solving smaller problems recursively. 
Furthermore, this trade-off is controllable: quantum dynamic programming can be employed in tandem with conventional memoryless strategies, enabling the optimisation of circuit dimensions for specific hardware platforms with constraints on maximum depth and width.

Combining our findings within resource theories and recursive quantum computation, this thesis provides a more holistic understanding of the advantages offered by auxiliary systems in quantum information processing and offers insights into their optimal utilisation. 
This understanding is expected to catalyse future research into protocols leveraging additional systems in fundamental problems such as resource interconversion or practical ones like optimising quantum circuit synthesis. 

\vspace*{\fill}
\newpage
\index{abstract}

\includepdf[pages=-]{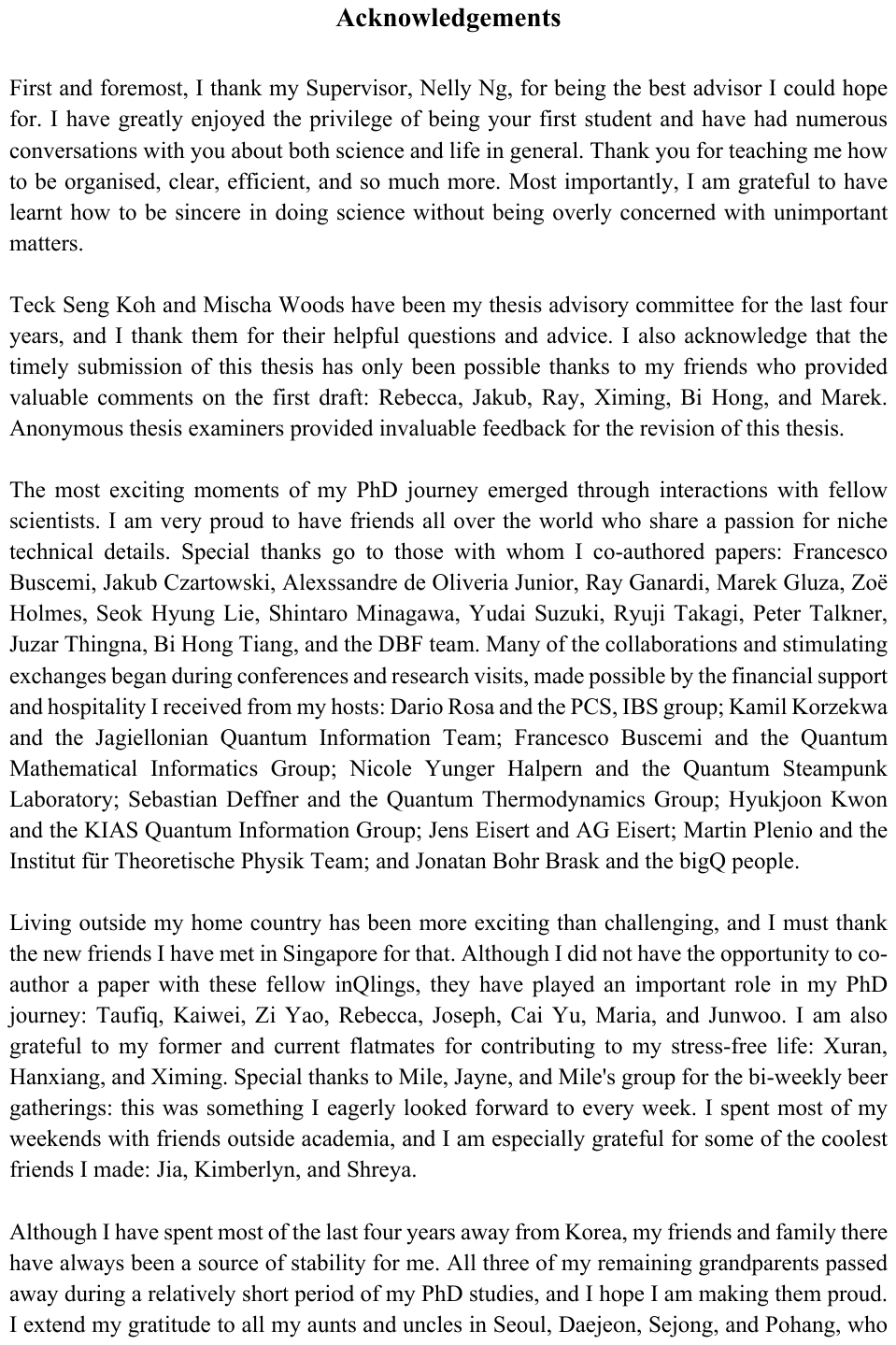}

\let\cleardoublepage\clearpage


\begingroup 
\pagenumbering{gobble}
\setlength{\textheight}{230\hscale} 

\etocstandarddisplaystyle 
\etocstandardlines 

{
  \hypersetup{linkcolor=black}
\tableofcontents* 
}


\newpage ~~
\thispagestyle{empty}
\endgroup

{
  \let\oldclearpage\clearpage 
  \let\clearpage\bigskip 


 
\mainmatter 
\setchapterstyle{kao} 

\pagelayout{wide} 
\setcounter{page}{0}
\chapter{Overview: auxiliary systems in quantum information processing}\label{chapter: overview}

Often, the primary impetus behind scientific research is the pursuit of universal laws.
Theorists approach this pursuit by formulating it as an optimisation problem: identifying the best strategy for a given task among all considered protocols.
The joy of being a theorist lies in the freedom to designate which protocols to consider for optimisation, chosen from all that are possible \emph{in potentia}, unencumbered by the time and budget constraints that bound experimentalists, but guided by physical laws such as symmetry or locality. 

A prominent example is thermodynamics. 
Modern thermodynamics began with Sadi Carnot's realisation that irrespective of specific details, the maximum achievable efficiency of heat engines is dictated by the temperatures of the heat baths.
This universal insight propelled theorists like Lord Kelvin and Rudolf Clausius to derive the famous upper bound on heat engine efficiency---valid for all engine operations with fixed hot and cold temperatures. 
Ultimately, this optimisation problem led to one of the most enduring universal laws: the second law of thermodynamics~\cite{MullerBook}. 

The history repeated itself in information theory~\cite{book_infotheory}, whose goal is to establish universal laws governing information processing and communication.
Coding theorems in (Shannon) information theory determine the optimal rate at which (noisy) communication channels can reliably transmit information, given all possible classical signals~\cite{Shannon1948}.
Theorists then extended the scope of this optimisation problem to include both quantum information sources and quantum channels, and new versions of coding theorems~\cite{Holevo1973, Schumacher1995Coding, Holevo1998} emerged as a result.
It is a great coincidence, or a fundamental connection, that information-theoretic generalisations of entropy inspired by thermodynamics (statistical mechanics to be specific) play central roles in both problems.

The extension to quantum systems had a more profound impact than simply generalising coding theorems: an entirely new field of quantum information theory has emerged as the amalgamation of information theory and quantum theory. 
In addition to the information transmission problem, this field encompasses a wide range of topics, including the studies on the information sources themselves.
While classical information theory also investigates the properties of information sources and how to manipulate them, quantum sources exhibit far richer and subtler behaviours.
Efforts to formulate universal laws governing the manipulation of quantum properties have led to the development of resource theories~\cite{Chitambar2019RTRev, Gour2024RTBook}, which form one of the two main themes of this thesis.

Another area in quantum information theory that has garnered a huge amount of interest is quantum computation. 
Information theory has long established fundamental limits on computation~\cite{Landauer1961, Kolmogorov1963}. 
Consequently, recognising the quantum nature of information processing presented a clear opportunity to explore a novel computational paradigm.
This quantum paradigm is believed to have the potential to provide computational advantages over classical counterparts. 
Evidence for this potential includes not only provable speed-ups for certain problems~\cite{DeutschJozsa1992, Cleve1998DJalgorithm, Bravyi2018Advantage, Bouland2019Advantage, Huang2021Advantage} but also algorithms offering efficient solutions to practically relevant problems~\cite{Grover96, Shor1997, Kitaev1995_QPE, Lloyd1996}.
At the same time, the quest for universality has driven efforts to establish fundamental bounds on computational power for various tasks and, conversely, to prove the optimality of certain algorithms~\cite{Bennett1997Strengths, Nielsen2006Geometry, Low2017signal}.
The second main theme of this thesis, dynamic programming for quantum algorithms, contributes to this theoretical endeavour by equipping quantum computation with a new toolkit. 

In both resource theories and quantum computation, the search for optimal protocols typically begins with fixing an initial system state. 
One approach is to determine the best attainable final state, while another is to identify the optimal pathway to a predefined final state. 
Both these setups are non-assisted, in the sense that no auxiliary system participates in the protocol, and they provide baseline results corresponding to the simplest implementable scenarios. 

However, introducing an auxiliary system often leads to more intriguing behaviours.
For example, milestones in quantum information theory include protocols like superdense coding and quantum teleportation~\cite{Bennett1992SuperdenseCoding, Bennett1993Teleportation}, where quantum communication or classical communication is strengthened by having a shared auxiliary state.
Computer memories are another example of an auxiliary system that plays a critical role, to the point that many complexity classes are defined by memory constraints~\cite{Papadimitriou2003compcomplexities}. 

Unfortunately, the advantage of introducing an auxiliary system comes with a significant growth in the complexity of the optimisation problem we need to solve. 
The origins for both the advantage and the complexity are the same: the vastly increased degrees of freedom in protocols. 
In simple terms, while the expanded set of protocols enabled by auxiliaries allows for improved optimal solutions, searching this larger space becomes significantly more demanding.

There are two primary reasons why auxiliary systems can significantly impact the complexity of finding the optimal solution.
Firstly, optimisation for a larger (system-auxiliary) composite system is generally harder than optimisation over individual subsystems combined, as the interplay between subsystems needs to be considered for composite systems. 
In quantum theories, the dimension grows multiplicatively: if the main system of interest is $d$-dimensional and the auxiliary system is $d'$-dimensional, the composite dimension becomes $dd'$, which can quickly become intractable even when $d$- and $d'$-dimensional objects are computationally manageable individually. 
Composite systems also include correlations between subsystems, which are generally notoriously hard to deal with in quantum information theory.  

Secondly, to fully exploit the use of an additional system, a good auxiliary system must be chosen first. 
This means that the aforementioned larger optimisation problem over composite systems needs to be repeated for different initial auxiliary systems, creating a new layer of optimisation.
Choosing an auxiliary system entails freedom in choosing the size of the auxiliary system (i.e. dimension $d'$), and it inherently leads to an unbounded search for the optimal auxiliary state. 
Moreover, this optimisation is further complicated by non-trivial constraints, often necessary because unrestricted auxiliary assistance would be too powerful.

Because of these difficulties, brute force optimisations for the best auxiliary system are infeasible, and the study of the utility of auxiliary systems has largely relied on specific examples or \emph{ad hoc} constructions without a systematic way of finding either.
The aim of this thesis is to catalyse a more comprehensive and methodical study of the use of auxiliary systems in quantum information theory. 
This thesis addresses this objective by first identifying the underlying mechanisms through which auxiliary systems provide assistance, and then leveraging these mechanisms to devise a powerful, general tool for applying auxiliary systems to a class of quantum computational problems.

\section{Structure of the thesis}

After reviewing the basic quantum theory in Chapter~\ref{chapter: preliminaries}, the contributions of this thesis are presented in two parts.

In the first part, auxiliary systems are studied within the framework of resource theories, where all questions boil down to the dichotomy of whether a process is feasible or not.
This dichotomy is a radical simplification ignoring that the difficulties in real world circumstances are in a continuous spectrum, not in a black and white binary. 
Hence, the resource theory framework is more useful as a sanitised laboratory to test and understand the power of auxiliary systems, than as a ready-to-use manual.  
Nevertheless, this controlled nature of resource theories facilitates the systematic study of advantages arising from adopting auxiliaries, as the advantages manifest by changing the binary feasibility question.

The rigidity of the dichotomy in resource theories also narrows down the types of auxiliary systems we can use. 
Because a process is either feasible or not feasible, a strict bookkeeping of resources is necessary.
If a resourceful auxiliary system is introduced to the process, and if it is allowed to degrade to a less resourceful state, this intricate bookkeeping would be disrupted.
In fact, it is shown that if an auxiliary state is permitted to change after the process, even by an arbitrarily small amount, there is a simple way to exploit the auxiliary for almost any process one wishes to enable~\cite{Leung2013coherent}; this phenomenon is called embezzlement~\cite{vanDam03_embezzle}, a name aptly capturing the essence of it. 

Therefore, we examine catalytic usages of auxiliaries, or catalyses, in resource theories; auxiliary systems, or catalysts, must return to their initial state at the end of the process so that they can be reused an indefinite number of times for the same process. 
The catalytic constraint is stringent as it forbids any change in the catalyst state, let alone any consumption of resources from the catalyst.
It is also exacting: to benefit from catalysis, the operation must non-trivially interrelate the system and the catalyst, while to meet the constraint, it must not leave any trace on the catalyst. 

Nevertheless, catalysis is perplexingly potent.
Somehow just having this additional state in the picture unlocks a wide range of operations that were not feasible without it, as observed in many resource theories~\cite{LipkaBartosik2023CatReview, Datta2023CatRev2}. 
The first research on catalysis~\cite{Jonathan99_PRL} displays a confounding advantage of employing a catalyst. 
When a quantum state is distributed to two spatially separated parties, it can only be manipulated locally, with the help of classical communication between two parties.
In such a setting, only a restricted set of final states can be attained from a given initial state. 
However, when there is another state, used as a catalyst, that is shared between these two parties, some final states that could not be obtained previously become attainable with the same set of operations: local manipulations and classical communications.
This advantage is gained for free, as the catalyst itself remains entirely unchanged.

However, the origin of this advantage is still elusive. 
The abovementioned example could be found by leveraging the clear necessary and sufficient condition for state transformations in spatially separate parties, as established in the resource theory of entanglement.
Even though the complete mathematical characterisation of catalysis in the same setting of spatially separate parties has been identified~\cite{Klimesh07_ineqs, Turgut_2007}, a clear physical interpretation explaining how an unchanged catalyst can enable such transformations remains absent.

The only insight the community has on this puzzling phenomenon is the connection between the catalytic and multi-copy processes.
In the latter process, multiple copies of the same state are collectively processed, which generally allow simultaneous transformations into a wider range of final states. 
It has been found that (with mild assumptions) any multi-copy process can be emulated using a setup with a single copy and a catalyst.
Moreover, a catalyst state needed for such emulations, known as Duan state~\cite{Duan2005_DuanState}, can be easily constructed. 
In other words, catalytic advantage includes multi-copy advantages.
A parallel result connecting an asymptotic transformation (from $n$ copies of the initial state to $n$ copies of the final state with the error vanishing as $n$ goes to infinity) and a catalytic one with the final correlation between the system and the catalyst has also been established~\cite{Takagi2022CorrCat}.
However, catalytic advantage cannot be fully explained this way, as exemplified by the existence of deterministic catalytic transformations that cannot be emulated by any deterministic multi-copy process~\cite{Feng2006multicopy}.
Chapter~\ref{chapter: resource theories} provides a concise summary of the necessary technical tools for resource theories and catalyses therein, including the aforementioned connection between multi-copy and catalytic advantages. 

In the first part of the thesis, we extend our analysis beyond this connection to the multi-copy processes.
The main contribution is the identification of three sources of power in catalysis: memory effect, fine-tuning, and resource broadcasting. 
In Chapter~\ref{chapter: CETO}, we focus on the memory effect.
Within thermodynamic resource theories, a hierarchy of different sets of feasible operations exists. 
The inner layers of the hierarchy consist of thermal operations---a popular operationally-inspired class---and other sets of operations that are decomposable due to the additional restrictions imposing different types of memory constraints.
This hierarchy is strict, as the larger set, i.e. thermal operations, includes operations that are not feasible in its subsets due to the additional constraints. 
We first examine the dynamics of catalysis---using small catalysts and decomposable operations, the change in catalyst state can be tracked stepwise. 
The resulting snapshots of the catalytic evolution portray how catalysts function as a temporary storage for the resource, which enables memory-constrained operations to emulate most of thermal operations using small catalysts as an additional memory.
Following this, we move on to demonstrate that catalysis indeed collapses this hierarchy completely: catalytic versions of all sets of operations in the hierarchy are proven to be equivalent.  
Moreover, this result is established by using catalysts that are resourceless.
Since these resourceless states are worthless for any other task, it evinces that the catalytic advantage solely comes from the memory effects they provide. 

In Chapter~\ref{chapter: resource broadcasting}, the other two sources, the importance of fine-tuning and the role of a catalyst as a resource broadcasting seed, are investigated. 
Catalysis is typically highly fine-tuned; the catalyst system, its state, and the operation applied all depend sensitively on the initial state of the system.
Hence, the initial states for the system and the catalyst must be prepared with high precision. 
We prove that this ability of carefully choose the catalyst according to the initial system state is one of the main factors that make catalysis possible. 
To be specific, we establish a no-catalysis theorem for a wide class of general resource theories when the catalyst needs to be insensitive to different initial system states.
Two highlights of this theorem are: i) it holds for general resource theories satisfying certain rules of composition, regardless of the specific resource of interest and ii) it clarifies that the fine-tuning of a catalyst state in the usual settings is extremely sensitive, by proving the equivalence between full insensitivity and arbitrarily weak insensitivity. 
Nevertheless, catalysts can activate non-trivial operations even without fine-tuning in some resource theories. 
We show that this state-agnostic class of catalysis corresponds to a process called resource broadcasting, where a seed state is broadcast into another system, while preserving its own state. 
Thus, the source of power for such catalyses is identified: it is the catalyst's capacity to act as the seed state to be broadcast.
Additionally, necessary and sufficient conditions for resource broadcasting are found for several classes of resource theories. 

In the second part of the thesis, we investigate auxiliary systems beyond resource theories and venture into quantum computation. 
In some algorithms, the desired circuit is fixed and known. 
The quantum Fourier transformation~\cite{Coppersmith2002QFT} is one such example; once the size of the system is determined, the desired unitary circuit is fixed.
In such cases, the synthesis problem is straightforward, as there exists a constructive algorithm to compile the circuit using elementary gates~\cite{dawson2006solovay}, although finding the optimal way to do this task is hard~\cite{Botea2018QCC}.
In contrast, for other classes of algorithms, the full information about the final circuit is not available, and only black-box access to oracles encoding some integral part of the desired circuit is given.
The synthesis problem for such algorithms thus becomes more non-trivial. 
An example is the Grover search algorithm~\cite{Grover96}, where the rotation operations around the initial and final states are given as a black box and synthesised to output a circuit finding the unknown final state. 
Our main focus is the latter class of algorithms.
In Chapter~\ref{chapter: circuit compilation}, we present existing methods to synthesise circuits without fully knowing them, using other unknown operations blindly.

A class of quantum algorithms, previously overlooked because of this non-trivial synthesis problem, is studied in Chapter~\ref{chapter: quantum recursions}. 
These are quantum recursions, defined as a recursive algorithm whose recursion steps consist of operations that explicitly depend on the result state of the previous recursion.
Since the algorithm is quantum, the resulting states on which the operations depend are also quantum states. 
Furthermore, if these states remain unmeasured (often the case, given the complexity of quantum measurement), the recursion steps must be synthesised dynamically based on unknown quantum outputs from prior recursions.
We find that such synthesis of quantum recursions is possible, but only with circuit depth growing exponentially with the number of recursion steps; perhaps this is the main reason why quantum recursions have been largely neglected thus far. 

Our new algorithm, introduced in Chapter~\ref{chapter: quantum recursions}, suggests that exponential circuit depth notwithstanding, quantum recursions can be a powerful algorithm for practical problems, such as ground state preparation. 
To be specific, we design a quantum imaginary-time evolution algorithm that effectively follows the path $e^{-\beta H}\ket{\psi}$ with increasing $\beta$ for any pure state $\ket{\psi}$, with proper renormalisation. 
This algorithm is compiled using only \emph{i)} real-time evolution by the Hamiltonian $H$, \emph{ii)} unitary operation preparing $\ket{\psi}$ from some fiducial state, and \emph{iii)} elementary unitary gates.
The performance guarantees we establish for this algorithm imply that even a small number of recursion steps, which prevents the circuit depth from being prohibitive, can already yield the solution for many problems to a satisfactory degree. 

In Chapter~\ref{chapter: QDP}, we improve the synthesis of quantum recursions using auxiliary systems.
This improvement is achieved through a quantum version of dynamic programming, or to be more precise, memoisation~\cite{Michie1968memoization}. 
In classical computing, memoisation works by introducing a memo function, i.e. memory states, that remembers the previous results of the past recursions and uses those stored values for future recursion steps. 
This technique can exponentially reduce the circuit depth of classical recursion problems compared to their memoryless counterpart.
Memoisation only demands a small amount of classical memory, as the previous result in the form of classical states can be written down and retrieved an indefinite number of times, once calculated.  

Equipped with the insights garnered from catalysts in resource theories, we design our auxiliaries for quantum dynamic programming (or quantum memoisation) as memory states that enable operations in a state-agnostic way.
In other words, we directly inject memories into the programme without reading them. 
This strategy circumvents the quintessential quantum restrictions that forbid the reading or reproduction of memories without destroying them. 
The state-agnostic utilisation still comes with a cost: because memories are injected to the programme and cannot be retrieved afterwards, they are one-time use. 
Hence, quantum dynamic programming must prepare many copies of resulting states in parallel and use them as memory states to reduce the circuit depth.
As a result, we obtain a trade-off: by using exponentially many initial states, i.e. requiring exponentially large circuit width, the circuit depth can be made linear in the number of recursion steps, achieving exponential depth reduction.
Quantum dynamic programming is also amenable to hybridisation with the memoryless recursion, providing a convenient way of controlling depth-width specifications of the algorithm. 
We highlight that quantum dynamic programming is also a perfect candidate for distributed quantum computing~\cite{Cacciapuoti2019Distributed}, where small quantum devices are connected by a quantum network to solve bigger problems.

We conclude this thesis with Chapter~\ref{chapter: conclusions}.
We briefly recapitulate our contributions, particularly regarding more systematic approaches to the use of auxiliary systems.
We suggest that the physical intuitions developed from this work could inform the design of improved protocols applicable to realistic settings beyond resource theories and dynamic programming. 
Building on this potential, we propose several specific directions for future research.
One is the extension to optical settings, where auxiliary systems are used either as a catalyst or a quantum memory for preparing valuable optical states.
The other further examines the composition of two systems, to understand not only the scenarios where an auxiliary system is attached to the main system of interest, but also ones where two systems cooperate and help surpass each system's individual limitations. 

\chapter{Preliminaries}\label{chapter: preliminaries}

We review the basic quantum formalism necessary for the remainder of the thesis. 
Naturally, we cannot provide a comprehensive text on quantum theory in a single chapter; instead, this serves as a brief tutorial on the concepts and terminology that are often used without explanation in contemporary research articles.
For a comprehensive and modern introduction to quantum theory, we recommend standard textbooks such as Refs.~\cite{Nielsen_and_Chuang2010, WildeBook}.

\section{Quantum states}\label{section: quantum states}

The starting point for the mathematical treatment of quantum mechanics is the superposition principle.
In his seminal monograph, Dirac states that~\cite{DiracQM} 
\begin{quote}
	[quantum theory] requires us to assume that between these states there exist peculiar relationships such that whenever the system is definitely in one state we can consider it as being partly in each of two or more other states.
\end{quote}
This property naturally leads to the notion of quantum states as vectors.
Moreover in the same monograph, the existence of distinct quantum states formed by the uniform superposition of two given states is shown, indicating that the vector space must be complex.

This leads to the question: what  type of vectors are these?
It has been shown~\cite{vonNeumann_book} that \emph{quantum states are normalised vectors living in a particular type of complex vector spaces, known as Hilbert space}.
In this thesis, we denote the Hilbert space as $\mH$ sometimes with a subscript to specify the system. 
A vector in a Hilbert space $\mH_{X}$ are denoted with the bra-ket notation as $\ket{\psi}_{X}$.
For a finite-dimensional system $X$, the vector $\ket{\psi}_{X}$ can be written as a column vector with complex number entries. 

Each Hilbert space comes with an inner product, which defines the norm of a vector.
For two vectors $\ket{\psi}$ and $\ket{\phi}$ belonging to the same Hilbert space $\mH$, their inner product is written as $\braket{\psi}{\phi}$ and satisfies the conjugate symmetry $\braket{\psi}{\phi} = \braket{\phi}{\psi}^{*}$, where $^{*}$ operation denotes complex conjugate.
Throughout this thesis, we assume that any vector $\ket{\psi}$ is normalised, so $\braket{\psi}{\psi} = 1$ unless specified otherwise. 
Note that any two quantum states $\ket{\psi},\ket{\phi}\in\mH$ are connected by some unitary operator $\ket{\psi} = U\ket{\phi}$.
When equality between states is claimed, it is understood to be up to a global phase, i.e. if $\ket{\psi} = e^{i\theta}\ket{\phi}$, for some real number $\theta$, we write $\ket{\psi} = \ket{\phi}$. 

In general, Hilbert spaces can be infinite-dimensional and certain quantum mechanical systems, such as bosonic modes, are indeed described by such spaces.
However, we limit our discussion to finite-dimensional Hilbert spaces, where the dimension can still be arbitrarily large.
Then a finite set of orthonormal basis vectors $\{\ket{i}\}_{i=1}^{\dim(\mH)}$ spans the Hilbert space $\mH$. 
The identity operator for this space is then defined as $\1 = \sum_{i}\dm{i}$, where $\dm{i}$ denotes the outer product of $\ket{i}$ with itself.

Given a quantum system, we are ultimately interested in its physical property.
In quantum theory, each observable has an associated operator $A$ selected from the set $\mL\coloneq\mL(\mH)$ denoting the linear operators 
from a Hilbert space (representing the system) to itself.
A Hermitian conjugate $A^{\dagger}$ of an operator $A\in\mL(\mH)$ is defined via the condition $\braket{\phi}{A\psi} = \braket{A^{\dagger}\phi}{\psi}$, $\forall \ket{\psi},\ket{\phi}\in\mH$, where $\ket{A\psi} = A\ket{\psi}$ and $\ket{A^{\dagger}\phi} = A^{\dagger}\ket{\phi}$.
If $A$ is associated to an observable, it needs to be a Hermitian operator, i.e. $A^{\dagger} = A$, and thus admits a spectral decomposition $A = \sum_{i}\alpha_{i}\dm{a_{i}}$ with real eigenvalues $\{\alpha_{i}\}_{i}$.
If the system is in the state $\ket{a_{i}}$, it is certain that it has the property corresponding to the definite value $\alpha_{i}$.
However, if the system is in a state $\ket{\psi} \neq \ket{a_{i}}$ for any $i$, an inherent statistical nature of quantum theory emerges: even the best possible measurement of the observable $A$ can only yield the probability of obtaining the outcome $i$, given by
\begin{align}
	P_{A}(i) =\lvert \braket{a_{i}}{\psi}\rvert^{2}.
\end{align}
Any moment of this observable is then written as $\expval{A^{n}} = \sum_{i}P_{A}(i)\alpha_{i}^{n} = \ev{A^{n}}{\psi}$.

Despite the probabilistic nature, knowing the quantum state $\ket{\psi}$ implies that the maximum achievable information about the system is obtained. 
However, this is often not the case; with limited access to information, one may only infer the probability $p_{i}$ that the system is in a state $\ket{\psi_{i}}\in\mH$.
Such statistical considerations can be represented~\cite{vonNeumann_book} by a \emph{density matrix} 
\begin{align}
\rho = \sum_{i}p_{i}\dm{\psi_{i}},
\end{align} 
i.e. a statistical mixture of rank-1 operators $\dm{\psi_{i}}\in\mL(\mH)$.
The trace of an operator $A$ can be defined using any orthonormal basis $\{\ket{i}\}_{i=1}^{\dim(\mH)}$ as $\Tr[\rho] \coloneq \sum_{i=1}^{\dim(\mH)} \bra{i}A\ket{i}$, and the trace of a density matrix $\Tr[\rho] = 1$ as $\Tr[\dm{\psi_{i}}] = 1$ and $\sum_{i}p_{i} = 1$. 
Furthermore, any $\rho$ is positive semi-definite as $p_{i}\geq0$ for all $i$.
The converse is also true. 
\begin{remark}
	If an operator $\rho\in\mL(\mH)$ is positive semi-definite and has unit trace, it is a density matrix. 
\end{remark}
Let us denote the set of all density matrices for a Hilbert space $\mH$ as $\mD(\mH)$ (or simply $\mD$ when the context is clear).
This set is convex and compact.
We allow ourselves some flexibility in nomenclature and refer to density matrices as quantum states.
A quantum state vector $\ket{\psi}$ and the corresponding density matrix $\dm{\psi}$ are called a \emph{pure state}.
Other density matrices, those with $\rank(\rho)>1$, are termed \emph{mixed states}. 

When a mixed state $\rho$ is given, predicting the observable property $A$ involves two layers of statistical considerations. 
First, $\rho$ is in a state $\ket{\psi_{i}}$ with probability $p_{i}$; for each possibility, the probability of obtaining the outcome $j$ for the observable $A$ is $P(j|i) = \lvert \braket{a_{j}}{\psi_{i}}\rvert^{2}$. 
Combined, the probability of obtaining outcome $j$ from $\rho$ is $\sum_{i}P(j|i)p_{i} = \Tr[\rho \dm{a_{j}}]$, where $\Tr$ denotes the trace operator.
The probabilities, which are ultimately the quantities measurable in experiments, depend solely on the projectors $\dm{a_{j}}$, termed \emph{effects}, and not on the values $\alpha_{j}$.
These effects can be generalised to (non-projective) operators~\cite{LudwigBook}---analogous to mixed states being a generalisation of pure states with projective density matrices~\cite{EffectsBook}---collected in a set $\mM = \{M_{i}\}_{i}$ to incorporate non-sharp measurement scenarios. The probability of obtaining outcome $i$ from $\rho$ then becomes
\begin{align}
	P_{\mM}(i) = \Tr\left[\rho M_{i}\right].
\end{align}
We adopt \emph{positive operator-valued measures (POVM)} as the most general set of effects: it is the set of operators $\{M_{i}\}_{i}\subset\mL(\mH)$, where each $M_{i}\geq0$ and they satisfy the completeness condition $\sum_{i}M_{i} = \1$.
We encourage interested readers to refer to a textbook by Wiseman and Milburn~\cite{BookWiseman} for extended discussion on this topic.

\subsection{Composition of quantum states}\label{subsection: composition of states}
Any quantum system---composite or not---is described with the associated Hilbert space.
If a composite system $XY$ consists of two subsystems $X$ and $Y$, the state vectors for the $XY$ system lives in a new Hilbert space $\mH_{XY}$. 
In this subsection, we explain how $\mH_{XY}$ is constructed from the subsystem spaces $\mH_{X}$ and $\mH_{Y}$, and discuss the implications of having access to the full space $\mH_{XY}$ rather than $\mH_{X}$ and $\mH_{Y}$ individually. 
This serves as an important prelude to the later sections of the thesis, where the power of auxiliary systems composed with the system of interest is studied. 

What does it mean to have a state for system $XY$? 
The simplest possibility is when the composite system is a collection of independent states in $X$ and $Y$. 
This is represented by $\ket{\psi}_{X}\otimes\ket{\psi'}_{Y}$ where $\ket{\psi}_{X}\in\mH_{X}$ and $\ket{\psi'}_{Y}\in\mH_{Y}$; therefore, $\ket{\psi}_{X}\otimes\ket{\psi'}_{Y}\in\mH_{XY}$ necessarily. 
Using the fact that each Hilbert space  $\mH_{X}$ and $\mH_{Y}$ has an orthonormal basis $\{\ket{i}_{X}\}_{i}$ and $\{\ket{i}_{Y}\}_{i}$, respectively, we identify the set of orthonormal vectors $\{\ket{i}_{X}\otimes\ket{j}_{Y}\}_{(i,j)}$ as a subset of all composite states in $XY$ that lives in $\mH_{XY}$.

On the other hand, the vector space spanned by the set $\{\ket{i}_{X}\otimes\ket{j}_{Y}\}_{(i,j)}$ (with complex coefficients) is defined as the tensor product of the two Hilbert spaces $\mH_{X}\otimes\mH_{Y}$, which itself forms a Hilbert space when equipped with the inner product 
\begin{align}
	(\bra{\psi}_{X}\otimes\bra{\psi'}_{Y})(\ket{\phi}_{X}\otimes\ket{\phi'}_{Y}) = \braket{\psi}{\phi}_{X}\braket{\psi'}{\phi'}_{Y}
\end{align} 
for any two vectors $\ket{\psi}_{X}\otimes\ket{\psi'}_{Y}$ and $\ket{\phi}_{X}\otimes\ket{\phi'}_{Y}$.
This choice turns out to be a correct way of defining the composite Hilbert space.
\begin{remark}
	The Hilbert space for the composite system $XY$ is  
	\begin{align}\label{eq: define composite Hilbert space}
		\mH_{XY} \coloneq \mH_{X}\otimes\mH_{Y}.
	\end{align} 
	In other words, any normalised vector in $\mH_{X}\otimes\mH_{Y}$ represents a quantum state of system $XY$.
\end{remark}
See, e.g. Ref.~\cite{LudwigBook} for a detailed justification. 
The rest of this subsection is a consequence of this composite structure. 

Now we can consider density matrices for the composite system $XY$. 
They are the linear, positive semi-definite operators from $\mH_{XY}$ to itself with unit trace. 
The tensor product $\rho_{X}\otimes\sigma_{Y}$ of any two states $\rho_{X}\in\mD_{X}\coloneq\mD(\mH_{X})$ and $\sigma_{Y}\in\mD_{Y}\coloneq\mD(\mH_{Y})$ defines a density matrix in $\mD_{XY} \coloneq \mD(\mH_{XY})$.
In particular, any state in $\mD_{XY}$ that can be expressed as a tensor product $\rho_{X}\otimes\sigma_{Y}$ for some $\rho_{X}\in\mD_{X}$ and $\sigma_{Y}\in\mD_{Y}$ is classified as a \emph{product state}, and these are the only states where systems $X$ and $Y$ are completely uncorrelated. 
Moreover, by the convexity of the set of density matrices, any convex combination of product states also forms a density matrix.
These states have a special name.
\begin{definition}[separable states~\cite{Werner1989separable}]\label{definition: separable states}
	A density matrix $\varrho_{XY}\in\mD_{XY}$ is called \emph{separable} for the partition $X|Y$ if it can be expressed as
	\begin{align}
		\varrho_{XY} = \sum_{i}p_{i}\rho^{(i)}_{X}\otimes\sigma^{(i)}_{Y},
	\end{align}
	where $\{p_{i}\}_{i}$ are non-negative coefficients and $\rho^{(i)}_{X}\in\mD_{X}$ and $\sigma^{(i)}_{Y}\in\mD_{Y}$ for all $i$.
	
	The set of all such density matrices is defined as 
	\begin{align}
		\SEP_{XY} \coloneq \conv\{\rho_{X}\otimes\sigma_{Y} \,\vert\, \rho_{X}\in\mD_{X},\sigma_{Y}\in\mD_{Y}\},
	\end{align}
	where \emph{conv} denotes the convex hull of a set. 
	The subscript $XY$ is sometimes omitted when there is no ambiguity. 
\end{definition}

A remarkable feature of the composition structure in quantum theory is that there exist valid quantum states that lie outside the convex hull of product states.
\begin{remark}
	If $\dim(\mH_{X}),\dim(\mH_{Y})\geq 2$, then
	\begin{align}
		\SEP_{XY} \subsetneq \mD_{XY}.
	\end{align}
	If a state $\rho_{XY}\in\mD_{XY}$ but $\rho_{XY}\notin\SEP_{XY}$, it is called \emph{entangled}.
\end{remark}
Intuitively, a separable state represents a statistical ensemble of product states, whereas entanglement signifies a more profound form of correlation.
Yet, determining if a given state is entangled or not is a hard problem in general~\cite{Gurvits2003EntanglementHardness, Ioannou2007EntanglementHardness, Gharibian2010EntanglementHardness}.

Beyond bi-partite systems, entanglement extends to multi-partite settings. 
Separable states in the $m$-partite settings are defined analogously: if a state is a convex combination of $m$-partite product states, it is separable. 
However, multi-partite entanglement has much more layers compared to its bi-partite counterpart. 
Even for the simplest case of three qubits, distinct classes of entanglement arise~\cite{Cirac2000}.
A recent introductory paper~\cite{Horodecki2024MultiPartite} provides an accessible guide to this topic.

Entanglement is often at the heart of quintessentially quantum phenomena, and its study remains one of the most significant subfields in quantum information theory.
We recommend the classic review by The Horodeckis~\cite{Horodecki4} as an entry point to this field.
Here, we focus only on results directly relevant to this thesis.

Let us begin with the simple case of bi-partite pure states. 
A pure state $\ket{\Psi}_{XY}\in\mH_{XY}$ can always be written as 
\begin{align}\label{eq: bipartite pure state decomposition}
	\ket{\Psi}_{XY} = \sum_{i,j}S_{ij}\ket{i}_{X}\otimes\ket{j}_{Y},
\end{align} 
with orthonormal bases of $\mH_{X}$ and $\mH_{Y}$.
We regard $S$ as a $\dim(\mH_{X})\times\dim(\mH_{Y})$ matrix, $S = \sum_{i}S_{ij}\ket{i}_{X}\bra{j}_{Y}$, which uniquely corresponds to each pure state $\ket{\Psi}_{XY}$.
Suppose that $\ket{\Psi}_{XY} = \ket{\psi}_{X}\otimes\ket{\psi'}_{Y}$ for some pure states $\ket{\psi}_{X}$ and $\ket{\psi'}_{Y}$, and choose unitary operators $U\in\mL_{X}$ and $V\in\mL_{Y}$, such that $U\ket{\psi}_{X} = \ket{1}_{X}$ and $V\ket{\psi'}_{Y} = \ket{1}_{Y}$. 
Then $USV^{\dagger} = \ket{1}_{X}\bra{1}_{Y}$, and we obtain singular value decomposition of $S$ revealing that $\rank(S) = 1$.
Conversely, if $\rank(S) = 1$, a similar singular value decomposition can be found, and the state $\ket{\Psi}_{XY}$ must be a product state. 
The rank of this matrix plays a crucial role even when the state is correlated. 
\begin{definition}[Schmidt rank~\cite{Schmidt1907}]\label{definition: Schmidt rank}
	The rank of the matrix $S$ defined by the elements $S_{ij}$ in Eq.~\eqref{eq: bipartite pure state decomposition} is called the Schmidt rank.
\end{definition}
\begin{proposition}[Schmidt decomposition~\cite{Schmidt1907}]\label{definition: Schmidt decomposition}
	Any bi-partite pure state $\ket{\Psi}_{XY}$ with Schmidt rank $r$ can be written in the Schmidt decomposition form
	\begin{align}
		\ket{\Psi}_{XY} = \sum_{i=1}^{r}\lambda_{i}\ket{\tilde{i}}_{X}\otimes\ket{\tilde{i}}_{Y},
	\end{align}
	where the \emph{Schmidt coefficients} $\lambda_{i}>0$ for all $i=1,\ldots,r$, and $\{\ket{\tilde{i}}_{X}\}_{i}$ and $\{\ket{\tilde{i}}_{Y}\}_{i}$ are some orthonormal bases for $\mH_{X}$ and $\mH_{Y}$, respectively. 
\end{proposition}
Proposition~\ref{definition: Schmidt decomposition} follows directly from the singular value decomposition argument above.
If a pure state $\ket{\Psi}_{XY}$ is separable, its density matrix takes the form $\dm{\Psi}_{XY}= \sum_{i}p_{i}\rho^{(i)}_{X}\otimes\sigma^{(i)}_{Y}$. 
Since $\dm{\Psi}_{XY}$ is a rank-1 operator and each $\rho^{(i)}_{X}\otimes\sigma^{(i)}_{Y}$ is positive semi-definite, the equality between can hold if and only if $\rho^{(i)}_{X}\otimes\sigma^{(i)}_{Y} = \dm{\Psi}_{XY}$ for all $i$ with $p_{i}\neq0$. 
In other words, a pure bipartite state is separable if and only if it is a product of two pure states.
\begin{proposition}[pure bi-partite state entanglement]
	A pure bi-partite state is separable if and only if its Schmidt rank is $1$. 
\end{proposition}
The Schmidt coefficients fully determine the degree of entanglement of a pure bipartite state, even when the state is entangled.
This follows from the fact that all bipartite pure states with the same Schmidt coefficients are locally unitarily equivalent: if $\ket{\Psi}_{XY}$ and $\ket{\Psi'}_{XY}$ share the same Schmidt coefficients, there always exist local unitary operators $U_{X}$ and $V_{Y}$ such that $\ket{\Psi'}_{XY} = (U_{X}\otimes V_{Y})\ket{\Psi}$.
Thus, for any observable of the form $A_{X}\otimes B_{Y}$, there exists an observable $U_{X}AU^{\dagger}_{X}\otimes V_{Y}B_{Y}V^{\dagger}_{Y}$, ensuring that the correlation measured in $\ket{\Psi}_{XY}$ is identical to that in $\ket{\Psi'}_{XY}$ with the new observable.

For mixed states, determining whether a state is entangled is significantly more cumbersome.
One necessary condition for a mixed state $\rho_{XY}$ to be separable is $\rho_{XY}^{\transp_{Y}}\geq0$, where $\transp_{Y}$ is the partial transpose operation defined as the linear map such that $(A_{X}\otimes B_{Y})^{\transp_{Y}} = A_{X}\otimes B_{Y}^{\transp}$ for all $A_{X}\in\mL_{X}$ and $B_{Y}\in\mL_{B}$~\cite{Peres1996PPT, Horodecki1996PPT}.
Various witnesses for entanglement---observables that yield a positive value for all separable states---have been extensively studied; see Refs.~\cite{Bruss2002Witness, Horodecki4, Chruscinski2014}.

\section{Quantum channels}\label{section: quantum channels}

Section~\ref{section: quantum states} introduced a static picture of quantum theory, i.e. how to represent a state at a point of time. 
We now turn to a dynamic picture, namely how quantum states can evolve in time. 

If a system is closed, i.e. if it is not interacting with external quantum system, a (pure) quantum state evolves according to Schrödinger's equation $\dv{t}\ket{\psi(t)} = -iH\ket{\psi(t)}$.
The Hamiltonian operator $H$ is Hermitian, implying that the evolution is unitary $\ket{\psi(t)} = U_{t}\ket{\psi(0)}$ for some unitary operator $U_{t}\in\mL$.
Since quantum theory is linear, a mixed state $\rho$ evolves by $\rho(t) = U_{t}\rho(0)U_{t}^{\dagger}$, which preserves the spectrum $\spec(\rho(t)) = \spec(\rho(0))$.

When the system $S$ interacts with an external environment $E$, the system-environment composite $SE$ undergoes the unitary evolution by some unitary operator $U_{SE}\in\mL_{SE}$.
Typically, environment $E$ is much larger than the system, and it is practically impossible (or unnecessary) to track the evolution of the composite $SE$.
Instead, we are interested in the effective description of the system evolution without any information about the environment. 
These evolutions must necessarily be linear maps from $\mD_{S}$ to $\mD_{S'}$ with $S'$ potentially different from $S$.
Furthermore, we would want this description to work when applied to any subsystem of a composite system that is in general correlated. 
These minimal requirements are encapsulated by a simple and elegant mathematical formulation
\begin{definition}[CPTP maps]\label{definition: CPTP maps}
	Let $\mE$ be a linear map $\mE$ from $\mL_{S}$ to $\mL_{S'}$.
	\begin{itemize}
		\item $\mE$ is completely positive if $\id_{R}\otimes\mE$ is a positive map---i.e. mapping all positive operators of $\mL_{SR}$ to other positive operators in $\mL_{S'R}$---for any additional system $R$, where $\id_{R}$ is the identity map from $\mL_{R}$ to itself.
		\item $\mE$ is trace-preserving if $\Tr \circ \mE = \mE \circ \Tr$. 
	\end{itemize}
	The set of completely positive and trace-preserving (CPTP) maps from $\mL_{S}$ to $\mL_{S'}$ is denoted as $\CPTP{S}{S'}$.
\end{definition}
We also use the name \emph{quantum channel} interchangeably with CPTP maps. 
Three classes of channels are worth mentioning: 
\begin{itemize}
	\item unitary channel $\mE(\cdot) = U(\cdot)U^{\dagger}\in\CPTP{S}{S}$ with a unitary operator $U\in\mL_{S}$,
	\item assigning map $\mA(\cdot) = \cdot \otimes \sigma_{E}\in\CPTP{S}{SE}$ with a density matrix $\sigma_{E}\in\mD_{E}$, and 
	\item partial trace $\Tr_{X}(\cdot)  = \sum_{i} \bra{i}_{X} (\cdot) \ket{i}_{X} \in \CPTP{SX}{S}$, where $\{\ket{i}\}_{i}$ is an orthonormal basis of $\mH_{X}$.
\end{itemize}
These three channels are singled out as they correspond to reversible evolution of a closed system, appending of an independent system, and discarding of a system, respectively.
In fact, it is known that any quantum channel can be obtained from the concatenation of these three in Theorem~\ref{theorem: unitary dilation} in Section~\ref{subsection: Kraus representation and unitary dilation}.

Having CPTP maps as the most general quantum evolution rules out a process to be non-physical, if it cannot be implemented via any CPTP map. 
One particularly interesting instance is the no cloning theorem. 
\begin{theorem}[no cloning theorem~\cite{Park1970nocloning, Wootters1982Nocloning, Dieks1982Nocloning}]\label{theorem: no cloning}
	Let $\dim(\mH_{S}) = \dim(\mH_{\bar{S}})$.
	There is no CPTP map $\mE\in\CPTP{S}{S\bar{S}}$, such that 
	\begin{align}
		\mE(\dm{\psi}_{S}) = \dm{\psi}_{S}\otimes\dm{\psi}_{\bar{S}},
	\end{align}
	for all $\ket{\psi}_{S}\in\mH_{S}$. 
\end{theorem}
A similar but more general result exists for mixed states. 
\begin{theorem}[no broadcasting theorem~\cite{Barnum1996Broad}]\label{theorem: no broadcasting}
	Let $\dim(\mH_{S}) = \dim(\mH_{\bar{S}})$.
	Suppose that a density matrix $\rho_{S}$ is chosen from the set $\{\sigma^{(1)},\sigma^{(2)}\}$.
	Then there is a CPTP map $\mE_{\in\CPTP{S}{S\bar{S}}}$, such that 
	\begin{align}
		\mE(\rho_{S}) &= \varrho_{S\bar{S}},\\
		\Tr_{\bar{S}}[\varrho_{S\bar{S}}] &= \rho_{S},\quad \Tr_{S}[\varrho_{S\bar{S}}] = \rho_{\bar{S}},
	\end{align}
	for both $\rho_{S} = \rho_{\bar{S}} = \sigma^{(1)},\sigma^{(2)}$ if and only if $[\sigma^{(1)},\sigma^{(2)}] = 0$.
\end{theorem}
Hence, even when we allow the correlation between our resulting states, copying information is not possible in general. 
These theorems severely limit the learning and manipulation of quantum information, given unknown states; see~\cite{Scarani2005CloningRev} for a detailed discussion. 

However, both CP and TP conditions entail making subtle assumptions.
We comment on the TP condition in the next subsection in relation to selective measurements; here, we focus on the CP condition. 
Complete positiveness of the dynamics is required when the system $S$ whose effective evolution is described starts from a state uncorrelated to the environment $E$.%
\footnote{Note that the initial correlation between $S$ and $R$ is allowed because it is assumed that $S$ and $R$ do not interact during the evolution. 
On the other hand, $S$ and $E$ interact and the correlation prior to this interaction affects the complete positivity of the dynamics.} 
If we want to describe the effective evolution from the point that the system $S$ is already correlated, the dynamics need not even be positive~\cite{Pechukas1994PnotCP, Alicki1995comment, Pechukas1995reply, Stelmachovic2001InitialEntanglement, Shaji2005PnotCP}. 
However, such initial states do not fit the spirit of quantum channels, which describe the process independent of the input state: assigning the $SE$ correlated state linearly for all $S$ states is impossible~\cite{Pechukas1994PnotCP}.

Complete positiveness is also closely related to the Markovianity of quantum processes~\cite{Stelmachovic2001InitialEntanglement, Breuer2009NonMarkovian, Rivas2010NonMarkovian, Buscemi2014NonMarkovianity}. 
For a process to be Markovian (or memoryless), there should be no backflow of information from the environment $E$ to the system $S$.
This is fulfilled when, during any interval of time $[t_{1},t_{2}]\subset [0,T]$ the dynamics can be described by a CPTP map, which can arise from the interaction with a new uncorrelated environment state.
It can also be used to define the non-Markovianity. 
\begin{definition}[Markovian quantum channel]\label{definition: Markovian quantum channel}
	A CPTP map $\mE\in\CPTP{S}{S}$ is Markovian if it is infinitesimally divisible~\cite{Wolf2008Divisibility}, i.e. for any $\epsilon>0$, there is a decomposition $\mE = \mE_{1}\circ\mE_{2}\circ\cdots\circ\mE_{N}$ with some $N$, such that each $\mE_{i}\in\CPTP{S}{S}$ is $\epsilon$-close to the identity map.  
\end{definition}
This definition is also equivalent to the existence of a GKSL master equation inducing the channel~\cite{GKS1976, Lindblad1976}.
See Ref.~\cite{Rivas2014NonMarkovianityReview} for a more in-depth review on this topic.

\subsection{Kraus representation and unitary dilation}\label{subsection: Kraus representation and unitary dilation}
A physical justification of using CPTP maps, defined as in Definition~\ref{definition: CPTP maps}, for describing generic quantum evolution can be made by looking at two representations that any CPTP map admits. 
We follow the narrative that Kraus presented in his book~\cite{EffectsBook}.
The first representation is often called \emph{Kraus representation} or operator-sum representation.
\begin{theorem}[the first representation theorem of Ref.~\cite{EffectsBook}]\label{theorem: Kraus representation}
	For each $\mE\in\CPTP{S	}{S'}$, there exists a set of operators $\{K_{i}\}_{i}$ with each $K_{i}: \mH_{S}\to\mH_{S'}$, such that
	\begin{align}
		\sum_{i}K_{i}(\cdot)K_{i}^{\dagger} &= \mE(\cdot) ,\\
		\sum_{i}K_{i}^{\dagger}K_{i} &= \1_{S}.\label{eq: Kraus normalisation}
	\end{align} 
	The converse is also true; hence, the set of CPTP maps is identical to the set of linear maps that have Kraus representation.
\end{theorem}
Note that originally the first representation theorem concerns \emph{CP trace non-increasing (CPTNI) operations}, also known as \emph{quantum instruments}, where $\Tr[\mE(\rho)]\leq\Tr[\rho]$. 
CPTNI operations can be interpreted as the state transformation resulting from applying a measurement and getting an outcome in a set $I$, i.e. a selective measurement. 
This can be written as $\mE(\cdot) = \sum_{i\in I}K_{i}\rho K_{i}^{\dagger}$, where the measurement effect $M_{i} = K_{i}^{\dagger}K_{i}$. 
The post-measurement state will be normalised to a unit trace state $\rho' = \frac{\mE(\rho)}{\Tr[\rho]}$, where the denominator $\Tr[\rho]$ is the probability of having outcomes in $I$. 
Note that the function $\frac{\mE(\cdot)}{\Tr[\mE(\cdot)]}$ is no longer linear. 
However, if we are unselective about the measurement outcome, i.e. if we forget or do not have access to the measurement outcomes, the post-measurement state becomes $\mE(\rho)$ with CPTP map $\mE$. 

The second representation, which is in fact chronologically precedes the first representation, has the name unitary dilation.
\begin{theorem}[the second representation theorem of Ref.~\cite{EffectsBook}]\label{theorem: unitary dilation}
	For each $\mE\in\CPTP{S	}{S'}$, there exist a density matrix $\sigma_{E}\in\mD_{E}$ and a unitary operator $U\in\mL_{SE}$, such that
	\begin{align}
		\mE(\cdot) = \Tr_{E'}\left[U(\cdot\otimes\sigma_{E})U^{\dagger}\right].
	\end{align}
\end{theorem}
Theorem~\ref{theorem: unitary dilation} can be proven using the Stinespring dilation theorem~\cite{Stinespring1955}.
The proof also implies that there always exists a choice where $\sigma_{E}$ is a pure state.   
Furthermore, the converse of the theorem is again true.
It means that any CPTP map can be described by a physical process where the system $S$ and some environment $E$ undergo a unitary evolution together, and vice versa.

\subsection{Choi–Jamiołkowski isomorphism}\label{subsection: Choi Jamiolkowski isomorphism}
We have introduced three different mathematical objects: state vectors (in a Hilbert space $\mH$), density matrices and other operators (linear maps from $\mH$ to $\mH'$), and quantum channels (linear maps from $\mL$ to $\mL'$).
The latter is also called superoperators. 
In fact, both operators and superoperators can also be regarded as vectors in some Hilbert spaces. 
However, unlike state vectors and operators which have simple representations of column vectors and matrices, superoperator representations are less intuitive. 
Choi–Jamiołkowski isomorphism, which connects a superoperator to a operator, thus becomes useful. 
\begin{theorem}[Choi's theorem~\cite{Choi1975}]\label{theorem: Choi theorem}
	Let us denote the maximally entangled state $\ket{\Omega}_{S\bar{S}} \coloneq \frac{1}{\sqrt{d}}\sum_{i=1}^{d}\ket{i}_{S} \otimes \ket{i}_{\bar{S}}$, where $d = \dim(\mH_{S}) = \dim(\mH_{\bar{S}})$.
	Then, $\Lambda$ from $\mL_{S}$ to $\mL_{S'}$ is a CP map \emph{if and only if} the operator 
	\begin{align}\label{eq: Choi state definition}
		C_{\Lambda} \coloneq (\Lambda\otimes\id_{\bar{S}})(\dm{\Omega}_{S\bar{S}})\in\mL_{S'\bar{S}},
	\end{align}
	known as \emph{Choi state}, is positive semi-definite. 
\end{theorem}
For a CPTP map $\mE$, the Choi state $C_{\mE}$ has another characterisation: $\Tr_{S'}[C_{\mE}] = \frac{\1_{\bar{S}}}{d}$.
It is also possible to retrieve the channel knowing the Choi state, since the following identity 
\begin{align}
	\Tr[A\mE(\rho)] = d\Tr[C_{\mE}(A\otimes\rho^{\transp})]
\end{align}
holds for any operator $A\in\mL_{S'}$.
Note that Jamiołkowski discovered a similar result~\cite{Jamiolkowski1972} with slightly different isomorphism before Ref.~\cite{Choi1975}.
See Ref.~\cite{Jiang2013Choi} for a more careful exposition and historical background on this matter.

\section{Distance measures and divergences}
For any quantitative study on (quantum) information processing, it is necessary to have a function that measures the degree of difference in two mathematical objects.
In this thesis, the main objects are vectors in $\mH$ (pure states), operators in $\mL$ (density matrices, unitary operators), and superoperators (CPTP maps).

The most immediate candidate would be the distance measures, which quantifies the distance between two objects. 
We define \emph{distance function} $d(\cdot,\cdot): \mA \times \mA \to \mbR$, where $\mA$ is a set (either $\mH$, $\mL$, or $\mathrm{CPTP}$ in this thesis), such that 
\begin{itemize}
	\item $d(\psi,\phi) = 0$ for $\psi,\phi\in\mA$ if and only if $\psi = \phi$,
	\item $d(\psi,\phi) = d(\phi,\psi)$ for any two $\psi,\phi\in\mA$, and
	\item $d(\psi,\phi) \leq d(\psi,\xi) + d(\xi,\phi)$ for any $\psi,\phi,\xi\in\mA$, also known as the triangle inequality. 
\end{itemize} 

In this thesis, distance functions are heavily used in the context of quantum computing and algorithm. 
Because they are symmetric and they satisfy triangle inequality, distance functions are the most suitable quantifiers for the error incurred during the algorithm implementation.

An easy way to construct a distance function is starting from the well-known norms. 
First consider $n$-dimensional vector spaces.
\begin{definition}[$p$-norms]\label{definition: p-norms}
	For $1\leq p < \infty$, the $p$-norm of a vector $x = (x_{1},x_{2},\cdots,x_{n})$ is 
	\begin{align}
		\| x\|_{p} \coloneq \left(\sum_{i}|x_{i}|^{p}\right)^{\frac{1}{p}},
	\end{align}
	where $|x_{i}| = \sqrt{x_{i}^{*}x_{i}}$.
	For $p = \infty$, we define
	\begin{align}
		\|x\|_{\infty} \coloneq \max_{i}|x_{i}|.
	\end{align}
\end{definition}
The distance between two vectors $x,y$ are then defined as $\| x - y \|_{p}$, and it follows all three axioms in the previous paragraph. 
Hence, for any two state vectors $\ket{\psi},\ket{\phi}\in\mH$, we use $\|\ket{\psi} - \ket{\phi}\|_{p}$ as the distance between them.  

Similarly, we can define norms and distances for operators. 
\begin{definition}[Schatten $p$-norms]\label{definition: Schatten p-norms}
	For $1\leq p < \infty$, the Schatten $p$-norm of an operator $A\in\mL$ is defined as
	\begin{align}
		\|A\|_{p} \coloneq \left(\Tr\left[ |A|^{p}\right]\right)^{\frac{1}{p}},
	\end{align}  
	where $|A| = \sqrt{A^{\dagger}A}$.
	For $p = \infty$, we define
	\begin{align}\label{eq: operator norm def}
		\|A\|_{\infty} \coloneq \max_{i}s_{i}(A),
	\end{align}
	where $s_{i}(A)$ are the singular values of $A$.
	Eq.~\eqref{eq: operator norm def} is defined to match the limit $\|A\|_{p\to\infty}$.
\end{definition}
Note that Schatten $p$-norms of an operator $A$ can also be considered as $p$-norms of the vector consists of singular values $\{s_{i}(A)\}_{i}$.
As a result, Schatten $p$-norms have unitary invariance, i.e. for any two unitary operators $U,V\in\mL$, 
\begin{align}
	\|A\|_{p} = \|UAV\|_{p}.
\end{align}
See Ref.~\cite{BhatiaBook} for the detailed properties of this and other norms. 
Distances between density matrices, unitary operators, etc., can then be defined by the norm of the difference $\frac{1}{2}\|A - B\|_{p}$ with $\frac{1}{2}$ factor included by convention. 
Three special cases are noteworthy. 
\begin{itemize}
	\item when $p = 1$, the Schatten $p$-norm becomes the trace norm $\Tr[|A|]$,
	\item when $p = 2$, the Schatten $p$-norm becomes the Hilbert-Schmidt norm $\sqrt{\Tr[A^{\dagger}A]}$,
	\item when $p = \infty$, the Schatten $p$-norm becomes the operator norm.
\end{itemize}

Finally, the distances between channels can be induced from those of operators. 
Here, we only show two of them. 
\begin{definition}\label{definition: channel distances} 
	For two quantum channels $\mE, \mF\in\CPTP{S}{S'}$, the \emph{trace norm distance} is defined as
	\begin{align}
			\frac{1}{2}\|\mE - \mF\|_{\Tr} \coloneq \frac{1}{2}\max_{\rho\in\mD_{S}}\|\mE(\rho) - \mF(\rho)\|_{1},
	\end{align}
	whereas the \emph{diamond norm distance}~\cite{Aharonov1998DiamondNorm} is defined as
	\begin{align}
		\frac{1}{2}\|\mE - \mF\|_{\diamond} \coloneq \frac{1}{2}\max_{\varrho\in\mD_{S\bar{S}}}\|(\mE\otimes\id_{\bar{S}})(\varrho) - (\mF\otimes\id_{\bar{S}})(\varrho)\|_{1},
	\end{align}
	where $\dim(\mH_{S}) = \dim(\mH_{\bar{S}})$.
\end{definition}
Note that we use $\|\cdot\|_{\Tr}$ for channels and $\|\cdot\|_{1}$ for operators. 
From the ranges of optimisation for two distance measures, $\frac{1}{2}\|\mE - \mF\|_{\diamond} \geq \frac{1}{2}\|\mE - \mF\|_{\Tr}$ is derived.

Another important quantity, although not a distance, is \emph{Uhlmann's fidelity}~\cite{Uhlmann1976fidelity}. 
\begin{definition}[fidelity of quantum states]\label{definition: fidelity}
	For two density matrices $\rho,\sigma\in\mD$,
	\begin{align}
		F(\rho,\sigma) \coloneq \left(\Tr\left[|\sqrt{\rho}\sqrt{\sigma}|\right]\right)^{2}.
	\end{align}
\end{definition}
Fidelity is a measure that has the value in $[0,1]$; its value is closer to $0$ when two states are more different, and to $1$ when two states are closer. 
For two pure states $\ket{\psi},\ket{\phi}\in\mH$, the fidelity coincides with the overlap $F(\ket{\psi},\ket{\phi}) = |\braket{\psi}{\phi}|^{2}$.
Moreover, in this case, the \emph{infidelity} coincides with the trace distance $1 - F(\ket{\psi},\ket{\phi}) = \frac{1}{2}\|\dm{\psi} - \dm{\phi}\|_{1}$.

In many cases, some of the three requirements of distance functions are unnecessary. 
For instance, if there is an asymmetry in the setting, i.e. if the roles of two objects are different, we would like a quantifier that is asymmetric to the two arguments in general. 
These measures that generalise distance functions are called divergences. 
The term divergence is defined in multiple non-identical ways~\cite{HayashiBook, AmariInfoGeometry, Gour2021Divergence}. 
We choose the loosest conditions as our definition of divergence: $\mbD(x|y)\geq 0$ for all $x,y$ and $\mbD(x|x)= 0$ for all $x$.
Note that divergences, as defined, have neither symmetry ($\mbD(x|y)\neq\mbD(y|x)$) nor triangle inequality ($\mbD(x|y)\nleq \mbD(x|z)+\mbD(z|y)$).
Furthermore, faithfulness ($\mbD(x|y)\geq 0$ if and only if $x = y$) and the data-processing inequality (as in Eq.~\eqref{eq: DPI classical} or Proposition~\ref{proposition: DPI for quantum Rényi divergences}), which are sometimes included as defining axioms for divergences, are not assumed \emph{a priori}.
We overview some the divergences that are used in this thesis. 

In this thesis, divergences are mainly used for resource theories, as many of them have direct interpretations in terms of state discrimination tasks~\cite{TomamichelBook, HayashiBook}, closely related to resource theoretic state transformation tasks.
In particular, divergences often can be used as a resource measure for any resource of interest; see Chapter~\ref{chapter: resource theories} Section~\ref{subsection: resource monotones} for detail. 

First is the family of Rényi divergences. 
\begin{definition}[Rényi divergences~\cite{Reyni1961}]\label{definition: reyni divergences}
	Suppose that $\pstate,\qstate$ are two probability vectors of the same dimension whose $i$th elements are $p_{i},q_{i}$, respectively. 
	For $\alpha\in(0,1)\cup(1,\infty)$, 
	\begin{align}
		D_{\alpha}(\pstate\|\qstate) \coloneq \frac{1}{\alpha-1}\log\left(\sum_{i}p_{i}^{\alpha}q_{i}^{1-\alpha}\right).
	\end{align}
	If $\alpha\to1$, the Kullback-Leibler divergence~\cite{KLdiv1951} 
	\begin{align}
		D_{\KL}(\pstate\|\qstate) \coloneq \sum_{i}p_{i}\log\left(\frac{p_{i}}{q_{i}}\right)
	\end{align}
	is recovered. 
	If $\alpha\to\infty$, 
	\begin{align}
		D_{\infty}(\pstate\|\qstate) \coloneq \log\left(\sup_{i}\frac{p_{i}}{q_{i}}\right),
	\end{align}
	and if $\alpha\to0$~\cite{Csiszar1995Reyni0},
	\begin{align}
		D_{0}(\pstate\|\qstate) \coloneq -\log\left(\sum_{i:p_{i}>0}q_{i}\right).
	\end{align}
\end{definition}
These divergences are derived from a set of axioms, such as continuity, unitary invariance, and additivity 
\begin{align}
	D_{\alpha}(\pstate_{1}\times \pstate_{2}\|\qstate_{1}\times \qstate_{2}) = D_{\alpha}(\pstate_{1}\|\qstate_{1}) +  D_{\alpha}(\pstate_{2}\|\qstate_{2}).
\end{align}
Note that $D_{0}$ does not qualify as a divergence in our definition, because there exists $\pstate\neq \qstate$, such that $D_{0}(\pstate\|\qstate) = 0$.
However, for $\alpha\in[0,\infty]$, a crucial property called data-processing inequality holds. 
For divergences $\mbD$ of classical probability vectors, data-processing inequality holds if for any probability vectors $\pstate,\qstate$ and any stochastic matrix $S$
\begin{align}\label{eq: DPI classical}
	\mbD(\pstate\|\qstate) \geq \mbD(S\pstate\|S\qstate).
\end{align}
Eq.~\eqref{eq: DPI classical} implies that post-processing of probability vectors cannot increase their distinguishability.
Ref.~\cite{vanErven2014Reyni} has a concise summary on the properties of Rényi divergences. 

Extending Rényi divergences to density matrices is not trivial. 
The set of axioms that uniquely defined Rényi divergences no longer does so for quantum states~\cite{TomamichelBook}. 
Here, we present two examples.
\begin{definition}[quantum Rényi divergences~\cite{TomamichelBook}]\label{definition: quantum Rényi entropies}
	For $\rho,\sigma\in\mD$, the minimal Rényi divergence is defined as
	\begin{align}
		\tilde{D}_{\alpha}(\rho\|\sigma) \coloneq \frac{1}{\alpha-1}\log\left(\Tr\left[\left(\sigma^{\frac{1-\alpha}{2\alpha}}\rho\sigma^{\frac{1-\alpha}{2\alpha}}\right)^{\alpha}\right]\right).	
	\end{align}
	This divergence converges to well-known divergences in the limiting cases: to max-relative entropy
	\begin{align}\label{eq: def max rel entropy}
		D_{\max}(\rho\|\sigma) \coloneq \log\inf\{\lambda: \rho\leq\lambda \sigma\},
	\end{align}
	when $\alpha\to\infty$, and to \emph{Umegaki relative entropy}~\cite{Umegaki1962}, also known as \emph{quantum relative entropy},
	\begin{align}\label{eq: def Umegaki}
		D(\rho\|\sigma) \coloneq \Tr\left[\rho(\log\rho - \log\sigma)\right],
	\end{align}
	for $\alpha\to1$.
	
	Petz quantum Rényi divergences are defined as
	\begin{align}
		\bar{D}_{\alpha}(\rho\|\sigma) \coloneq \frac{\sgn(\alpha)}{\alpha-1}\log\left(\Tr\left[\rho^{\alpha}\sigma^{1-\alpha}\right]\right).
	\end{align}
	For $\alpha\to1$, $\bar{D}_{\alpha}$ also converges to quantum relative entropy; for $\alpha\to0$, to min-relative entropy
	\begin{align}
		D_{\min}(\rho\|\sigma) \coloneq -\log\left(\Tr\left[\Pi_{\rho}\sigma\right]\right),
	\end{align}
	where $\Pi_{\rho}$ is the projector onto $\supp(\rho)$.
\end{definition}
Except for the min-relative entropy, divergences defined in Definition~\ref{definition: quantum Rényi entropies} are faithful, i.e. $\mD(\rho\|\sigma) = 0$ if and only if $\rho = \sigma$. 

For certain ranges of $\alpha$, these quantum Rényi divergences have data-processing inequality. 
\begin{proposition}[data-processing inequalities for quantum Rényi divergences]\label{proposition: DPI for quantum Rényi divergences}
	For $\alpha\geq\frac{1}{2}$,
	\begin{align}
		\tilde{D}_{\alpha}(\rho\|\sigma) \geq \tilde{D}_{\alpha}(\mE(\rho)\|\mE(\sigma)),
	\end{align}
	for any $\rho,\sigma\in\mH_{S}$ and for any $\mE\in\CPTP{S}{S'}$.
	For $\alpha \in[0,2]$,
	\begin{align}
		\bar{D}_{\alpha}(\rho\|\sigma) \geq \bar{D}_{\alpha}(\mE(\rho)\|\mE(\sigma)),
	\end{align}
	for any $\rho,\sigma\in\mH_{S}$ and for any $\mE\in\CPTP{S}{S'}$.
\end{proposition}
See Ref.~\cite{TomamichelBook} for the proof. 
Data processing inequalities for density matrices imply that quantum channels cannot be used to make two states more distinguishable.

We introduce one peculiar divergence that plays an important role in Chapter~\ref{chapter: resource broadcasting}.
Suppose that $\mM = \{M_{i}\}_{i}$ is a POVM with effects $M_{i}\in\mL_{S}$. 
We use $\mM(\rho)$ to denote a vector with elements $(\mM(\rho))_{i} = \Tr[M_{i}\rho]$, i.e. a vector of outcome probabilities when measuring $\rho\in\mD_{S}$ with POVM $\mM$.
\begin{definition}[measured relative entropy]\label{definition: measured relative entropy}
	The measured relative entropy between two states $\rho,\sigma\in\mD_{S}$ is defined as 
	\begin{align}
		D_{\mbM}(\rho\|\sigma) \coloneq \sup_{\mM: \text{POVM}}D_{\KL}(\mM(\rho)\|\mM(\sigma)),
	\end{align}
	where the supremum is over all POVMs.%
	\footnote{It is also possible to define the function with supremum over a set of POVMs that might not be the entire set of POVMs. However, to make this function a proper divergence, i.e. to make $D_{\mbM}(\rho\|\sigma) = 0$ if and only if $\rho = \sigma$, the set of POVMs must be informationally complete~\cite{Piani2009MRelEnt}.}
\end{definition}
This notion of measured divergence can be generalised for $\alpha$-divergences.
Moreover, these generalisations are useful for proving data-processing inequalities for other divergences, as the supremum in the definition of the measured divergences guarantees the data-processing inequality for them~\cite{TomamichelBook}.

\addpart{Catalysis in resource theories}
\chapter{Resource theories and catalysis: the framework}\label{chapter: resource theories}

Resource theory is a framework for studying the limit of achievable processes under the assumption that a particular resource is scarce.  
Two complementary but seemingly different pictures are adopted for the mathematical treatment of this framework. 
The first is a more traditional and operationally inspired approach as in Refs.~\cite{Chitambar2019RTRev, Gour2024RTBook}.
In this picture, resource theories define a set of quantum channels, known as \emph{free operations} $\fro{S}{S'}\subset\CPTP{S}{S'}$, deemed to be free between any two systems $S$ and $S'$.
The minimum requirement for these operations is that the resource, which motivates the theory, cannot increase after a free operation. 
Since the sets of quantum channels are the starting point of this approach, we exactly know what the feasible physical operations are, regardless of the input states.  
Often these sets are defined operationally, meaning that even the recipes to implement free operations are given. 
Then, at least in principle, it is possible to determine whether a transformation $\rho_{S}\to\rho'_{S'}$ is feasible by some CPTP map in $\fro{S}{S'}$.
From this, we can construct a set of all pairs $(\rho_{S},\rho'_{S'})$, such that $\rho_{S}\to\rho'_{S'}$.
Conversely, to show $\rho_{S}\to\rho'_{S'}$ one needs to identify an operation enabling this transition.
Moreover the requirement that free operations must be quantum channels imposes a certain restriction on the way we construct the pairs. 

The second picture, on the other hand, starts from the pairs of states. 
Then the resource theory is defined as the set of feasible state transitions $\rho_{S}\to\rho'_{S'}$ without much consideration on the operations that induce such transitions. 
This is an approach taken by category theoretical studies in resource theories~\cite{Coecke2016RT}.
In this picture, less direct state transitions, e.g. asymptotic or catalytic transformations, can naturally be included by just defining $\rho_{S}\to\rho'_{S'}$ when it is achievable by some asymptotic or catalytic process. 
Although this approach may seem more attractive for our purpose of using auxiliary systems, the lack of a guiding principle of free operations makes it difficult to construct a physically relevant resource theory solely from state transformations.

We therefore take the first picture, i.e. the paradigm of free operations, for the rest of this thesis. 
However, we pay particular attention when introducing compositions between different systems, albeit without the mathematical rigour of category theory. 

\section{Resource theories from free operations}\label{section: resource theories}
To define a resource theory, we classify quantum channels into ones that are feasible (\emph{free operations} $\fro{S}{S'}\subset\CPTP{S}{S'}$) and those that are not. 
We impose several constraints on this classification, which originate from operational scenarios deemed natural. 
\begin{definition}[proper free operations for resource theories]\label{definition: free operations}
	A set of free operations $\fro{S}{S'}\subset\CPTP{S}{S'}$ is proper if it satisfies the three defining properties. 
	\begin{enumerate}
		\item The identity channel is free: $\id_{S}\in\fro{S}{S}$.
		\item Concatenation of free channels is free: $\mE\circ\mF\in\fro{S}{S'}$ if $\mE\in\fro{S''}{S'}$ and $\mF\in\fro{S}{S''}$.
		\item Discarding a (sub)system is free: $\Tr_{R}\in\fro{SR}{S}$.
	\end{enumerate}
\end{definition}
Note that these three axioms are almost identical to those in Ref.~\cite{Gour2024RTBook}, except that we also allow partial traces, not only the full trace operation, to be free. 
The first property is required to ensure that it is possible to leave the system as it is. 
Also, it renders the relation $\to$ reflexive, i.e. $\rho\to\rho$ for any $\rho$.
The second property is intuitive, although it might exclude scenarios where the number of operations is relevant, as in the resource theory of uncomplexity~\cite{YungerHalpern2022Uncomplexity}.  
Concatenations immediately establish the transitivity of the $\to$ relation: if $\rho\to\rho'$ and $\rho'\to\rho''$, then $\rho\to\rho''$.
However, the relation is not necessarily strongly connected as there exist pairs of \emph{incomparable} states $(\rho,\rho')$ such that $\rho\not\to\rho'$ and $\rho'\not\to\rho$.
These properties define a nice mathematical structure for the alternative picture of state transitions. 
\begin{remark}
	Resource theories define a \emph{preorder} between density matrices with the binary relation $\to$.	
\end{remark}

Property 3 is the only one that concerns the composition, or rather decomposition, of systems.
The partial trace $\Tr_{R}$ corresponds to separating system $R$ from the rest and forgetting about it. 
One also needs to be careful that this discarded state does not interact with any other auxiliary systems that might interact with the system of interest in the future. 
Nevertheless, this operation is easily achievable in most physical scenarios.
At the same time, any $\fro{SR}{S}$ operation can be implemented with the $\fro{SR}{SR}$ operation followed by the partial trace $\Tr_{R}$ in most resource theories. 
\begin{observation}
	Suppose that appending a state $\mA(\cdot) = (\cdot)\otimes\gamma_{R}\in\CPTP{S}{SR}$ is a free operation. 
	Then, any free channel $\mE\in\fro{SR}{S}$ can be constructed as $\mE = \Tr_{R}\circ\mF$, where $\mF\in\fro{SR}{SR}$ is free.
\end{observation}
\begin{proof}
The proof is simple but somewhat circular. 
By setting $\mF = \mA\circ\mE$, we obtain $\mE = \Tr_{R}\circ\mF$.
This observation hints that $\Tr_{R}$ is the only new ingredient that is needed for the decomposition process $\fro{SR}{S}$ given the set of free operations from a system to itself. 
\end{proof}

From the set of free operations, we can define states that are free. 
\begin{definition}[free states]\label{definition: free states}
	The set of free states is defined as 
	\begin{align}
		\frs{S} \coloneq \{\gamma_{S} \,|\, \forall\rho_{S}\in\mD_{S},\, \exists\mE\in\fro{S}{S},\ \text{s.t.}\ \mE(\rho_{S}) = \gamma_{S}\}.
	\end{align}
\end{definition}
Here, our definition deviates from that of Ref.~\cite{Gour2024RTBook}, which uses $\tilde{\mathcal{S}}_{S} \coloneq \{\mE(1) \,|\, \mE\in\fro{1}{S} \}$, where $1$ represents the trivial one-dimensional state in a one-dimensional space $1$.
We choose this alternative definition to encompass interesting resource theories that are not adequately captured by the conventional definition. 
The conventional definition of $\tilde{\mathcal{S}}_{S}$ as a free state set is more stringent than Definition~\ref{definition: free states} when free operations are defined first and subsequently induce free states.
Specifically, if $\gamma_{S}\in\tilde{\mathcal{S}}_{S}$, then $\gamma_{S}\in\frs{S}$ because we can consider the process $\rho_{S}\to1\to\gamma_{S}$, which corresponds to a concatenation of free operations.
However, the converse is not necessarily true under our axioms. 
This is because the composition process $\fro{A}{AB}$ is not guaranteed to include any channel \emph{a priori}.
For instance, we could define $\fro{A}{AB}=\emptyset$, effectively disallowing the composition of system $B$ with $A$.

Concrete examples are variants of thermodynamic resource theories---elementary thermal operations (Definition~\ref{definition: ETO}) and Markovian thermal operations (Definition~\ref{definition: MTO})---which are the main focus of Chapter~\ref{chapter: CETO}.
Adopting Definition~\ref{definition: free states}, both theories have the Gibbs state with respect to the ambient temperature (Eq.~\eqref{eq: Gibbs state def}) as a sole free state for each system.
However, with the conventional definition of $\tilde{\mathcal{S}}_{S}$, both theories cannot have any free state. 
We believe it is more physical that Gibbs states are free for these theories. 

Switching to the state transition perspective, the third axiom for free operations---that the (partial) trace operation is free---sets the trivial one-dimensional state $1$ as the bottom element of the preorder defined by $\to$, meaning that for any state $\rho$, we have $\rho\to1$.
Using the conventional definition $\tilde{\mathcal{S}}_{S}$ would imply that all free states are bottom elements within the preordered set $\bigcup_{X}\mD_{X}$.
However, with our Definition~\ref{definition: free states}, free states in $\frs{X}$ are only bottom elements within $\mD_{X}$, the set of states on system X, and not necessarily across all systems.

This modification in Definition~\ref{definition: free states} means that a fundamental property often assumed in resource theories cannot be directly derived from our initial definitions. 
Therefore, we explicitly define this property and assume that all resource theories of interest satisfy it.
\begin{definition}[the golden rule of resource theories]\label{definition: golden rule of resource theories}
	Resource theories follow the golden rule if free operations, when applied to free states, invariably result in a free state, i.e.
	\begin{align}\label{eq: golden rule}
		\mE(\gamma_{S})\in\frs{S'},\quad \forall\gamma_{S}\in\frs{S},\ \forall\mE\in\fro{S}{S'}.
	\end{align}
\end{definition}
It is possible to construct a theory consistent with Definitions~\ref{definition: free operations} and~\ref{definition: free states} that violates this golden rule.%
\footnote{Note that this is not possible with the conventional definition of free states $\tilde{\mathcal{S}}$.
If $\gamma_{S}\in\tilde{\mathcal{S}}_{S}$, then $1\to\gamma_{S}$ by definition. 
Hence, $1\to\gamma_{S}\to\mE(\gamma_{S})$, implying $\mE(\gamma_{S})\in\tilde{\mathcal{S}}_{S'}$ for any $\mE\in\fro{S}{S'}$.}
Suppose that $S = S'R$ and $S'$ 
represents an inaccessible system, meaning its only free operation is the identity ($\fro{S'}{S'} = \{\id_{S'}\}$), and there are no free operations from $S'$ to any other non-trivial system $X$ (i.e., $\fro{S'}{X}=\emptyset$ for $X\neq1,S'$).
In this case, the set of free states for $S'$ is empty because no state $\rho_{S'}$ can be transformed into a different state $\rho'_{S'}$ via a free operation. 
Now, suppose there somehow exists a free state $\gamma_{S'R}\in\frs{S'R}$. 
Since the partial trace operation is free, if we take $\mE = \Tr_{R}$ in Eq.~\eqref{eq: golden rule}, we would obtain $\Tr_{R}[\gamma_{S'R}]$, which cannot be free as $\frs{S'}$ is empty.
Although such a theory is mathematically consistent with the initial three properties of free operations, the existence of a free state $\gamma_{S'R}\in\frs{S'R}$ when $S'$ is inaccessible appears unnatural. 
Therefore, we will only consider resource theories that adhere to the golden rule as defined in Definition~\ref{definition: golden rule of resource theories}.

In addition to Definitions~\ref{definition: free operations} and~\ref{definition: golden rule of resource theories}, we consider several properties of the resource theory that facilitate the mathematical treatment.
\begin{definition}[tensor product structure]\label{definition: tensor product structure}
	A resource theory has a tensor product structure if
	\begin{enumerate}[start=4]
		\item relabelling of systems is free
		\item if an operation is free, it is completely free: $\mE\otimes\id_{R}\in\fro{SR}{S'R}$ for all $\mE\in\fro{S}{S'}$ and for all systems $R$.
	\end{enumerate}
\end{definition}
This definition matches that of Ref.~\cite{Gour2024RTBook}. 
Note that properties 4 and 5 concern composite systems. 
In particular, property 5, when combined with property 2 in Definition~\ref{definition: free operations} indicates that the independent application of free operations in each subsystem is free: $\mE\otimes\mF\in\fro{S_{1}S_{2}}{S_{1}'S_{2}'}$ when $\mE\in\fro{S_{1}}{S_{1}'}$ and $\mF\in\fro{S_{2}}{S_{2}'}$.
This composition rule for free operations is the most natural condition to demand for composite systems, as it does not necessitate any interaction between the two systems. 

However, the tensor product structure is not universally present in resource theories studied in the literature.
To see this, let us consider constructions that begin with free states rather than free operations.
\begin{definition}[resource non-generating (RNG) operations]\label{definition: RNG}
	Suppose that the set of free states $\frs{S}$ is given for each system $S$.
	An RNG operation is defined as
	\begin{align}
		\RNG{S}{S'} \coloneq \{\mE\in\CPTP{S}{S'} \,|\, \mE(\gamma_{S})\in\frs{S'},\ \forall\gamma_{S}\in\frs{S}\}.
	\end{align}
\end{definition}
In other words, $\RNG{S}{S'}$ encompasses all channels that satisfy the golden rule.
The other defining properties of free operations can be verified for RNG operations.
Property 1, $\id_{S}\in\RNG{S}{S}$, holds because $\id_{S}(\gamma_{S})\in\frs{S}$ for any $\gamma_{S}\in\frs{S}$.
Property 2 is also satisfied: if $\mF\in\fro{S}{S''}$ and $\mE\in\fro{S''}{S'}$, then for any $\gamma_{S}\in\frs{S}$, $\mF(\gamma_{S}) \in\frs{S''}$ and thus $\mE(\mF(\gamma_{S}))\in\frs{S'}$, by the golden rule.
Finally, Property 3 can be made compatible by requiring that the set of free states is such that $\Tr_{R}[\gamma_{SR}]\in\frs{S}$ for all $\gamma_{SR}\in\frs{SR}$.

An instructive example can be found in the resource theory of entanglement.
\begin{example}[non-entangling operations]\label{example: non-entangling operations}
	Suppose that each system is defined as a bi-partite composite system, $S_{i} = A_{i}B_{i}$, where either $A_{i}$ or $B_{i}$ can be trivial. 
	Define the set of free states $\frs{A|B}$ as the set of all separable states $\SEP_{AB}$, as defined in Definition~\ref{definition: separable states}. 
	Then, the set of RNG operations $\RNG{S}{S'}$ is referred to as non-entangling operations.
\end{example}
This definition yields a valid resource theory, studied in works such as Refs.~\cite{Brandao2010Entanglement, Lami2023No2ndlaw}.
However, the non-entangling resource theory lacks the tensor product structure because the swap operation is considered free.
For any separable state $\gamma_{S} = \sum_{i}p_{i}\rho^{(i)}_{A}\otimes\sigma^{(i)}_{B}$, applying the swap operation results in $\gamma'_{S'} =  \sum_{i}p_{i}\rho^{(i)}_{B}\otimes\sigma^{(i)}_{A}$, which remains separable. 
Now, consider the system $S_{1}S_{2} = A_{1}A_{2}|B_{1}B_{2}$. 
For any (potentially entangled) state $\rho_{A_{1}A_{2}}$ and $\sigma_{B_{1}B_{2}}$, the product state $\rho_{A_{1}A_{2}}\otimes\sigma_{B_{1}B_{2}}$ is separable, with respect to the $A_{1}A_{2}|B_{1}B_{2}$ partition.
However, the swap operation acting between subsystems $A_{1}$ and $B_{1}$ yields the output state $\rho_{B_{1}A_{2}}\otimes\sigma_{A_{1}B_{2}}$, which is an entangled state for $A_{1}A_{2}|B_{1}B_{2}$ partition when either $\rho$ or $\sigma$ is entangled itself.  

Imposing the tensor product structure therefore excludes certain RNG resource theories like Example~\ref{example: non-entangling operations}.
However, this exclusion does not posit a significant conceptual problem. 
RNG theories are primarily considered because they represent the most relaxed definition of free operations, which is valuable for proving the impossibility of certain processes across all more restrictive resource theories~\cite{Lami2023No2ndlaw}.
Consequently, further constraining the theory by requiring a tensor product structure does not alter the conclusions of such fundamental no-go results.

The maximally relaxed operations consistent with the tensor product structure (Definition~\ref{definition: tensor product structure}) can also be defined.
\begin{definition}[completely resource non-generating (CRNG) operations]\label{definition: CRNG}
	Suppose that the set of free states $\frs{S}$ is given for each system $S$.
	CRNG operation is defined as
	\begin{align}
		\CRNG{S}{S'} \coloneq \{\mE\in\CPTP{S}{S'} \,|\, (\mE\otimes\id_{R})(\gamma_{SR})\in\frs{S'R},\ \forall\gamma_{SR}\in\frs{SR},\ \forall R\}.
	\end{align}
\end{definition}
CRNG for entanglement theory, analogous to Example~\ref{example: non-entangling operations}, is termed separable operations~\cite{Vedral1997SEP, Rains1998SEP}.
Other examples of CRNG operations include Gibbs-preserving operations~\cite{Faist2015GP}, which correspond to scenarios where the set of free states contains only a single Gibbs state at a fixed temperature. 
Covariant operations~\cite{Keyl1999Cov, Gour2008Cov} are another example, where the free states are those that remain invariant under operations obeying a particular symmetry. 
Finally, maximally incoherent operations~\cite{Aberg2006MIO} arise when the free states are density matrices that are diagonal in a fixed basis. 
In these latter three cases, CRNG operations are also RNG operations.

We now consider a geometric property of the set of free operations
\begin{definition}[convex resource theories]\label{definition: convex RTs}
	A resource theory is convex if the set $\fro{S}{S'}$ is convex.
\end{definition}
Note that if $\fro{S}{S'}$ is convex, then the set of free states $\frs{S'}$ is also convex, but the converse is not true. 
Furthermore, the convexity enables the derivation of universal results that hold for general resource theories~\cite{Takagi2019Convex, Takagi2019Convex2}.%
\footnote{More recently, similar results have also been established for non-convex theories~\cite{Kuroiwa2024nonConvex}.}

The majority of resource theories studied are convex. 
This is partly because the set of free operations can be made convex by allowing the use of classical randomness, i.e. choosing which operation to apply by rolling a die and forgetting about the result. 
Yet, there are notable counterexamples, such as the theories of non-Gaussianity~\cite{Zhuang2018nonGaussianity} and randomness~\cite{Lie2023Delocalized}.
In Chapters~\ref{chapter: CETO} and~\ref{chapter: resource broadcasting}, we will always assume the convexity of the theories under consideration.

%

\subsection{Resource monotones}\label{subsection: resource monotones}
As the name suggests, resource theories are concerned with the quantification and manipulation of resources. 
However, the axioms presented in Definition~\ref{definition: free operations} do not explicitly define what constitutes a resource.
Nevertheless, in most resource theories of interest, the definition of free operations is motivated by a specific resource. 
Broadly speaking, an operation is considered free if its implementation does not require the resource in question.
Conversely, the notion of a resource can be inferred from the set of free operations.
Specifically, a resource can be understood as a property of a quantum state that cannot be increased by any free operation.

Resource monotones provide a formal framework for this concept.
\begin{definition}[resource monotones]\label{definition: resource monotones}
	A function $\mbP: \bigcup_{X}\mD_{X} \to \mbR$ is a resource monotone if 
	\begin{enumerate}
		\item $\mbP(\rho_{S})\geq\mbP(\rho'_{S'})$, whenever $\rho_{S}\to\rho'_{S'}$ and
		\item $\mbP(\gamma_{S}) = 0$, whenever $\gamma_{S}\in\frs{S}$
	\end{enumerate}
\end{definition}
The first condition implies that a resource monotone is non-increasing under free operations, thus respecting the preorder defined by $\to$.
The second condition is a normalisation, stipulating that free states, which should intuitively possess zero resource, are assigned a value of zero by the monotone. 
It is important to note that while $\mbP(\rho_{S})\geq\mbP(\rho'_{S'})$ is a necessary condition for the transformation $\rho_{S}\to\rho'_{S'}$ to be possible, it is generally not a sufficient condition on its own.
Therefore, resource monotones quantify the amount of resource in a state, although this quantification is often incomplete.

For any given resource theory, there can exist multiple resource monotones.
If a family of monotones $\mR$ completely characterises the preorder, i.e.
\begin{align}
	\rho\to\rho' \quad\Leftrightarrow\quad \mbP(\rho)\geq \mbP(\rho'),\ \forall \mbP\in\mR,
\end{align}
then $\mR$ is termed a complete set of monotones.
Such a family can be constructed in principle~\cite{Coecke2016RT, Takagi2019Convex2}, although it is hard to actually calculate them, in general. 
Interestingly, if a single monotone completely characterises all possible state transformations within a resource theory, then the preorder relation $\to$ defines a \emph{total order}.
This means that for any pair of states $(\rho,\rho')$, either $\rho\to\rho'$ or $\rho'\to\rho$ (or both), and the converse is also true~\cite{Datta2023Monotones}.

We now define some properties that can be useful.
The first property we consider is the converse of the second condition in Definition~\ref{definition: resource monotones}.
\begin{definition}[faithfulness]\label{definition: faithfulness monotone}
	A monotone $\mbP$ is faithful if
	\begin{align}
		\mbP(\gamma_{S}) = 0 \quad\Leftrightarrow\quad \gamma_{S}\in\frs{S}.
	\end{align}
\end{definition}
Thus, the existence of a faithful monotone provides a direct and convenient method to determine whether a given quantum state is a free state or not.

A useful feature of resource monotones is that they provide a means to compare the resource content of states belonging to different systems.
We now introduce properties that relate the resourcefulness of composite systems to that of their constituent subsystems.
First, observe that for any bipartite state $\varrho_{XY}\in\mD_{XY}$, the following inequalities hold:
\begin{align}
	\mbP(\varrho_{XY})\geq \max\left\{ \mbP(\Tr_{Y}[\varrho_{XY}]),\mbP(\Tr_{X}[\varrho_{XY}])\right\},
\end{align}
This is because the partial trace operation is always considered free.
In other words the resource content of a composite system, as quantified by any resource monotone, is always greater than or equal to the resource content of either of its marginal systems.
However, the relationship between the resource of the composite system and the sum of the resources of its marginals is not fixed and depends on the specific monotone.
\begin{definition}[additivity of resource monotones]\label{definition: additivity of resource monotones}
	A resource monotone $\mbP$ is 
	\begin{itemize}
		\item \emph{strongly additive} if $\mbP(\varrho_{XY}) = \mbP(\Tr_{Y}[\varrho_{XY}])+\mbP(\Tr_{X}[\varrho_{XY}])$, $ \forall \varrho_{XY}\in\mD_{XY}$,
		\item \emph{strongly super-additive} if $\mbP(\varrho_{XY}) \geq \mbP(\Tr_{Y}[\varrho_{XY}])+\mbP(\Tr_{X}[\varrho_{XY}])$, $ \forall \varrho_{XY}\in\mD_{XY}$,
		\item \emph{strongly sub-additive}, if $\mbP(\varrho_{XY}) \leq \mbP(\Tr_{Y}[\varrho_{XY}])+\mbP(\Tr_{X}[\varrho_{XY}])$, $ \forall \varrho_{XY}\in\mD_{XY}$.
	\end{itemize}
	More relaxed versions of these conditions can be defined for product states. 
	A resource monotone $\mbP$ is 
	\begin{itemize}
		\item \emph{additive} if $\mbP(\rho_{X}\otimes\sigma_{Y}) = \mbP(\rho_{X})+\mbP(\sigma_{Y})$, $ \forall \rho_{X}\in\mD_{X}$, $\forall \sigma_{Y}\in\mD_{Y}$.
		\item \emph{super-additive} if $\mbP(\rho_{X}\otimes\sigma_{Y}) \geq \mbP(\rho_{X})+\mbP(\sigma_{Y})$, $ \forall \rho_{X}\in\mD_{X}$, $\forall \sigma_{Y}\in\mD_{Y}$.
		\item \emph{sub-additive}, if $\mbP(\rho_{X}\otimes\sigma_{Y}) \leq \mbP(\rho_{X})+\mbP(\sigma_{Y})$, $ \forall \rho_{X}\in\mD_{X}$, $\forall \sigma_{Y}\in\mD_{Y}$.
	\end{itemize}
\end{definition}
Strong (super-/sub-) additivity always implies (super-/sub-) additivity.
In addition, if a monotone is (strongly) additive, it is also (strongly) super-additive and (strongly) sub-additive. 
A monotone may be neither (strongly) super-additive nor (strongly) sub-additive. 
For example, a monotone might satisfy $\mbP(\rho_{X}\otimes\sigma_{Y})\geq \mbP(\rho_{X})+\mbP(\sigma_{Y})$ for some states $\rho_{X},\sigma_{Y}$, but for other states $\rho'_{X},\sigma'_{Y}$, it might satisfy $\mbP(\rho'_{X}\otimes\sigma'_{Y})\leq \mbP(\rho'_{X})+\mbP(\sigma'_{Y})$. 

These additivity properties place constraints on how resources behave under the composition of physical systems.
For instance, the existence of a strongly super-additive monotone is a crucial ingredient in proving the no-go theorem presented in Chapter~\ref{chapter: resource broadcasting}.
Nevertheless, the absence of a particular type of additivity property does not directly imply any result, as monotones provide necessary but not sufficient conditions for state transformations to be feasible. 

A convenient way to construct resource monotones is by using divergences between quantum states that satisfy the data-processing inequality, as exemplified by Proposition~\ref{proposition: DPI for quantum Rényi divergences} and Definition~\ref{definition: measured relative entropy}.
\begin{proposition}[monotones from data-processing inequality]\label{proposition: faithful forward and reverse}
	Suppose that $\mbD$ is a divergence between quantum states that follows the data-processing inequality and faithfulness. 
	Then 
	\begin{align}
		\mbP(\rho_{S}) &\coloneq \inf_{\gamma_{S}\in\frs{S}} \mbD(\rho_{S}\|\gamma_{S}),\\
		\mbPrev(\rho_{S}) &\coloneq \inf_{\gamma_{S}\in\frs{S}} \mbD(\gamma_{S}\|\rho_{S}),\label{eq: reversed divergence}
	\end{align}
	are faithful resource monotones. 
\end{proposition}
\begin{proof}
	Consider a free operation $\mE\in\fro{S}{S'}$ that transforms $\rho_{S}\to\rho'_{S'}$.
	Let $\tilde{\gamma}_{S}\in\frs{S}$ be a free state that achieves the infimum in the definition of $\mbP(\rho_{S})$, such that $\mbP(\rho_{S}) = \mbD(\rho_{S}\|\tilde{\gamma}_{S})$.
	Then by the data-processing inequality, we have
	\begin{align}
		\mbP(\rho_{S})\geq\mbD(\rho_{S}\|\tilde{\gamma}_{S})\geq \mbD(\mE(\rho_{S})\|\mE(\tilde{\gamma}_{S})).
	\end{align}
	The golden rule guarantees that $\mE(\tilde{\gamma}_{S})\in\frs{S'}$ and thus 
	\begin{align}
		\mbD(\mE(\rho_{S})\|\mE(\tilde{\gamma}_{S})) = \mbD(\rho'_{S'}\|\mE(\tilde{\gamma}_{S}))\geq \inf_{\gamma_{S'}\in\frs{S'}} \mbD(\rho'_{S'}\|\gamma_{S'}),
	\end{align}
	which proves the monotonicity of $\mbP$. 
	The faithfulness arises from that of the divergence.
	The proof for $\mbPrev$ follows the same argument. 
\end{proof}

Finally, we introduce several examples of resource monotones that will be used in the subsequent chapters of this thesis.
\begin{example}\label{example: resource monotone examples}
	When divergences are adapted as resource monotones, we denote them with the symbol $R$ in place of $D$ in the original notation.
	\begin{itemize}
		\item $R(\rho_{S}) \coloneq \inf_{\gamma_{S}\in\frs{S}}D(\rho_{S}\|\gamma_{S})$ is the \emph{relative entropy of resource}.
		\item $R_{\max (\min)}(\rho_{S}) \coloneq \inf_{\gamma_{S}\in\frs{S}}D_{\max (\min)}(\rho_{S}\|\gamma_{S})$ is the \emph{max (min) relative entropy of resource}.
		\item $R_{\mbM}(\rho_{S})\coloneq \inf_{\gamma_{S}\in\frs{S}}D_{\mbM}(\rho_{S}\|\gamma_{S})$ is the \emph{measured relative entropy of resource}. 
	\end{itemize}
\end{example}

\section{Catalysis}\label{section: catalysis}
Catalysis is a process involving an auxiliary system that returns to its original state at the end.
Fascinatingly, the very presence of such an auxiliary system, dubbed a catalyst, often enables transformations that would otherwise be impossible.
Within the framework of resource theories and the $\to$ relation, we can formally define strict catalysis as follows:
\begin{definition}[strict catalysis]\label{definition: strict catalysis}
	Strict catalysis is a process defined by a free operation $\mE\in\fro{SC}{S'C}$, initial state $\rho_{S}\in\mD_{S}$, and a catalyst $\tau_{C}\in\mD_{C}$, such that
	\begin{align}\label{eq: strict catalysis}
		\mE(\rho_{S}\otimes\tau_{C}) = \rho'_{S}\otimes\tau_{C}.
	\end{align}
\end{definition}
We use the term \emph{strict catalysis} following Ref.~\cite{LipkaBartosik2023CatReview} to distinguish this version of catalysis from other more generalised ones.
Note that if $\rho'_{S'} = \mF(\rho_{S})$ with some free operation $\mF\in\fro{S}{S'}$, it is always possible to find a catalysis $\mE(\rho_{S}\otimes\tau_{C}) = \rho'_{S'}\otimes\tau_{C}$ with any catalyst state $\tau_{C}\in\mD_{C}$. 
This can be achieved by choosing a free operation $\mE = \mF\otimes\id_{C}\in\fro{SC}{S'C}$.

However, only the non-trivial catalysis where $\rho_{S}\not\to\rho'_{S'}$ but $\rho_{S}\otimes\tau_{C}\to\rho_{S'}\otimes\tau_{C}$ are usually of interest. 
If we choose $\mE$ in Eq.~\eqref{eq: strict catalysis} to be a tensor product $\mE_{1}\otimes\mE_{2}$ with $\mE_{1}\in\fro{S}{S'}$ and $\mE_{2}\in\fro{C}{C}$ both free, it is impossible to get a non-trivial catalysis. 
This is because $\mE_{1}(\rho_{S}) = \rho'_{S'}$ without the help of a catalyst.
Therefore, finding a catalysis necessitates the study of composite free operations $\fro{SC}{S'C}$ that are not merely derived from the tensor product structure (Definition~\ref{definition: tensor product structure}). 
Another layer of difficulty in studying catalysis arises from the freedom of choosing the system $C$ and the catalyst state $\tau_{C}$.
Practically, these choices must be constrained by considerations regarding the difficulty of preparing and operating on the catalyst state $\tau_{C}$.
Nonetheless we are often interested in the ultimate limit of catalysis and would like to learn whether a desired transformation is feasible with the help of some catalyst $\tau_{C}$, regardless of its size or complexity. 
This goal corresponds to the characterisation of a new binary relation $\scto$ defined as 
\begin{align}
	\rho_{S} \scto \rho'_{S'} \quad\Leftrightarrow\quad \exists \tau_{C},\ \text{s.t.}\ \rho_{S}\otimes\tau_{C}\to\rho'_{S'}\otimes\tau_{C}. 
\end{align}

This new relation $\scto$ again defines a preorder.
The reflexivity $\rho\scto\rho$ follows from $\rho\to\rho$.
The transitivity can be proven explicitly: if $\mE(\rho_{S}\otimes\tau_{C}) = \rho'_{S'}\otimes\tau_{C}$ for some free operation $\mE\in\fro{SC}{S'C}$ and $\mF(\rho'_{S'}\otimes\tau'_{C'}) = \rho''_{S''}\otimes\tau'_{C'}$ for some free operation $\mF\in\fro{S'C'}{S''C'}$, the concatenation gives
\begin{align}
	[\mF\otimes\id_{C}]\circ[\mE\otimes\id_{C'}](\rho_{S}\otimes\tau_{C}\otimes\tau'_{C'}) = \rho''_{S''}\otimes\tau_{C}\otimes\tau'_{C'}.
\end{align}
Furthermore, if the preorder defined with $\to$ relation is convex, i.e. $\rho\to\rho'$ and $\rho\to\rho''$ implies $\rho\to \lambda\rho'+(1-\lambda)\rho''$ for any $\lambda\in[0,1]$, the preorder defined with $\scto$ also inherits the convexity.

Catalysis is also constrained by some of the monotones. 
$\rho\scto\rho'$ is equivalent to $\rho\otimes\tau\to\rho'\otimes\tau$ for some $\tau$, which in turn implies $\mbP(\rho\otimes\tau)\geq\mbP(\rho'\otimes\tau)$ for any resource monotone $\mbP$. 
If $\mbP$ is additive and if $\mbP(\tau)\neq\infty$, the monotone for the usual $\to$ relation also respects the preorder defined by $\scto$, i.e.
\begin{align}
	\rho\scto\rho' \quad\Rightarrow\quad \mbP(\rho)\geq\mbP(\rho').
\end{align}
In many theories, there are additive resource monotones. 
Consequently, strict catalysis cannot enhance the resource (as quantified by these monotones), but it can facilitate transitions from states possessing at least the same amount of resource to states with less resource by relaxing other constraints.

Indeed, the advantage of using strict catalysis has been reported in theories of entanglement~\cite{Jonathan99_PRL, Eisert2000Catalysis, Daftuar01_trumping, Klimesh07_ineqs}, coherence~\cite{Du2015CoherenceCatalysis, Bu16_coherence, Xing2020CoherenceCat}, thermodynamics~\cite{Brandao2015_2ndlaws, Woods19_engines}, and magic~\cite{Campbell2011MagicCat}. 
Yet, the origin of such advantages is still not understood clearly, despite having the full characterisation of $\scto$ in some cases~\cite{Klimesh07_ineqs, Brandao2015_2ndlaws}.
The main motivation for the results in Chapters~\ref{chapter: CETO} and~\ref{chapter: resource broadcasting} is this absence of understanding that hinders more systematic studies on catalysis. 

Nonetheless, one conceptual connection between catalysis and composite state transformation has been established.  
Suppose that a multi-copy transformation 
\begin{align}\label{eq: multi copy transf}
	\rho^{\otimes n}_{S_{1}\cdots S_{n}} \to \sigma^{\otimes n}_{S_{1}\cdots S_{n}}
\end{align}
with some finite $n$ is feasible.
Then there is a catalyst $\tau_{C}$, such that
\begin{align}
	\rho_{S}\otimes\tau_{C} \to \sigma_{S}\otimes\tau_{C},
\end{align}
i.e. $\rho_{S}\scto\sigma_{S}$, if one can conditionally apply the free operation, conditioned on the state of some ancillary system $A$. 
That is, if one can measure the system $A$ by some basis $\{\dm{i}\}_{i}$ and depending on the result, apply a free operation $\mE^{(i)}_{S}\in\fro{S}{S'}$, and then forget the measurement outcome $i$.
The necessity of the conditional free operation would become clear after we explain the protocol below.

This state $\tau$ is called the Duan state~\cite{Duan2005_DuanState}, and it is a mixture of $(n-1)$ copies of initial and final states
\begin{align}\label{eq: Duan state}
	\tau = \frac{1}{n}\sum_{i=1}^{n}\rho^{\otimes i-1}_{S_{2}\cdots S_{i}}\otimes\sigma^{\otimes n-i}_{S_{i+1}\cdots S_{n}}\otimes\dm{i}_{A}.
\end{align}
Here, system $A$ represents the classical flag, where $\dm{i}_{A}\to\dm{j}_{A}$ for any $i,j$ by mere relabelling.
The free operation is as follows: 
\begin{enumerate}
	\item measure $A$ system in a basis $\{\dm{i}_{A}\}_{i}$; when the outcome is $n$, apply a free operation that achieves the transformation in Eq.~\eqref{eq: multi copy transf} to $\rho_{S_{1}}\otimes\tau$, which yields a state
	\begin{align}
		\frac{1}{n}\sum_{i=1}^{n-1}\rho^{\otimes i}_{S_{1}\cdots S_{i}}\otimes\sigma^{\otimes n-i}_{S_{i+1}\cdots S_{n}}\otimes\dm{i}_{A} + \sigma^{\otimes n}_{S_{1}\cdots S_{n}}\otimes\dm{n}_{A},
	\end{align}
	\item apply permutation $S_{i}\to S_{i+1}$ and $S_{n}\to S_{1}$ to obtain
	\begin{align}
		\frac{1}{n}\sum_{i=1}^{n-1}\sigma_{S_{1}}\otimes\rho^{\otimes i}_{S_{2}\cdots S_{i+1}}\otimes\sigma^{\otimes n-i-1}_{S_{i+2}\cdots S_{n}}\otimes\dm{i}_{A} + \sigma^{\otimes n}_{S_{1}\cdots S_{n}}\otimes\dm{n}_{A},
	\end{align}
	\item apply relabelling $\dm{i}_{A}\to\dm{i+1}_{A}$ and $\dm{n}_{A}\to\dm{1}_{A}$, which yields $\sigma_{S_{1}}\otimes \tau_{C}$, as desired.
\end{enumerate}
Essentially, the catalyst allows us to `conceal' multiple copies of the initial and final states within it.
However, it is known that the converse is not always true~\cite{Feng2006multicopy}, at least for exact state transformation cases. 
This also reveals that the power of catalytic processes is no less powerful than the alternative composite approaches that rely on the collective processing of multiple copies.

In the beginning, we loosely defined catalysis to be the process where a catalyst returns to its original state. 
For strict catalysis, this return is interpreted strictly; that is, it must be uncorrelated to the system of interest as in the beginning of the process. 
Sometimes, this strict recovery may not be necessary. 
We define another type of catalysis. 
\begin{definition}[correlated catalysis]\label{definition: correlated catalysis}
	Correlated catalysis is a process defined by a free operation $\mE\in\fro{SC}{S'C}$, initial state $\rho_{S}\in\mD_{S}$, and a catalyst $\tau_{C}\in\mD_{C}$, such that
	\begin{align}\label{eq: correlated catalysis}
		\mE(\rho_{S}\otimes\tau_{C}) = \varrho_{S'C}, \quad \Tr_{S'}[\varrho_{S'C}] = \tau_{C},
	\end{align}
	which achieves the transformation $\rho_{S}\ccto\rho'_{S'} \coloneq \Tr_{C}[\varrho_{S'C}]$.
\end{definition}
Note that in Eq.~\eqref{eq: correlated catalysis}, only the reduced state of the catalyst needs to be recovered, but it must be recovered \emph{exactly} without any error. 
The binary relation $\ccto$ corresponding to the correlated catalytic transformation is defined analogously to $\scto$.
Despite the correlation established between $S$ and $C$, this catalyst can be reused for the same process when an identical but independent copy of the initial state is given.
Observe that
\begin{align}
	(\mE_{\bar{S}C}\otimes\id_{S'})(\rho_{\bar{S}}\otimes\varrho_{S'C}) = \varsigma_{S'\bar{S}'C},
\end{align}
where $\Tr_{S'\bar{S}'}[\varsigma_{S'\bar{S}'C}] = \tau_{C}$ and $\Tr_{\bar{S}'C}[\varsigma_{S'\bar{S}'C}] = \Tr_{S'C}[\varsigma_{S'\bar{S}'C}]$.
Yet, the correlation can be problematic if the system $S'$ or $\bar{S}'$ are used afterwards. 
As a result of this accumulated correlation, manipulating system $S'$ might affect both $\bar{S}'$ and $C$, potentially spoiling the desired final state in the other system or the catalyst. 
While this correlation can be made arbitrarily small, it typically requires a catalyst with diverging resources~\cite{Rubbolo2022CorrCat}. 

By definition, a strict catalysis is also a correlated catalysis, and $\rho\scto\rho'$ implies $\rho\ccto\rho'$.
Hence, correlated catalysis is usually adopted to consider the maximal advantage that can be obtained from catalysis. 
In Chapter~\ref{chapter: resource broadcasting}, we mostly work with correlated catalyses, because if the no-go theorems can be established there, they are automatically valid for weaker definitions such as strict catalysis. 

Correlated catalysis is typically far more powerful than its strict counterpart. 
First, it imposes strong constraints on all relevant monotones, because it is always possible to have a transformation~\cite{LipkaBartosik2023CatReview} 
\begin{align}
	\varrho_{S_{1}S_{2}}\ccto\Tr_{S_{2}}\left[\varrho_{S_{1}S_{2}}\right]\otimes\Tr_{S_{1}}\left[\varrho_{S_{1}S_{2}}\right]. 
\end{align}
To see this, let the catalyst state in system $\bar{S}_{2}$ have the same density matrix as $\Tr_{S_{1}}[\varrho_{S_{1}S_{2}}]$. 
Then, a simple swap operation between $S_{2}$ and $\bar{S}_{2}$ outputs a state $\varsigma_{S_{1}S_{2}\bar{S}_{2}} = \varrho_{S_{1}\bar{S}_{2}}\otimes\Tr_{S_{1}}[\varrho_{S_{1}S_{2}}]$, such that $\Tr_{\bar{S}_{2}}[\varsigma_{S_{1}S_{2}\bar{S}_{2}}] = \Tr_{S_{2}}[\varrho_{S_{1}S_{2}}]\otimes\Tr_{S_{1}}[\varrho_{S_{1}S_{2}}]$ and $\Tr_{S_{1}S_{2}}[\varsigma_{S_{1}S_{2}\bar{S}_{2}}] = \Tr_{S_{1}}[\varrho_{S_{1}\bar{S}_{2}}]$, satisfying Definition~\ref{definition: correlated catalysis}.
Therefore, it must always be true that
\begin{align}
	\mbP(\varrho_{S_{1}S_{2}}) \geq \mbP\left(\Tr_{S_{2}}\left[\varrho_{S_{1}S_{2}}\right]\otimes\Tr_{S_{1}}\left[\varrho_{S_{1}S_{2}}\right]\right),
\end{align}
for $\mbP$ to be a monotone for \emph{correlated catalytic} transformations. 

A similar connection to multi-copy can also be established for correlated catalysis. 
In fact, correlated catalysis allows a broader class of multi-copy transformation, namely the approximate multi-copy transformation:
\begin{align}
	\rho^{\otimes n}_{S_{1}\cdots S_{n}} \to \varsigma_{S_{1}\cdots S_{n}} \overset{\epsilon}{\sim} \sigma^{\otimes n}_{S_{1}\cdots S_{n}},
\end{align}
where the symbol $\overset{\epsilon}{\sim}$ indicates that two operators have a trace distance smaller than $\epsilon$.
Then, the transformation $\rho\ccto\sigma^{\epsilon}$ for some $\sigma^{\epsilon}\overset{\epsilon}{\sim}\sigma$ can be achieved when the permutation between identical systems and classical conditioning are free. 
The catalyst state needed is similar to the Duan state Eq.~\eqref{eq: Duan state}, with multiple copies of the final state $\sigma^{\otimes n-i}_{S_{i+1}\cdots S_{n}}$ replaced by $\Tr_{S_{1}\cdots S_{i}}[\varsigma_{S_{1}\cdots S_{n}}]$~\cite{Shiraishi2021GP}. 
A few remarks are in order.
Firstly, although this process implements an approximate multi-copy transformation, the catalyst reduced state is always recovered \emph{exactly}, without any approximation.
Secondly, approximate multi-copy transformations are qualitatively different from their exact counterparts; even when $\epsilon\to0$, the approximate transformation reaches a vastly larger set of states~\cite{Bennett1996concentrating, Winter2016CoherenceAsymptotic, Wilming2021AsymptoticApproximate, Shiraishi2024CoherenceCat}.
Furthermore, when $n$ can be arbitrarily large, the setting becomes that of the usual approximate asymptotic transformation.
The equivalence between correlated catalysis and asymptotic has been demonstrated in many resource theories~\cite{Shiraishi2024CoherenceCat, Kondra2024Asymmetry, Ganardi2024CorrCatEnt}, which leads to the characterisation of $\ccto$ with a single complete monotone~\cite{Muller18_corr, Lipka-Bartosik21_tele, Kondra2021Corr,Wilming2021AsymptoticApproximate, Shiraishi2021GP, Char2023CorrCatCoherence}, usually given by the relative entropy of resource in Example~\ref{example: resource monotone examples}.  

Despite all such positive results accumulated for catalytic processes, the framework has some serious issues to be used in practice. 
First, the preparation of catalysts is typically hard, e.g. when the Duan state construction Eq.~\eqref{eq: Duan state} is used, which requires a catalyst system whose size is comparable to multiple copies of the original system and whose state already includes multiple copies of the desired final state. 
Another problem is the fragility of catalysis in the presence of noise due to its fine-tuned nature.
We discuss and (partially) resolve these problems in Chapters~\ref{chapter: CETO} and~\ref{chapter: resource broadcasting}.

\chapter{Overcoming Markovianity in thermal processes using catalysts}\label{chapter: CETO}

In this chapter, we investigate catalytic processes in resource theories modelling quantum thermodynamics. 
In particular, we focus on the theories with certain degree of Markovian restrictions. 
As described in Definition~\ref{definition: Markovian quantum channel}, Markovianity is closely related to the decomposability of the operations. 
Leveraging the decomposability, we identify the memory effect that catalysis provides. 
Furthermore, we prove that catalysis completely negates the additional restrictions imposed by Markovianity in thermodynamic resource theories, showcasing the power of catalysts as a non-Markovian auxiliary.

My original results in this section include Theorems~\ref{theorem: nice order_main}, \ref{theorem: GCETO is TO}, \ref{theorem: CETO is CTO} and Figures~\ref{fig: ETO cone}, \ref{fig: CETO tracking}, \ref{fig: local free Es}, \ref{fig: CETO tracking alphas}, \ref{fig: higher dim catalysis thermal to thermal}.

\section{Background: a hierarchy of thermodynamic resource theories}

There are several classes of thermal processes studied in thermodynamic resource theories. 
These classes form a hierarchy: smaller classes are strictly contained within larger ones. 
Because choosing a different class in this hierarchy yields distinct descriptions of thermodynamic processes, it has been an active open problem whether there exists a setting in which all classes become equivalent. 
In the following subsections, the catalytic setting is considered as a potential candidate. 
This subsection gives a brief introduction to these classes of thermal processes, summarising results from the literature; it contains no original results.

\subsection{Free operations}

Thermodynamic resource theories, also known as resource theories of athermality, are motivated by the typical thermodynamic settings, where the system of interest is embedded in a thermal environment with a single fixed temperature. 
In these settings, a natural free operation is the full thermalisation channel mapping the system to a thermal equilibrium state. 
For quantum systems with Hamiltonian $H_{S}$, the thermal equilibrium state at the inverse temperature $\beta$ is the Gibbs state
\begin{align}\label{eq: Gibbs state def}
	\gbs_{S} \coloneq \frac{1}{Z_{S}}e^{-\beta H_{S}},
\end{align}
where the partition function $Z_{S}\coloneq \Tr[e^{-\beta H_{S}}]$.

Gibbs states are special for several reasons. 
Firstly, when the entropy is defined as the von Neumann entropy $S(\rho) \coloneq -\Tr[\rho\log\rho]$, the free energy
\begin{align}\label{eq: nonequilibrium free E def}
	F(\rho_{S}) \coloneq \Tr[\rho_{S}H_{S}] - \beta^{-1}S(\rho_{S}),
\end{align}
is minimised when $\rho_{S} = \gbs_{S}$.
Secondly, Gibbs states can be uniquely identified by a few physical axioms. 
Passive states are defined as the states from which no energy can be extracted unitarily; if $n$ copies of the state $\rho^{\otimes n}$ is also passive for all $n$, the state $\rho$ is completely passive.
Gibbs states of inverse temperature $0\leq\beta\leq\infty$ are the only states that are completely passive~\cite{Pusz1978CompletelyPassive}.
Alternatively, Gibbs states are the only passive states that are structurally stable and consistent, i.e. small perturbations in Hamiltonians only give small perturbations in Gibbs states, and Gibbs states of a composite system is the tensor product of Gibbs states of each subsystem~\cite{Lenard1978GibbsState}. 

Therefore, Gibbs state, with respect to the environment inverse temperature (or ambient inverse temperature) $\beta$, is the most natural choice for a free state.
Indeed, if we assume that the full thermalisation is a free operation, Gibbs state is always the free state by Definition~\ref{definition: free states}.
Resource theories of athermality are then a family of theories sharing the same set of free states $\frs{S} = \{\gbs_{S}\}$, but with different free operations.
We introduce five notable sets of thermodynamic free operations that form a hierarchy. 
For simplicity, we only discuss operations from $S$ to itself, with a fixed ambient inverse temperature $\beta$ and a fixed Hamiltonian $H_{S}$.

\begin{definition}[Gibbs-preserving operations (GP)~\cite{Faist2015GP}]\label{definition: GP}
	Given a system $S$ and its Hamiltonian $H_{S}$, the set of Gibbs-preserving operations from $S$ to itself is defined as
	\begin{align}
		\GP{S}{S} \coloneq \left\{\mE\in\CPTP{S}{S} \,|\, \mE(\gbs_{S}) = \gbs_{S}\right\}.
	\end{align}
\end{definition}
This set is by definition an RNG and in fact also a CRNG; see Definitions~\ref{definition: RNG} and~\ref{definition: CRNG}.
The measure of resourcefulness most familiar to thermodynamicsts is the non-equilibrium free energy defined as in Eq.~\eqref{eq: nonequilibrium free E def}~\cite{Kawai2007NonequilF, Esposito2010NonequilF}.
Interestingly, it is easy to show that~\cite{Landi2021EntProd} non-equilibrium free energy corresponds directly to the relative entropy of resource (Example~\ref{example: resource monotone examples}), a resource measure popular in the resource theory community. 
The two are related by the factor $\beta^{-1}$ as
\begin{align}\label{eq: resource monotone free E}
	R(\rho_{S}) = \beta^{-1}F(\rho_{S}) - \beta^{-1}F(\gbs_{S}) \eqcolon \beta^{-1}\Delta F(\rho_{S}).
\end{align}
Therefore, $\Delta F$ is monotonic under GP, and this monotonicity extends to all subsets of GP that will be introduced below. 

GP, however, can be too general to capture all thermodynamic restrictions. 
In fact, any (C)RNG free operations with respect to singleton free state sets can be regarded as GP with some temperature and Hamiltonian, as $\beta H_{S} = \log(\gbs_{S})$.

To capture another aspect of thermodynamics, let us consider coherence between different energy levels. 
The resource theory of asymmetry with respect to U(1) group generated by the Hamiltonian precisely tackles that problem.
There, the set of free operations are covariant operations~\cite{Keyl1999Cov} 
\begin{align}\label{eq: covariant operation def}
	\COV{S}{S} \coloneq \left\{\mE\in\CPTP{S}{S} \,|\, \mE(e^{-i\theta H_{S}} (\cdot) e^{i\theta H_{S}} ) = e^{-i\theta H_{S}} \mE(\cdot) e^{i\theta H_{S}},\quad \forall\theta\in\mbR\right\}.
\end{align}
Covariant operations also correspond to the CRNG of symmetric states
\begin{align}\label{eq: symm state def}
	\SYM{S} \coloneq \left\{\rho_{S}\in\mD_{S} \,|\, U(\theta)\rho_{S} U(\theta)^{\dagger} = \rho_{S},\  \forall\theta\in\mbR\right\}.
\end{align}
For the U(1) group with the representation $U(\theta) = e^{-i\theta H_{S}}$, Eq.~\eqref{eq: symm state def} is equivalent to $\SYM{S} \coloneq \{\rho_{S}\in\mD_{S} \,|\, [\rho_{S},H_{S}] = 0\}$.

Covariant operations also have an operational meaning: $\COV{S}{S}$ is exactly the set of operations that can be written with a dilation 
\begin{align}\label{eq: free dilation asymmetry}
	\mE(\cdot) = \Tr_{R}[U (\cdot \otimes \gamma_{R}) U^{\dagger}],
\end{align}
with some system $R$ with a Hamiltonian $H_{R}$, free ancilla state $\gamma_{R}\in\frs{R}$, and \emph{energy-preserving unitary} $U$, satisfying 
\begin{align}\label{eq: energy preserving unitary def}
	[U, H_{S}\otimes\1_{R} + \1_{S}\otimes H_{R}] = 0.
\end{align}
Eq.~\eqref{eq: energy preserving unitary def} has a physical interpretation that $U$ strictly preserves the energy of any energy eigenstate of $SR$.
This also implies that the energy statistics of the $SR$ state before and after the unitary are identical. 
Note that the energy of each subsystem can change after an energy-preserving unitary. 

To incorporate this global energy-preserving behaviour, we impose covariance (sometimes called time-translation symmetry), Eq.~\eqref{eq: covariant operation def}, in addition to the existing Gibbs-preserving condition.

\begin{definition}[Gibbs-preserving covariant operations (GPC)~\cite{Cwiklinski2015GPC, Lostaglio15_PRX_coh}]\label{definition: GPC}
	Gibbs-preserving covariant operations from $S$ to itself is defined as
	\begin{align}
		\GPC{S}{S} \coloneq \GP{S}{S} \cap \COV{S}{S}.
	\end{align}
\end{definition}
This set is also known as \emph{enhanced thermal operations} or \emph{thermal processes}. 
Note that $\GPC{S}{S}$ is a strict subset of $\GP{S}{S}$ in general: a simple example of a GP that creates asymmetric state from a symmetric state is found in Ref.~\cite{Faist2015GP}.
In fact, there exist GP channels that can be implemented with GPC channels only with the injection of infinite coherence auxiliaries~\cite{Tajima2025GPInfinite}. 

Both GP and GPC are defined axiomatically; because the axioms only concern mathematical structure of the channels, the implementation of them are not considered. 
The remaining three sets of operations are defined bottom-up: these are the channels that can be constructed from implementation recipes.
We first introduce the one defined as the set of operations with free dilations, similar to Eq.~\eqref{eq: free dilation asymmetry}.
However, the set of free states for the theory of athermality is a singleton including only the Gibbs state; hence, the free ancilla is also fixed to be the Gibbs state $\gbs_{R}$. 
\begin{definition}[thermal operations (TO)~\cite{Janzing00_TO}]\label{definition: TO}
	Consider a set of channels $\mE$ satisfying
	\begin{align}\label{eq: TO dilation}
		\mE(\cdot) = \Tr_{R}\left[U\left(\cdot\otimes\gbs_{R}\right)U^{\dagger}\right],
	\end{align}
	for some system $R$ with some Hamiltonian $H_{R}$ and some energy-preserving unitary $U$ with the condition Eq.~\eqref{eq: energy preserving unitary def}.
	Thermal operations from $S$ to itself $\TO{S}{S}$ is defined as the closure of this set. 
\end{definition}
In some literature, the set without the closure is denoted TO and the closure of it is separately named as CTO, but in this thesis, we always define TO following Definition~\ref{definition: TO}.

TO is a subset of GPC, as $\gbs_{R}\in\SYM{R}$ and thus channel in the form Eq.~\eqref{eq: TO dilation} is a covariant channel.
The converse is true when $S$ is a qubit system~\cite{Cwiklinski2015GPC}, but not in general: for dimension larger than 2, there exists a GPC that is not a TO; in fact there even exists a state transformation that is feasible in GPC but not in TO~\cite{Ding21_EnTO}.

Although TO is operationally motivated, implementing arbitrary energy-preserving unitaries between system and (potentially infinitely large) environment is not realistic~\cite{YungerHalpern17_Realization}. 
Another hindrance arises from the structure of the theoretical development, which tends to focus on the possibility of state transformations, instead of providing a construction of concrete, simple heat baths together with corresponding interaction Hamiltonians underlying the process.%
\footnote{See Refs.~\cite{Hu2019_singlemode, Shiraishi2020_construction, Ende2022_bath} for rare examples where the concrete construction of TO has been investigated.} 
As a result, even when a transformation is known to be possible, it remains non-trivial to construct the protocol that implements it, obscuring the dynamical description of the process. 
Hence, restrictions of TO inspired by more experimentally accessible setups have been defined. 

\begin{definition}[elementary thermal operations (ETO)~\cite{Lostaglio_18_ETO}]\label{definition: ETO}
	Let $\eTO{S}{S}$ be a subset of $\TO{S}{S}$ that can be written with the dilation Eq.~\eqref{eq: TO dilation}, where the energy-preserving unitary $U$ acts non-trivially on at most two energy eigenstates of $S$, i.e. there exists an energy-eigenbasis $\{\ket{i}_{S}\}_{S}$, such that
	\begin{align}
		\bra{i}_{S}U\ket{i}_{S} = \1_{R},
	\end{align}
	for all $i$ \emph{except at most two}. 
	Then, $\ETO{S}{S}$ is defined as the closed convex hull of all concatenations of channels in $\eTO{S}{S}$.
\end{definition}

ETO by definition has a decomposition that offers a natural way to prescribe a process to the experimenter.
Furthermore, any extreme point of $\eTO{S}{S}$ can be realised with a single bosonic mode bath and the intensity-dependent Jaynes-Cummings interaction~\cite{Buzek89_intensityJC}. 
With the addition of a pure dephasing operation, any $\eTO{S}{S}$ can be implemented~\cite{Hu2019_singlemode}.
A good portion of these operations can also be achieved via more experiment-friendly models~\cite{Lostaglio_18_ETO, Hu2019_singlemode}, such as the collision model~\cite{Rau63_collision} or the Jaynes-Cummings model~\cite{JCmodel}.
Compared to the baths proposed in Refs.~\cite{Horodecki13_fundamental, Shiraishi2020_construction}, using a single-mode bath for each step is much more feasible. 

Since $\TO{S}{S}$ is a set closed under concatenation and convex combination, ETO is a subset of TO; in fact it is a strict subset, i.e. $\ETO{S}{S}\subsetneq\TO{S}{S}$ except when $S$ is a qubit system or $\beta = 0$~\cite{Lostaglio_18_ETO}.
This is rather surprising, given the universality achieved in $n$-qubit unitaries into two-qubit ones~\cite{DiVincenzo1995Universal}, $n$-level unitaries into two-level ones~\cite{Reck94TwoLvl}, or $n$-mode Gaussian unitaries into two-mode ones~\cite{Reck94TwoLvl, Braunstein2005}.%
\footnote{We note that another type of unexpected non-universality has been observed: when $k$-local and symmetric operations are composed together, some of the global symmetric operations cannot be achieved~\cite{Marvian2022_locality, Marvian2024Rotationally, Marvian2024Abelian, Lastres2024NonUniverality}.}
Moreover, a seemingly much easier task of decomposing TO into series of TO$_{d-1}$---defined as a subset of TO involving at most all but one system levels at a time---is also shown to be impossible~\cite{Mazurek2018_Decomp}. 

Another realistic restriction is requiring Markovianity of the bath.
Similarly to ETO, Markovianity introduces the decomposition, providing a pathway to implement an operation with smaller and easier steps. 
In particular, when the full Markovianity is imposed, these steps can be infinitesimally small and continuous.
\begin{definition}[Markovian thermal operations (MTO)~\cite{Spaventa22_MTO}]\label{definition: MTO}
	A channel $\mE\in\TO{S}{S}$ is an MTO if it is Markovian in the sense of Definition~\ref{definition: Markovian quantum channel}, i.e. if it can be written as a concatenation of TO channels that are $\epsilon$-close to the identity channel with arbitrarily small $\epsilon>0$.
\end{definition}
This is equivalent to the condition that there exists a Markovian master equation, such that the evolution from $t = 0$ to $t = T$ yields the channel $\mE$, while for any time slices $0\leq t_{1}\leq t_{2}\leq T$, the evolution from $t = t_{1}$ to $t = t_{2}$ is TO. 
Note that Refs.~\cite{Lostaglio2022MTO1, Korzekwa2022MTO2} first considered such Markovian restrictions, but they considered a Markovian version of GPC not TO.

By definition, MTO is a subset of TO, and it is indeed a strict subset even for qubit systems~\cite{Lostaglio2022MTO1}. 
The simplest example is the cooling of a qubit excited state $\dm{1}_{S}$. 
With TO, it is known that the transformation $\dm{1}_{S}\to\dm{0}_{S}$ is possible even when the ambient inverse temperature $\beta<\infty$, i.e. a complete cooling is attainable.
When Markovianity is imposed, $\dm{1}_{S}$ must be evolved into $\dm{0}_{S}$ continuously. 
Since TO is covariant, the state always remain incoherent, i.e. it is described by a one-parameter evolution of $\rho(t) = p(t)\dm{1}_{S} + (1-p(t))\dm{0}_{S}$ with $p(0) = 1$ and $p(T) = 0$.
However, as soon as $\rho(t)$ reaches the Gibbs state, it cannot evolve to any other state via TO. 
Hence, if $\gbs_{S}\neq\dm{0}_{S}$, i.e. if $\beta<\infty$, MTO cannot completely cool down an excited state. 

We summarise the hierarchy among these free operations; see also Figure~\ref{fig: hierarchy}. 
\begin{remark}[Hierarchies of thermodynamic resource theories]\label{remark: hierarchies}
	Suppose that $S$ has dimension higher than two. 
	Then,
	\begin{align}
		\MTO{S}{S}, \ETO{S}{S} \subsetneq \TO{S}{S} \subsetneq \GPC{S}{S} \subsetneq \GP{S}{S}.
	\end{align}
\end{remark}

\begin{figure}[t!]
	\includegraphics[width=0.8\columnwidth]{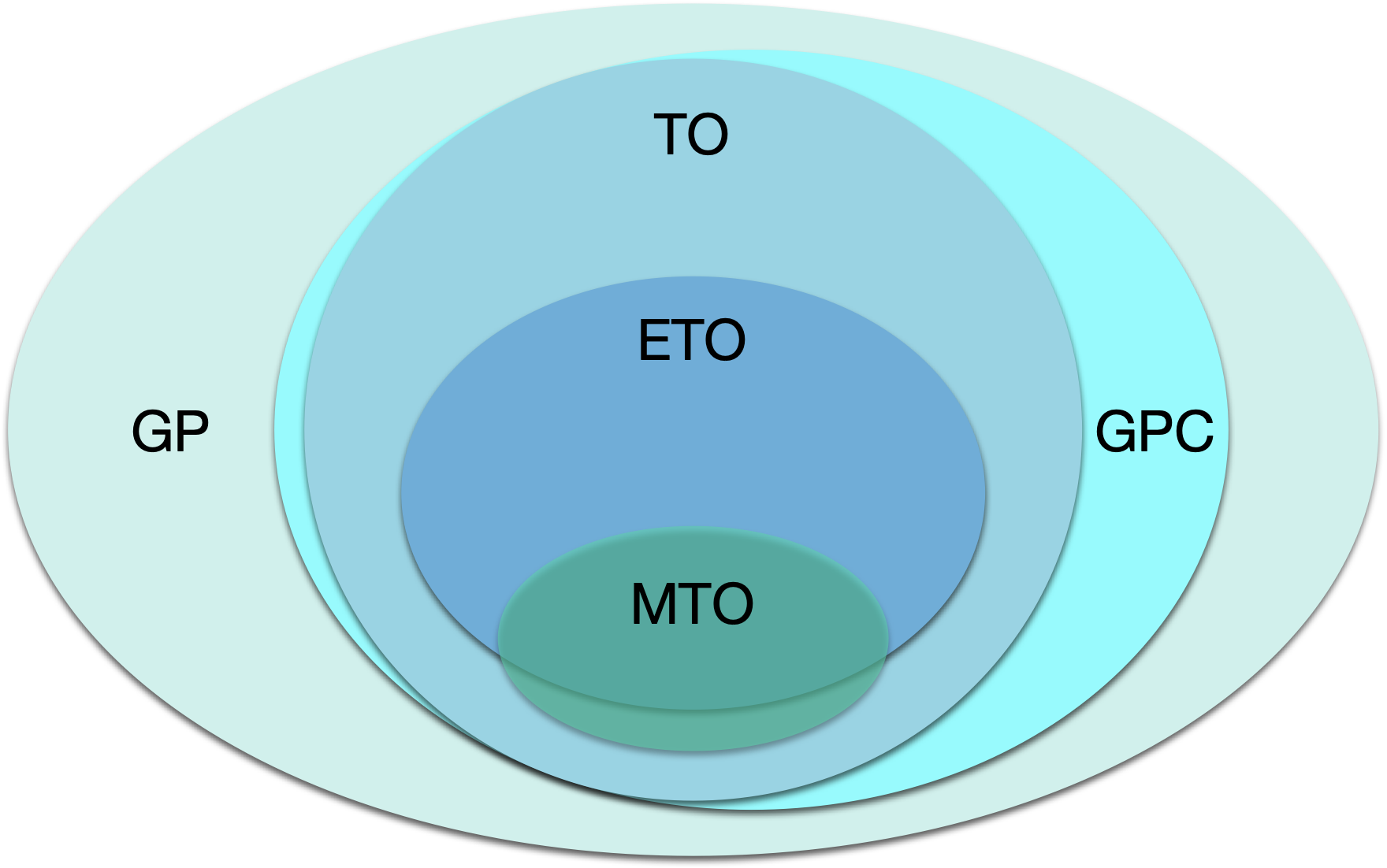}
	\caption{Illustration showing the hierarchy between thermodynamic resource theories in Remark~\ref{remark: hierarchies}.
	Note that the existence of ETO channels that are not MTO is known, but it is not known whether all MTO are ETO.} 
	\label{fig: hierarchy}
\end{figure}

\subsection{State transformations}

A hierarchy can also be formed at the level of state transformations. 
The inclusion relations in Remark~\ref{remark: hierarchies} would again hold for state transformations, yet the strict inclusions may not necessarily hold.
For example, when the initial states for the state transformations are restricted, the gap between different free operations may not manifest in the set of reachable final states. 

Hence, we restrict the initial states to be symmetric states in $\SYM{S}$. 
This choice brings a huge simplification.
With the exception of the largest set GP, all free operations are covariant; that is, if the initial state is in $\SYM{S}$, the final state must also be in $\SYM{S}$.
Thus both initial and final states are block-diagonal in the energy-eigenbasis, with each block corresponding to degenerate-energy subspaces. 
It is therefore more convenient to represent states not as $d \times d$-dimensional density matrices, but as $d$-dimensional vectors whose elements are the energy-eigenvalues of those density matrices. 

\begin{definition}[population vectors]\label{definition: propulation vectors}
	Suppose that $\rho_{S}\in\SYM{S}$, i.e. $[\rho_{S},H_{S}] = 0$. 
	Then, there exists an energy-eigenbasis $\{\ket{i}_{S}\}_{i}$, such that $H\ket{i}_{S} = E_{i}\ket{i}_{S}$ and $\rho_{S}\ket{i}_{S} = p_{i}\ket{i}_{S}$ for all $i$.
	The ordering is given by $E_{i}\leq E_{j}$ whenever $i\leq j$.
	We define a population vector corresponding to $\rho_{S}$ as $\pstate$, where its $i_\text{th}$ element is defined as $(\pstate)_{i} = p_{i}$.
	We denote a set of all $d$-dimensional probability vectors as $\probspace_{d}$.
\end{definition}

There might exist multiple density matrices corresponding to one population vector $\pstate$ when the Hamiltonian has degeneracy.
However, these density matrices are connected by energy-preserving unitary operations, because they all have the same energy eigenvalues. 
Since energy-preserving unitary is free in any set of free operations we consider, all density matrices with the same population vector are fully equivalent in the preorder given by thermodynamic resource theories. 
This means that we can define the preorder $\pstate\to\qstate$ using the one defined with the corresponding density matrices $\rho_{S}\to\rho'_{S}$ and vice versa.
We sometimes denote different binary relations stemming from the set of free operations $X$ as $\xrightarrow{X}$.

Since state transformations between symmetric states are of interest, we define the set of population vectors that can be reached from a given initial population vectors. 
\begin{definition}[set of reachable states]\label{definiton: set of reachable states}
	The set of reachable states from $\pstate$ via some class of operation $X$ is defined as
	\begin{align}
		\mT_{X}(\pstate) \coloneq \left\{\qstate \,|\, \pstate \xrightarrow{X} \qstate \right\},\quad X = \mathrm{GPC}, \mathrm{TO}, \mathrm{ETO}, \mathrm{MTO}.
	\end{align}
	We denote the extreme points of this set as $\extr[\mT_{X}(\pstate)]$.
\end{definition}

Analogously to the mapping of density matrices into population vectors, any GPC channel can be represented as a $d \times d$-dimensional matrix that preserves the Gibbs population vector and the sum of probability vector elements. 
\begin{definition}[Gibbs-stochastic matrices]\label{definition: Gibbs stochastic}
	A $d \times d$-dimensional matrix $E$ is Gibbs-stochastic, if
	\begin{align}
		\estate^{\transp} E = \estate^{\transp},\quad E\gbs = \gbs,
	\end{align}
	for the uniformly distributed probability vector $\estate = (\frac{1}{d},\cdots,\frac{1}{d})$ and the population vector $\gbs$ corresponding to the Gibbs state $\gbs_{S}$.
\end{definition}
The set of Gibbs-stochastic matrices is well-characterised with explicit recipes to find all extreme points, using the techniques developed for transportation matrices~\cite{Mazurek19_channels}.

Now we establish the correspondence between GPC and TO channels with Gibbs-stochastic matrices. 
\begin{proposition}[Ref.~\cite{Janzing00_TO}, Theorems~5 and~6]\label{proposition: GPC TO equiv GStochastic}
	Suppose that $\mE\in\GPC{S}{S}$ for $d$-dimensional system $S$.
	Then there exists a $d \times d$-dimensional Gibbs-stochastic matrix $E$, such that 
	\begin{align}\label{eq: TO GPC and GStochastic}
		\mE(\rho_{S}) = \rho'_{S} \quad \Leftrightarrow \quad E\pstate = \qstate,
	\end{align}
	for any $\rho_{S},\rho'_{S}\in\SYM{S}$, where $\pstate$ and $\qstate$ are population vectors corresponding to $\rho_{S}$ and $\rho'_{S}$, respectively. 
	
	Furthermore, for any Gibbs-stochastic matrix $E$, there exists a TO channel $\mE\in\TO{S}{S}$, satisfying Eq.~\eqref{eq: TO GPC and GStochastic}.
\end{proposition}
This proposition also implies that $\setTO(\pstate) = \mT_{\mathrm{GPC}}(\pstate)$ for all population vectors $\pstate\in\probspace_{d}$.

For ETO and MTO, a special class of Gibbs-stochastic matrices assumes such a role. 
To construct them, we first consider how $\eTO{S}{S}$ operations effectively act on population vectors. 
Suppose that $\rho_{S}$ and $\rho'_{S}$ have corresponding population vectors $\pstate$ and $\qstate$.
If there exists $\mE\in\eTO{S}{S}$ acting on energy levels $j\leq k$, such that $\mE(\rho_{S}) = \rho'_{S}$, then there exists a Gibbs-stochastic matrix $M^{(j,k)}_{\lambda}$ defined as  
\begin{align} \label{eq: Mswap def}
	M^{(j,k)}_{\lambda} \coloneq \begin{pmatrix}
		1-\lambda \Delta_{jk} && \lambda \\
		\lambda \Delta_{jk} && 1-\lambda
	\end{pmatrix} \oplus \mathbb{1}_{\setminus (j,k)},\quad \Delta_{jk} \coloneq \frac{\gbs_{k}}{\gbs_{j}},
\end{align}
such that $ \qstate = M^{(j,k)}_\lambda\pstate$.
When the order between $j$ and $k$ is not certain, we may write $M^{(j,k)}_{\lambda}$ even if $k< j$; in such cases, $M^{(j,k)}_{\lambda}$ must be understood as $M^{(k,j)}_{\lambda}$.
For $M^{(j,k)}_{\lambda}$ to be Gibbs-stochastic the parameter $\lambda\in[0,1]$. 

When $\lambda = 0$, the matrix $M^{(j,k)}_{0} = \1$. 
The other extremal case of $\lambda=1$ is named a $\beta$-swap (sometimes we refer to it simply as a swap),
\begin{align}\label{eq: beta swap def}
	\beta^{(j,k)} \coloneq M^{(j,k)}_{1} = \begin{pmatrix}
		1- \Delta_{jk} & 1 \\
		\Delta_{jk} & 0
	\end{pmatrix} \oplus \mathbb{1}_{\setminus (j,k)}.
\end{align} 
Another notable special case is the full two-level thermalisation swap (or T-swap for short) corresponding to $\lambda = \frac{1}{1+\Delta}$, denoted as
\begin{align}\label{eq: T swap def}
	T^{(j,k)} \coloneq M^{(j,k)}_{\frac{1}{1+\Delta}}.
\end{align}
As the name suggests, 
\begin{align}
	\frac{(T^{(j,k)}\pstate)_{k}}{(T^{(j,k)}\pstate)_{j}} = \Delta_{jk},
\end{align}
for any initial population vector $\pstate$ i.e. two levels $j$ and $k$ are fully thermalised relative to each other. 

Finally, we define an ordering of energy eigenbasis labels for each population vector. 
Intuitively, this ordering shows which levels are more populated, compared to the Gibbs state. 
It is thus a special case of $d$-majorisation (also known as relative majorisation) where the weight vector is the Gibbs state population. 

\begin{definition}[$\beta$-ordering]\label{definition: beta-ordering}
	Given $\pstate\in\probspace_{d}$ and a Gibbs population vector $\gbs\in\probspace_{d}$, we define the element-wise ratio of the two vectors as 
	\begin{align}\label{eq: slopes}
		\slope(\pstate)_i \coloneq \frac{p_{i}}{\gbs_{i}}.
	\end{align}
	We define the $\beta$-order $\pi_{\pstate}$ as  a permutation of $(1,\cdots,d) $, such that the ratios according to this ordering is non-increasing, i.e. 
	\begin{align}\label{key}
		\slope(\pstate)_{\pi_{\pstate}(k)} \geq  \slope(\pstate)_{\pi_{\pstate}(k+1)},\quad \forall k\leq d-1,
	\end{align}
	where we omit $\gbs$ in the argument when it is obvious from context.
	We also denote $\pi_{\pstate}$ as a vector $(\pi_{\pstate}(1),\cdots,\pi_{\pstate}(d))$.
\end{definition}
Note that the $\beta$-ordering of a population vector $\pstate$ is not unique when there are two identical element-wise ratios.
We accept all possible $\beta$-orderings of a vector as valid $\beta$-orders.

Previous results on state transformation characterisations and our new improvements for the ETO case is in Appendix~\ref{chapter: ETO cone characterisation}.

\section{Unravelling catalytic operations with small catalysts and step-wise operations}\label{section: unravelling CETO}

One great advantage of ETO is that it opens up the opportunity to analyse intermediate states, which are found by partially applying the swap sequence---providing a time-resolved description of the system dynamics, rarely possible in other resource theories.

\begin{figure}[t!]
	\includegraphics[width=0.92\columnwidth]{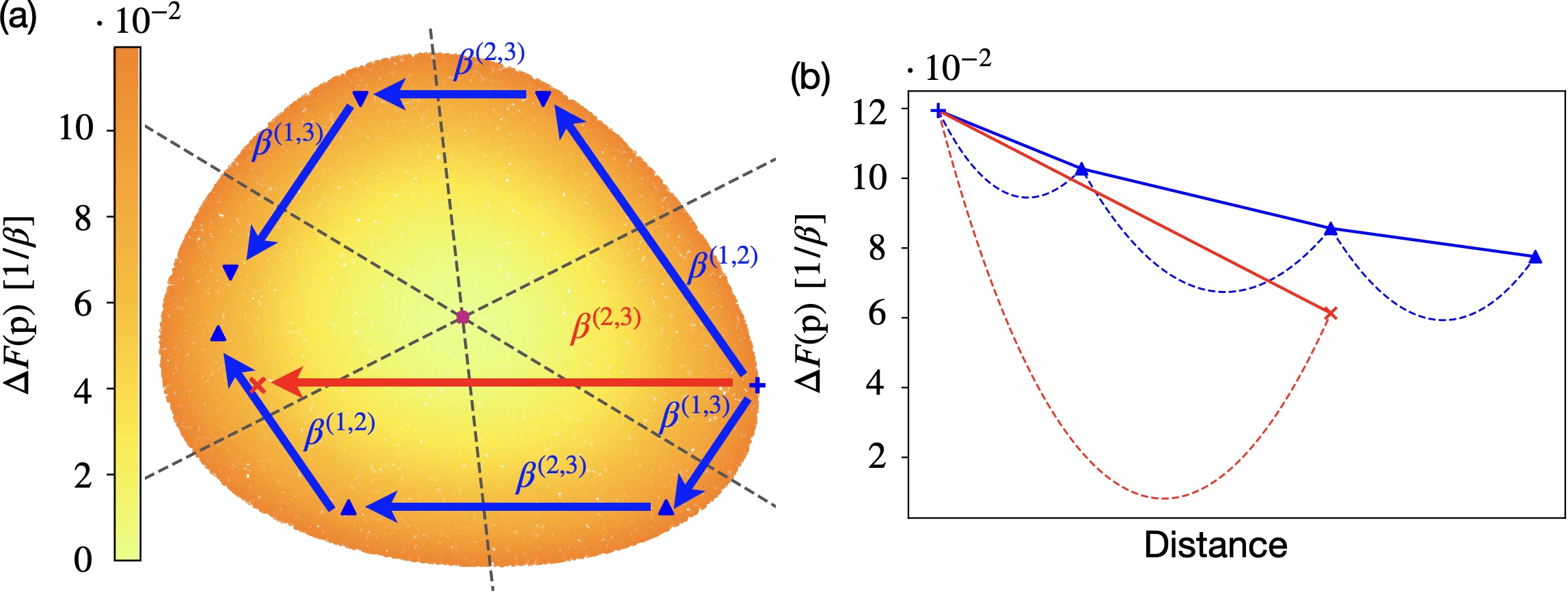}
	\caption{Initial $ d=3 $ state $\pstate_{1} = (0.35,0.55,0.1)$ (blue plus marker) undergoing ETO, where its Hamiltonian is fixed as $\beta\mathrm{H}_S = (0,0.2,0.5)$, and all energies are scaled in a unit of $1/\beta$.
		Panel (a): ETO transformations in barycentric representation, mapping the probability simplex $\probspace_{3}$ into an equilateral triangle and its interior following $(p_{1},p_{2},p_{3}) \mapsto \left(\frac{\sqrt{3}}{2}(p_{2}-p_{3}), p_{1} - \frac{1}{2}(p_{2}+p_{3})\right)$. 
		This transforms the initial state $\pstate_{1}$ to have the barycentric coordinate close to $(0.390, 0.025)$, while the two leftmost points have coordinates $(-0.184, 0.097)$ (upward blue triangle) and $(-0.171 ,0.180)$ (downward blue triangle). 
		Only the relevant part of the entire equilateral triangle (representing vectors $ \qstate $ with $F (\qstate) \leq F (\pstate_{1})$) is displayed in colour.
		Triangles label the extreme points of $\setETO(\pstate_{1})$, achieved by $ \beta $-swaps indicated by blue arrows. The red X marker is the state $\beta^{(2,3)}\pstate_{1}$. Different $\beta$-order cells are separated by black dashed lines connecting pure states and the thermal state. 
		Panel (b): the free energy difference (Eq.~\eqref{eq: resource monotone free E}), in two different paths corresponding to (a). Dashed lines show the continuous values of $ \Delta F $ from points on arrow paths in (a). Straight lines connect the discrete values obtained from endpoints denoted with triangle and X symbols. $ x $-axis is the total length of the path taken from the initial state as plotted in (a), e.g. $ x $-coordinate of the second triangle is the summation of the first and the second blue arrow lengths starting from the plus symbol.
		Figure adapted from Figure~4 of Ref.~\cite{Son2024_CETO}.} 
	\label{fig: ETO cone}
\end{figure}

For example, see Figure~\ref{fig: ETO cone}, which displays the real time evolution of the state during ETO transitions. 
By construction, only two levels of the state undergo change in time, i.e. all the other populations are fixed during that period, imposing the system to follow the straight lines in the barycentric representation as in (a). 
In (b), dashed lines correspond to the free energies during the evolution.
Suppose we apply $ \beta^{(j,k)} $ to a state $ \pstate $.
The state evolves as $ \qstate(\lambda(t)) = M^{(j,k)}_{\lambda(t)}\pstate $, where $ \qstate(\lambda = 1) = \beta^{(j,k)}\pstate $ with some monotonic function $\lambda(t)$ reaching $1$ at some $t = T$. 
Since the evolution is continuous, the system passes through a state $ \qstate(\lambda^{*}) = T^{(j,k)}\pstate$ during the evolution. 
This state is the closest to being thermal, and achieves minimal generalised free energies---minima of dashed lines in (b)---among states $ \qstate(\lambda(t))$.
In other words, during a single $\beta$-swap, free energy decreases until the minimum point $\qstate(\lambda^*)$ is reached, and then increases until the end of the $ \beta $-swap. 
On the other hand, if only the endpoints values are considered (solid lines of (b)), free energy cannot increase after each ETO step. 
These intermediate increases within a single swap evince the non-Markovian effect of thermal reservoirs at each step, differentiating ETO from MTO. 

\subsection{Qubit catalysts}

Now we move on to the catalytic case. 
When the system is symmetric and two-dimensional, thermal operations do not benefit from strict catalysis~\cite{Czartowski2024Catalytic}.
Hence, we consider the simplest non-trivial case: qutrit system and qubit catalyst. 
The composite state $\pstate\otimes\catstate$ lives in a six-dimensional probability space $ \probspace_{6} $. 
Our goal is to construct a set attainable by catalytic elementary thermal operations with a qubit catalyst (CETO$_2$)
\begin{align}
	\setCETOqb(\pstate) \coloneq \{\qstate \,|\, \exists \catstate\in\probspace_2\quad \text{s.t.}\quad \pstate\otimes\catstate\toeto\qstate\otimes\catstate \}.
\end{align}
It is easy to show that $\setETO(\pstate)\subset \setCETOqb(\pstate) $, but qubit catalytic advantage exists if and only if $\setCETOqb(\pstate)\not\subset\setETO(\pstate)$. 
For a number of limited cases, given a fixed catalyst state $\catstate = (c_{1},1-c_{1})$, some parts of the set
\begin{align}
	\setCETO(\pstate;\catstate) \coloneq \{ \qstate \,|\, \pstate\otimes\catstate\toeto~\qstate\otimes\catstate \}
\end{align}
can be evaluated analytically.
Nonetheless, $ \setCETO(\pstate;\catstate) $ is in general constructed by numerically finding the extreme points of $\setETO(\pstate\otimes\catstate)$ and imposing exact catalyst recovery conditions. 
The set $\setCETOqb(\pstate)$ is then given by iterating the process for different values of $c_{1}$.
In this section, we focus on the strict catalysis, where $S$ remains uncorrelated to $C$ to affirm that catalytic advantages exist even in the most conservative setting. 

\begin{figure}[t!]
	\centering
	\includegraphics[width=0.9\columnwidth]{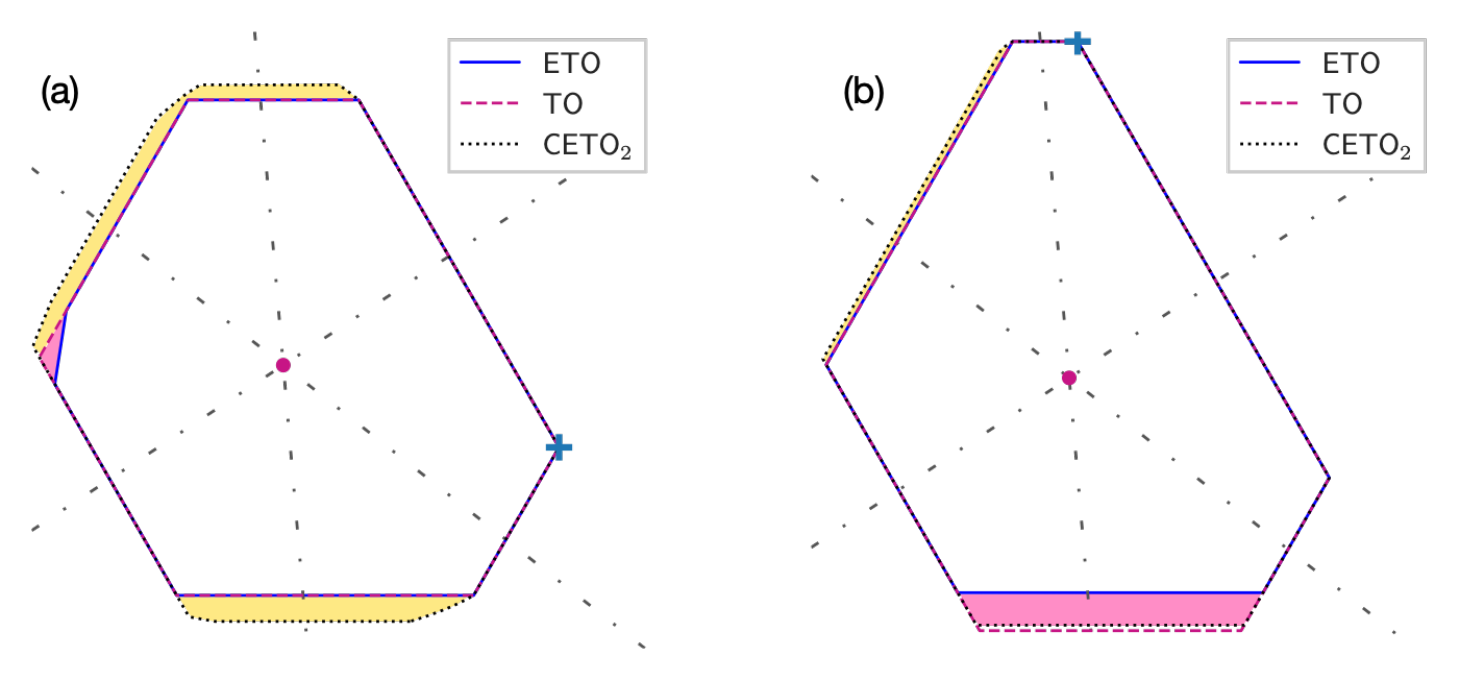}
	\caption{Barycentric representation of $\setETO$ (blue solid lines), $\setTO$ (violet dashed lines), and $\setCETOqb$ (black dotted lines), for the initial states (a) $ \pstate_{1} $ in Figure~\ref{fig: ETO cone} and (b) $\pstate_2 = (0.7,0.2,0.1)$, such that $\pi_{\pstate_{2}} = (1,2,3)$. 
	The initial state, thermal state $ \gbs $, and $ \beta $-order regimes are denoted similarly as to Figure~\ref{fig: ETO cone}. 
	Pink shaded areas mark catalytic advantage within $\setTO$, while yellow shaded areas are states that can be reached via ETO+qubit catalyst, but not by TO without catalysis.
	Figure adapted from Figure~5 of Ref.~\cite{Son2024_CETO}.}
	\label{fig: ext points 213}
\end{figure}

In Figure~\ref{fig: ext points 213}, we present two $\setCETOqb$ sets corresponding to two different initial states that have distinct $\beta$-orders. 
The sets are displayed in comparison with non-catalytic sets of reachable states $\setETO$ and $\setTO$. 
Notice that in Figure~\ref{fig: ext points 213} (a), where the initial state is set to have $\pi = (2,1,3)$, we observe that $ \setTO\subsetneq\setCETOqb $. 
This feature persists for a number of randomly chosen initial states with $\beta$-orders $\pi = (2,1,3)$ and $(3,1,2)$. 
When this happens, we can on one hand reproduce every TO transition using ETO with a single qubit catalyst; and on the other hand, combat some of the finite-size effects and enable a larger set of transitions than previously allowed by an arbitrary TO. 
Likewise, consider Figure~\ref{fig: ext points 213} (b), for an initial state $\pi = (1,2,3)$. 
Here, $\setCETOqb$ overlaps almost entirely with $\setTO$, but neither is fully contained by the other. 
This qualitative characteristic is again present for different initial states with the same $ \beta $-ordering.

In Figure~\ref{fig: ext points 213}, the set of states that go beyond $\setTO$ (yellow) highlights the additional advantage brought forward by CETO$_2$. 
In addition, the set of states between $\setTO$ and $\setETO$ (pink) also has operational merits---there exist states which require genuine multi-level TO to achieve, but can be obtained by an alternative pathway that involves only basic, Jaynes-Cummings-like interactions needed for ETO, when a catalyst is present.

Now that we have established the existence of the catalytic advantage, we leverage the decomposability of ETO and track changes in the system free energy throughout the catalytic process. 
Specifically, we analyse a simple series of ETO swaps that leads to the transformation of an initial state $\pstate$ into an extreme point of $\setCETOqb(\pstate)$.

We tackle the problem by the following procedure:
\begin{enumerate}
	\item \textbf{Construction of $\setETO(\pstate\otimes\catstate)$ for each $\catstate$.} 
	Given the initial state $\pstate\in \probspace_{3}$, we choose a qubit catalyst $\catstate\in\probspace_{2}$.
	Furthermore, the catalyst Hamiltonian is assumed to be degenerate%
	\footnote{It is believed that any catalytic transformation can be done with some catalyst with a degenerate Hamiltonian.
	For example, Ref.~\cite{LipkaBarosik2021_universal} shows that any large enough catalyst can be universal for any transformation (approximately).
	The same assumption has been made in Refs.~\cite{Czartowski2023ThermalRecall}.}, 
	meaning that a choice of the population $ \catstate = (c_{1},1-c_{1}) $ completely determines the catalyst. 
	The set $ \setETO(\pstate\otimes\catstate) $ is found by identifying all its extreme points $\extr[\setETO(\pstate\otimes\catstate)]$ by using necessary and sufficient conditions for ETO state transformations described in Appendix~\ref{chapter: ETO cone characterisation}.
	
	\item \text{Finding $\setCETOqb(\pstate;\catstate)$ from $\setETO(\pstate\otimes\catstate)$.}
	The resulting set $\setETO(\pstate\otimes\catstate)$ also includes states $p'\in\probspace_{6}$ that are correlated.
	Final states of the form $ \qstate\otimes\catstate $ can be distilled by imposing the catalytic condition. 
	This is done numerically by applying half-space intersections on the full set $\setETO(\pstate\otimes\catstate)$.
	
	\item \text{Iteration.}
	This procedure is iterated for different choices of $\catstate$, by varying $c_{1}$ in a sufficiently fine-grained manner. As a result, the desired set $\setCETOqb(\pstate)$ can be obtained. 
\end{enumerate}

\begin{figure}[t!]
	\includegraphics[width=\columnwidth]{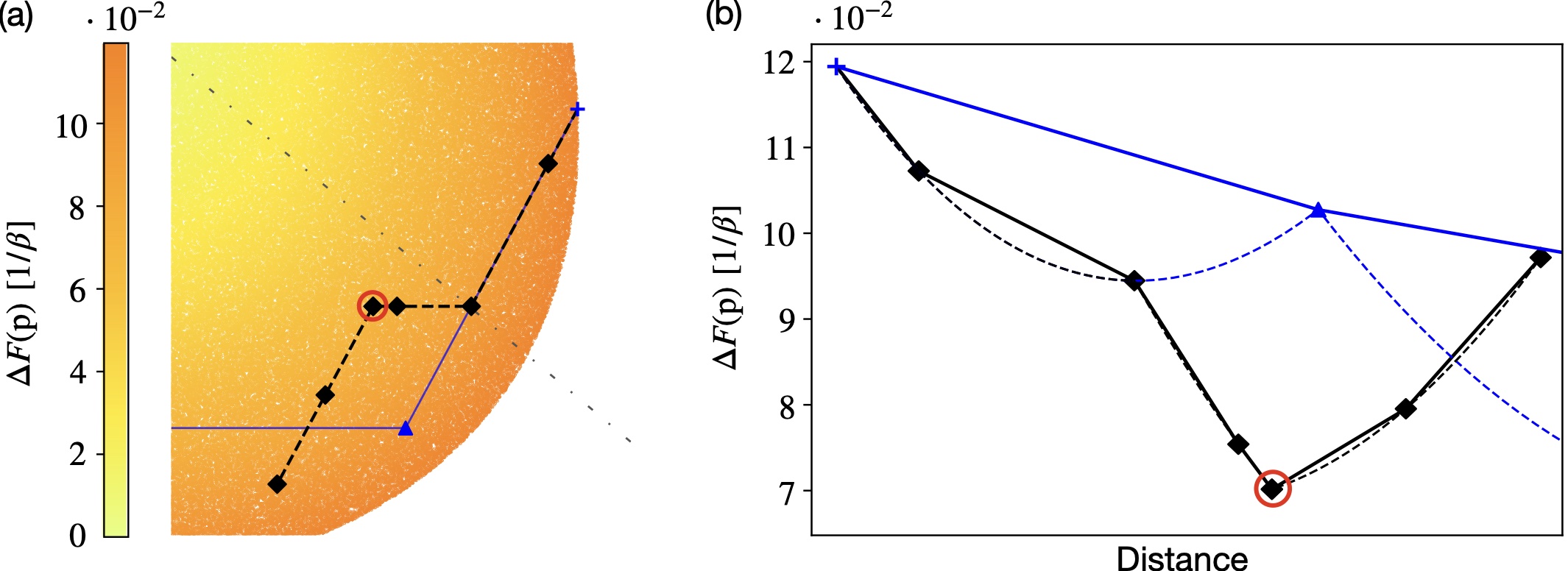}
	\caption{(a): Illustration of system reduced state evolutions in a CETO process (diamond markers and dashed lines) starting from an initial state $\pstate_{1}$ with final state lying beyond $\setETO(\pstate_{1})$. 
		This state is chosen to be an extreme point of $\setCETOqb(\pstate_{1})$. 
		The rest of the plot is the zoomed-in view of Figure~\ref{fig: ETO cone} (a) around $\beta$-orders $(2,1,3)$ and $(2,3,1)$. 
		(b): Free energy differences from the equilibrium state $\gbs$. 
		Solid lines connect the values after each swap and dashed lines mark continuous change between them.
		The fifth swap applied during the catalytic evolution does not alter the system reduced state, making the point before and after that swap not distinguishable in the system reduced picture of (a) and (b). 
		These points are marked with red circles in both (a) and (b). Initial state $\pstate_{1}$ and its Hamiltonian $\beta\mathrm{H}_S$ are the same to those of Figure~\ref{fig: ETO cone}.
		Figure adapted from Figure~6 of Ref.~\cite{Son2024_CETO}.}
	\label{fig: CETO tracking}
\end{figure}

In general, $\qstate\in\extr[\setCETO(\pstate)]$ does not guarantee $\qstate\otimes\catstate\in\extr[\setETO(\pstate\otimes\catstate)]$ in system-catalyst composite set. 
Therefore, even the extremal catalytic ETO may not be written as a single $\beta$-swap series.
Nevertheless, it is easy to find the convex combination of $\beta$-swap series that gives rise to the transformation $\pstate\otimes\catstate\toeto\qstate\otimes\catstate$, as it can be formulated as a linear programming.

Now we examine one specific extreme point of $ \setCETOqb(\pstate;\catstate) $, plotted in Figures~\ref{fig: CETO tracking}, \ref{fig: local free Es}, and \ref{fig: CETO tracking alphas}, following the procedure described above. 
This example is also an extreme point of $ \setCETOqb(\pstate) $ having the smallest ground state population among reachable states.

\begin{kaobox}[frametitle =An extreme point of $ \setCETOqb $]
	The initial state $ \pstate = (0.35,0.55,0.1) $ and the Hamiltonian $\beta\mathrm{H}_S = (0,0.2,0.5)$, giving the $\beta$-order $\pi_{\pstate} = (2,1,3)$. The initial catalyst distribution is fine-tuned to be $\catstate = (c_{1},1-c_{1})$, with
	\begin{align}
		c_{1} = \frac{-p_3+\sqrt{p_3^2+8\Delta_{13}p_{1}p_3}}{4\Delta_{13}p_{1}} \simeq 0.3816,
	\end{align}
	which provides the maximum advantage when minimising the ground state population. 
	Then the composite state $\beta$-order becomes
	\begin{align}
		\pi_{\pstate\otimes\catstate} = (2*2,2*1,1*2,1*1,3*2,3*1),
	\end{align}
	where $a*b$ indicates the energy eigenstate $\ket{a}_{S}\ket{b}_{C}$ of a system (S) plus catalyst (C) state.
	
	From $ \setETO(\pstate\otimes\catstate) $, we obtain new extreme points of $ \setCETOqb(\pstate;\catstate) $, including our example point $ \qstate $. 
	Furthermore, by solving a linear programming problem, four extreme points $\qstate^\prime_{1,2,3,4}\in\probspace_6$ of the set $\setETO(\pstate\otimes\catstate)$, such that  
	\begin{align}
		\qstate\otimes\catstate = \sum_{i=1}^{4}\alpha_i\qstate^\prime_i,\quad \sum_{i=1}^{4}\alpha_i = 1,\quad \alpha_i\geq 0,
	\end{align}
	can be found. 
	In addition, the corresponding $\beta$-swap series $ \qstate^\prime_i = \vec{\beta}_i(\pstate\otimes\catstate) $ are also identified.
	These series $ \vec{\beta}_i $ differ only slightly from each other (differences marked red),
	\begin{align}
		\vec{\beta}_{1} &= \beta^{(1*2,3*1)} \beta^{(1*1,3*1)}\beta^{(2*1,3*2)}\beta^{(1*2,3*2)}\beta^{(1*1,3*2)},\\
		\vec{\beta}_2 &= \beta^{(1*2,3*1)} \beta^{(1*1,3*1)}{\color{red!65!black}\beta^{(2*1,2*2)}}\beta^{(2*1,3*2)}\beta^{(1*2,3*2)}\beta^{(1*1,3*2)},\\
		\vec{\beta}_3 &= \beta^{(1*2,3*1)} \beta^{(1*1,3*1)}{\color{red!65!black}\beta^{(2*2,3*2)}}\beta^{(2*1,3*2)}\beta^{(1*2,3*2)}\beta^{(1*1,3*2)},\\
		\vec{\beta}_4 &= \beta^{(1*2,3*1)} \beta^{(1*1,3*1)}{\color{red!65!black}\beta^{(2*1,2*2)}\beta^{(2*2,3*2)}}\beta^{(2*1,3*2)}\beta^{(1*2,3*2)}\beta^{(1*1,3*2)},
	\end{align}
	facilitating the recombination into
	\begin{align}\label{eq: example CETO path}
		\qstate\otimes\catstate = \beta^{(1*2,3*1)} \beta^{(1*1,3*1)} M_{\lambda_2}^{(2*1,2*2)} M_{\lambda_{1}}^{(2*2,3*2)}\beta^{(2*1,3*2)}\beta^{(1*2,3*2)}\beta^{(1*1,3*2)}(\pstate\otimes\catstate),
	\end{align}
	with $ \lambda_{1} = \alpha_3/(\alpha_{1}+\alpha_3) = \alpha_4/(\alpha_2+\alpha_4) $ and $ \lambda_2 = \alpha_2/(\alpha_{1}+\alpha_2) = \alpha_4/(\alpha_3+\alpha_4) $. 
	
	Given Eq.~\eqref{eq: example CETO path}, we can analyse system-catalyst interplay during the catalytic evolution, starting from
	\begin{align}
		\pstate\otimes\catstate \simeq (0.1336, 0.2164, 0.2099, 0.3401, 0.0382, 0.0618).
	\end{align}
	The swap series can be grouped into three phases:
	\begin{enumerate}
		\item The first four swaps $M_{\lambda_{1}}^{(2*2,3*2)}\beta^{(2*1,3*2)}\beta^{(1*2,3*2)}\beta^{(1*1,3*2)}$ all involve the $3*2$ population. 
		The first and the third swaps ($ \beta^{(1*1,3*2)} $ and $ \beta^{(2*1,3*2)} $) work to shift population from the first level of the catalyst to the second level and thus intensify the non-uniformity of the catalyst reduced state, as reflected in corresponding steps of Figure~\ref{fig: local free Es} (b). 
		The ratio between $3*2$ and $3*1$ populations is increasing more rapidly than the one between levels $ 2 $ and $ 1 $ of catalyst reduced state, correlating the system and catalyst as shown in the mutual information.
		The fourth swap is chosen to be $ \lambda\neq1 $ swap to prevent $3*2$ population to become too large to recover $ \catstate $.
		
		Since both catalyst local free energy and mutual information increase, system local free energy should always decrease at this stage to keep the total free energy non-increasing. 
		The composite population vector and its marginals after these swaps are
		\begin{align}
			\rstate^\prime_{1} &\simeq (0.1144, 0.1662, 0.1857, 0.3323 , 0.0382, 0.1633),\\ 
			\Tr_{C}[\rstate^\prime_{1}] &\simeq (0.2806, 0.5180, 0.2015),\quad \Tr_{S}[\rstate^\prime_{1}] \simeq (0.3382, 0.6618)
		\end{align}
		
		\item The fifth swap $M_{\lambda_2}^{(2*1,2*2)}$ balances catalyst marginal distribution in the degenerate block $\ket{2}_S$ while system marginal populations are fixed, yielding
		\begin{align}
			\rstate^\prime_2 &\simeq (0.1144, 0.1662, 0.1977, 0.3203 , 0.0382, 0.1633),\\ 
			\Tr_{C}[\rstate^\prime_2] &\simeq (0.2806, 0.5180, 0.2015),\quad \Tr_{S}[\rstate^\prime_2] \simeq (0.3502, 0.6498).
		\end{align}
		Catalyst local free energy decreases as a result of mixing, while correlation increases. 
		
		\item Last two swaps $\beta^{(1*2,3*1)} \beta^{(1*1,3*1)}$ increase $3*1$ level population to recover the original ratio between $3*1$ and $3*2$, while at the same time further reducing the system ground state population. 
		For this particular choice of catalyst, we have a simplification in the sense that these swaps also balance $1*1$ and $1*2$, leading to 
		\begin{align}
			\qstate\otimes\catstate &\simeq (0.0832, 0.1348, 0.1977, 0.3203, 0.1008,	0.1633), \\
			\qstate &\simeq (0.2179, 0.5180, 0.2641),
		\end{align}
		with vanishing correlation and retrieval of the original catalyst. 
		The system free energy increases here, since level $\ket{1}_S$, which already has the lowest slope, loses more population and the new state thermomajorises the old state.
	\end{enumerate}
\end{kaobox}

\begin{figure}[t!]
	\includegraphics[width=\columnwidth]{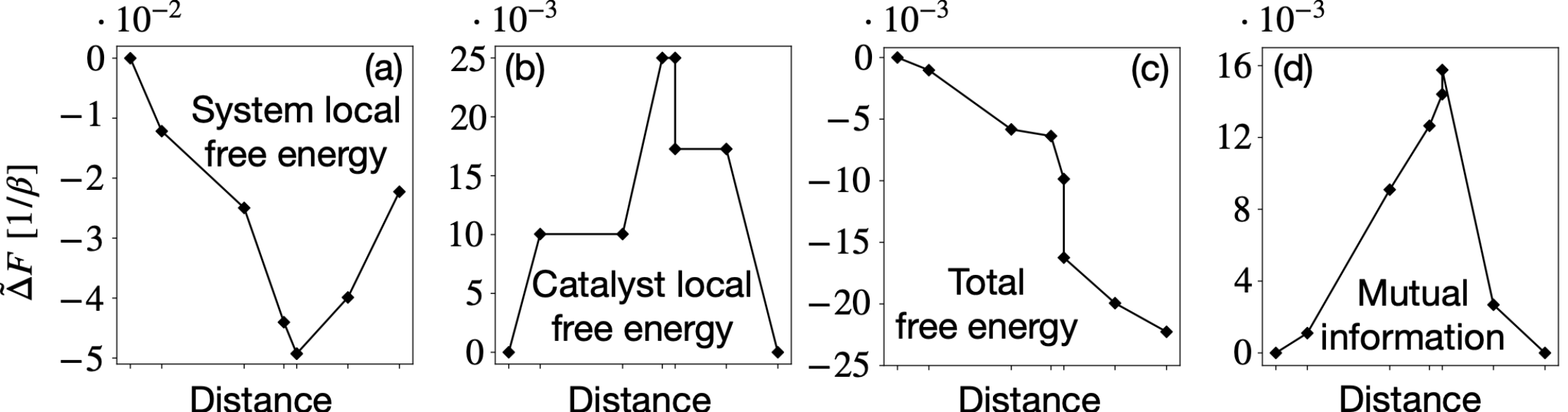}
	\caption{Non-equilibrium free energies of CETO evolution described in Figure~\ref{fig: CETO tracking}. 
		The symbol $\tilde{\Delta}$ indicates the difference from the values of the initial state. 
		(a): system local free energy identical to black lines in (b) of Figure~\ref{fig: CETO tracking}; (b): catalyst local free energy; (c): total free energy; (d): mutual information which, in this case, is identical to the difference between the total free energy and the sum of two local free energies. 
		$ x $-coordinates of points are identical to ones in Figure~\ref{fig: CETO tracking} (b).
		Figure adapted from Figure~7 of Ref.~\cite{Son2024_CETO}.
	}
	\label{fig: local free Es}
\end{figure}

The above example showcases a typical strategy for constructing catalytic transformations: 
\begin{enumerate}
	\item exploits expanded dimensionality to swap a particular energy level with more numbers of levels, storing free energy in the form of i) temporary correlations and ii) local variations on catalyst; 
	\item resolves correlations by mixing in the degenerate energy subspace; and
	\item recovers original catalyst distribution while increasing system local free energy.
\end{enumerate}

The most striking observation from this example is the increase of system non-equilibrium free energy in the later steps of the process. 
This behaviour of increasing free energy after swaps is strictly forbidden in non-catalytic setting, and allowed in this case by sacrificing correlation and catalyst free energy stored from previous operations. 

Since strict catalysis, the most conservative form of catalysis, is assumed, catalyst local free energy goes back to its original level, and mutual information also returns to zero, as shown in (b) and (d) of Figure~\ref{fig: local free Es}. 
Hence, the catalyst's role is restricted to temporary free energy storages. 
Furthermore, the total free energy always decreases after each swap, obeying the monotonicity of free energies under ETO.

Similar behaviours are observed for other monotones, such as generalised free energies with $\alpha\neq 1$ constructed from the $\alpha$-Rényi divergence (Definition~\ref{definition: reyni divergences}) as $\Delta F_{\alpha}(\pstate) \coloneq \beta D_{\alpha}(\pstate\|\gbs)$.  
In Figure~\ref{fig: CETO tracking alphas}, $\alpha = 0.5$ and $2$ are presented as representative examples. 
In both cases, $F_{\alpha}$ for a system reduced state (see (c) and (d) of Figure~\ref{fig: CETO tracking alphas}) shares the decreasing/increasing trend, albeit with different slopes. 

\begin{figure}[t!]\centering
	\includegraphics[width=\columnwidth]{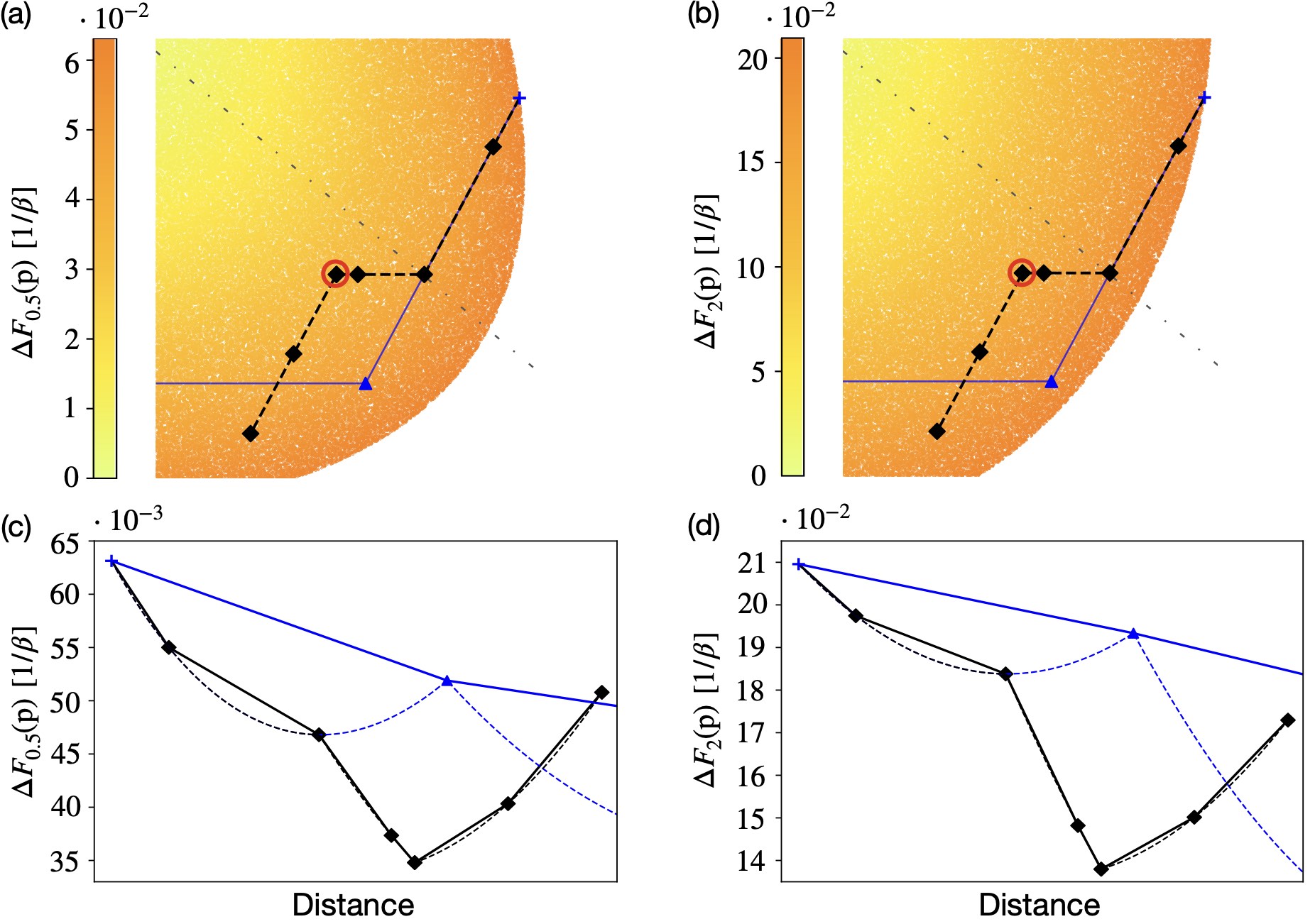}
	\caption{Replica of Figure~\ref{fig: CETO tracking} with generalised free energies defined with $\alpha$-Rényi divergences (Definition~\ref{definition: reyni divergences}).
	Panels (a) and (c) display $\alpha = 0.5$ quantities, while panels (b) and (d) show $\alpha = 2$ cases.
	Figure adapted from Figure~F1 of Ref.~\cite{Son2024_CETO}.
	}
	\label{fig: CETO tracking alphas}
\end{figure}

\subsection{Intermediate-size catalysts}

After examining qubit catalysis, a natural question follows: how does the gap between CETO$_d$ and ETO change with $d$ being the dimension of the allowed catalyst? 
In general, identifying the whole set of reachable states $\setCETOqd{d}(\pstate)$ with $d$-dimensional catalysts is highly challenging, even when $\pstate$ is three-dimensional and $d = 3$, because constructing the set $\setETO(\pstate\otimes\catstate)$ requires calculating $(3d)^{2(3d)!}$ different final states in the worst case.
However, by developing a theoretical tool, namely Theorem~\ref{theorem: nice order_main}, we show that the numerical cost is dramatically reduced for initial states $ \pstate $ whose $ \beta $-orders are monotonic in energy levels, i.e. $\pstate$ such that $\pi_{\pstate} = (1,2,\cdots,d)$ or $(d,d-1,\cdots,1)$. 
The extreme state of $\setETO(\pstate)$ corresponding to each $\beta$-ordering is uniquely determined by Theorem~\ref{theorem: nice order_main}, hence reducing $(md)^{2(md)!}$ potential candidates for extreme points into $(md)!$ for $m$-dimensional catalyst and $d$-dimensional system. 
See Theorem~\ref{theorem: nice order} in Appendix~\ref{chapter: ETO cone characterisation} for the reiteration of the theorem and its proof.

\begin{theorem}[original result]\label{theorem: nice order_main} 
	If $\pi_{\pstate} $ is monotonic in energy, extreme points of $\setETO(\pstate)$ are achieved if and only if the corresponding $\beta$-swap series that produce them are
	\begin{enumerate}
		\item always acting on two consecutive energy levels of the $\beta$-ordering, 
		\item containing no repetition of each swap.
	\end{enumerate} Furthermore, when $ \vec\beta_{1}\pstate, \vec\beta_{2}\pstate \in \extr[\setETO(\pstate)] $ and $ \pi_{\vec\beta_{1}\pstate} = \pi_{\vec\beta_{2}\pstate}$ for such $ \pstate $, the two series are identical ($ \vec\beta_{1} = \vec\beta_{2} $). 
\end{theorem}

A particularly important class of states that satisfy this property is the set of Gibbs states with temperatures different from the ambient temperature $\beta^{-1}$. 
When $ \beta_{h}<\beta $, the state $ \gamma^{\beta_{h}} $ is hotter than the environment with temperature $ \beta^{-1} $, and the $ \beta $-order $ \pi_{\gamma^{\beta_{h}}} = (d,d-1,\cdots,1) $. 
Similarly, colder states with $\beta_{c}>\beta$ have the order $\pi_{\gamma^{\beta_{c}}} = (1,2,\cdots,d)$. 
If we further employ a catalyst $\catstate$ from the set of states which are sufficiently thermal, the monotonicity of $\beta$-order would be preserved, i.e. $\pi_{\pstate\otimes\catstate}$ is again monotonic in the total energy. 
In such cases, the analysis of higher-dimensional catalysts becomes computationally tractable. 
We will refer to such catalyst states as \emph{minimally-disturbing catalysts}, in the sense that they do not disturb the $\beta$-ordering of the system-catalyst composite.
This setup has an additional merit that catalysts are low-resource states, as they are chosen to be closer to the thermal state. 

To demonstrate the benefit of this reduction, let us consider the cooling process via (C)ETO starting from a high temperature thermal state $ \gamma^{\beta_{h}} $ with $\beta_{h}<\beta$ to a colder thermal state $\gamma^{\beta_{c}}$.
For any temperature $\beta_{h}^{-1}$, it is always possible to reach $\beta_{c} = \beta$ by a full thermalisation with the environment. 
However, with ETO and TO, colder temperatures $\beta_{c}>\beta$ can be achieved.
To corroborate the effectiveness of small catalysts with practicable procedures and investigate the scaling of catalytic advantage with respect to catalyst size, we apply Theorem~\ref{theorem: nice order_main} to find the limits of the cooling performance for a qutrit when using catalysts of varying dimensions, ranging from two to thirty. 
The catalyst Hamiltonian is again set to be trivial for simplicity. 

\begin{figure}[t]
	\centering
	\includegraphics[width = 0.7\columnwidth]{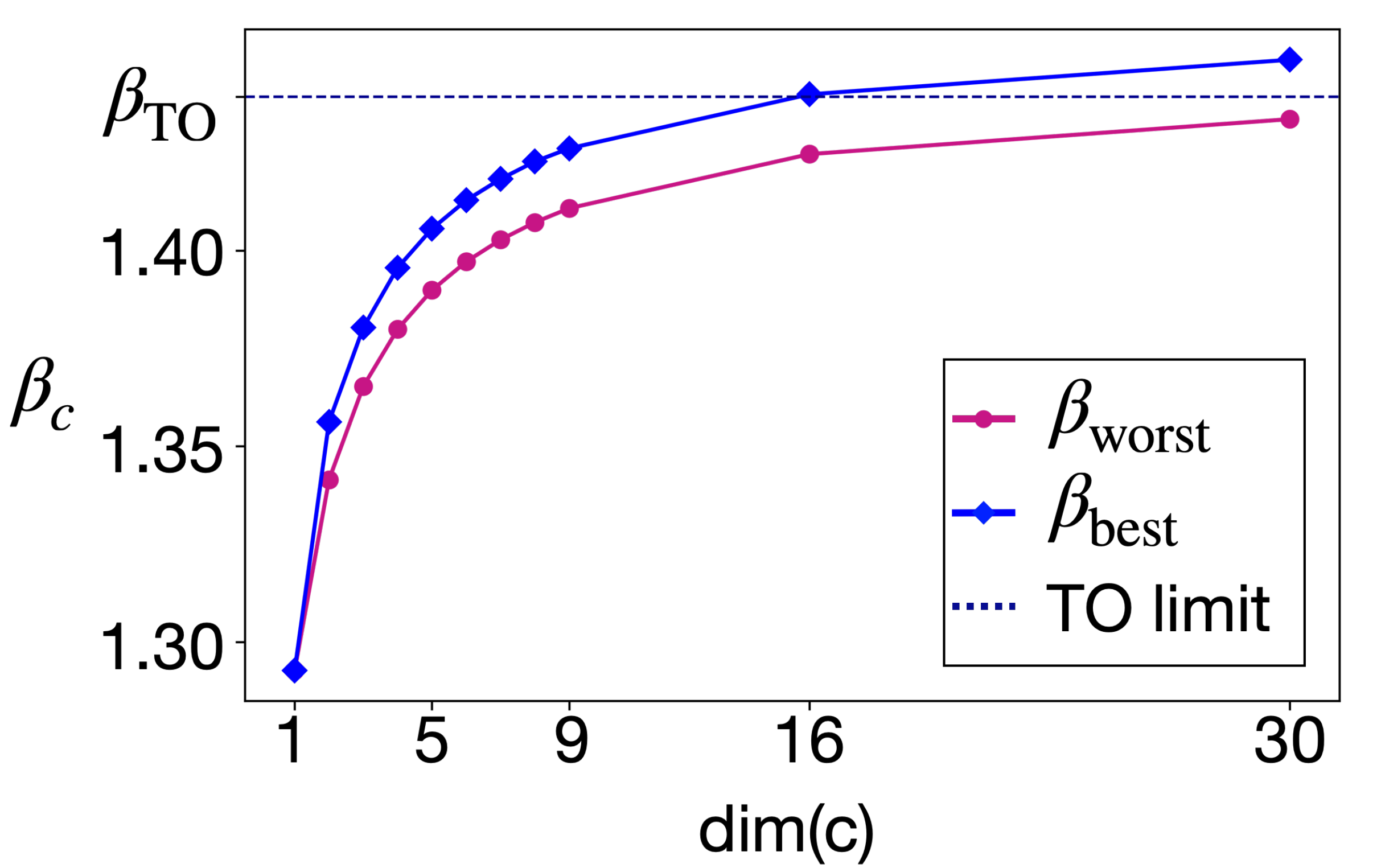}
	\caption{The cooling performance of CETO from a thermal state to another thermal state, quantified by final inverse temperature $ \beta_{c} $ attainable when the catalyst state is fixed. 
	We set the initial system inverse temperature to be $ \beta_{h} = 0.5 $, and the ambient temperature to be $ \beta = 1 $. 
	The system of interest is three-dimensional with energy levels $ (0,0.4,0.5) $ and catalysts of dimension $ \dim(\catstate) $ from two to nine, sixteen, and thirty are used. 
	$ \dim(\catstate) = 1 $ case corresponds to non-catalytic ETO, and the blue dashed line above marks the inverse temperature $ \beta_\mathrm{TO} $ that non-catalytic TO can achieve. 
	We searched over catalyst state distributions among minimally-disturbing catalysts. 
	The results from the worst performing catalysts in each dimension are marked with purple circles, while blue diamonds are from the best catalysts in each dimension.
	Figure adapted from Figure~8 of Ref.~\cite{Son2024_CETO}.}
	\label{fig: higher dim catalysis thermal to thermal}
\end{figure}

Figure~\ref{fig: higher dim catalysis thermal to thermal} shows the coldest achievable $ \beta_{c} $ from minimally-disturbing catalysts, where the worst and the best cases are marked with purple circles and blue diamonds, respectively.
Even with qubit catalysts, almost half of the gap between the TO limit (dashed line) and the ETO limit ($ \dim(\catstate) = 1 $) is covered. 
The maximal catalytic advantage (blue solid line) gradually increases with the catalyst size, and at $ \dim(\catstate) = 16 $, best catalysts among the sample surpass TO limit, whilst at $ \dim(\catstate) = 30 $, most of the samples perform better than TO. 

Note that we have limited the range of catalyst distributions to fix the initial composite state $\beta$-order; hence there might exist (not minimally-disturbing) catalyst states that activate a better cooling process than the ones marked in Figure~\ref{fig: higher dim catalysis thermal to thermal}. 
Also, even the worst case catalysts provide some advantage for the same reason.
Usually, if the catalyst is pure or almost pure, catalytic advantage vanishes, but minimally-disturbing ones are typically far from pure. 
Nevertheless, our results give an efficiently computable lower bound to the achievable amount of cooling when any $ d $-dimensional catalyst is allowed.

One heuristic approach for exploring catalyst distributions that are not minimally-disturbing, is to search over sequences of $\beta$-swaps satisfying the conditions of Theorem~\ref{theorem: nice order_main}. 
This method is computationally efficient and yields a subset of $\setETO(\pstate\otimes\catstate)$, typically strictly smaller than the full set. 
Nonetheless, we have observed instances where catalytic advantage is still present even within this restricted set.

Overall, the results in this section demonstrate that small catalysts do provide substantial advantage in the setting of ETO, where simple two-level swaps are sufficient to execute the procedure. 
Furthermore, we leverage the step-wise structure of ETO to track and analyse catalytic evolutions, by capturing snapshots of states after each ETO step. 
This approach opens up a new avenue for understanding the underlying origins of catalytic advantage. 
In our example, the catalyst's role was to receive the free energy flowing out from the system, either through reduced state population changes or correlations with the system. 
Without the catalyst, all changes in system free energy would dissipate into the surrounding bath, which thermalises after each swap. This interpretation could potentially be extended to catalysts in different resource theories. 
For instance, it would be intriguing to further investigate the snapshots of correlated catalysis in ETO to observe whether the memory effect is still the primary reason behind the catalytic advantage. 
	
\section{A hierarchy of thermodynamic resource theories collapses under catalysis}\label{section: a hierarchy collapses}

Interestingly, regardless of the catalyst dimension, the worst catalysts in Figure~\ref{fig: higher dim catalysis thermal to thermal} are given by maximally mixed states which are the Gibbs states for the trivial Hamiltonian that is assumed.
This is striking because Gibbs states are free states in thermodynamic resource theories.
In GP, GPC, and TO, appending any free state is free, i.e. $\rho_{S}\to\rho_{S}\otimes\gbs_{R}$.
Hence, catalytic operations using $\gbs_{R}$ as a catalyst can be done without using it catalytically, as it can be used freely. 

However, MTO assumes that the bath is fully Markovian, i.e. the bath loses its memory much faster than the timescale of the system dynamics.
In such cases, Gibbs states can activate state transitions as a catalyst by providing additional non-Markovianity, as studied in Refs.~\cite{Korzekwa2022MTO2, Czartowski2023ThermalRecall}.
This is reminiscent of the discussion in the qubit catalyst case, where the catalyst functions as a temporary storage during the evolution.

In general, ETO is not fully Markovian and thus distinct from MTO; nevertheless, an innate Markovianity is also embedded in the definition of ETO. 
ETO is written as sequences of two-level operations $\mE_{i}\in\eTO{S}{S}$ (and their convex combinations), where each two-level operation is implemented with a \emph{fresh} bath $ \gbs_{R_{i}} $.
Intuitively, the irreversible loss of information into bath $R_i$ in each channel $ \mE_{i} $ produces an in-built Markovian behaviour, although each bath $ \gbs_{R_{i}} $ goes out of equilibrium during individual evolutions $ \mE_{i} $ as can be seen in Figure~\ref{fig: ETO cone} (b).
We understand the inclusion $\setETO(\rho) \subsetneq \setTO(\rho)$ with a strict gap from this in-built Markovianity of ETOs.

Is the gap between thermal operations and its simpler counterparts solely attributed to their Markovianity? 
If this were the case, the injection of sufficient memory states should enable the implementation of any complex, multi-level thermal operation using only sequential two-level operations.
We answer this question affirmatively---indeed, any thermal operation can be accomplished with this strategy. 
In other words, we discover that the use of Gibbs catalysts is the decisive factor in closing the gap between thermal operations and their elementary counterparts (Theorem~\ref{theorem: GCETO is TO}). 
More significantly, we show that Gibbs catalysts also close the gap between \emph{catalytic} versions of TO and ETO (Theorem~\ref{theorem: CETO is CTO}). 

To address the irreversible loss of information to the bath after each step in ETO, we must be able to retrieve the information from the bath.
This necessitates the controllability over certain parts of the used baths. 
We do so by modelling such parts of the baths as catalysts; we thus retain the baths throughout the process without tracing them out midway. 
This naturally leads to the formulation of operations supplemented with Gibbs state catalysts, i.e. controllable thermal baths; see Figure~\ref{fig: GCETO illustration} for the GC-ETO illustration. 
\begin{definition}[Gibbs-catalytic operations]\label{definition: GCETO}
	A transformation $\rho_{S}\rightarrow\sigma_{S}$ is achievable by Gibbs-catalytic X (GC-X) if there exists a Gibbs state  $ \gbs_{C}$ such that 
	\begin{align}
		\rho_{S}\otimes \gbs_{C} \xrightarrow{\text{X}}\sigma_{S}\otimes \gbs_{C}
	\end{align}
	for X = TO, ETO, MTO.
	In other words, the existence of the catalyst Hamiltonian $H_{C}$ such that $\sigma_{S}\otimes \gbs_{C} \in \setXTO(\rho_{S}\otimes \gbs_{C})$ is equivalent to $\sigma_{S}\in\setGCXTO(\rho_{S})$.
\end{definition}

Note that Gibbs-catalytic ETO or MTO are different from the usual ETO or MTO, because $\rho_{S}$ cannot be freely transformed into $\rho_{S}\otimes\gbs_{C}$ even though $\gbs_{C}$ is a free state; see Definition~\ref{definition: free states}.

\begin{figure}[t!]
	\centering
	\includegraphics[width=0.92\columnwidth]{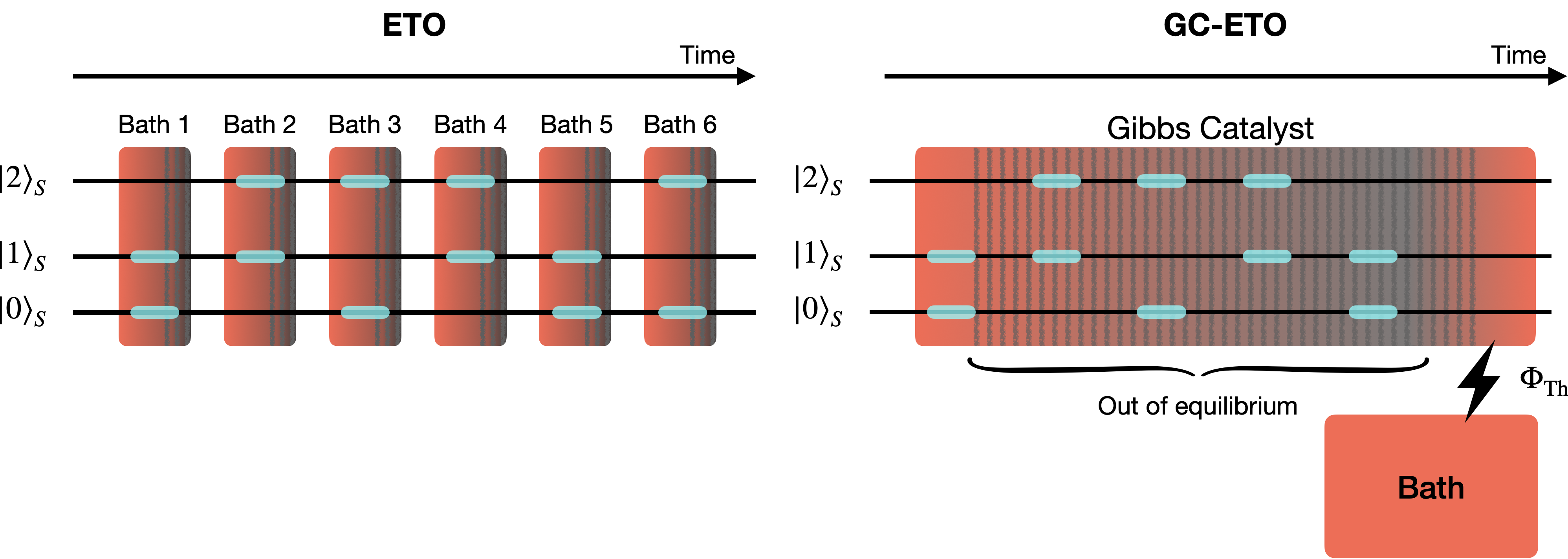}
	\caption{Comparison between ETO and GC-ETO. 
		The left diagram depicts an ETO sequence applied to a qutrit. 
		At each step, two levels (highlighted in light blue) interact with a refreshed thermal bath (red), which becomes athermal at the end (grey). 
		When two new levels are chosen, a fresh bath is also chosen. 
		The right part portrays a GC-ETO process with a catalyst starting from a Gibbs state (in red). 
		During the process, the catalyst goes out of equilibrium (in grey with stripes), but at the end of the process, it is rethermalised via the thermalising channel $ \mE_\mathrm{Th}\in\ETO{C}{C}$.
		Figure adapted from Figure~1 of Ref.~\cite{Son2024hierarchy}.
		}
	\label{fig: GCETO illustration}
\end{figure}

Importantly, the full thermalisation can be performed by a sequence of ETO (or MTO), with the aid of an additional bath. 
Since the full thermalisation channel transforms every density matrix into a single fixed-point, it also eliminates any correlation between the catalyst and the system. 
Therefore, whenever a Gibbs state is used, the fulfilment of the strict catalytic condition is free, i.e.
\begin{align}
	\rho_{SC} \to \Tr_{C}[\rho_{SC}]\otimes\gbs_{C}
\end{align} 
via ETO or MTO, for any potentially correlated final state $\rho_{SC}$.

We already established that TO does not benefit from using Gibbs states as a catalyst. 
In other words, GC-TO is exactly equivalent to TO. 
Since ETO is a subset of TO, any transformation $\rho_{S}\otimes\gbs_{C}\toeto\sigma_{S}\otimes\gbs_{C}$ implies that $\rho_{S}\otimes\gbs_{C}\toto\sigma_{S}\otimes\gbs_{C}$.
This in turn means that
\begin{align}\label{eq: GCETO is in TO}
	\setGCETO(\rho_{S}) \subset \setTO(\rho_{S}),
\end{align}
for any input state $ \rho_{S} $.
Similarly, we can establish the relation $ \setGCMTO(\rho_{S}) \subset \setTO(\rho_{S}) $ for any input state $ \rho_{S} $.

Our first result shows that the converse of Eq.~\eqref{eq: GCETO is in TO} is also true. 
We emphasise that this theorem holds for arbitrary \emph{asymmetric} initial states that may not commute with the system Hamiltonian, as opposed to other majorisation-based approaches that are only valid for symmetric state transitions.
Such generality is achieved thanks to the emulation of TO channels themselves via Gibbs-catalytic operations. 

\begin{theorem}[original result]\label{theorem: GCETO is TO} 
	$ \setTO(\rho_{S})= \setGCETO(\rho_{S}) = \setGCMTO(\rho_{S}) $ for all states $\rho_{S}\in\mD_{S}$. 
\end{theorem}

\begin{proof}
	For the proof, we only need to show that any TO channel can be recast into a GC-ETO (or GC-MTO) channel.
	Consider a TO channel $\mE\in\TO{S}{S}$ that admits the dilated form $\mE(\rho_{S}) = \Tr_{R}[U(\rho_{S}\otimes\gbs_{R})U^\dagger]$,
	with some bath Hamiltonian $ H_R\in\mL_{R} $ and an energy-preserving unitary $ U\in\mL_{SR} $, such that $ [U,H_{0}] = 0 $, where we denote the non-interacting Hamiltonian $ H_{0} = H_{S}\otimes\1_{R}+\1_{S}\otimes H_{R} $. 
	
	The language of Lie group and Lie algebra, briefly summarised in Appendix~\ref{chapter: Lie groups}, is useful for our proof.
	The set of all energy-preserving unitaries $ G $ given the Hamiltonian $ H_{0} $ [Eq.~\eqref{eq: Lie group def} in Appendix~\ref{chapter: Lie groups}] forms a compact connected Lie group, where the corresponding Lie algebra $ \mathfrak{g} $ is the set of energy preserving Hamiltonians (times $ -i $). 
	Let us choose a set of linearly independent anti-Hermitian operators $ \{K_{1},K_{2},\cdots,K_{L}\} $ that generates the Lie algebra $ \mathfrak{g}$. 
	Then from Lemma~\ref{lemma: Lie controllability} in Appendix~\ref{chapter: Lie groups}, any energy-preserving unitary $ U $ is a product of a finite number of exponentials of the form $ \exp(Kt) $, where $ K\in \{K_{1},K_{2},\cdots,K_{L}\} $ and $ t\in\mathbb{R} $.
	
	Now we show that each $ K_{i} $ in the set can be chosen as an operator acting on at most two levels. 
	Any Hermitian operator $ H_{\mathrm{int}} \in\mL_{SC}$ can be expanded in the energy eigenbasis as
	\begin{equation}\label{eq:H_interaction}
		H_{\mathrm{int}} = \sum_{k,l,E,E^{\prime},g,g^{\prime}}H^{(klEE^{\prime} gg^{\prime})},
	\end{equation}
	with each term being $ H^{(klEE^{\prime}gg^{\prime})} \propto \ketbra{k}{l}_{S}\otimes\ketbra{E,g}{E^{\prime},g^{\prime}}_{C}  $, where $\{\ket{k}_{S}\}_{k}$ is the eigenbasis of $H_{S}$, such that $H_{S}\ket{k}_{S} = E_{k}\ket{k}_{S}$ and $\{\ket{E,g}_{C}\}_{E,g}$ that of $H_{C}$ with $H_{C}\ket{E,g}_{C}= E\ket{E,g}_{C}$. 
	Here, we have chosen the catalyst to be the Gibbs state $ \gbs_{C} $ with Hamiltonian $H_{C}$ identical to the bath Hamiltonian $H_{R} $.
	Now observe that
	\begin{equation}\label{eq:each_term}
		[H^{(klEE^{\prime} gg^{\prime})}, H_{0}] = (E_{k}+E-E_{l}-E^{\prime})H^{(klEE^{\prime} gg^{\prime})}.
	\end{equation}
	Eq.~\eqref{eq:each_term} vanishes only when $ E_{k}+E = E_{l}+E^{\prime} $, i.e. when $ \ket{k}_{S}\ket{E,g}_{C} $ and $ \ket{l}_{S}\ket{E^{\prime},g^{\prime}}_{C} $ are in the same energy subspace. 
	The energy-preserving condition $ [H_{\mathrm{int}},H_{0}] = 0 $ indicates that Eq.~\eqref{eq:each_term} is zero for all terms in Eq.~\eqref{eq:H_interaction}, as $H^{(klEE^{\prime} gg^{\prime})}$ for different superscripts are all linearly independent.
	 
	As a result, any energy-preserving Hamiltonian $ H_{\mathrm{int}} $ is a linear combination of (genuine) two-system-level terms 
	\begin{align}
		\ketbra{k}{l}_{S}&\otimes\ketbra{E-E_{k},g}{E-E_{l},g^{\prime}}_{C} +\mathrm{h.c.},\label{eq:basis1}\\
		i\ketbra{k}{l}_{S}&\otimes\ketbra{E-E_{k},g}{E-E_{l},g^{\prime}}_{C} +\mathrm{h.c.},\label{eq:basis2}
	\end{align}
	with $ k\neq l $ and one-system-level terms with $ g\neq g^{\prime} $,
	\begin{align}
		\dm{j}_{S}&\otimes\ketbra{E-E_{j},g}{E-E_{j},g^{\prime}}_{C} + \mathrm{h.c.},\label{eq:basis3}\\
		i\dm{j}_{S}&\otimes\ketbra{E-E_{j},g}{E-E_{j},g^{\prime}}_{C} + \mathrm{h.c.},\label{eq:basis4}\\
		\dm{j}_{S}&\otimes\dm{E-E_{j},g}_{C}.\label{eq:basis5}
	\end{align} 
	By collecting each of these terms (times $ -i $), we obtain the basis set $ \{K_{1},K_{2},\cdots,K_{L}\} $ of the Lie algebra $\mathfrak{g}$.
	Basis set is also a generating set; therefore, any energy-preserving unitary can be written as a finite product of at-most-two-level unitaries.
	
	Since we regard the Gibbs state as a catalyst in GC-ETO framework, a unitary channel with the unitary operator $ \exp(K_{i}t) $ is itself an ETO channel; the product of these unitaries is then an ETO sequence. 
	After this sequence the resulting intermediate state is $\sigma_{SC}$, whose reduced state on $S$ is already the target state $\sigma_{S}$. 
	Another ETO sequence that implements the fully thermalising channel on $C$ yields the final product state $\sigma_{S}\otimes\gbs_{C}$.
	This concludes the proof of $ \setTO(\rho_{S})\subset\setGCETO(\rho_{S}) $.
	
	We remark that the rank-1 projectors of the form Eq.~\eqref{eq:basis5}, which act on a single system-catalyst level instead of two, generate valid ETO channels, according to the definition of ETO allowing \emph{at most} two energy levels to be manipulated simultaneously.%
	\footnote{In addition, it is always possible to choose a different generating set $\{K'_{1},K'_{2},\cdots,K'_{L}\}$, where all elements are rank-2 operators}
	
	Finally, note that the same energy-preserving unitary channels generated by $ K_{i} $ are also MTO channels since the system-catalyst composite is regarded as a controllable system. 
	Furthermore, fully thermalising channel is also in MTO. 
	Therefore, we also conclude that $ \setTO(\rho_{S})\subset\setGCMTO(\rho_{S}) \cap \setGCETO(\rho_{S}) \subset \setGCMTO(\rho_{S}) $.
\end{proof}

This proof can be easily generalised to other subsets of thermal operations besides ETO and MTO. 
Suppose that a set X consists of step-wise operations and in each step within an operation, the system and the bath are coupled via an interaction Hamiltonian $H_{j}$, selected from some restricted set $\mathrm{X} = \{H_{i}\}_{i}$, whose elements commute with the total Hamiltonian $H_{0}$ (e.g. for ETO, this set is the set of interaction Hamiltonians that act non-trivially on at most two system energy levels). 
Because the operation is step-wise decomposable, the bath is rethermalised after each interaction by $H_{j}$.
Then, the same proof employed for Theorem~\ref{theorem: GCETO is TO} can be used to prove $\setTO(\rho_{S})= \setGCXTO(\rho_{S})$ for any $\rho_{S}$.

Lemma~\ref{lemma: Lie controllability} in Appendix~\ref{chapter: Lie groups} provides a powerful tool for the decomposition of energy-preserving unitary operators. 
When a target Hamiltonian $H_{\mathrm{int}}$ can be generated by linear combinations and commutations of the accessible Hamiltonians $\{H_{i}\}$, it is possible to \emph{exactly} decompose $e^{-itH_{\mathrm{int}}}$ into a \emph{finite} product of unitaries $e^{-it_{i}H_{i}}$ generated from individual Hamiltonians $H_{i}$.
Therefore, we obtain the desired result $\setTO(\rho_{S})\subset\setGCXTO(\rho_{S})$.
The condition imposed on $\mathrm{X} = \{H_{i}\}_{i}$ in Lemma~\ref{lemma: Lie controllability} can also be understood as X achieving universality (for energy-preserving unitaries). 
ETO corresponds exactly to the case where X is the set of two-system-level interaction Hamiltonians.  

Due to the rethermalisation, even when the set X achieves universality for all energy-preserving interactions, concatenations (and even convex combinations) of X channels cannot achieve a general TO channel. 
Remarkably, even with the freedom of choosing a bath different from the one used for the thermal operation, some TO channels are still not decomposable into X channels.
The main idea of our proof is that employing Gibbs states as catalysts simplifies the problem of decomposing TO channels into that of decomposing energy-preserving unitary operators. 
As long as such X operations still allow for full thermalisation, one can use Gibbs states catalytically, i.e. they can be restored to their original state by the use of additional thermal baths.

The above Theorem~\ref{theorem: GCETO is TO} states that for any TO channel, there always exists a GC-X process that emulates it, yet the explicit construction is not granted by the proof.  
Nevertheless, an exact and explicit construction can be established for the decomposition by GC-ETO.
In Ref.~\cite{Reck94TwoLvl}, a decomposition of \emph{generic} $d\times d$ unitary operator into a product of length $ \frac{d(d-1)}{2} $ two-level unitaries is given.
We observe that the argument also translates to \emph{energy-preserving} unitaries. 
This is true, because any energy-preserving unitary can be first written as a direct sum of unitaries $\bigoplus_i U_i$ with each $U_i$ acting on a fully degenerate energy subspace. 
Using the construction for generic unitaries, we can decompose each $U_i$ into a series of two-level unitaries. Each of these two-level unitaries would be energy-preserving, and therefore the full decomposition corresponds to an ETO sequence.

Building on Theorem~\ref{theorem: GCETO is TO}, we formulate our second main result of this section: when arbitrary catalysts are allowed, the hierarchy of thermodynamic free operations---MTO, ETO, and TO, which were previously studied independently---collapses. 
Such a statement can be subtle due to the various types of catalysis~\cite{LipkaBartosik2023CatReview}, leading to significantly different state transition conditions. 
In particular, without the exact emulation of TO in Theorem~\ref{theorem: GCETO is TO}, there is no guarantee that the catalyst can be exactly recovered, even when the GC-ETO or GC-MTO approximates TO arbitrarily well.
Nevertheless, since our result provides a direct and exact decomposition of the unitary corresponding to TO, it holds for any catalytic type (strict or correlated), as long as the definition remains consistent across different thermodynamic free operations. 

\begin{theorem}[original result]\label{theorem: CETO is CTO} 
	$\setCTO(\rho_{S}) = \setCETO(\rho_{S}) = \setCMTO(\rho_{S})$ for all $\rho_{S} \in\mD_{S}$, where CX may stand for either strict or correlated catalytic X operations.
\end{theorem} 

\begin{proof}
	By the inclusion in Remark~\ref{remark: hierarchies}, we immediately have $ \setCETO(\rho_{S}), \setCMTO(\rho_{S}) \subset\setCTO(\rho_{S})$.
	For the other direction, consider $ \sigma_{S}\in\setCTO(\rho_{S}) $: this implies the existence of a catalyst state $ \tau_C $, such that $ \sigma_{SC} \in \setTO(\rho_{S}\otimes\tau_C) $ and $\Tr_{C}[\sigma_{SC}] = \sigma_{S} $. 
	Furthermore, the catalyst recovery condition, which depends on the definition of catalysis, is imposed; see Definitions~\ref{definition: strict catalysis} and~\ref{definition: correlated catalysis}.
	Using Theorem~\ref{theorem: GCETO is TO}, we find that $ \sigma_{SC}\in \setGCETO(\rho_{S}\otimes\tau_{C}) $, or equivalently, $ \sigma_{SCC'}\coloneq\sigma_{SC}\otimes\gbs_{C'}\in \setETO(\rho_{S}\otimes\tau_C\otimes\gbs_{C'}) $ for some $ H_{C^{\prime}} $. 
	Any catalyst recovery condition imposed on $\sigma_{SC}$ is satisfied by $\sigma_{SCC'}$.
	Therefore, we have that $ \sigma_{S}\in\setCETO(\rho_{S}) $, and hence $ \setCTO(\rho_{S})\subset\setCETO(\rho_{S}) $. The same proof strategy applies to MTO.
\end{proof}

The equivalence for correlated catalysis is significant for the programme of using correlating catalysis to make thermodynamic resource theories reversible~\cite{Shiraishi2025CorrCat}.
First, it is known that GP with correlated catalysis achieves reversibility, i.e. state transformation is always determined by a single complete monotone~\cite{Shiraishi2021GP}.

In addition, GPC with correlated catalysis is shown to be almost reversible~\cite{Shiraishi2024GPCCat}.
To be precise, suppose that $H_{S} = \sum_{i}E_{i}\dm{i}_{S}$ with all energy eigenvalues $E_{i}$ being commensurable. 
Then there exists $\mathsf{E}$, such that all $E_{i}$ are integer multiples of $\mathsf{E}$.
In such cases, for each quantum state $\rho_{S}$ there is a set of durations, called period, such that the state returns to itself after these durations, i.e. $\tau$ is a period of $\rho_{S}$ if $e^{-i\tau H_{S}}\rho_{S}e^{i\tau H_{S}} = \rho_{S}$.
For symmetric states, any $\tau$ works; for any asymmetric state, $\tau = \frac{2\pi}{\mathsf{E}}$ is a period.
If the smallest such period for $\rho_{S}$ is $\frac{2\pi}{\mathsf{E}}$, we have $\mT_{\mathrm{CGPC}}(\rho_{S}) = \mT_{\mathrm{CGP}}(\rho_{S})$, which is the set of all states with the free energy smaller than that of $\rho_{S}$.

On the other hand, the other part of the hierarchy, namely $\MTO{S}{S}, \ETO{S}{S} \subsetneq \TO{S}{S}$ collapses by Theorem~\ref{theorem: CETO is CTO}.
If, furthermore, $\setCTO(\rho_{S}) = \mT_{\mathrm{CGPC}}(\rho_{S})$ can be shown for correlated catalysis, all thermodynamic resource theories in the hierarchy Remark~\ref{remark: hierarchies} would (almost) become catalytically equivalent and reversible. 

\section{Concluding remarks}
We have demonstrated catalytic advantages in action within various resource theories of thermodynamics.
Furthermore, fulfilling our initial objective, we identified the memory-effect as a key mechanism underpinning auxiliary system assistance.
This insight arises from two key observations: firstly, snapshots of the catalytic evolution, and secondly, our proof that catalysis enables memory-restricted operations to achieve the full scope of thermodynamic transformations.

The collapse of a hierarchy (or part of the full hierarchy stated in Remark~\ref{remark: hierarchies}) has been an important open problem in quantum thermodynamics, particularly for those adopting resource-theoretic approaches. 
This stems from the fact that the choice of free operations yields inequivalent thermodynamic theories, especially in cases involving energy coherence, i.e. beyond the semi-classical setting. 
Hence, finding a setting in which these discrepancies vanish amounts to reconciling different paradigms of quantum thermodynamics. 
Interestingly, the other part of the hierarchy (between GP and GPC) is also (almost) resolved using catalysts~\cite{Shiraishi2024GPCCat}, leaving only the gap between TO and GPC unaddressed.

One understated feature of our results is that GC-ETO and GC-MTO implement TO channels catalytically, not merely TO state transformations. 
In other words, the emulation of TO by GC-ETO (or GC-MTO) works regardless of the input system state. 
By contrast, GPC achieves GP state transformations catalytically but cannot implement GP channels themselves. 
This impossibility is most apparent from a recent result~\cite{Tajima2025GPInfinite} showing that some GP channels require an infinite amount of energy coherence in the ancilla, even when that ancilla need not be catalytic. 
The distinction between catalytic channels and catalytic state transformations is a central topic in Chapter~\ref{chapter: resource broadcasting}.

In Chapter~\ref{chapter: QDP}, we identify other underlying mechanisms of catalysis and leverage these to design a novel algorithmic paradigm for quantum computing. 
The memory-effects investigated in this chapter will prove particularly instructive, as our paradigm employs auxiliaries as a memory---storing the operations performed upon them---much like catalysis within ETO and MTO.

\chapter{The importance of fine-tuning and catalysts as a resource broadcasting seed}\label{chapter: resource broadcasting}

In this chapter we identify two additional factors that explain catalytic advantages, i.e. how catalysts can enable state transformations or channels that are impossible without them. 
These factors correspond to two paradigms of catalysis: processes that implement state transformations and processes that implement channels. 
Section~\ref{section: fragility of catalysis} introduces these two paradigms. 
A catalytic implementation of a channel always yields a catalytic state transformation, but not vice versa.
The main result of Section~\ref{section: fragility of catalysis} shows that the converse---that every catalytic state transformation implements a catalytic channel---holds if we impose an additional noise-robustness requirement. 
This demonstrates that one source of catalytic power is the ability to fine-tune the catalyst state based on exact knowledge of the system state, a capability that can make catalytic state transformations strictly stronger than catalytic channels.
In Section~\ref{section: resource broadcasting} we show that catalytic channels (under certain assumptions) necessarily enable resource broadcasting: the catalyst acts as a seed that can be broadcast, which constitutes the second power of catalyst we identified.
This connection to resource broadcasting also serves as a critical technical tool for establishing main theorems in the next section. 
Section~\ref{section: composition and resource broadcasting} then presents a no-go theorem for resource broadcasting---and thereby for catalytic channels---in one class of resource theories and an existence theorem in another. 
To obtain these results we develop a new perspective on resource theories that focuses on composing the subsystem resource theories into a theory for the composite system. 
This viewpoint lets us prove the no-go and existence theorems in a very general setting, independent of the specific physical resource.
We close the chapter with concluding remarks in Section~\ref{section: concluding remarks robust catalysis}.

My original results in this section are: Theorems~\ref{theorem: rc is catchan}, \ref{theorem: catchan and broadcasting}, \ref{theorem: mincomp no catchan}, \ref{theorem: broadcasting possible}, Remarks~\ref{remark: broadcasting to cat chan}, \ref{remark: minmax compositions}, Lemma~\ref{lemma: RNG CRNG}, Observation~\ref{observation: limited subspace theories}, and Proposition~\ref{proposition: strict robust catalysis no-go}.
 
\section{Fragility of catalysis and catalytic channels}\label{section: fragility of catalysis}

So far, catalysis is defined as \emph{catalytic transformations} as in Definitions~\ref{definition: strict catalysis} and~\ref{definition: correlated catalysis}, i.e., as a process with three interconnected elements: initial system state $\rho_{S}$, catalyst $\tau_{C}$, and a free channel $\mE\in\fro{SC}{S'C}$, which output the final state with the catalyst state intact. 
For example, consider strict catalysis in Definition~\ref{definition: strict catalysis} which outputs $\rho'_{S}\otimes\tau_{C} = \mE(\rho_{S}\otimes\tau_{C})$. 
Although the final catalyst state $\tau_{C}$ is retrieved as an independent system, tensored with $\rho'_{S}$, the process $\mE$ generally acts non-trivially and collectively on both systems $S$ and $C$.
That is because the operation of the form $\mE = \mE_{S}\otimes\mE_{C}$ can only prepare $\rho'_{S} = \mE_{S}(\rho_{S})$ regardless of the catalyst state used. 
This intricacy of the process renders catalysis sensitive to the noise affecting the initial setup. 
The effective operation acting on the catalyst reads 
\begin{align}\label{eq: catalyst effective channel}
	\tau_{C} \mapsto \mE_{C}(\tau_{C}) \coloneq \Tr_{S'}\left[\mE(\rho_{S} \otimes \tau_{C})\right],
\end{align}
and the process can be made catalytic by choosing the catalyst state $\tau_{C}$ to be the fixed-point of the map $\mE_{C}$~\cite{boes2020passing, LipkaBartosik2023Temperature, AdOJ2024CavityQED, SonEntGen}.
Since $\mE$ is not decomposable into local operations, Eq.~\eqref{eq: catalyst effective channel} is affected by any change in any of the three elements in the initial setup: initial system state $\rho_{S}$, catalyst $\tau_{C}$, and the operation $\mE$.

We can upper-bound the error in the final catalyst state using the linearity of quantum channels. 
The desired outcome is $\Tr_{S'}[\mE(\rho_{S}\otimes\tau_{C})] = \tau_{C}$, for an initial system state $\rho_{S}$, catalyst state $\tau_{C}$, and a free operation $\mE\in\fro{SC}{S'C}$.
Now consider the case where a noisy channel is used instead of $\mE$, which corresponds to panel (b) in Figure~\ref{fig: chapter 5 illustration}. 
Suppose that $\mE_{\epsilon}\in\CPTP{SC}{S'C}$ is a channel such that $\frac{1}{2}\| \mE - \mE_{\epsilon}\|_{\Tr}\leq \epsilon$ for some $\epsilon>0$, i.e. the one that is $\epsilon$-close to $\mE$.
Here, we use the trace norm $\|\cdot\|_{\Tr}$, but any other channel norm, including the diamond norm $\|\cdot\|_{\diamond}$, may alternatively be used; see Definition~\ref{definition: channel distances} for definitions. 
Then, the upper bound 
\begin{align}\label{eq: catalyst error when channel error}
	\frac{1}{2}\left\| \Tr_{S'}\left[ \mE(\rho_{S}\otimes\tau_{C})\right] - \Tr_{S'}\left[\mE_{\epsilon}(\rho_{S}\otimes\tau_{C})\right]\right\| \leq \frac{1}{2}\left\|  \mE(\rho_{S}\otimes\tau_{C})- \mE_{\epsilon}(\rho_{S}\otimes\tau_{C})\right\|_{1}\leq \epsilon,
\end{align}
follows from the data-processing inequality for the trace norm.
The LHS of Eq.~\eqref{eq: catalyst error when channel error} is the trace distance between the initial (ideal) catalyst state $\tau_{C}$ and the final state obtained after applying the noisy channel $\mE_{\epsilon}$, and it is upper-bounded by the noise in the channel itself. 
This scenario appears somewhat hopeless to salvage. 
Even with the simplest noise model---small, independent noise occurring after the ideal channel $\mE$---the catalyst state inevitably suffers disturbance. 
Furthermore, without appropriate error correction, this disturbance tends to accumulate over multiple uses, eventually rendering the catalyst ineffective. 
Currently, there is no generally applicable method to prevent this error propagation.

Next, consider a catalytic process starting from a faulty catalyst $\tau^{\epsilon}_{C}$ [panel (d) in Figure~\ref{fig: chapter 5 illustration}], such that $\frac{1}{2}\|\tau_{C} - \tau^{\epsilon}_{C}\|_{1}\leq\epsilon$. 
Data-processing inequality gives
\begin{align}\label{eq: catalyst error when catalyst error}
	\frac{1}{2}\left\| \Tr_{S'}\left[ \mE(\rho_{S}\otimes\tau_{C})\right]  - \Tr_{S'}\left[ \mE(\rho_{S}\otimes\tau^{\epsilon}_{C})\right] \right\|_{1} \leq \frac{1}{2}\left\| \tau_{C}  - \tau^{\epsilon}_{C} \right\|_{1},
\end{align}
by considering $\Tr_{S'}[ \mE(\rho_{S}\otimes\cdot)] $ as the data-processing process.
Again, the final error is upper-bounded by the initial error.
However, compared to the noisy channel $\mE$, this noisy initial catalyst scenario is easier to handle. 
The key difference lies in the fact that when catalysis is repeated, the same catalyst system is reused without a fresh source of noise.
After one round of catalysis, the initial noisy catalyst state $\tau^{\epsilon}_{C}$ is transformed into a new state $\tilde{\tau}^{\epsilon}_{C}$, and the latter is no noisier than the former, by Eq.~\eqref{eq: catalyst error when catalyst error}.
If the new catalysis begins with $\tilde{\tau}^{\epsilon}_{C}$, the resulting catalyst state would still be at least as good as the initial one. 
In other words, the error does not accumulate, and after each round of catalysis, the final system state can be obtained with $\epsilon$ error without further degradation. 
Therefore, even without additional measures, catalysis is robust against initial error in the catalyst state. 

\begin{figure}
	\centering
	\includegraphics[width=\linewidth]{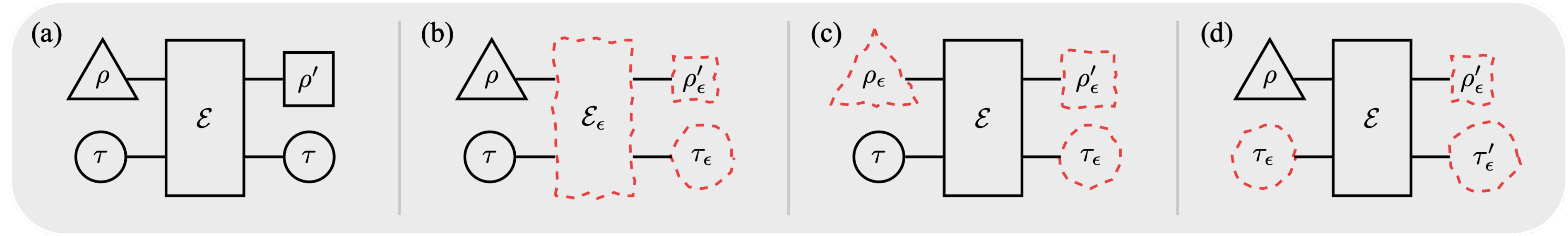}
	\caption{Illustrations of noisy catalysis.
		Panel (a) depicts the ideal scenario, where a catalyst ($\tau$) facilitates a quantum process (transforming $\rho$ into $\rho'$) to its exact initial state at the end of the process. 
		The other three panels suppose that either the channel [panel (b)], the system initial state [panel (c)], or the catalyst initial state [panel (d)] is noisy.
		In all cases, the initial noise affects the final states of both system and catalyst, albeit to a degree not exceeding the initial noise.
		Figure adapted from Figure~1 of Ref.~\cite{SonRCRB}.  
}\label{fig: chapter 5 illustration}
\end{figure}

Finally, when the initial system state is noisy as in panel (c) of Figure~\ref{fig: chapter 5 illustration}, data-processing inequality can again be used to yield
\begin{align}\label{eq: catalyst error when system error}
	\frac{1}{2}\left\| \Tr_{S'}\left[ \mE(\rho_{S}\otimes\tau_{C})\right]  - \Tr_{S'}\left[ \mE(\rho^{\epsilon}_{S}\otimes\tau_{C})\right] \right\|_{1} \leq \epsilon,
\end{align}
for any $\rho^{\epsilon}_{S}$, such that $\frac{1}{2}\|\rho_{S} - \rho^{\epsilon}_{S}\|_{1}\leq \epsilon$, which implies that the noise in the system state translates to the catalyst state. 

This last scenario is the most interesting.
In this case, the catalyst state effectively experiences a channel $\Tr_{S'}[\mE(\rho^{\epsilon}_{S}\otimes\cdot)]$ at each round of catalysis, and the error accumulates, potentially driving the catalyst far from its original state. 
In the worst case, the accumulated error on the catalyst may grow linearly with the number of repetitions, leading to eventual breakdown of catalysis.

\begin{example}[accumulating errors on the catalyst] 
	It suffices to demonstrate that catalyst continues to degrade after the first round of catalysis. 
	Consider a qutrit system and a qubit catalyst undergoing a joint unitary evolution. 
	Denote the eigenvalues of the system and catalyst to be:
	\begin{equation}
		p_S = (p_1,p_2,p_3), \qquad q_C = (q_1,q_2).
	\end{equation}
	The unitary swaps the eigenstates corresponding to $p_1q_1 \leftrightarrow p_2q_2$ and $p_2q_1 \leftrightarrow p_3q_2$. 
	This operation is catalytic whenever $(p_1+p_2)q_1 = (p_2+p_3)q_2$. 
	Such toy examples are useful for illustrations, and have been used, e.g. in Appendix B of \cite{boes2020passing}. 
	
	Now, suppose that in the first round, we have a noisy system state $p_\varepsilon = (p_1-\varepsilon,p_2,p_3+\varepsilon)$ for some $\varepsilon >0$. 
	This leads to a final degraded catalyst $q' = (q_1',q_2')$, such that
	\begin{equation}
		q_1' = p_2q_2+(p_3+\varepsilon)q_2+ (p_3+\varepsilon)q_1 = q_1 + \varepsilon.
	\end{equation}
	From normalisation, we also have that $q_2' = q_2 - \varepsilon$. In other words, the full amount of error $\varepsilon$ has propagated into the catalyst. 
	Next, suppose that in a second round, we have another noisy system state $p_{-\varepsilon} = (p_1+\varepsilon,p_2,p_3-\varepsilon)$. 
	Under the action of the same catalytic unitary, the catalyst further degrades into $q''_C = (q''_1, q''_2)$, where
	\begin{align}
		q''_1 = p_2q_2' + (p_3-\varepsilon)q_2' + (p_3-\varepsilon)q_1' &= q_1 - \varepsilon (p_2+p_3+q_2+q_1-p_3)\nonumber\\ 
		&= q_1 - \varepsilon(1+p_2).
	\end{align}
	In summary, the error accumulated almost linearly during two rounds of catalysis, as we anticipated. 
\end{example}

To prevent this error accumulation, a robustness condition needs to be established and imposed.
We define robust catalysis as a catalytic process that is robust against small system-preparation noise, which is the most relevant and interesting type of noise, as explained above.
The formal definition is as follows. 
\begin{definition}[robust catalysis]\label{definition: robust catalysis}
	Given an initial state $\rho_{S}$ and a robustness parameter $\epsilon>0$, a channel $\mE\in\fro{SC}{S'C}$ implements $(\rho,\epsilon)$-\textit{robust catalysis} if there exists a catalyst state $\tau_{C}\in\mD_{C}$ such that, for all system states $\sigma_{S}$ with $\frac{1}{2}\| \sigma_{S} - \rho_{S}\|_{1}\le\epsilon$, the catalyst is recovered exactly, i.e. $\Tr_{S'}[\mE(\sigma_{S}\otimes\tau_{C})] = \tau_{C}$. 
	We say that $(\rho,\epsilon)$-robust catalysis is \textit{strict} when no correlation is established between the system and the catalyst, i.e., $\mE(\sigma_{S}\otimes\tau_{C}) = \tilde{\sigma}_{S'}\otimes\tau_{C}$.
\end{definition}
Definition~\ref{definition: robust catalysis} relaxes the catalyst's fine-tuned dependence on the initial system state, addressing a key conceptual weakness in the theory of catalysis. 
The definition makes no assumption about the system's state \emph{after} the process, although we already know that it will be at most $\epsilon$-far from the desired final state.
Notably, the robustness parameter $\epsilon$, capturing the degree of initial state preparation errors, can be arbitrarily small. 
The only truly stringent requirement is that the catalyst be returned to its initial state without error. 
This is a common requirement in the literature on catalytic transformations, necessary to prevent embezzlement~\cite{vanDam03_embezzle,leung2014characteristics,ng2015limits}. 

The extreme case of the $(\rho,\epsilon)$-robust catalysis in Definition~\ref{definition: robust catalysis} is when $\epsilon = 1$, which becomes completely agnostic to the input state $\rho$. 
This input-agnostic approach fundamentally transforms how catalysis is defined, i.e. from \emph{catalytic transformations} to \emph{catalytic channels}.
In other words, we can define a set of catalytically free operations similarly to the set of free operations, unlike catalytic transformations, which must be investigated only at the level of state transformation, i.e. pairs of states $(\rho,\rho')$ such that $\rho\to\rho'$. 
The concept of catalytic channels has been studied in the literature for a setting where the set of free operations $\fro{SC}{SC}$ is the set of all unitary operations~\cite{Boes2018CatRandom, Lie2021Generecity, Lie2021CatalyticRandomness}.
Here, we generalise this notion to any set of free operations $\fro{SC}{S'C}$.
\begin{definition}[catalytic channel]\label{definition: catalytic channels}
	A channel $\tilde{\mE}\in\CPTP{S}{S'}$ is a catalytic channel if there exists a dilation $\mE\in\fro{SC}{S'C}$ and a catalyst state $\tau_{C}\in\mD_{C}$, such that 
	\begin{align}
		\Tr_{C}[\mE(\rho_{S}\otimes\tau_{C})] &= \tilde{\mE}(\rho_{S}),\\
		\Tr_{S'}[\mE(\rho_{S}\otimes\tau_{C})] &= \tau_{C},
	\end{align}
	for all input states $\rho_{S}\in\mD_{S}$.
\end{definition}

Surprisingly, this extreme limit in Definition~\ref{definition: catalytic channels} turns out to be equivalent to a seeming much less stringent definition of robust catalysis with any $\epsilon >0$. 
\begin{theorem}[original result]\label{theorem: rc is catchan}
	For any $\rho_S\in\mD_{S}$ and any $\epsilon>0$, a channel $\mE\in\fro{SC}{S'C}$ implements a $(\rho,\epsilon)$-robust catalysis with catalyst's state $\tau_{C}$, if and only if $\tilde{\mE}(\cdot_{S}) = \Tr_{C}[ \mE (\cdot_{S}\otimes\tau_{C})]$ is a catalytic channel.  
\end{theorem}
\begin{proof}
	If $\tilde{\mE}$ is a catalytic channel, it implements a robust catalysis, for any state $\rho_S$ and any robustness parameter $\epsilon>0$. 
	We thus only need to prove the converse, and we do it by contradiction.
	Suppose that $\mE$ implements a $(\rho,\epsilon)$-robust catalysis, but $\tilde{\mE}$ is not a catalytic channel. 
	This means that there exists at least one state, say $\eta_{S}\in\mD_{S}$, such that
	\begin{align}
		\tau_C=\Tr_{S'}[ \mE (\rho_{S}\otimes\tau_{C})] \neq \Tr_{S'}[ \mE (\eta_{S}\otimes\tau_{C})]\eqcolon\tau_C'.
	\end{align}
	Let us then consider the state $\tilde\rho_{S} \coloneq (1-\epsilon)\rho_{S} + \epsilon\eta_{S}$. 
	By construction,  we have $\frac{1}{2}\| \tilde\rho_{S} - \rho_{S}\|_{1}=\frac{1}{2}\| \epsilon\eta_{S}- \epsilon\rho_{S}\|_{1}\le\epsilon$. 
	Nevertheless, by linearity, $\Tr_{S'}[ \mE (\tilde\rho_{S}\otimes\tau_{C})] = (1-\epsilon)\tau_{C} + \epsilon\tau'_{C} \neq\tau_{C}$, contradicting the assumption that $\mE$ implements an $\epsilon$-robust catalysis for $\rho_S$.
\end{proof}
Theorem~\ref{theorem: rc is catchan} clarifies that all forms of catalysis, other than catalytic channels, are inevitably fine-tuned to a very specific initial state of the system, and risk degrading the catalyst whenever the system's state is not prepared with strictly infinite precision.
It also offers a mathematically streamlined pathway towards robust catalysis, as one may directly examine the structure of catalytic channels rather than inspect different catalyst states and fine-tuned strategies for state transformations.

Even though both catalytic channels and universal catalysts share universality with respect to input system states, they should not be conflated.
The latter concept refers to a catalyst state that can be used for any catalytic transformation, assuming the operation can be appropriately fine-tuned. 
The intuition behind such catalysts is simple: the universal state just needs to subsume all catalyst states that may be required.
Specifically, one can coarse-grain over all catalytic transformations and collect all associated catalysts into a single, large universal catalyst~\cite{Datta2024_universal}.
Alternatively, the asymptotic reversibility of a resource theory---if it exists---can be leveraged.
By selecting sufficiently many copies of some quantum state to be a universal catalyst, this catalyst can be pre-processed into many copies of the initial system state, transformed into many copies of the desired final state, and post-processed back into the original catalyst state, all without dissipating (almost) any resources~\cite{LipkaBarosik2021_universal}. 
In both approaches, however, the process is not robust against noise in the preparation of the initial system state.%
\footnote{To make such universal catalysts robust, one may envisage a protocol where the initial state can be known by some free operations. 
In Refs.~\cite{Watanabe2024BlackBox, Watanabe2025WorkExt}, an interesting protocol is presented. 
Given asymptotically many copies of the initial state unknown to us, an optimal resource extraction can be performed by measuring some of the copies to learn the initial state and extracting the resource from the rest of the copies.
This process is indeed independent of different choices of the initial state.
However, the measurement sampling cost is negligible only when working within the asymptotic setting, which negates the need of catalysis in asymptotically reversible theories that these universal catalysts are considered.}
Once the operation is fixed, any deviation from the intended input can break the catalytic behaviour.
Catalytic channels are therefore anticipated to be more reliably implementable, as they guarantee reusability of the catalyst even in the presence of such errors.

Theorem~\ref{theorem: rc is catchan} prompts us to examine whether catalytic channels can provide any meaningful advantage. 
Indeed, special cases have been studied in the literature. 
Initially, catalytic channels were defined and analysed for the unitary theory, i.e. when $\fro{X}{X}$ is the set of unitary channels for system $X$~\cite{Boes2018CatRandom, Lie2021Generecity, Lie2021CatalyticRandomness}.
In this setting, the most straightforward catalytic channels are mixed unitary channels. 
Suppose that the catalyst $\tau_{C} = \sum_{i}p_{i}\dm{i}_{C}$ for some basis $\{\ket{i}_{C}\}_{i}$.
By choosing the unitary operator $U = \sum_{i}U_{i}\otimes\dm{i}_{C}$, the catalytic channel 
\begin{align}\label{eq: catalytic channel mixed unitaries}
	\Tr_{C}\left[U \left( \rho_{S}\otimes\tau_{C} \right)U^{\dagger}\right] = \sum_{i}p_{i} U_{i}\rho_{S} U_{i}^{\dagger}, 
\end{align}
is obtained with the guarantee of the exact recovery $\Tr_{S}[U ( \rho_{S}\otimes\tau_{C} )U^{\dagger}] = \tau_{C} $ for any $\rho_{S}\in\mD_{S}$.
In other words, any mixed unitary channel (i.e. a channel that can be written as a convex combination of unitary channels) can be implemented using a catalyst state with its eigenvalues coinciding with the convex coefficients. 

It turns out that catalytic channels can also implement channels that are not in the form of Eq.~\eqref{eq: catalytic channel mixed unitaries}~\cite{HierarchyManuscript}. 
They form a set that is strictly larger than the set of mixed unitaries and strictly smaller than the set of strongly factorisable channels (also known as noisy operations~\cite{Gour15_TRTreview}).%
\footnote{Interestingly, the free unitary $U\in\fro{SC}{SC}$ can implement a catalytic channel if and only if its partial transpose $U^{\transp_{S}}$ is also unitary.
When the dimensions for $S$ and $C$ are the same, this class of unitary is equivalent to dual unitaries (multiplied by a SWAP operator)~\cite{Bertini2019DualUnit}, or sometimes called $\Gamma$-dual unitaries~\cite{Rather2024GammaDualUnit}, which are actively studied in the context of chaos and operator spreading.} 

Catalytic channels similar to Eq.~\eqref{eq: catalytic channel mixed unitaries} can be constructed more generally; if the set of free operations is not convex but not necessarily unitary, and if controlled operations are free, catalysts can be used as a source of randomness that is encoded in its spectrum; see Ref.~\cite{Lie2023Delocalized} for more general discussion on this topic. 

Alternatively, catalytic channels also emerge naturally when the catalyst is a free state; in fact, this catalysis is also strict. 
By Definition~\ref{definition: free states}, there always exists a free operation $\mF\in\fro{C}{C}$, which prepares a fixed free state $\gamma_{C}\in\frs{C}$.
By applying $\mF$ to the catalyst after any operation on system and catalyst, the resulting operation becomes catalytic, i.e. $(\id_{S}\otimes\mF)\circ\mE (\rho_{S}\otimes\gamma_{C}) = \rho'_{S}\otimes\gamma_{C}$ for any $\rho_{S}\in\mD_{S}$.
Typically, the catalytic channel $\mE(\cdot \otimes\gamma_{C})$ is a free channel when $\gamma_{C}\in\frs{C}$ is a free state and $\mE\in\fro{SC}{SC}$ is a free operation.
However, this is not always the case, as observed in some resource theories of athermality~\cite{Korzekwa2022MTO2, Czartowski2023ThermalRecall, Son2024_CETO, Son2024hierarchy}; see Section~\ref{section: a hierarchy collapses} how even Gibbs states can expand the set of free operations.
In such cases, catalytic channels exploit the memory effect of the catalyst, overcoming the inherent Markovianity of the resource theory.

\section{Resource broadcasting and catalytic channels}\label{section: resource broadcasting}

To study more intricate advantages from catalytic channels, beyond the aforementioned special cases of catalytic randomness and non-Markovianity boost, a few additional axioms are imposed on resource theories of interest.  
\begin{kaobox}[frametitle = Axioms for convex resource theories with free tensor product and partial trace]
	\begin{enumerate}[topsep=3pt,itemsep=0ex,leftmargin=*,label={(\makebox[1.4em][c]{A\arabic*})}]
		\item $\rho_{A}\otimes\rho_{B}\in\frs{AB}$ is a free state whenever $\rho_{A}\in\frs{A}$ and $\rho_{B}\in\frs{B}$ are free states
		\item If $\rho_{AB}\in\frs{AB}$ is a free state, then $\Tr_{B}[\rho_{AB}]\in\frs{A}$ and $\Tr_{A}[\rho_{AB}]\in\frs{B}$ are free states
		\item the set of free states $\frs{X}$ is a convex set for any system $X$
	\end{enumerate}
\end{kaobox}
(A1) and (A2) set basic rules for composing free state sets in larger spaces.
(A2) is implied by the tensor product structure of resource theories: it follows from the third axiom in the definition of free operations (Definition~\ref{definition: free operations}) and the second axiom in the tensor product structure (Definition~\ref{definition: tensor product structure}).
(A1) serves to exclude a phenomenon known as (super-)activation~\cite{Shor2003Superactivation, Navasecues2011Activation, Palazuelos2012Superactivation}, whereby even free states can activate catalytic channels. 
Similarly, the convexity assumption (A3) ensures that any classical randomness is already incorporated into the definition of free operations, negating the need for catalytic randomness.  

These axioms do not uniquely characterise the composite free states $\frs{AB}$ given the subsystem sets $\frs{A}$ and $\frs{B}$. 
Nevertheless, they are standard assumptions that hold in a broad range of quantum resource theories, including those of entanglement, athermality, coherence, asymmetry, and magic.
Furthermore, these axioms leave considerable flexibility in the choice of free operations, the only constraint being the golden rule (Definition~\ref{definition: golden rule of resource theories}).
A canonical way of defining free operations from free states is via completely resource non-generating (CRNG) operations (Definition~\ref{definition: CRNG}), which comprise the full set of channels that cannot generate any resource from free states, even when acting on a marginal state of a larger free state. 
An additional merit of CRNG operations is that they are maximal under assumptions (A1)--(A3) and the golden rule.
For these reasons, we will focus of CRNG operations from now on. 

Since catalysis refers to the activation of processes that would otherwise be impossible, robust catalysis must implement a catalytic channel $\tilde{\mE}\in\CPTP{S}{S'}$ that is not free. 
To ascertain that $\tilde{\mE}$ is not free, it suffices to find a process that maps a free state $\gamma_{S}$ into a resourceful state $\tilde{\mE}(\gamma_{S})\notin\frs{S'}$. 
Conversely, if $\tilde{\mE}$ maps free states $\frs{S}$ to free states $\frs{S'}$, it is resource non-generating (RNG). 
Furthermore, if every catalytic channel $\tilde{\mE}$ is RNG, so are their extensions $\id_{A} \otimes \tilde{\mE}$, as these are themselves catalytic channels.
This in turn implies that all $\tilde{\mE}$ are \emph{completely} resource non-generating.
In this case, if the set of free operations is chosen to be CRNG, catalytic channels offer no advantage.

To examine the power of catalytic channels, let us introduce a related concept with the name \emph{resource broadcasting}.
The name broadcasting comes from the fact that the channel propagates some resource from one system to another system, while leaving the original one completely intact.
This notion is inspired by, but not identical to, quantum state broadcasting (Theorem~\ref{theorem: no broadcasting}); see also Refs.~\cite{superbroad1,superbroad2,Barnum2007Broad, Piani2008Broad, Parzygnat2024VirtualBroad}.
Resource broadcasting is an interesting phenomenon that has attracted attention from the community~\cite{Lu2013FisherBroad, Marvian2019Broadcasting, Lostaglio2019Broadcasting, Yang_2021Broad, Zhang2024_magicnobroad}.
Nevertheless, in this thesis, this concept is introduced primarily as a technical tool for establishing results for robust catalysis; the connection between the two will become more apparent in Theorem~\ref{theorem: catchan and broadcasting} below. 

There are variations in how resource broadcasting is defined in the literature. 
Here, we take the most lenient one as our definition.
\begin{definition}[resource broadcasting]\label{definition: resource broadcasting}
	A broadcasting channel $\mB\in\fro{A}{AB}$ is a free operation such that 
	\begin{align}
		\Tr_{B}[\mB(\rho_{A})] &= \rho_{A},\\
		\Tr_{A}[\mB(\rho_{A})]&\notin\frs{B},
	\end{align}
	i.e. producing a non-free state in system $B$ for some state $\rho_{A}\in\mD_{A}$.
\end{definition}

Setting $A$ as the catalyst system $C$ and $B$ as the system of interest $S$ highlights the similarity to catalytic channels. 
Consider the application of the catalytic channel $\tilde{\mE}$ to a free system state $\gamma_{S}\in\frs{S}$, yielding a resourceful state $\tilde{\mE}(\gamma_{S}) = \Tr_{C}[\mE(\gamma_{S}\otimes\tau_{C})]\notin\frs{S'}$; this process can be viewed as broadcasting the catalyst $\tau_{C}$.
We establish that, under some assumptions, the two concepts accompany one another.

 \begin{theorem}[original result]\label{theorem: catchan and broadcasting}
	Suppose that a resource theory having all CRNG operations as the free operation satisfies axioms (A1)--(A3). 
	If there exists a non-free catalytic channel $\tilde{\mE}\notin\fro{S}{S'}$ with some catalyst $\tau_{C}$, there must exist a free broadcasting channel $\mB\in\fro{C}{S'C}$ defined as in Definition~\ref{definition: resource broadcasting}.
	Conversely, given a broadcasting channel $\mB\in\fro{C}{S'C}$, it is possible to construct a non-free catalytic channel $\Tr_{C}\circ\mB\circ\Tr_{S}[\cdot\otimes\tau_{C}]\notin\fro{S}{S'}$.
\end{theorem}
\begin{proof}
	Suppose that $\tilde{\mE}\notin\fro{S}{S'}$ is a non-free catalytic channel with a dilation into free channel $\mE\in\fro{SC}{S'C}$ and a catalyst $\tau_{C}$.
	Since the set of free operations $\fro{S}{S'}$ is defined to be CRNG, $\tilde{\mE}$ must be able to transform a free state into a resourceful one, i.e. $\exists\gamma_{S}\in\frs{S}$ such that the output $\rho_{SC} = \mE(\gamma_{S}\otimes\tau_{C})$ satisfies the catalytic condition, $\Tr_{S}[\rho_{SC}] = \tau_{C}$ and the resourcefulness of the final state $\Tr_{C}[\rho_{SC}]\notin\frs{S'}$.
	Now, define the free channel $\mB\in\fro{C}{S'C}$ by $\mB(\cdot) = \mE(\gamma_{S}\otimes\cdot)$.
	This construction makes $\mB$ a broadcasting channel that maps $\tau_{C}$ into $\rho_{SC}$.
	
	For the converse direction, suppose that the free channel $\mB\in\fro{C}{S'C}$ broadcasts $\tau_{C}$. 
	Then the channel $\mE = \mB\circ\Tr_{S}\in\fro{SC}{S'C}$ is free, as it is a concatenation of two free operations.
	A catalytic channel can be constructed as $\tilde{\mE}(\rho_{S}) = \Tr_{C}[\mE(\rho_{S}\otimes\tau_{C})]$ and it is not CRNG since $\tilde{\mE}(\gamma_{S}) = \Tr_{C}[\mB(\tau_{C})]\notin\frs{S'}$ is not free.
\end{proof}

Despite the equivalence in the existences of two phenomena, catalytic channels and resource broadcasting exhibit subtle differences. 
In real experimental setups, implementing the broadcasting channel $\mB\in\fro{C}{S'C}$ requires an auxiliary state $\gamma_{S'}\in\frs{S'}$, which is assumed to be fixed. 
Unlike catalytic channels, the broadcasting process may fail to preserve the original state $\tau_{C}$ if the auxiliary state $\gamma_{S'}$ is perturbed in any way. 
This point does not play a role in the proof of Theorem~\ref{theorem: catchan and broadcasting}, because we assume that the channel $\mB\in\fro{C}{S'C}$ can be implemented faithfully. 

In fact, the implication from the existence of broadcasting channels to the existence of catalytic channels holds even without some of the assumptions.
\begin{remark}[original result]\label{remark: broadcasting to cat chan}
	As long as the resource theory follows the Definitions~\ref{definition: free operations} and~\ref{definition: golden rule of resource theories}, i.e. as long as it is a proper resource theory according to our definitions, given a broadcasting channel $\mB\in\fro{C}{S'C}$, it is possible to construct a non-free catalytic channel $\Tr_{C}\circ\mB\circ\Tr_{S}[\cdot\otimes\tau_{C}]\notin\fro{S}{S'}$.
\end{remark}
This remark follows immediately from the construction of the non-free catalytic channel.

On the other hand, the implication from the existence of catalytic channels to that of resource broadcasting requires all assumptions in Theorem~\ref{theorem: catchan and broadcasting}.
As an example, suppose that the set of free operations is strictly smaller than the entire CRNG operations.
Then, there may exist robust catalysis that implements a catalytic channel $\tilde{\mE}$ that is not free, but still CRNG. 
The example below presents one such case. 

\begin{example}\label{example: non-CRNG RC does not imply non-CRNG RB}
	Elementary thermal operations (ETO, Definition~\ref{definition: ETO}) and Markovian thermal operations (MTO, Definition~\ref{definition: MTO}) admit robust catalysis, with the Gibbs catalyst state $\gbs_{C}$ as shown in Chapter~\ref{chapter: CETO}.
	Both operations have the Gibbs state as the only free state, but it is known that ETO or MTO with any catalysis are bound to be Gibbs-preserving.
	The latter can be inferred from Theorem~\ref{theorem: CETO is CTO} and the fact that Gibbs-preserving operations (Definition~\ref{definition: GP}) cannot broadcast.
	Hence, resource theories whose free operations are ETO and MTO do not have broadcasting channels.
\end{example}

Another subtle relationship emerges for robust catalysis in theories having free operations strictly smaller than CRNG versus the full CRNG set. 
Firstly, even if there exists robust catalysis in the latter, it is possible to construct a smaller set of free operations that excludes this process. 
On the other hand, even if the impossibility of robust catalysis is proved with the full set of CRNG operations, this result cannot be extended to its subsets. 
An example below shows the contrapositive of this statement. 

\begin{example}\label{example: non-CRNG RC does not imply CRNG RC}
	Consider the theory where the set of free operations $\fro{S}{S}$ is always the set of all unitary channels from $S$ to itself, regardless of the system $S$.
	These sets are strict subsets of the CRNG operation, which is the set of unital channels (i.e. channels that preserve the identity $\1_{S}$).  
	Eq.~\eqref{eq: catalytic channel mixed unitaries} shows that robust catalysis exists when $\fro{S}{S}$ is defined in this way. 
	However, it is known that catalytic channels implemented via unitary channels $\fro{SC}{SC}$ are always unital~\cite{Lie2021Generecity}.
	Furthermore, from no-broadcasting of unital channels implies that there is no robust catalysis for the unital theory. 
\end{example}

The full relationships between robust catalysis and resource broadcasting for theories having free operations strictly smaller than CRNG versus the full CRNG set are summarised in Figure~\ref{fig: chapter 5 diagram}.

\begin{figure}
	\centering
\begin{tikzpicture}[]
	\draw (0, 2) node[draw] (a) {CRNG + RC};
	\draw (0, 0) node[draw] (b) {non-CRNG + RC};
	\draw (8, 0) node[draw] (c) {non-CRNG + RB};
	\draw (8, 2) node[draw] (d) {CRNG + RB};
	\node [left] at (-0.8,1) {Example~\ref{example: non-CRNG RC does not imply CRNG RC}};
	\node [below] at (4,-0.3) {Example~\ref{example: non-CRNG RC does not imply non-CRNG RB}};
	
	\draw[implies-implies, double equal sign distance, thick] (a) -- (d) node[midway, above] {Theorem~\ref{theorem: catchan and broadcasting}};
	\draw[->, thick] ($(a.south)+(0.2, 0)$) -- ($(b.north)+(0.2, 0)$) node[pos=0.5,rotate=135] {$/$};
	\draw[->, thick] ($(b.north)-(0.2, 0)$) -- ($(a.south)-(0.2, 0)$) node[pos=0.5,rotate=135] {$/$};
	\draw[->, thick] ($(c.north)-(0.2, 0)$) -- ($(d.south)-(0.2, 0)$);
	\draw[->, thick] ($(d.south)+(0.2, 0)$) -- ($(c.north)+(0.2, 0)$) node[pos=0.5,rotate=135] {$/$};
	\draw[->, thick] ($(b.east)-(0, 0.15)$) -- ($(c.west)-(0, 0.15)$) node[pos=0.5, ,rotate=135] {$/$};
	\draw[->, thick] ($(c.west)+(0, 0.15)$) -- ($(b.east)+(0, 0.15)$);
\end{tikzpicture}
	\caption{Diagram explaining the relationships between robust catalysis (RC) and resource broadcasting (RB).
	Figure adapted from Figure~2 of Ref.~\cite{SonRCRB}.}
	\label{fig: chapter 5 diagram}
\end{figure}

\section{How composition of local resource theories determines the existence of resource broadcasting}\label{section: composition and resource broadcasting}

As illustrated in Figure~\ref{fig: chapter 5 diagram}, the existence of resource broadcasting map can always probed by working with the set of free operations defined as all CRNG operations. 
Hence, for the rest of this chapter, we assume the set of free operations to be the set of CRNG operations. 

Moreover, we impose the fourth axiom for the resource theories of our interest.
\begin{kaobox}[frametitle = Additional axioms for convex resource theories with free tensor product and partial trace]
	\begin{enumerate}[topsep=3pt,itemsep=0ex,leftmargin=*,label={(\makebox[1.4em][c]{A\arabic*})}]
		\setcounter{enumi}{3}
		\item There always exists a full-rank free state $\gamma_{X}\in\frs{X}$ for any system $X$
	\end{enumerate}
\end{kaobox}
Axiom (A4) is introduced to eliminate the possibility of having a state with infinite resources. 
The below example demonstrates that when (A4) does not hold, i.e. when there exists a catalyst state that has infinite resources, any state can be obtained from broadcasting. 

\begin{observation}[limited subspace theories; original result]\label{observation: limited subspace theories}
	Suppose that all free states are non-full rank, while axioms (A1)--(A3) still hold; we call such theories as limited subspace theories.
	Due to the convexity assumption (A3), the non-full rank condition implies the existence of a subspace orthogonal to the space spanned by free states.
	Then the non-zero projectors for each of these two subspaces can be defined: $\Pi_{0}$ is the projector onto the subspace spanned by free states $\bigcup_{\gamma_{C}\in\frs{C}}\supp(\gamma_{C})$, and $\Pi_{1} = \1_{C} - \Pi_{0}$. 
	Since $\Pi_{1}\neq0$, a state $\tau_{C}$ can be chosen in the latter space, i.e. $\tau_{C}\Pi_{0} =0$.
		
	Then the robust catalysis mapping $\rho_{S}\to\rho'_{S'}$ for any pair $(\rho_{S},\rho'_{S'})$ can be constructed:
	\begin{enumerate}
		\item the catalyst $C$ is measured with effects $\Pi_{0}$ and $\Pi_{1}$;
		\item if the outcome is $0$ (corresponding to $\Pi_{0}$), prepare a $S'C$ free state;
		\item otherwise, prepare $\rho'_{S'}\otimes\tau_{C}$, where $\rho'_{S'}$ is any resourceful system state and $\tau_{C}$ is the initial catalyst state.
	\end{enumerate}
	Combined, this measure-and-prepare channel is written as 
	\begin{align}
		\mE(\rho_{SC}) = \gamma_{S'C}\Tr[\rho_{SC}(\1_{S}\otimes\Pi_{0})] +  \rho'_{S'}\otimes\tau_{C}\Tr[\rho_{SC}(\1_{S}\otimes\Pi_{1})], 
	\end{align}
	where $\gamma_{S'C}\in\frs{S'C}$ is some free state.
	When any $SC$ free state is input, this channel always outputs $\gamma_{S'C}$, i.e. $\mE\in\fro{SC}{S'C}$ is free.
	Furthermore, $\mE(\rho_{S}\otimes \tau_{C}) = \rho'_{S'}\otimes\tau_{C}$ for any $\rho_{S}$, achieving the strict robust catalysis.
	
	The broadcasting map can be constructed similarly as 
	\begin{align}
		\mB(\rho_{C}) = \gamma_{S'C}\Tr[\rho_{C}\Pi_{0}] +  \rho'_{S'}\otimes\tau_{C}\Tr[\rho_{C}\Pi_{1}],
	\end{align}
	for any $\rho'_{S}$.
	Then $\mB(\tau_{C}) = \rho'_{S'}\otimes\tau_{C}$, as promised. 
	Note that the broadcasting is also strict, i.e. the final state is uncorrelated.
	
	This catalyst state $\tau_{C}$ can be interpreted as the infinite resource state, because any state within the subspace spanned by free states cannot have a finite overlap with $\tau_{C}$ after any free operation. 
	Moreover, divergence-based monotones (Example~\ref{example: resource monotone examples}) diverge for this state, as $\mbP(\tau_{C}\|\gamma_{C}) = \infty$, whenever $\supp(\tau_{C})\nsubset\supp(\gamma_{C})$.
	
	The simplest example is the resource theory of athermality at temperature $T=0$, where the only free state is the ground state $\gamma_{X} = \dm{0}_{X}$ for any $X$.
	Suppose that $C$ is a qubit system with $\1_{C} = \dm{0}_{C} + \dm{1}_{C}$. 
	Then by setting $\Pi_{0} = \dm{0}_{C}$ and $\Pi_{1} = \dm{1}_{C}$, any final state can be generated from a free initial state using the catalytic channel with a catalyst state $\tau_{C} = \dm{1}_{C}$.
	A similar construction was used in Ref.~\cite{Kuroiwa2020catreplication}.
\end{observation}

We address the central question again with the additional axiom (A4): when does robust catalysis offer a net advantage?
It turns out that the answer hinges on an additional degree of freedom that has not been explicitly considered before: how the set of free states $\frs{AB}$ is composed given $\frs{A}$ and $\frs{B}$.
In typical resource theories, this composition is usually defined operationally, depending on the particular resource at hand.
For instance, the subset of separable states (Definition~\ref{definition: separable states}) can be clearly defined for any set of density matrices, given a partition in entanglement theory; the set of symmetric states is determined by the unitary representation of the symmetry group in each Hilbert spaces in the theory of asymmetry; etc. 

However, in abstract resource theories, where no operational definition exists a priori, any composition rule can be adopted, as long as it does not cause inconsistencies with the rest of the theory.
In particular, we can establish lower and upper bound of the composition. 
\begin{remark}[minimal and maximal compositions; original result]\label{remark: minmax compositions}
	Define minimal and maximal compositions of free state sets
	\begin{align}
		\mincomp{A}{B} &\coloneq \conv\left\{ \rho_{A}\otimes\rho_{B} \,\vert\, \rho_{A}\in\frs{A},\ \rho_{B}\in\frs{B} \right\},\label{eq: min comp}\\
		\maxcomp{A}{B} &\coloneq \left\{ \rho_{AB} \,\vert\, \Tr_{B}[\rho_{AB}]\in\frs{A},\Tr_{A}[\rho_{AB}]\in\frs{B}\right\}.\label{eq: max comp}
	\end{align}
	for given free state sets $\frs{A}$ and $\frs{B}$.
	Then for any free state set $\frs{AB}$ satisfying axioms (A1)-(A3), 
	\begin{align}
		\mincomp{A}{B} \subset \frs{AB} \subset \maxcomp{A}{B}.
	\end{align}
\end{remark}
We note that Ref.~\cite{Pinske2024Censorship} defines $\mincomp{A}{B}$ as their composition rule in the context of resource censorship.

The notations $\otimes_{\textrm{min}}$ and $\otimes_{\textrm{max}}$ are inspired by minimal and maximal tensor products of convex cones~\cite{Aubrun_2021,deBruyn_2022,aubrun2022entanglement}, often discussed in the context of general probabilistic theories~\cite{janotta2014generalized,plavala2023general}. 
\begin{kaobox}[frametitle = A digression on minimal and maximal tensor products]
	We define convex cones and their duals following Refs.~\cite{boyd2004convex,aliprantis2007cones}.
	Let $\mV$ be a vector space.
	A non-empty subset $\mC\subset\mV$ is a convex cone if it is convex and closed under positive scalar multiplication.
	We also assume that $\mC$ is closed and $\mC\cap(-\mC)=\{0\}$.
	Let $\mV^*$ be the set of linear functionals on $\mV$ with the duality $\langle,\rangle:\mV^*\times\mV\to\mathbb R$.
	The dual cone of $\mC$ is then given by $\mC^{*} = \Bqty{x^* \in\mV^*\,|\, \langle x^*, z \rangle \geq 0 \textrm{ for all } z \in \mathcal{C}}$.
	The minimal and maximal tensor products of two cones $\mathcal{C}_{A}$ and $\mathcal{C}_{B}$ are defined as
	\begin{align}
		\mC_{A}\!\otimes_{\textrm{min}}\!\mC_{B} &\coloneq \conv\left\{ z_{A} \otimes z_{B} \,|\, z_{A} \in \mC_{A}, z_{B} \in \mC_{B}\right\},\label{eq: min tensor product def} \\
		\mC_{A}\!\otimes_{\textrm{max}}\!\mC_{B} &\coloneq \left\{z \,|\, \langle x^*\otimes y^*, z \rangle \geq 0,\, x^* \in \mC^{*}_{A},\,y^* \in \mC^{*}_{B} \right\}. \label{eq: max tensor product def}
	\end{align}
	
	Consider the case where cones are defined on the space of linear Hermitian operators acting on Hilbert spaces, with the duality $\langle,\rangle$ given by the Hilbert-Schmidt inner product.
	We now observe that for each free state set $\frs{A}$, the associated cone can be defined as $\mC(\frs{A}) = \bigcup_{\lambda\geq0}\lambda\frs{A}$.
	Then, it follows that $\mC(\frs{A})\!\otimes_{\textrm{min}}\!\mC(\frs{B}) = \mC(\mincomp{A}{B})$, i.e. the minimal tensor product is equivalent to the minimal composition in Eq.~\eqref{eq: min comp}.
	
	For the maximal counterparts, we have $\mC(\frs{A})\!\otimes_{\textrm{max}}\!\mC(\frs{B}) \subset \mC(\maxcomp{A}{B})$.
	To see this, note that choosing $ x^{*} = \1_{A}\in\mC^{*}(\frs{A})$ in Eq.~\eqref{eq: max tensor product def}, ensures the $B$ marginal of $\gamma_{AB}\in\mC(\frs{A})\!\otimes_{\textrm{max}}\!\mC(\frs{B})$ is always in $\mC(\frs{B})$.
	Choosing $y^{*} = \1_{B}\in\mC^{*}(\frs{B})$ implies the same for the $A$ marginal.
	The converse of this inclusion does not hold in general, as one can check by constructing counter-examples: consider qubit systems $A$ and $B$ with $\frs{A} = \frs{B} = \{\frac{\1}{2}\}$.
	The maximum tensor product turns out to be $\mC(\frs{A})\!\otimes_{\textrm{max}}\!\mC(\frs{B}) = \{\lambda\1_{A}\otimes \1_{B} \,|\, \lambda\geq0\} $, which is strictly smaller than $\mC(\maxcomp{A}{B})$.
\end{kaobox}

Minimal and maximal compositions encapsulate the limits of allowed correlations in free states.
For instance, the resource theories of athermality, where the tensor product $\frac{e^{-\beta (H_{A}+H_{B})}}{Z_{AB}} = \frac{e^{-\beta H_{A}}}{Z_{A}}\otimes\frac{e^{-\beta H_{B}}}{Z_{B}}$ of subsystem Gibbs states $\frac{e^{-\beta H_{A}}}{Z_{A}}$ and $\frac{e^{-\beta H_{B}}}{Z_{B}}$ is a free state, and coherence~\cite{Streltsov2017CoherenceReview}, where diagonal states are free, follow the minimal composition rule.
On the other hand, the maximal composition of athermality includes the thermofield double state $\sum_{i}\frac{e^{-\beta E_{i}/2}}{\sqrt{Z}}\ket{i}_{A}\ket{i}_{B}$~\cite{Israel1976Thermofield} as a free state, when $H_{A} = H_{B} = \sum_{i}E_{i}\dm{i}$.
While $\maxcomp{A}{B}$ may appear contrived, it can be interpreted as a theory concerned with \emph{local}, rather than global, resources.

Perhaps the most problematic aspect of maximal composition is that entanglement between subsystems is considered free, which may seem counter-intuitive. 
We define another composition rule precisely to address this issue.
Separable composition
\begin{align}\label{eq: sep comp}
	\sepcomp{A}{B}\coloneq (\maxcomp{A}{B}) \cap \mathrm{SEP},
\end{align}
where $\mathrm{SEP}$ denotes all separable states across the $A \vert B$ partition, explicitly excluding entanglement between subsystems as a free resource. 

These composition rules streamline the analysis by making the set of CRNG operations identical to RNG operations, eliminating the need to consider resource-generating effects on larger Hilbert spaces.
\begin{lemma}[original result]\label{lemma: RNG CRNG}
	If the set of free state $\frs{AB} $ is either $\mincomp{A}{B}$, $\sepcomp{A}{B}$, or $\maxcomp{A}{B}$, then RNG = CRNG.
\end{lemma}
\begin{proof}
	First consider the case $\frs{AB} = \mincomp{A}{B}$.
	For any system $B$, a free state $\gamma_{AB}\in\frs{AB}$ can be written as $\gamma_{AB} = \sum_{i}p_{i}(\gamma^{(i)}_{A}\otimes\gamma^{(i)}_{B})$, where $\gamma^{(i)}_{B}\in\frs{B}$ is another free state.
	Let $\mE\in\CPTP{A}{A'}$ be an RNG channel.
	Then the extension
	\begin{align}
		\mE\otimes\id_{B}(\gamma_{AB}) = \sum_{i}p_{i}(\tilde{\gamma}^{(i)}_{A'}\otimes\gamma^{(i)}_{B}),
	\end{align} 	
	where each $\tilde{\gamma}^{(i)}_{A'} = \mE(\gamma^{(i)}_{A}) \in\frs{A'}$ is a free state.
	This implies that $\mE\otimes\id_{B}$ is an RNG channel and thus $\mE$ is a CRNG channel.
	
	Next, we prove the case $\frs{AB} = \maxcomp{A}{B}$. 
	Let $\gamma_{AB}\in\frs{AB}$ be any free state and $\mE$ be any RNG channel from $A$ to $A'$. 
	The extension $ \tilde{\gamma}_{AB} = \mE\otimes\id_{B}(\gamma_{AB}) $ is also free if and only if its reduced states are free. 
	Since $\mE$ is RNG, the $A$ reduced state $\Tr_{B}[\tilde{\gamma}] = \mE(\Tr_{B}[\gamma_{AB}])$ is free.
	The $B$ reduced state $\Tr_{A}[\tilde{\gamma}] = \Tr_{A}[\gamma_{AB}]$ is free because $\gamma_{AB}\in\frs{AB}$ is free.
	Therefore, $\mE$ is a CRNG channel. 
	
	The proof is very similar for $\frs{AB} = \sepcomp{A}{B}$. 
	The free state $\gamma_{AB} = \sum_{i}p_{i}(\xi_{A}^{(i)}\otimes\zeta_{B}^{(i)})$ for $\xi_{A}^{(i)}\in\mD_{A}$, $\zeta_{B}^{(i)}\in\mD_{B}$, $p_{i}\geq 0$ for all $i$ and 
	$\sum_{i}p_{i}\xi_{A}^{(i)}\in\frs{A}$, $\sum_{i}p_{i}\zeta_{B}^{(i)}\in\frs{B}$ are free.
	The final state after the extended channel becomes
	\begin{align}
		\mE\otimes\id_{B}(\gamma_{AB}) = \sum_{i}p_{i}(\tilde{\xi}^{(i)}_{A'}\otimes\zeta^{(i)}_{B}),
	\end{align} 
	where $\tilde{\xi}^{(i)}_{A'}\in\mD_{A'}$ for all $i$. 
	Furthermore, since $\mE$ is an RNG channel, we have that $\sum_{i}p_{i}\tilde{\xi}_{A'}^{(i)}\in\frs{A'}$ is free.
	Hence, $\mE\otimes\id_{B}(\gamma_{AB}) $ is separable and its reduced states are free, making it a free state in $\sepcomp{A'}{B}$. 
\end{proof}

After establishing these composition rules for sets of free states, one might wonder whether composition rules for free operations can be constructed in a similar fashion.
The minimal composite free operation can be defined straightforwardly: convex combinations of tensor products of free operations are natural candidates.
Unfortunately, there is no inclusion between $\mathrm{CRNG}(\mincomp{A}{B})$ and $\mathrm{CRNG}(\maxcomp{A}{B})$, and taking the union of the two (and concatenations within the union set) trivialises the theory, as all quantum channels can be constructed in this way. 
Therefore, a better method is needed to define the maximal composite free operation.
We leave this problem for future work. 

\subsection{The impossible: no-broadcasting in minimal composition and other theories}

With this newly developed categorisation of compositions in quantum resource theories, we establish a no-go theorem for robust catalysis when the composition restricts correlations between partitions of free states, i.e. when the free state set is minimal. 
\begin{theorem}[original result]\label{theorem: mincomp no catchan}
	Suppose that a resource theory satisfies axioms (A1)--(A4) and takes CRNG as its free operations. 
	If the composite free state sets are minimal, i.e. $\frs{AB} = \mincomp{A}{B}$ for any $A$ and $B$, then the theory allows neither resource broadcasting nor robust catalysis.
\end{theorem}

While no-broadcasting has previously been established in specific theories, such as asymmetry under connected Lie groups~\cite{Marvian2019Broadcasting, Lostaglio2019Broadcasting} and, more recently, stabilizer operations~\cite{Zhang2024_magicnobroad}, our result provides the first known sufficient condition that guarantees no-broadcasting across generic classes of resource theories.

The proof boils down to constructing an inequality that resembles the strong super-additivity of resource monotones. 
The strategy is to show that resource broadcasting increases the total amount of resources, which must not be possible.  
If there exists a faithful, strongly super-additive resource monotone $\mbP$ (see Definitions~\ref{definition: faithfulness monotone} and~\ref{definition: additivity of resource monotones}), it is easy to prove no-broadcasting; if the broadcasting channel $\mB\in\fro{A}{AB}$ maps $\rho_{A}$ to $\mB(\rho_{A}) = \varrho_{AB}$ with $\Tr_{B}[\varrho_{AB}] = \rho_{A}$, the monotonicity and super-additivity of $\mbP$ implies
\begin{align}\label{eq: no broadcasting from superadditivity}
	\mbP(\rho_{A}) \geq \mbP(\varrho_{AB}) \geq \mbP(\rho_{A}) + \mbP(\Tr_{A}[\varrho_{AB}]).
\end{align}
When $\mbP(\rho_{A})<\infty$, which is imposed by axiom (A4), $\mbP(\Tr_{A}[\varrho_{AB}]) = 0$ and the faithfulness of $\mbP$ dictates that $\Tr_{A}[\varrho_{AB}]\in\frs{B}$, indicating the failure of broadcasting. 

In some resource theories, the existence of such monotones is known. 
Entries in the upper right cell of Table~\ref{table: resource broadcasting}, except for the theory of asymmetry with respect to connected Lie group symmetries, are the examples. 
The asymmetry theory is an interesting exception: it is proven that no faithful strongly super-additive resource monotone exists for the theory, yet broadcasting is still not possible.
Hence, no-broadcasting does not imply the existence of faithful strongly super-additive monotones. 

In general, the existence of a faithful and super-additive monotone is not known.%
\footnote{For the theories satisfying the assumptions of Theorem~\ref{theorem: mincomp no catchan}, measured relative entropy is found to be faithful and strongly super-additive~\cite{Brandao2020Superadditive}.
This result can provide a simpler alternative proof for Theorem~\ref{theorem: mincomp no catchan}.} 
Nevertheless, our proof establishes an inequality similar to Eq.~\eqref{eq: no broadcasting from superadditivity} for resource broadcasting. 
First, we introduce a lemma necessary for the proof.
\begin{lemma}[Ref.~\cite{Piani2009MRelEnt}, Thm.~1]\label{lemma: Piani}
	Suppose that assumptions (A2) and (A3) hold.
	Furthermore, assume that for any POVM $\mM = \{M_{i}\}_{i}$ on $X$ and for all $\gamma_{XY}\in\frs{XY}$, the post-measurement $Y$ marginal state $\frac{\Tr_{X}[(M_{i}\otimes\1_{Y})\gamma_{XY}]}{\Tr[(M_{i}\otimes\1_{Y})\gamma_{XY}]}\in\frs{Y}$ is free for any effect $M_{i}$.
	Then, for any $\rho_{XY}\in\mD_{XY}$, 
	\begin{align}\label{eq: quasi super additivity}
		R(\rho_{XY}) \geq R_{\mbM}(\Tr_{Y}[\rho_{XY}]) +  R(\Tr_{X}[\rho_{XY}]),
	\end{align}
	where the relative entropy of resource $R$ and the measured relative entropy of resources $R_{\mbM}$ are defined in Example~\ref{example: resource monotone examples} with (measured) relative entropy defined in Definitions~\ref{definition: quantum Rényi entropies} and~\ref{definition: measured relative entropy}.
\end{lemma}
Intuitively, Eq.~\eqref{eq: quasi super additivity} establishes a quasi-strong-super-additivity for the monotone $R$.
In the original paper~\cite{Piani2009MRelEnt}, Lemma~\ref{lemma: Piani} is proven for a more general case, where the measured relative entropy of resource is optimised over a set of measurements, not necessarily all POVMs. 
For our proof of Theorem~\ref{theorem: mincomp no catchan}, the canonical choice $R_{\mbM}$ of optimisation over all POVMs is sufficient. 

\begin{proof}[Proof for Theorem~\ref{theorem: mincomp no catchan}]
	First, we state some properties of $R_{\mbM}$:
	\begin{enumerate}
		\item $R_{\mbM}(\rho_{X})\leq R(\rho_{X})$ for any $\rho_{X}$, by the data-processing inequality for Umegaki relative entropy;
		\item $R_{\mbM}$ is faithful, i.e. $ R_{\mbM}(\rho_{X}) \geq 0$ with the equality if and only if $\rho_{X}\in\frs{X}$ is free; 
		\item $R_{\mbM}$ is a monotone, i.e. $ R_{\mbM}( \mE(\rho_{X}))\leq R_{\mbM}(\rho_{X})$ for any $\rho_{X}$ and any free operation $\mE\in\fro{X}{Y}$.
	\end{enumerate}
	
	Now consider a catalyst $\tau_{C}$ and a free operation $\mE\in\fro{SC}{S'C}$ inducing a catalytic channel $\tilde{ \mE}$.
	For any free system state $\gamma_{S}\in\frs{S}$, denote $\chi_{S'C} =  \mE(\gamma_{S}\otimes\tau_{C})$, where $\Tr_{C}[\chi_{S'C}] = \tilde{ \mE}(\gamma_{S})$ and $\Tr_{S'}[\chi_{S'C}] = \tau_{C}$. 
	By monotonicity of $ R$, 
	\begin{align}\label{eq: SC dataprocessing}
		R(\tau_{C}) =  R(\gamma_{S}\otimes\tau_{C}) \geq  R(\chi_{S'C}),
	\end{align}
	where the first equality follows from both appending and discarding a free state $\gamma_{S}$ being a free operation.
	
	To apply Lemma~\ref{lemma: Piani}, set the set of free states $\frs{S'C}=\mincomp{S'}{C}$. 
	\begin{itemize}
		\item The first requirement---assumptions (A2) and (A3)---are already assumed as the axioms.
		
		\item The second requirement follows from the structure of $\mincomp{S'}{C}$: note that any free state $\gamma_{S'C}\in\mincomp{S'}{C}$ can be written as 
		\begin{align}
			\gamma_{S'C} = \sum_{i}p_{i}(\gamma_{S'}^{(i)}\otimes\tilde{\gamma}_{C}^{(i)}).
		\end{align}
		Then for any POVM effect $M_{i}$ that acts on system $S'$, 
		\begin{align}\label{eq: post measurement state free}
			\frac{\Tr_{S'}\left[(M_{i}\otimes\1_{C})\gamma_{S'C}\right]}{\Tr\left[(M_{i}\otimes\1_{C})\gamma_{S'C}\right]} = \frac{\sum_{i}p_{i}\Tr[M_{i}\gamma_{S'}^{(i)}]\tilde{\gamma}_{C}^{(i)}}{\sum_{i}p_{i}\Tr[M_{i}\gamma_{S'}^{(i)}]} \eqcolon \sum_{i}\tilde{p}_{i}\tilde{\gamma}_{C}^{(i)},
		\end{align} 
		where $\{\tilde{p}_{i}\}_{i}$ are valid convex coefficients. 
		By convexity of the free state set, the resulting state in Eq.~\eqref{eq: post measurement state free} remains free.
	\end{itemize}
	
	Now we are ready to use Lemma~\ref{lemma: Piani}, which gives
	\begin{align}
		R(\chi_{S'C}) \geq  R_{\mbM}(\tilde{ \mE}(\rho_{S})) +  R(\tau_{C}).
	\end{align}
	Combined with Eq.~\eqref{eq: SC dataprocessing}, it follows that  
	\begin{align}
		R(\tau_{C}) \geq  R_{\mbM}(\tilde{ \mE}(\rho_{S})) +  R(\tau_{C}),
	\end{align}
	reminiscent of Eq.~\eqref{eq: no broadcasting from superadditivity}.
	Whenever $ R(\tau_{C})<\infty$, which is guaranteed by axiom (A4), 
	\begin{align}
		0 \geq  R_{\mbM}(\tilde{ \mE}(\gamma_{S})).
	\end{align}
	The faithfulness of $R_{\mbM}$ leads to the conclusion $\tilde{ \mE}(\gamma_{S})\in\frs{S'}$ is free for any free state $\gamma_{S}\in\frs{S}$.
	In other words, catalytic channels for resource theories with minimal composition are always free operations.
	By Theorem~\ref{theorem: catchan and broadcasting}, this also proves the no-broadcasting statement. 
\end{proof}

Note that in Ref.~\cite{Takagi2022CorrCat} it has been shown that the super-additive monotone, if exists, also restricts marginal or correlated catalysis that are not robust. 
Our result then implies that theories with the minimal composition cannot be trivialized via (fine-tuned) marginal or correlated catalysis.

\begin{table}
	\begin{tabular}{|c|c|c|}
		\hline
		\multirow{2}{*}{} & \multicolumn{2}{c|}{\textbf{Robust Catalysis and Resource Broadcasting}} \\
		\cline{2-3} 
		&  &  \\
		& \textbf{Yes} & \textbf{No} \\
		&  &  \\
		\multirow{7}{*}{\textbf{\makecell{full CRNG\\resource\\ theories}}} & Athermality ($T=0$)~\cite{Kuroiwa2020catreplication} & {Athermality ($T>0$)} [Thm.~\ref{theorem: mincomp no catchan}] \cite{Wilming2017FreeE} \\
		& Imaginarity~\cite{Takagi2017Imaginarity}  & {MIO Coherence} [Thm.~\ref{theorem: mincomp no catchan}] \cite{Xi2015CoherenceSuperAdd}\\
		& Asymmetry (finite groups)~\cite{Marvian2013GCov} &  Entanglement~\cite{Christandl2004Squashed, Piani2009MRelEnt} \\
		& Theories in Theorem~\ref{theorem: broadcasting possible}  &PPT entanglement~\cite{Ganardi2024CorrCatEnt}\\
		& \multirow{2}{*}{\makecell{Limited subspace theories\\ $[$Observation~\ref{observation: limited subspace theories}$]$}} & Magic~\cite{Zhang2024_magicnobroad} \\
		&   & \multirow{2}{*}{\makecell{Asymmetry (connected Lie groups)\\ \cite{Marvian2019Broadcasting, Lostaglio2019Broadcasting}}} \\
		&   &  \\
		&  & Optical nonclassicality~\cite{Ferrari2023CVResources} \\
		&   & \\
		\hline \hline
		\multirow{2}{*}{} & \multicolumn{2}{c|}{\textbf{Robust Catalysis}} \\
		\cline{2-3}
		&  &  \\
		& \textbf{Yes} & \textbf{No} \\
		& &  \\
		\multirow{3}{*}{\textbf{\makecell{smaller than\\CRNG\\resource\\ theories}}} & Elementary TO~\cite{Son2024_CETO, Son2024hierarchy}  &  GPC operations ($T>0$) \\
		&Markovian TO~\cite{Korzekwa2022MTO2,Czartowski2023ThermalRecall,Son2024hierarchy}& Thermal operations ($T>0$)~\cite{HierarchyManuscript}\\
		&Unitary operations~\cite{Boes2018CatRandom, Lie2021CatalyticRandomness, Lie2021Generecity}&  \\
		&  &  \\
		\hline
	\end{tabular}
	\caption{Entries in the no-broadcasting column have been shown either to prohibit broadcasting directly or have a strongly super-additive monotone, as demonstrated in the corresponding references.
	The references in the other column contain examples of robust catalysis. 
	It is harder to show the non-existence of robust catalysis for free operations smaller than CRNG.
	As such, it remains unknown whether notable theories such as LOCC or stabiliser operations have robust catalysis.
	However, from their full CRNG counterparts, the possibility of resource broadcasting is already ruled out; see Figure~\ref{fig: chapter 5 diagram}.}
	\label{table: resource broadcasting}
\end{table}

Furthermore, many significant theories that lie beyond the minimal composition class also lack robust catalytic advantage and do not permit resource broadcasting.
In Table~\ref{table: resource broadcasting}, entanglement, PPT entanglement, magic, asymmetry (connected Lie groups) and optical nonclassicality are such examples. 
Particularly interesting cases are Gibbs-preserving covariant (GPC) operations (Definition~\ref{definition: GPC}) and thermal operations (Definition~\ref{definition: TO}).

GPC operation is an example that is not itself full CRNG, but is an intersection of multiple CRNG sets for different resources; in GPC's case, these are Gibbs-preserving operations (athermality) and covariant operations (asymmetry).
In such cases, if each constituent CRNG set does not allow broadcasting of its respective resource, then their intersection also inherits the no-broadcasting property and yields no advantage from robust catalysis.

For thermal operations, the no-broadcasting result already follows from that of Gibbs-preserving operations.
However, showing that no advantageous robust catalysis exists is a trickier problem that requires an entirely different set of techniques.
This is done in Ref.~\cite{HierarchyManuscript}.

Thus far, all our results apply to robust catalysis scenarios in which the catalyst may retain correlations with the system post-operation.
A stricter condition can be imposed: the catalyst must be recovered without any correlation (strict robust catalysis, as defined in Definition~\ref{definition: robust catalysis}).
By definition, strict robust catalysis cannot offer any advantage whenever general robust catalysis fail to do so.
We demonstrate that the strict setting indeed imposes even more stringent requirements for the existence of robust catalysis.
\begin{proposition}[original result]\label{proposition: strict robust catalysis no-go}
	In any resource theory having all CRNG operations as the free operation satisfying axioms (A1)-(A4), there is no strict robust catalytic advantage if the catalyst state is full-rank, i.e. if $\rho_{S}\otimes\tau_{C}\to\rho'_{S}\otimes\tau_{C}$ with a full-rank $\tau_{C}$, then $\rho_{S}\to\rho'_{S}$ is always possible without catalysts. 
\end{proposition}
\begin{proof}
	We begin by defining the reversed relative entropy of resource, first used in Ref.~\cite{Eisert_2003} in the context of entanglement theory,
	\begin{align}
		\revrelent{\rho_{S}} \coloneq \inf_{\gamma_{S} \in \frs{S}} D(\gamma_{S} \| \rho_{S}).
	\end{align}
	This is a special case of the measure defined in Eq.~\eqref{eq: reversed divergence}, with the choice $\mbD = D$.
	Furthermore, Proposition~\ref{proposition: faithful forward and reverse} proves that it is faithful, and also it is finite, i.e. $\revrelent{\rho_{S}}<\infty$ when $\rho_{S}$ is full rank.
	
	Another important property is the additivity
	\begin{align}
		\revrelent{\rho_{S}\otimes\omega_{S'}} = \revrelent{\rho_{S}} + \revrelent{\omega_{S'}},
	\end{align}
	which is derived using the property of the quantum relative entropy: $D(\gamma_{SS'} \| \rho_{S}\otimes\omega_{S'}) \geq D(\Tr_{S'}[\gamma_{SS'}] \| \rho_{S}) + D(\Tr_{S}[\gamma_{SS'}] \| \omega_{S'}) $, with the equality if and only if $\gamma_{SS'}$ is an uncorrelated state. 
	For any free state $\gamma_{SS'} \in \frs{SS'}$, the uncorrelated state $\Tr_{S}[\gamma_{SS'}]\otimes\Tr_{S'}[\gamma_{SS'}]\in\frs{S'}$ is free, and the infimum of $ D(\gamma_{SS'} \| \rho_{S}\otimes\omega_{S'})$ over free states $\gamma_{SS'} \in \frs{SS'}$ is always obtained when $\gamma_{SS'}$ is uncorrelated; hence, we obtain the claim.
	
	Now, suppose a free state $\gamma_{S}\in\frs{S}$ can be transformed into another state $\sigma_{S'}$ via strict robust catalysis with a full rank catalyst $\tau_{C}$. 
	Then there exists a free channel $\mE\in\fro{SC}{S'C}$, such that 
	\begin{align}
		\mE(\gamma_{S}\otimes\tau_{C}) = \sigma_{S'}\otimes\tau_{C}.
	\end{align}
	From the additivity and the monotonicity, 
	\begin{align}
		\revrelent{\tau_{C}} = \revrelent{\gamma_{S}\otimes\gamma_{C}} \geq \revrelent{ \sigma_{S'}\otimes\tau_{C}} = \revrelent{\sigma_{S'}} + \revrelent{\tau_{C}}.
	\end{align}
	By assumption, $\tau_{C}$ is full-rank, so $\revrelent{\tau_{C}}$ is finite, which implies that $0\geq \revrelent{\sigma_{S'}}$, or equivalently $\sigma_{S'}\in\frs{S'}$ is free from the faithfulness. 
\end{proof}

\subsection{The possible: examples of resource broadcasting}

Now we move on to the cases, where resource broadcasting is possible. 
Surprisingly, larger composition rules enable useful robust catalysis and resource broadcasting.
Let us first illustrate this with the example of the resource theory of imaginarity~\cite{Hickey2018Imaginarity, Wu2021ImaginarityPRL, Wu2021Imaginarity}.

The resource theory of imaginarity is defined by the set of free state
\begin{align}
	\frs{X} = \left\{ \rho \,|\, \rho\in\mD_{X},\ \bra{i}\rho\ket{j}\in\mbR,\ \forall i,j \right\},
\end{align}
where $\{\ket{i}\}_{i}$ is a fixed basis for system $X$ prescribed by some restrictions.
The composition rule for this theory is not minimal: a Bell state 
\begin{align}
	\dm{\Psi^{+}}_{AB} = \frac{1}{2}\begin{pmatrix}
		0 & 0 & 0 & 0\\
		0 & 1 & 1 & 0\\
		0 & 1 & 1 & 0\\
		0 & 0 & 0 & 0
	\end{pmatrix},
\end{align}
is a real state in $\frs{AB}$, even though it cannot be written as a convex combination of tensor products of $\frs{A}$ and $\frs{B}$ states. 

This choice of composition rule, which is larger than minimal, also includes certain entangling operations as free operations.
The controlled-$iY$ gate 
\begin{align}\label{eq: CiY definition}
	\mathrm{CiY} \coloneq \dm{0}_{S}\otimes\1_{C} + \dm{1}_{S}\otimes(|0\rangle\langle1|_{C} - |1\rangle\langle0|_{C}) \eqcolon \dm{0}_{S}\otimes\1_{C} +\dm{1}_{S}\otimes iY_{C} ,
\end{align}
is an example, along with controlled-$X$ and controlled-$Z$ gates defined analogously.
We use this operation to demonstrate robust catalysis and resource broadcasting following the scheme in Ref.~\cite{Takagi2017Imaginarity} Figure~8. 

Suppose that the maximally imaginary state $\ket{\hat{+}}_{C} = \frac{1}{\sqrt{2}}(\ket{0}_{C} + i\ket{1}_{C})$ is provided as a catalyst.
For any pure state $\ket{\psi}_{S}$, CiY gate Eq.~\eqref{eq: CiY definition} does not change the catalyst state.
\begin{align}
	\mathrm{CiY}(\ket{\psi}_{S}\otimes\ket{\hat{+}}_{C}) &= \frac{1}{\sqrt{2}}\left(\braket{0}{\psi}\ket{0}_{S}\otimes(\ket{0}_{C}+i\ket{1}_{C}) +\braket{1}{\psi}\ket{1}_{S}\otimes(i\ket{0}_{C}-\ket{1}_{C}) \right) \nonumber\\
	&= \sqrt{Z}_{S}\ket{\psi}_{S}\otimes\ket{\hat{+}}_{C},\label{eq: imaginarity global}
\end{align}
where the unitary operator 
\begin{align}\label{eq: sqrt Z def}
	\sqrt{Z}_{S} \coloneq \dm{0}_{S} + i\dm{1}_{S} = \begin{pmatrix}
		1 & 0 \\ 0 & i
	\end{pmatrix}.
\end{align}
Eq.~\eqref{eq: imaginarity global} implies that this process applies $\sqrt{Z}_{S}$ to the system state while preserving the catalyst state $\ket{\hat{+}}_{C}$, regardless of the initial state $\ket{\psi}_{S}$.
Furthermore, the resulting catalytic channel from the unitary operator $\sqrt{Z}_{S}$ is clearly not a free operation, as can bee seen its matrix representation in Eq.~\eqref{eq: sqrt Z def}.

The broadcasting version of this channel $\mB\in\fro{C}{SC}$, defined as
\begin{align}
	\mB(\rho_{C}) = \mathrm{CiY} (\dm{0}_{S}\otimes\rho_{C}) (\mathrm{CiY})^{\dagger},
\end{align}
yields $\mB(\dm{\hat{+}}) = \dm{\hat{+}}^{\otimes 2}$, which is a special case of resource broadcasting, known as catalytic replication in Ref.~\cite{Kuroiwa2020catreplication}.
To develop more intuition, we invoke Proposition~1 of Ref.~\cite{Wu2021Imaginarity}, which states that any pure state $\ket{\psi}$ can be transformed into an effectively qubit pure state $\ket{\phi} = \alpha\ket{0} + i\beta\ket{1}$ with $\alpha,\beta\in\mathbb{R}$, via some real unitary operation. 
Since the operation is unitary, its inverse is also a real operation. 
In other words, any pure state $\ket{\psi}$ is equivalent to a qubit pure state $\ket{\phi}$ in terms of imaginarity. 
This also implies that no pure state can possess more imaginarity resource than the maximally imaginary pure qubit state $\ket{\hat{+}}$, even if the former state consists of multiple copies of the latter. 

On the other hand, this non-extensiveness does not indicate that the catalytic replication $\rho\to\rho^{\otimes 2}$ is always possible using real operations. 
Ref.~\cite{Zhang_2024a} shows that the maximally imaginary state is the only state that admits catalytic replication among qubit states or pure states.

Beyond specific resource theories, the composition structure once again allows us to establish a general result applicable to abstract resource theories; this time, the existence of resource broadcasting can be proven.
Moreover, the necessary and sufficient condition for resource broadcasting is given in terms of the max-relative entropy of the resource, $R_{\max}$, introduced in Example~\ref{example: resource monotone examples}, based on the max-relative entropy in Eq.~\eqref{eq: def max rel entropy}.
\begin{theorem}[original result]\label{theorem: broadcasting possible}
	Suppose that the set of free operations is the set of all CRNG operations, and that it satisfies axioms (A1)-(A4). 
	Furthermore, assume that the free state set of system $C$ is a singleton $\frs{C} = \{\gamma_{C}\}$ and the composite free state set $\frs{SC}$ is either $\maxcomp{S}{C}$ or $\sepcomp{S}{C}$.
	Then there exists a state $\tau_{C}\in\mD_{C}$ and a broadcasting channel $\mB\in\fro{C}{SC}$, such that the state $\sigma_{S} = \Tr_{C}[\mB(\tau_{C})]$ can be prepared in $S$, if and only if
	\begin{align}\label{eq: Dmax condition}    
		\sup_{\rho_{C}\in\mD_{C}} R_{\max} (\rho_{C}) \geq R_{\max}(\sigma_{S}).
	\end{align}
\end{theorem}
We offer a few remarks before presenting the proof. 
Firstly, the theorem holds for any $\frs{S}$, even if it is not a singleton. 
Furthermore, using the broadcasting channel $\mB$, we can construct a free channel $\mE \coloneq \mB\circ\Tr_{S}\in\fro{SC}{SC}$ that induces robust catalysis with catalyst $\tau_{C}$. 
The theorem then implies that robust catalysis can transform any $\rho_{S}\in\mD_{S}$ into $\sigma_{S}$ if $\sigma_{S}$ satisfies Eq.~\eqref{eq: Dmax condition}. 
Additionally, when the free state set $\frs{S} = \{\gamma'_{S}\}$, where $\gamma'_{S}$ has a smallest eigenvalue greater than or equal to that of $\gamma_{C}$, transformations between arbitrary states become feasible, effectively trivialising the theory.
We note that robust catalysis from $\rho_{S}$ to $\sigma_{S}$ may still be possible even if $\sigma_{S}$ violates Eq.~\eqref{eq: Dmax condition}, since there exist catalytic channels that cannot be written as $\mB\circ\Tr_{S}$.%
\footnote{A rather trivial example is when $\sigma_{S} = \rho_{S}$ and $R_{\max}(\rho_{S})>\sup_{\tau_{C}\in\mD_{C}} R_{\max} (\tau_{C})$. 
It violates Eq.~\eqref{eq: Dmax condition} but $\rho_{S}\to\sigma_{S}$ is clearly possible.}
The possibility of resource broadcasting extends beyond this setting to analogous theories such as those of local coherence and local entanglement.

\begin{proof}
	We first show the necessity of Eq.~\eqref{eq: Dmax condition}.
	Suppose that there exists a broadcasting channel $\mB\in\fro{C}{SC}$, such that $\Tr_{S}[\mB(\tau_{C})] = \tau_{C}$ and $\Tr_{C}[\mB(\tau_{C})] = \sigma_{S}$.
	Since $R_{\max}$ is a resource monotone, it does not increase after any free operation, i.e.
	\begin{align}
		R_{\max}(\tau_{C})\geq R_{\max}(\Tr_{C}\circ\mB(\tau_{C})) = R_{\max}(\sigma_{S}).
	\end{align}
	Hence, for any $\tau_{C}$, Eq.~\eqref{eq: Dmax condition} is satisfied.
	
	To prove sufficiency, note that $\max_{\tau_{C}\in\mD_{C}}R_{\max} (\tau_{C}) = -\log(\lambda_{1})$ is attained for $\psi_{C} = \dm{\psi}_{C}$; here, $\ket{\psi}_{C}$ is the eigenvector of the free state $\gamma_{C}$ in a catalyst system, corresponding to the smallest eigenvalue $\lambda_{1}$.
	Consider the measure-and-prepare channel
	\begin{align}
		\mB(\rho_{C}) \coloneq \Tr[\psi_{C}\rho_{C}] (\sigma_{S}\otimes\psi_{C}) + \Tr[(\1_{C}-\psi_{C})\rho_{C}] (\omega_{S}\otimes\zeta_{C}),
	\end{align}
	which broadcasts $\psi_{C}\to\sigma_{S}\otimes\psi_{C}$.
	To complete the proof, it remains to verify that $\mB\in\fro{C}{SC}$ is free.
	From Lemma~\ref{lemma: RNG CRNG}, it is sufficient to show that $\mB$ is RNG.
	When the free state $\gamma_{C}$ is input,
	\begin{align}
		\mB(\gamma_{C}) = \lambda_{1}(\sigma_{S}\otimes\psi_{C}) + (1-\lambda_{1})(\omega_{S}\otimes\zeta_{C}).
	\end{align}
	The $C$ reduced state can always be made free by choosing $(1-\lambda_{1}) \zeta_{C} = \gamma_{C} - \lambda_{1}\psi_{C}$, while the $S$ reduced state is free if $\lambda_{1}\sigma_{S}+(1-\lambda_{1})\omega_{S}\in\frs{S}$.
	The latter is equivalent to the fact that there exists a free state $\gamma_{S}\in\frs{S}$, such that $\gamma_{S}-\lambda_{1}\sigma_{S}\geq0$, i.e. $D_{\max}(\sigma_{S}\Vert\frs{S})\leq-\log(\lambda_{1}) = D_{\max}(\psi_{C}\Vert\frs{c})$.
	If that is the case, $\mB\in\fro{C}{SC}$ when the set of free channels $\fro{C}{SC}$ is defined by the maximal composition $\fro{C}{SC} = \textrm{CRNG}(\frs{C} \to \maxcomp{S}{C})$.
	Furthermore, $\mB(\gamma_{C})$ is a separable operation.
	Hence, when $\fro{C}{SC}$ is defined by the separable composition $\fro{C}{SC} = \textrm{CRNG}(\frs{C} \to \sepcomp{S}{C})$, $\mB$ is also free. 
\end{proof}

\section{Concluding remarks}\label{section: concluding remarks robust catalysis}

We have showcased the power of fine-tuning in catalysis by examining robust catalysis, the alternative scenario in which fine-tuning is not allowed.
Surprisingly, most quantum resources do not admit catalytic advantages without fine-tuning, even though many benefit from fine-tuned catalysis.
An alternative interpretation of fine-tuning is in terms of knowledge of the quantum state: the catalyst state can be tailored to the system state only when the latter is known.
If fine-tuning---which may demand infinite precision, as shown in Theorem~\ref{theorem: rc is catchan}---is required, then perfect knowledge of the system state must precede.

On the other hand, our results reveal a deep connection between catalysis without fine-tuning and the phenomenon of resource broadcasting, thereby identifying a distinct mechanism through which a catalyst can be useful.
Resource broadcasting is only possible when the resource is not strongly extensive, as seen in the example of imaginarity.
This highlights the importance of composition rules for free state sets, since the composition structure determines the extensiveness of a resource.
Our work identifies new classes of non-extensive resource theories by finding compositions that permit broadcasting, and establishes a necessary and sufficient condition for obtaining an outcome state, as given in Theorem~\ref{theorem: broadcasting possible}.

This study leaves several open questions.
As our focus was on CRNG-free operations, the potential for robust catalytic advantage using non-CRNG operations remains unexplored.
For example, it is unclear whether robust catalysis could implement separable operations via LOCC, or stabiliser-preserving operations via stabiliser operations.
Nevertheless, any potential advantage within such operationally defined theories cannot exceed the limitations imposed by their corresponding CRNGs.
Moreover, if CRNG operations prohibit broadcasting, then all of their subsets must also be no-broadcasting---thereby proving the no-broadcasting property for LOCC, stabiliser operations, thermal operations, etc.
Our findings thus establish an upper bound on resource broadcasting applicable to any well-defined resource theory.
Conversely, observing resource broadcasting within a non-CRNG framework would indicate the existence of a robust catalytic advantage.
Beyond implementing non-free operations, catalytic channels may provide other types of advantages, such as the dimensional advantage of quantum catalysts over classical randomness observed in Ref.~\cite{Boes2018CatRandom}.

Robust catalysis also paves an alternative pathway for investigating channel catalysis, where a free channel induces a non-free one through the assistance of a catalyst channel.
Although resource theories of channels have been actively studied~\cite{Devetak2008Dynamical, Chiribella2008Supermap, Rosset2018Dynamical, Liu2019Dynamical, Gour2019Dynamical, Takagi2020Dynamical, Gour2020Dynamical}, their catalytic versions remain largely unexplored~\cite{LipkaBartosik2023CatReview,Datta2023CatRev2}.
Our work makes initial strides towards filling this gap: robust catalysis can be seen as a particular form of channel catalysis, where the catalyst channel is simply a state-preparation channel with fixed output.
In this case, channel catalysis becomes equivalent to a catalytic channel, as described in Theorem~\ref{theorem: catchan and broadcasting}.
Generic channel catalysis, on the other hand, is not immediately precluded by the absence of robust catalysis, as demonstrated in Ref.~\cite{Vidal_catchannel}, where a non-free channel can be transformed to another non-free channel only when a catalyst is present. 
This does not fit our definition of robust catalysis as the system-catalyst channel is not free; however, it is robust under errors in initial state preparation. 
Given the input-agnostic nature of channel catalysis, a deeper connection between channel catalysis and robust catalysis seems plausible.

Lastly, we identify a hierarchy of composition rules for free states in resource theories, revealing a spectrum that ranges from the impossibility to the possibility of robust catalysis. 
This hierarchy effectively delineates the types of correlations permitted under free operations. 
We anticipate that this framework will foster new strategies for extending or even hybridising resource theories, offering novel approaches to exploring complex resource interactions.
\addpart{Memory states in quantum computing}
\chapter{Synthesising quantum circuits with black-boxes}\label{chapter: circuit compilation}

In the second part of this thesis, we transition from resource theories to investigate another form of quantum information processing. 
At the core of this investigation is quantum recursion, a process wherein a quantum state evolves stepwise, with each step depending on the system's current state. 
This state-dependent nature of the operation imparts non-linearity to the evolution, distinguishing it from the conventional linear evolution governed by fixed unitary operations. 
A detailed exposition of the implementation of such quantum recursions is laid out in Chapters~\ref{chapter: quantum recursions} and~\ref{chapter: QDP}.
This chapter reviews the requisite techniques for these implementations. 
More specifically, we shall introduce subroutines designed to effect higher-order transformations of unknown Hamiltonian evolutions and to implement circuits instructed by quantum states.

The motivation for these higher-order transformations arises from inherent challenges in synthesising certain quantum operations. 
In the context of quantum circuit synthesis, we assume that access to quantum computational resources is unrestricted. 
That is, any physical evolution of a finite-dimensional quantum system can, in principle, be approximated with arbitrary precision. 
This capability of quantum computers is termed universality.

Standard gate-based quantum computers are typically conceptualised as comprising multiple qubits; henceforth, we shall assume all systems under consideration are multi-qubit systems. 
It has been discovered that universal sets of unitary gates exist, typically acting on at most two qubits at a time~\cite{DiVincenzo1995Universal, Barenco1995Universal, Deutsch1995Universal, Lloyd1995Universal}. 
This means that by combining these elementary gates, any unitary operation acting on an arbitrary number of qubits can be compiled. 
A canonical example of such a set includes all single-qubit operations, supplemented with the CNOT gate. 
Moreover, explicit algorithms have been developed to compile any unitary operation from these universal gate sets~\cite{Kitaev1997SolovayKitaev, dawson2006solovay, Bouland2021SolovayKitaev}.

Nevertheless, despite universality, certain processes remain unrealisable---not due to a deficiency in control capabilities, but owing to a lack of specific knowledge.
An illuminating example is the synthesis of a controlled unitary gate. 
When a unitary gate $U$ is fully known, its controlled counterpart, $\dm{0}_{A}\otimes\1 + \dm{1}_{A}\otimes U$, is likewise fully known. 
From the universality of the gate set, the latter is readily compilable. 
Even if only partial information regarding $U$ is available, namely an eigenstate $\ket{\phi}$, such that $U\ket{\phi} = \ket{\phi}$, a controlled-$U$ operation can still be compiled using elementary gates and a single application of $U$~\cite{Kitaev1995_QPE}.

However, when designing higher order processes (transforming unitaries into other unitaries, not states into other states), the unitary $U$ needs to be assumed as a black box, meaning no prior information about it is available.
Then, its controlled version cannot be realised merely by dressing $U$ with auxiliary qubits and operations~\cite{Araujo2014BlackBoxControl}. 
Indeed, it has been shown that a controlled-$U$ operation cannot be synthesised even with an arbitrary number of black-box applications of $U$ and its inverse $U^{\dagger}$%
\footnote{Surprisingly, it is also impossible to implement the controlled-$U$ by learning $U$ with multiple queries of it, e.g. via process tomography~\cite{Chuang1997ProcessTomography, Poyatos1997ProcessTomography}.
This impossibility arises because process tomography estimates the action $(\cdot) \mapsto U(\cdot)U^{\dagger}$ and not the operator $U$ itself.
See Ref.~\cite{Gavorova2024BlackBoxControlled} for a detailed discussion on this point.}
Similarly, the synthesis of an arbitrary fractional power of a unitary $U^{q}$ from black-box access to $U$ is known to be impossible in general~\cite{Sheridan2009FractionalU}.
This impossibility extends to the specific case of the $d$th root, $U^{\frac{1}{d}}$, where $U$ is a $d\times d$-dimensional unitary operator~\cite{Gavorova2024BlackBoxControlled}.
If the reader is interested in further transformations of black-box unitaries, the recent review paper Ref.~\cite{Taranto2025HigherOrder} would be helpful. 

\section{Higher-order transformation of Hamiltonians}\label{section: higher order transformation}

As discussed above, limiting access to a unitary operator as a mere black box imposes substantial restrictions. 
A more lenient, and arguably more physically motivated, model arises when we consider the typical means of implementing unitary evolutions: the application of a specific Hamiltonian for a controlled duration. 
Consequently, rather than assuming black-box access to a unitary $U$ itself, we instead presuppose the ability to realise \emph{any evolution} $e^{-iHt}$ for an arbitrary time $t\geq0$, governed by an unknown Hamiltonian $H$. 
This paradigm is equivalent to possessing access to $U^{r}$ for any real number $r \in \mbR$, thereby including arbitrary fractional powers of the unitary. 
Such a capability stands in contrast to the black-box unitary model, under which generating general fractional unitaries is impossible.

The other task previously shown to be impossible, namely synthesising the controlled operation of the black-box, also becomes feasible. 
Before we describe the algorithm, let us define what we mean by $\epsilon$-approximation when our outcome is probabilistic. 
\begin{definition}[$\epsilon$-approximation for randomised algorithms~\cite{Odake2025UHET}]\label{definition: epsilon approx randomised}
	Suppose that we aim to approximate a channel $\mU$ by implementing a channel $\mE_{j}$ with probability $p_{j}$ for each $j$.
	We succeed at $\epsilon$-approximation if the distance of the average channel $\|  \sum_{j}p_{j}\mE_{j} - \mU \|_{\diamond}\leq \epsilon$.
	See definitions of channel distance $\|\cdot\|_{\diamond}$ in Definition~\ref{definition: channel distances}.
\end{definition}
The condition $\|  \sum_{j}p_{j}\mE_{j} - \mU \|_{\diamond}\leq \epsilon$ bounding the error of the average channel implies an upper bound for the average of errors $\sum_{j}p_{j}\|\mE_{j} - \mU \|_{\diamond}^{2} \leq 2\epsilon$; see Ref.~\cite{Odake2023higherorder} Supplemental Materials Section B. 

We now present the promised algorithm, which is named controlisation.
\begin{proposition}[controlisation~\cite{Dong2021ControlledU}]\label{proposition: controlisation}
	Suppose that $\Tr[H] = 0$ and $\|H\|_{\infty}\leq1$; the operator norm $\|\cdot\|_{\infty}$ is defined in Definition~\ref{definition: Schatten p-norms}.
	Then by applying the unitary operator $e^{-i\frac{t}{N}H}$ $N$ times and randomly chosen controlled-Pauli operators $2N$ times, there is an algorithm that implements $\epsilon$-approximation (à la Definition~\ref{definition: epsilon approx randomised}) of the unitary channel $\mU(\cdot) = U(\cdot)U^{\dagger}$, such that 
	\begin{align}
		U = \dm{0}_{A}\otimes e^{-itH} + \dm{1}_{A}\otimes \1.
	\end{align} 
	This can be done for any $\epsilon>0$ given a number of queries 	
	\begin{align}
		N \geq  \max\left\{\frac{10t^{2}}{\epsilon},\frac{5t}{2}\right\}. 
	\end{align}
\end{proposition}
The proposition implements $\dm{0}_{A}\otimes e^{-itH} + \dm{1}_{A}\otimes \1$, but the standard controlled evolution $\dm{0}_{A}\otimes \1 + \dm{1}_{A}\otimes e^{-itH}$ can be synthesised with two additional applications of bit-flips.

We describe the algorithm without a proof.
Let us denote Pauli operators 
\begin{align}
	\sigma_{0} = \begin{pmatrix}
		1 & 0 \\ 0 & 1
	\end{pmatrix},\quad \sigma_{1} = \begin{pmatrix}
	0 & 1 \\ 1 & 0
	\end{pmatrix},\quad \sigma_{2} = \begin{pmatrix}
	0 & -i \\ i & 0
	\end{pmatrix},\quad \sigma_{3} = \begin{pmatrix}
	1 & 0 \\ 0 & -1
	\end{pmatrix}.
\end{align}
for each qubit.
We further denote strings of Pauli operators 
\begin{align}
	\sigma_{\vec{v}} = \sigma_{v_{1}}\otimes\sigma_{v_{2}}\otimes \cdots \otimes \sigma_{v_{n}},
\end{align}
where $\vec{v}\in\{0,1,2,3\}^{n}$.
The main idea comes from the identity
\begin{align}
	\sum_{\vec{v}\in\{0,1,2,3\}^{n}} \frac{1}{4^{n}} \left(\dm{0}_{A}\otimes\1 + \dm{1}_{A}\otimes\sigma_{\vec{v}}\right)\left(\1_{A}\otimes H\right) \left(\dm{0}_{A}\otimes\1 + \dm{1}_{A}\otimes\sigma_{\vec{v}}\right)\nonumber\\ 
	= \dm{0}_{A}\otimes H,
\end{align}
for any $n$-qubit traceless Hamiltonian $H$.
Then, we can write the desired unitary as
\begin{align}\label{eq: exponential of Pauli dressed sum}
	e^{-i\frac{t}{4^{n}}\sum_{\vec{v}\in\{0,1,2,3\}^{n}}  \left(\dm{0}_{A}\otimes\1 + \dm{1}_{A}\otimes\sigma_{\vec{v}}\right)\left(\1_{A}\otimes H\right) \left(\dm{0}_{A}\otimes\1 + \dm{1}_{A}\otimes\sigma_{\vec{v}}\right)}
	=  \dm{0}_{A}\otimes e^{-itH} + \dm{1}_{A}\otimes\1.
\end{align}

There are multiple ways of implementing the LHS of Eq.~\eqref{eq: exponential of Pauli dressed sum}.
The first is to use Lie product formula (Proposition~\ref{proposition: Lie product formula} in Appendix~\ref{chapter: Lie groups}).
Then $M$ repetitions of a product of exponentials $e^{-i\frac{t}{4^{n}M} \left(\dm{0}_{A}\otimes\1 + \dm{1}_{A}\otimes\sigma_{\vec{v}}\right)\left(\1_{A}\otimes H\right) \left(\dm{0}_{A}\otimes\1 + \dm{1}_{A}\otimes\sigma_{\vec{v}}\right)}$ for each $\vec{v}\in\{0,1,2,3\}^{n}$ yield controlled-$e^{-itH}$.
Furthermore, using another identity 
\begin{align}
	U e^{itH} U^{\dagger} = e^{itUHU^{\dagger}},
\end{align}
for any unitary $U$, we arrive at the formula
\begin{align}\label{eq: Trotter for controlisation}
	\left[\prod_{\vec{v}\in\{0,1,2,3\}^{n}}\left(\dm{0}_{A}\otimes\1 + \dm{1}_{A}\otimes\sigma_{\vec{v}}\right)e^{-i\frac{t}{4^{n}M}(\1_{A}\otimes H)}\left(\dm{0}_{A}\otimes\1 + \dm{1}_{A}\otimes\sigma_{\vec{v}}\right)\right]^{M} \nonumber\\
	= \dm{0}_{A}\otimes e^{-itH} + \dm{1}_{A}\otimes\1 + O\left(\frac{t^{2}}{4^{n}M}\right).
\end{align}
Here we used the big O notation where $f(x) = O(g(x))$ indicates that there exists positive numbers $C,x_{0}$ such that $|f(x)|\leq C|g(x)|$ for any $x\geq x_{0}$. 
In other words, $f(x)$ does not grow faster than $g(x)$ asymptotically. 
Another, more efficient method uses the idea of the random compiler~\cite{Campbell2019qDRIFT}.
Instead of choosing all $4^{n}$ exponentials for each different $\vec{v}$ as in Eq.~\eqref{eq: Trotter for controlisation} and repeat for $M$ times, $N$ uniformly random choices of $\vec{v}$ collected as the set $\mR$ yields 
a uniformly random choice of $\vec{v}$ 
\begin{align}\label{eq: random Trotter for controlisation}
	\prod_{\vec{v}\in\mR}
	\left(\dm{0}_{A}\otimes\1 + \dm{1}_{A}\otimes\sigma_{\vec{v}}\right)e^{-i\frac{t}{4^{n}M}(\1_{A}\otimes H)}\left(\dm{0}_{A}\otimes\1 + \dm{1}_{A}\otimes\sigma_{\vec{v}}\right)
\end{align}
that approximates the controlled operation often more efficiently. 
Furthermore, the number of controlled-Pauli operations can be reduced to $N+1$ by combining two consecutive applications of controlled-Pauli operations into one. 

This randomisation idea underlies more general syntheses achieving higher-order transformations.  
The essence of these techniques is the capability to construct an effective evolution $e^{-itf(H)}$ with multiple queries to $e^{-itH}$ for specific classes of functions $f$.
As stated in the beginning of this chapter, the ultimate goal is the implementation of quantum recursions wherein the unitary operation is conditional upon the quantum state on which it acts.
Employing higher-order transformations, the complex task of realising a general state-dependent evolution $e^{-itf(\rho)}$ can be reduced to that of implementing a more elementary state-dependent evolution $e^{-it\rho}$, which is typically much easier. 
See unfolding implementation in Section~\ref{section: quantum recursions definition} for more discussion. 

\begin{proposition}[Hermitian-preserving linear maps~\cite{Odake2023higherorder}]\label{proposition: higher order linear}
	Suppose that $f$ is a Hermitian-preserving linear map, such that $f(\1)\propto\1$ and $\|H\|_{\infty}\leq 1$. 
	An $\epsilon$-approximation (à la Definition~\ref{definition: epsilon approx randomised}) of the unitary channel corresponding to $e^{-itf(H)}$ can be synthesised from $N$ applications of $e^{-i\frac{\beta t}{N}H}$ dressed with elementary gates and an auxiliary qubit. 
	The number of queries scales as $N = O(\beta^{2}t^{2}n\epsilon^{-1})$, where $\beta$ is a quantity that only depends on the map $f$.
\end{proposition}

\begin{proposition}[smooth functions~\cite{Odake2025UHET}]\label{proposition: higher order smooth}
	Suppose that $f$ is a $C^{3}$ smooth function $[-1,1]\to\mbR$ and $\|H\|_{\infty}\leq 1$.
	An $\epsilon$-approximation (à la Definition~\ref{definition: epsilon approx randomised}) of the unitary channel corresponding to $e^{-itf(H)}$ can be synthesised from $N$ applications of $e^{\pm i\tau H}$ for different $\tau>0$ dressed with elementary gates and an auxiliary qubit.
	The number of queries scales as $N = O(\beta^{2}t^{2}n\epsilon^{-1})$, where $\beta$ is the quantity that only depends on the map $f$.
\end{proposition}
Note that for Proposition~\ref{proposition: higher order smooth} requires the reverse time evolution $e^{i\tau H}$ in addition to the forward one. 
However, this reverse evolution can also be synthesised with Proposition~\ref{proposition: higher order linear} as $f(H) = -H$ is a linear map. 

Both Propositions~\ref{proposition: higher order linear} and~\ref{proposition: higher order smooth} provide an algorithm to synthesise $e^{-itf(H)}$ with a black-box access to $e^{-itH}$.
The error scales inversely proportional to the circuit depth, which is desirable. 
However, the constant factor $\beta^{2}$ can become prohibitively large (at worst scaling with the dimension of the system $2^{n}$) for some functions $f$, rendering the synthesis impractical in the worst case. 

\section{Quantum circuits instructed by quantum states}\label{section: state instructed circuits}

For the implementation of $e^{-itf(\rho)}$, a synthesis paradigm alternative to Propositions~\ref{proposition: higher order linear} and~\ref{proposition: higher order smooth} exists. 
This approach involves the direct injection of multiple copies of the state $\rho$ to instruct the circuit, thereby averting the need for access to an evolution of the form $e^{-it\rho}$.
In the quantum recursion scenario, represented by the transformation $\rho \mapsto e^{-itf(\rho)}\rho e^{itf(\rho)}$, the non-linearity of the operation with respect to $\rho$ is resolved by starting with multiple copies of $\rho$.
This is because the resulting transformation that yields $e^{-itf(\rho)}\rho e^{itf(\rho)}$ (or its approximation) can be rendered linear to $\rho^{\otimes M}$ for some $M$.
We describe how these techniques are applied to quantum recursions in Chapter~\ref{chapter: QDP}.

The first of its kind is a subroutine known as density matrix exponentiation.

\begin{proposition}[density matrix exponentiation (DME)~\cite{Lloyd2014quantum, Go2024DME}]
	Let us define a unitary channel
	\begin{align}\label{eq: exact exponentiation def}
		\Eopch{\tau}{\rho}(\sigma_{S}) \coloneq e^{-i\tau\rho_{S}}\sigma_{S}e^{i\tau\rho_{S}},
	\end{align} 
	and a channel
	\begin{align}\label{eq: DME def}
		\hE{\tau}{\rho}(\sigma_{S}) \coloneq \Tr_{\bar{S}}\left[e^{-i\tau\SWAP_{\bar{S}S}}\left(\rho_{\bar{S}}\otimes\sigma_{S}\right)e^{i\tau\SWAP_{\bar{S}S}} \right]
	\end{align}
	that requires an ancillary system $\bar{S}$ with the same size as $S$. 
	For $\tau = \frac{t}{M}$ with some positive integer $M$, 
	\begin{align}\label{eq: DME sample complexity}
		\frac{1}{2}\left\| \Eopch{t}{\rho} - \left(\hE{\tau}{\rho}\right)^{M} \right\|_{\diamond} \leq \frac{4t^{2}}{M}.
	\end{align}
\end{proposition}

Similarly to the algorithms in Section~\ref{section: higher order transformation}, the approximation error scales inversely proportionally to the circuit depth, which is $M$ times that of the $\SWAP$ operation.

To approximate the unitary evolution $e^{-it\rho}$, Eq.~\eqref{eq: DME def} employs auxiliary systems prepared in the state $\rho$, which encapsulates the information about the evolution.
It then applies the $\SWAP$ operation between these systems.  
The most important observation here is that the operation that needs to be exerted on the system ($\SWAP$ operation in this case) does not depend on the desired final operation $e^{-it\rho}$, as it is completely independent of $\rho$.
Hence, we do not need to know anything about the state $\rho$, provided that $M$ copies of this state are somehow available. 

Notably, the approximation using Eq.~\eqref{eq: DME def} is much more efficient compared to the alternative, learning-and-compiling method that compiles the evolution $e^{-i\rho t}$ by learning the density matrix $\rho$ through tomography. 
The latter method requires the sample complexity
\begin{align}\label{eq: tomography scaling}
	M = O\left(\frac{Cdr(t-\epsilon)^{2}}{\epsilon^{2}\log(dt/r\epsilon)} + \frac{t^{2}}{\epsilon^{2}}\right),
\end{align}
where $d = 2^{n}$ is the dimension of $\rho$ and $r = \rank(\rho)$~\cite{Kimmel2017DME_OP}. 

More general evolutions, analogous to Proposition~\ref{proposition: higher order linear}, can also be synthesised within this paradigm of injecting instruction quantum states. 
\begin{proposition}[Hermitian-preserving map exponentiation (HME)~\cite{Wei2023hermpreserving}]\label{proposition: HME}
	Let $f: \mL_{S}\to\mL_{S'}$ be a linear Hermitian-preserving map with a Choi matrix $C_{f}$ as defined in Definition~\ref{theorem: Choi theorem}.
	Let $F\coloneq d(C_{f})^{\transp_{\bar{S}}}$, where the partial transpose $\transp_{\bar{S}}$ acts only on the system $\bar{S}$ and $d$ is the dimension of $S$.
	We also define a unitary channel
	\begin{align}\label{eq: exact exponentiation HP def}
		\Eopch{\tau}{f,\rho}(\sigma_{S'}) \coloneq e^{-i\tau f(\rho_{S})}\sigma_{S'}e^{i\tau f(\rho_{S})},
	\end{align} 
	and a channel
	\begin{align}\label{eq: HME def}
		\hE{\tau}{f, \rho}(\sigma_{S}) \coloneq \Tr_{\bar{S}}\left[e^{-i\tau F}\left(\rho_{\bar{S}}\otimes\sigma_{S'}\right)e^{i\tau F} \right].
	\end{align}
	With $\tau = \frac{t}{M}$ for any $\epsilon>0$, there exits an $M =  O(\frac{\| F\|_{\infty}^{2}t^{2}}{\epsilon})$ such that
	\begin{align}
		\frac{1}{2}\left\| \Eopch{t}{f, \rho} - \left(\hE{\tau}{f, \rho}\right)^{M} \right\|_{\diamond} \leq \epsilon,
	\end{align}
	with the diamond norm $\|\cdot\|_{\diamond}$ defined in Definition~\ref{definition: channel distances}.
\end{proposition}
Note that HME reduces to DME when $f = \id$ and $F = \SWAP$.
Moreover, the external control $e^{-i\tau F}$ again does not depend on the instruction state $\rho$, and thus HME can be run agnostic to $\rho$.

Another generalisation to polynomial functions can be made.
\begin{proposition}[polynomial function exponentiation (PFE)~\cite{Kimmel2017DME_OP}]\label{proposition: PFE}
	Let $f$ be a Hermitian-preserving polynomial of matrices $\{\rho_{j}\}_{j}$ with maximum degree $\mathtt{d}$ for each variable.
	That is, $f(\{\rho_{j}\}_{j})$ is a sum of concatenations $A_{1}\rho_{i_{1}}A_{2}\rho_{i_{2}}\cdots A_{d}$ with constant operators $A_{1},\cdots,A_{d}$ for some finite $d$ and indices $i_{1},i_{2},\cdots,i_{d-1}$ selected from the indices of $\rho_{j}$. 
	A unitary channel
	\begin{align}\label{eq: exact exponentiation PF def}
		\Eopch{\tau}{f,\{\rho_{j}\}_{j}}(\sigma_{S'}) \coloneq e^{-i\tau f(\{\rho_{j}\}_{j})}\sigma_{S'}e^{i\tau f(\{\rho_{j}\}_{j})},
	\end{align} 
	can be approximated with a channel
	\begin{align}\label{eq: PFE def}
		\hE{\tau}{f, \{\rho_{j}\}_{j}} \coloneq \hE{\tau}{g, \bigotimes_{j}\rho_{j}^{\otimes \mathtt{d}}},
	\end{align}
	where $g$ is a linear Hermitian-preserving map such that $g(\rho_{j}^{\otimes \mathtt{d}}) = f(\{\rho_{j}\}_{j})$.
\end{proposition}

Finally, we note that non-unitary evolutions described by Lindblad master equations can also be implemented using a similar strategy~\cite{Patel2023WML1, Patel2023WML2}.

\section{Mixedness reduction subroutines}\label{section: mixedness reduction}

Digressing slightly from the topics of Sections~\ref{section: higher order transformation} and~\ref{section: state instructed circuits} we now turn our attention to another subroutine that manipulates an unknown state.
Specifically, given many copies of 
\begin{align}\label{eq: rho spectral form}
	\rho = \sum_{i=1}^{d}p_{j}\dm{\psi_{j}},\quad p_{j}\geq p_{k},\ \forall 1 \leq j \leq k \leq d,
\end{align} 
with unknown coefficients $\{p_{j}\}_{j}$ and unknown eigenbasis $\{\ket{\psi_{j}}\}_{j}$, the goal is to prepare a state closer to the pure state $\ket{\psi_{1}}$, namely the one corresponding to the principal component.
The degree of mixedness of the state is quantified by the parameter $x = 1 - p_{1}$, which we term the \emph{mixedness parameter}. 

This subroutine is introduced to address an issue arising from the procedures described in Section~\ref{section: state instructed circuits}, wherein unitary channels are approximated by channels that are not strictly unitary. 
Consequently, at least for pure state evolutions, this process of mixedness reduction can mitigate the errors engendered by such approximations. 
This technique is mainly used for Theorem~\ref{theorem: qdp pure recursion} in Chapter~\ref{chapter: QDP}.

The subroutine resembles the swap test.
Therefore, we first denote the controlled-swap operation 
\begin{align}
	\mathrm{CSWAP} = \dm{0}_{A}\otimes\1_{\bar{S}S} + \dm{1}_{A}\otimes\SWAP_{\bar{S}S}.
\end{align}

\begin{proposition}[swap-test based mixedness reduction protocol]\label{proposition: swap test mixedness reduction}
	Suppose that two copies of a state $\rho$ are given in $S$ and $\bar{S}$, and an auxiliary qubit prepared in the state $\ket{+}_{A} = \frac{1}{\sqrt{2}}(\ket{0}_{A}+\ket{1}_{A})$.
	A round of the swap-test based protocol
	\begin{enumerate}
		\item applies $\mathrm{CSWAP}$ to $\bar{S}S$, controlled by $A$,
		\item measure the auxiliary qubit in the basis $\{\dm{+}_{A}, \dm{-}_{A}\}$, and
		\item trace out $\bar{S}$.
	\end{enumerate}

	If the original state has the spectral decomposition as Eq.~\eqref{eq: rho spectral form} with the mixedness parameter $x = 1 - p_{j}$, then
	\begin{enumerate}
		\item the probability of obtaining the outcome $+$ is 
		\begin{align}
			p_{+} = \frac{1}{2}(1+\sum_{j}p^{2}_{j}),
		\end{align}
		which satisfies $p_{+}>1-x$ and
		
		\item the corresponding outcome state has the same eigenvectors $\psivec{j}$ and the new mixedness parameter 
		\begin{align}\label{eq: x ratio upperbound}
			x' \leq \frac{(1+x)x}{2-2x+x^{2}},
		\end{align}
		which becomes $x' = \frac{1}{2}x + O(x^{2})$ when $x\ll1$.
	\end{enumerate}
\end{proposition}

We used the same swap-test based protocol as in Refs.~\cite{Cirac1999_purification, Childs2024purification}, but unlike them, our analysis in Eq.~\eqref{eq: x ratio upperbound} is not restricted to qubits~\cite{Cirac1999_purification} or uniformly depolarised states~\cite{Childs2024purification}, and applicable for general density matrices. 

\begin{proof}
	The described process results in the \emph{unnormalised} state 
	\begin{align}\label{eq: two copies after the swap test}
		\Tr_{A\bar{S}}\left[(\dm{+}\otimes\1_{\bar{S}S})\mathrm{CSWAP}\left(\dm{+}_{A}\otimes\rho^{\otimes 2}_{\bar{S}S}\right)\mathrm{CSWAP}^{\dagger}\right] = \frac{1}{2}(\rho+\rho^{2}),
	\end{align}
	after the measurement outcome $+$.
	The probability of getting this outcome is the trace of this unnormalised state, i.e. 
	\begin{align}
		p_{+} = \frac{1}{2}\Tr[\rho+\rho^{2}] = \frac{1}{2}(1+\sum_{j}p^{2}_{j}),
	\end{align}
	as stated in the Proposition. 
	In addition, 
	\begin{align}
		p_{+} = \frac{1}{2}(1+\sum_{j}p^{2}_{j}) \geq \frac{1}{2}(1+p^{2}_{1}) \geq p_{1} = 1-x. 
	\end{align}
	
	Using Eq.~\eqref{eq: rho spectral form}, we calculate the normalised post-measurement state as 
	\begin{align}
		\rho' = \frac{\rho+\rho^{2}}{\Tr[\rho+\rho^{2}]} = \sum_{j}\frac{p_{j}(1+p_{j})}{1+\sum_{j}p_{j}^{2}}\dm{\psi_{j}} \eqcolon \sum_{j}p'_{j}\dm{\psi_{j}},
	\end{align}
	where the eigenbasis $\{\ket{\psi_{j}}\}_{j}$ remains intact after the protocol.
	Each eigenvalue $p_{j}$ of $\rho$ is multiplied by the factor $\frac{(1+p_{j})}{1+\sum_{j}p_{j}^{2}}$ for $\rho'$.
	This factor is larger for larger eigenvalues $p_{j}$, i.e. it exacerbates the separation between different eigenvalues.
	In particular, for the largest eigenvalue $p_{1}$, it is guaranteed that
	\begin{align}
		p'_{1} = p_{1}\frac{(1+\sum_{j}p_{1}p_{j})}{1+\sum_{j}p_{j}^{2}} > p_{1},
	\end{align}
	unless $p_{j} = \frac{1}{d}$ for all $j$.
	
	Finally, the new mixedness parameter can be bounded as
	\begin{align}
		1-p'_{1} = 1 - \frac{(1-x)+(1-x)^{2}}{1+(1-x)^{2}+\sum_{j\neq1}p_{j}^{2}} = \frac{x+\sum_{j\neq1}p_{j}^{2}}{1+(1-x)^{2}+\sum_{j\neq1}p_{j}^{2}} \leq \frac{x(1+x)}{1+(1-x)^{2}},
	\end{align}
	where we use $\sum_{j\neq1}p_{j}^{2}\leq (\sum_{j\neq1}p_{j})^{2} = x^{2}$ and $0 \leq \sum_{j\neq1}p_{j}^{2}$ for the numerator and the denominator for the last inequality. 
	This concludes the proof.
\end{proof}

Indeed, the swap test protocol as in Proposition~\ref{proposition: swap test mixedness reduction} can be used iteratively to amplify the largest eigenvalue $p_{1}$; this scheme was first developed for qubits in Ref.~\cite{Cirac1999_purification} and later generalised to higher-dimensional systems in Ref.~\cite{Childs2024purification}.
It is in general hard to calculate the exact performance of this protocol, but the simple case of $\rho = (1-\delta)\dm{\psi_{d}} + \frac{\delta}{d}\1$ is well-studied in Ref.~\cite{Childs2024purification}.
The mixedness parameter $\frac{d-1}{d}\delta$ evolves as
\begin{align}\label{eq: swap test mixedness reduction asymptotics}
	\delta_{n} \leq \frac{1}{2^{n}(1-2\delta_{0})+2\delta_{0}}\delta_{0},
\end{align} 
where $\delta_{0} = \delta\leq \frac{1}{2}$ and $\delta_{n}$ is the corresponding parameter after $n$ rounds of the mixedness reduction as in Proposition~\ref{proposition: swap test mixedness reduction}. 
When $\delta_{0}$ is small, Eq.~\eqref{eq: swap test mixedness reduction asymptotics} effectively halves the mixedness parameter after each round. 
The success probability at the $m$th round is given as $1 - \frac{d-1}{d}\delta_{m-1} + \frac{d-1}{2d}\delta_{m-1}^{2}$, which is almost 1 when $\delta_{m-1}$ is small, and lower-bounded by $\frac{1}{2}$ for any $\rho$. 

For qubit density matrices, the protocol in Proposition~\ref{proposition: swap test mixedness reduction} has been proven to be optimal when given two copies of the same state, yet protocols better than this iterative method exist when given $N$ copies~\cite{Cirac1999_purification}.
Recently, the optimal strategy for $N$ copies for any higher-dimensional density matrix with the form $\rho = (1-\delta)\dm{\psi_{d}} + \frac{\delta}{d}\1$ has been found using the permutation symmetry properties of the $N$-copy states~\cite{Li2024PurityAmplification}.
However, we use Proposition~\ref{proposition: swap test mixedness reduction} for our algorithm in Chapter~\ref{chapter: QDP} for its simplicity and the fact that only two copies of a state interact at a time, without the need for multi-copy operations.  

\chapter{Quantum recursions and quantum imaginary-time evolution as a special case}\label{chapter: quantum recursions}

The goal of this chapter is twofold. 
First, in Section~\ref{section: quantum recursions definition} we introduce a class of recursive algorithms that consist of multiple steps, each of which depends on the results of previous steps. 
Such recursions in classical computing can be efficiently dealt with by a technique called memoisation~\cite{Michie1968memoization}, i.e. by remembering the previous results and using them for later steps; see Chapter~\ref{chapter: QDP} for more details on memoisation and its quantum generalisation. 
In this section, however, we outline a general approach to solve these quantum recursions without memoisation using techniques introduced in Chapter~\ref{chapter: circuit compilation}.
Section~\ref{section: DBQITE} then proceeds to focus on one special case of quantum recursion, which implements an effective quantum imaginary-time evolution to a pure state. 
Importantly, effective quantum imaginary-time evolution prepares the ground state of the Hamiltonian, which is regarded as one of the most important applications of quantum computers. 
Finally, Section~\ref{section: concluding remarks dbqite} concludes the chapter with some remarks. 

My original results in this section include Theorems~\ref{theorem: fluctuation refrigeration relation}, \ref{theorem: fidelity convergence DBQITE}, Corollary~\ref{corollary: DBQITE depth scaling}, and Lemmas~\ref{lemma: more compact form of DBQA states}, \ref{lemma: characteristic function bound}.

\section{Definition and unfolding implementation}\label{section: quantum recursions definition}

This subsection defines and presents a general method for solving a class of algorithms we term \emph{quantum recursions}.
Quantum recursions are recursive quantum algorithms in which each recursion step is a unitary operation that depends on the state it acts upon. 
As a result, each recursion step, and thereby the whole algorithm, induces non-linear evolution of the state, diverging from the usual quantum‑computing paradigm of linear evolution. 
This non-linearity precludes conventional implementation methods; instead, the techniques developed for black-box unitaries (reviewed in Chapter~\ref{chapter: circuit compilation}) prove more effective. 
We dub this method the unfolding implementation and summarise it in the box below.
To the best of my knowledge, the material in this section has not previously appeared in the literature, including my own. 
The contents of this section, to the best of my knowledge, has not been presented in existing works, including the ones this thesis is based on. 
However, some instances of quantum recursions (e.g. nested fixed-point Grover search in Ref.~\cite{Yoder2014Grover}) have been an important part of the quantum algorithm literature, and they are solved using the unfolding implementation, although not under that name.
Our contribution here is the systematic categorisation and study of quantum recursions beyond specific examples.

\begin{definition}[quantum recursions]\label{definition: quantum recursions}
	Consider a unitary operator that depends on a quantum state $\rho$, defined as 
	\begin{align}\label{eq: Upsirec def}
		\Upsirec{\rho} \coloneq e^{if(\rho)},
	\end{align}
	with some Hermitian-preserving function $f$.
	The corresponding unitary channel is denoted as
	\begin{align}\label{eq: Upsirecch def}
		\Upsirecch{\rho}(\cdot) \coloneq e^{if(\rho)}(\cdot)e^{-if(\rho)}.
	\end{align}
	Then quantum recursions are defined as a recursive algorithm iterating
	\begin{align}\label{eq: quantum recursion def}
		\rho_{k} \mapsto \rho_{k+1} = \Upsirecch{\rho_{k}}(\rho_{k}),
	\end{align}
	starting from a base case $\rho_{0}$.
\end{definition}
Channels defined in Eqs.~\eqref{eq: exact exponentiation def} and~\eqref{eq: exact exponentiation HP def} in Chapter~\ref{chapter: circuit compilation} are examples of unitary channels of the form Eq.~\eqref{eq: Upsirecch def}.
In a quantum recursion, $(k+1)$th recursion step $\Upsirecch{\rho_{k}}$ depends on the result of $k$th recursion $\rho_{k}$.
Importantly, unless one learns the resulting state $\rho_{k}$ after each step (which requires a prohibitively large sample complexity) or classically simulates all recursion steps (which requires a prohibitively large computational time), the resulting states $\rho_{k}$ are unknown to us. 
Therefore, the synthesis of $\Upsirecch{\rho_{k}}$ is highly non-trivial.

We assume that the base case $\rho_{0}$ is completely known, which indicates that any unitary of the form $\Upsirec{\rho_{0}}$ can be compiled using elementary gates. 
Furthermore, we always assume that the function $f$ is known. 
Then, an approximate synthesis of $\Upsirecch{\rho_{k}}$ for any $k$ becomes possible, using the technique we call \emph{unfolding implementation}.

\begin{kaobox}[frametitle = Unfolding implementation]
	First step is the approximation of the unitary $\Upsirec{\rho}$ into the one that is easier to synthesise. 
	Let us define
	\begin{align}\label{eq: L memory calls}
		\Upsirec{\{f_{i}\}_{i}, \rho} \coloneq V_{L}e^{if_{L}(\rho)}V_{L-1}\cdots V_{1}e^{if_{1}(\rho)}V_{0},
	\end{align}
	where $\{V_{i}\}_{i = 0}^{L}$ are constant unitary operators independent of $\rho$, and $\{f_{i}(\rho)\}_{i}$ are linear or polynomial Hermitian-preserving functions of $\rho$.
	We refer to each $e^{if_{L}(\rho)}$ as \emph{memory-call} unitary, as it requires a call to the state $\rho$. 
	The operator $\Upsirec{\{f_{i}\}_{i}, \rho}$ is then a quantum recursion unitary with $L$ memory-calls.
	
	It is known that for any recursion unitary channel $\Upsirecch{\rho}$ and for any $\epsilon>0$, there is a unitary channel $\Upsirecch{\{f_{i}\}_{i}, \rho}$, such that 
	\begin{align}
		\| \Upsirecch{\rho} - \Upsirecch{\{f_{i}\}_{i}, \rho} \|_{\diamond} \leq \epsilon, \quad \forall\rho,
	\end{align}
	with the diamond norm $\|\cdot\|_{\diamond}$ defined in Definition~\ref{definition: channel distances}.
	One crude way of finding this approximation is to expand the function $f$ in Taylor series and use the Lie product formula to find $\Upsirec{\{f_{i}\}_{i}, \rho}$.
	This approximation is likely to be inefficient. 
	However, we assume that there exists a reasonably efficient approximation that is given or that can be easily found.
	Without such good approximations, using an unfolding implementation would be infeasible.
	
	Next, we synthesise each memory-call $e^{if_{i}(\rho)}$ using black-box access to the evolution $e^{it\rho}$.
	Since $f_{i}$ is a polynomial function, this synthesis can be done using Proposition~\ref{proposition: higher order linear} or~\ref{proposition: higher order smooth}.
	
	Finally, we use the identity 
	\begin{align}\label{eq: covariance of unitary def}
		e^{it\rho_{k}} = \Upsirec{\rho_{k-1}}e^{it\rho_{k-1}} (\Upsirec{\rho_{k-1}})^{\dagger},
	\end{align}
	which allows us to synthesise $e^{it\rho_{k}}$ using black-box access to $e^{it\rho_{k-1}}$. 
	This reduction $k \to k-1$ can be recursively performed until $\Upsirec{\rho_{k}}$ is written as a product of constant unitary operators and $e^{it\rho}$.
\end{kaobox}

Note that the concept of quantum recursions can be extended.
For instance, the recursion unitary $\Upsirec{\rho}$ might vary with the recursion step $k$, or the update rule could depend on the entire history of states $\{\rho_{i}\}_{i=0}^{k}$ not solely on the state $\rho_{k}$ immediately preceding the current step.

The most severe shortcoming of unfolding implementation is its circuit depth growth with the number of recursion steps $k$.
Suppose that each $e^{if_{i}(\rho)}$ is always approximated using $M$ black-box queries to $e^{it\rho}$, which means that each recursion step $\Upsirec{\rho}$ requires $LM$ queries to $e^{it\rho}$. 
Then, each query to $e^{it\rho_{k}}$ in Eq.~\eqref{eq: covariance of unitary def} needs $2LM+1$ queries to $e^{it\rho_{k-1}}$: $LM$ queries for $\Upsirec{\rho_{k-1}}$, one query for $e^{it\rho_{k-1}}$, and $LM$ queries for $(\Upsirec{\rho_{k-1}})^{\dagger}$.
Consequently, each recursion step $\Upsirec{\rho_{k}}$ would include $D_{k} = LM(2LM+1)^{k}$ queries to $e^{it\rho_{0}}$, and the entire algorithm for $N$ recursion steps demands 
\begin{align}\label{eq: unfolding exponential circuit depth growth}
	\sum_{k = 0}^{N-1} D_{k} = \frac{(2LM+1)^{N} - 1}{2} = O((2LM+1)^{N}),
\end{align}
exponential to the number of total steps $N$.

Although this exponential scaling appears discouraging, there exist practically relevant algorithms with performance guarantees that suggest recursions up to small number of steps $N$ would be sufficient. 
We present one such example in the next section.

\section{Double-bracket quantum imaginary-time evolution}\label{section: DBQITE}

\subsection{Motivation}
Preparing ground states of Hamiltonians is a fundamental task in quantum computation with wide-ranging applications.
However, ground state preparation is not only NP-hard~\cite{barahona1982computational} but also QMA-complete~\cite{GottesmanQMA,QMA,Kempe-QIC-2003}, and thus is a challenging problem even for quantum computers~\cite{Osborne2012HamiltonianComplexity,gharibian2015quantum}, let alone classical ones. 
Nevertheless, improvements on the computational efficiency over classical simulation is anticipated to be feasible with quantum processors~\cite{Cade2020Strategies, wu2024variational}. 

To date, various quantum algorithms have been proposed for ground state preparation both for fault-tolerant \cite{Kitaev1995_QPE, Brassard2002ampamp, temme2011quantum, ge2019faster, Lin2020NearOptimal, motta2020determining, dong2022ground, fragmented_QITE2024} and near-term quantum computers~\cite{mcArdle2019, larose2019variational, kokail2019self, yeter2020practical, google2020hartree, nishi2021, yeter2021benchmarking, Zeng_2021, LinQPE, Wiersema2023RiemannianFlow}.
Among these a promising family of protocols take a thermodynamics-inspired approach and use imaginary-time evolution (ITE) to cool an initial state $\psivec{0}$ with respect to a Hamiltonian $H$ via 
\begin{align}\label{eq: QITE}
	\ket{\psi(\tau)} = \frac{e^{-\tau H} \psivec{0}}{\| {e^{-\tau H}}\psivec{0} \|_{2} }.
\end{align}
Here $\tau$ is the ITE duration and the normalisation involves the norm defined for any vector $\ket{\omega}$ by $\|\ket{\omega}\|_{2} = \sqrt{\braket{\omega}{\omega}}$ is the 2-norm of a vector defined in Definition~\ref{definition: p-norms}.  

The effectiveness of ITE for ground state preparation can be easily shown. 
First, let us assume the following: 
\begin{itemize}
	\item the Hamiltonian $H = \sum_{j=0}^{d-1}\lambda_{j}\dm{\lambda_{j}}$, where $d$ denotes the dimension of the system, and energy eigenvalues $\lambda_{j}$ are ordered such that $\lambda_{j}\leq\lambda_{k}$ for any $j\leq k$,
	\item the ground state $\ket{\lambda_{0}}$ is unique and has the eigenvalue $\lambda_{0} = 0$, 
	\item the spectral gap, i.e. the gap between the ground and the first excited state eigenvalues, $\Delta = \lambda_{1}>0$, and
	\item the operator norm (Definition~\ref{definition: Schatten p-norms}) $\| H \|_{\infty} \geq 1$.
\end{itemize}
These assumptions do not entail significant physical consequences apart from the constant shift and the rescaling of energies; the only non-trivial assumptions are the uniqueness of the ground state and the existence of a non-zero spectral gap $\Delta$.

ITE is then guaranteed to converge to the ground state $\ket{\lambda_{0}}$ of $H$ in case $\braket{\lambda_{0}}{\psi_{0}}\neq0$.
It is straightforward to see this convergence: 
\begin{align}
	F(\tau) \coloneq |\braket{\lambda_{0}}{\psi(\tau)}|^{2} = \frac{F_{0}}{\bra{\psi_{0}}e^{-2\tau H}\psivec{0}} = \frac{F_{0}}{F_{0} + \sum_{j=1}^{d-1}e^{-2\tau \lambda_{j}}|\braket{\lambda_{j}}{\psi_{0}}|^{2}},
\end{align}
where we defined $F_{0} \coloneq |\braket{\lambda_{0}}{\psi_{0}}|^{2}$ and assumed $\lambda_{0} = 0$.
From the inequality $e^{-2\tau \Delta} \geq e^{-2\tau \lambda_{j}}$ for all $j\geq 1$, 
\begin{align}\label{eq: ITE fidelity convergence}
	F(\tau) \geq \frac{F_{0}}{F_{0}+e^{-2\tau\Delta}(1-F_{0})} \geq 1 - \left(\frac{1-F_{0}}{F_{0}}\right)e^{-2\tau\Delta}.
\end{align}

The change in average energy in the ITE trajectory can also be explicitly calculated. 
ITE defined in Eq.~\eqref{eq: QITE} satisfies
\begin{align}\label{eq: GML}
	\partial_{\tau} \ket{\psi(\tau)} = -(H  -  E(\tau))\ket{\psi(\tau)}
\end{align}
with the energy $E(\tau) \coloneq \bra{\psi(\tau)}H\ket{\psi(\tau)}$.
From Eq.~\eqref{eq: GML}, a direct computation gives 
\begin{align}\label{eq: fluctuation refrigeration}
	\partial_{\tau} E(\tau) &= \bra{\psi(\tau)}(E(\tau) - H)H\ket{\psi(\tau)} + \bra{\psi(\tau)}H(E(\tau) - H)\ket{\psi(\tau)}\nonumber\\
	&= -2 V(\tau),    
\end{align}
where $V(\tau) \coloneq \bra{\psi(\tau)} (H-E(\tau))^{2} \ket{\psi(\tau)}$ is the energy fluctuation (the operator variance of the Hamiltonian).
It follows that higher energy fluctuations in the state lead to a faster energy decrease. 
We call Eq.~\eqref{eq: fluctuation refrigeration} a \emph{fluctuation-refrigeration relation}, and we will show that an analogous relation holds for the algorithm we propose. 

We make a distinction between ITE and \emph{quantum imaginary-time evolution (QITE)}~\cite{motta2020determining, mcArdle2019, nishi2021, yeter2020practical, yeter2021benchmarking, fragmented_QITE2024} in that ITE is defined by the normalised action of a non-unitary propagator and QITE is the implementation of ITE by explicitly using a unitary $Q_\tau$ such that $\ket{\psi(\tau)} = Q_{\tau} \psivec{0}$.
Finding the unitaries that implement the ITE states $\ket{\psi(\tau)}$ is not straightforward.
One family of approaches use a hybrid quantum-classical optimisation loop to learn $Q_\tau$ ~\cite{motta2020determining, mcArdle2019, nishi2021, yeter2020practical, yeter2021benchmarking}. 
This can yield compressed circuits by fine-tuning to individual input instances, but scaling to large problem sizes is generally inhibited by growing requirements on measurement precision~\cite{Larocca2024BarrenPlateaus, Cerezo2024BarrenPlateaus}.
Another approach is to extend the system size and approximate the non-unitary propagator with qubitisation~\cite{fragmented_QITE2024}. However, the overheads of implementing so-called block-encodings preclude flexible near-term experiments.
In other words, constructing efficient circuits for QITE remains an open problem. 

In Section~\ref{subsection: DBQITE definition and synthesis}, we offer a resolution to this problem by drawing on the observation that ITE is a solution to well-studied differential equations known as \emph{double-bracket flows (DBF)}~\cite{Brockett1991DBF, Bloch1990DoubleBracket, Moore1994DiscreteDBI, Smith_Thesis, HelmkeMoore1994Book}.
DBFs are appealing for quantum computation because their solutions arise from unitaries, a feature recently exploited for quantum circuit synthesis by means of double-bracket quantum algorithms (DBQA)~\cite{Gluza2024DBI, SonDBINumerics, Xiaoyue2024DBQA, SonDBQSP, Zander2025RiemannianDBQITE, SonGrover}. 
DBF comes with local optimality proofs in that they implement gradient flows and these are known to converge exponentially fast to their fixed points~\cite{Karimi2016Convergence}.

Our DB-QITE algorithm implements steps of gradient flows \textit{coherently} on a quantum computer, without resorting to classical computations, heuristic hybrid quantum-classical variational methods, or block encodings. 
Instead, the DBQA approach produces recursive quantum circuits which at every step approximates the DBF of ITE, as sketched in Fig.~\ref{fig: DBQITE summary}.

\begin{figure}[t]
	\centering
	\includegraphics[width=\textwidth]{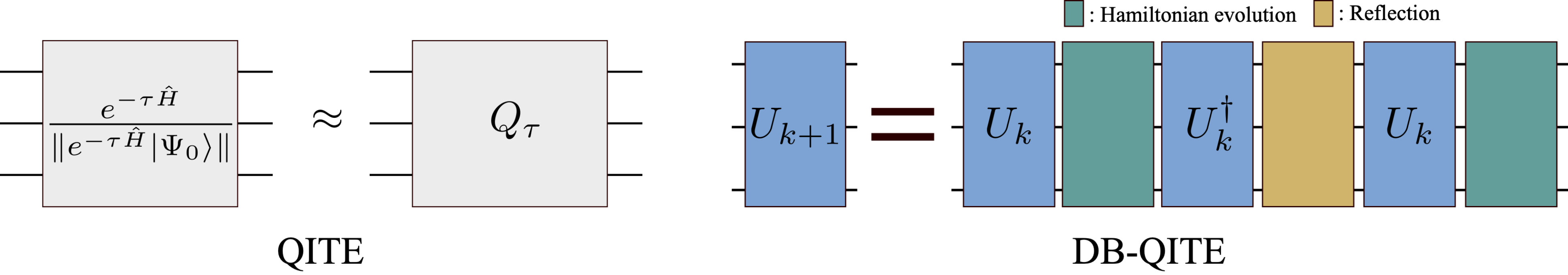}
	\caption{\textbf{Double-bracket Quantum Imaginary-Time Evolution (DB-QITE).}
		We propose a new quantum algorithm to implement imaginary-time evolution (ITE). To implement the Quantum Imaginary-Time Evolution (QITE) unitary $Q_{\tau}$, we utilise a Double-Bracket Quantum Algorithm (DBQA) and show that QITE can be recursively compiled using Hamiltonian evolution and reflection gates. 
		Figure adapted from Figure~1 of Ref.~\cite{SonDBQITE}.}
	\label{fig: DBQITE summary}
\end{figure}

In Section~\ref{subsection: DBQITE performance guarantees}, we show that DB-QITE inherits the cooling properties of ITE, as summarised in Table~\ref{table: DBQITE summary}. 
Namely, we provide rigorous guarantees that DB-QITE systematically lowers the energy of a state and increases its fidelity with the ground state.
As we will see, DB-QITE has a similar cooling rate as ITE, with the rate larger for quantum states with high energy fluctuations.
Moreover, a single step of DB-QITE is guaranteed to increase the fidelity with the ground state by an amount proportional to the initial fidelity with the ground state and the spectral gap of the target Hamiltonian. 
Furthermore, as our bounds hold for any input state (in contrast to Ref.~\cite{Anshu2021Improved}) this argument can be applied iteratively to show that the DB-QITE unitaries yield states which converge exponentially fast in the number of iterations to the ground state. 
Given the recursive structure of these additional iterations the circuit depths required to implement DB-QITE grow exponentially with each additional step but can be controlled by keeping the number of iterations moderate. 

\begin{table}[t]
	\centering
	\begin{tabular}[c]{|c||c|c|}
		\hline
		 & ITE (Eq.~\eqref{eq: QITE}) & DB-QITE (Eq.~\eqref{eq: DBQITE  main equation})\\
		 \hline
		 \multirow{2}{*}{Energy decrease} & \multirow{2}{*}{\makecell{$\partial_\tau E(\tau) = -2 V(\tau)$ \\ $[$Eq.~\eqref{eq: fluctuation refrigeration}$]$}} &  \multirow{2}{*}{\makecell{$E_{k+1} \leq E_{k} - 2s_{k} V_{k} + O (s_{k}^{2})$\\$[$Thm.~\ref{theorem: fluctuation refrigeration relation}$]$}}\\
		 &&\\
		 \hline
		 \multirow{2}{*}{Fidelity convergence} & \multirow{2}{*}{\makecell{$F(\tau) \geq 1-O(e^{-2\tau\Delta})$\\$[$Eq.~\eqref{eq: ITE fidelity convergence}$]$}} & \multirow{2}{*}{\makecell{$F_{k} \geq 1-O(q^{k})$\\$[$Thm.~\ref{theorem: fidelity convergence DBQITE}$]$}}\\
		 &&\\
		 \hline
	\end{tabular}
	\caption{\textbf{Energy reduction guarantees for DB-QITE}.
				It is known that the ITE can lower the energy $E(\tau)$ by an amount proportional to the energy fluctuation $V(\tau)$. We call this a fluctuation-refrigeration relation.
				ITE is also guaranteed to improve the fidelity to the ground state exponentially in the imaginary-time $\tau$. 
				Similarly to ITE, DB-QITE agrees with the amount of energy reduction $E_{k+1}-E_{k}$ up to corrections of order $O(s_{k}^{2})$, and improve the fidelity exponentially fast in the number of algorithm iterations $k$. }
	\label{table: DBQITE summary}
\end{table}

\subsection{Definition and synthesis}\label{subsection: DBQITE definition and synthesis}

The unitary synthesis of ITE using double-bracket flow, i.e. DB-QITE, is inspired by the observation that ITE Eq.~\eqref{eq: GML} can be rewritten as
\begin{align}\label{eq: GML DBF}
	\partial_{\tau} \ket{\psi(\tau)} = [\psi(\tau),H] \ket{\psi(\tau)},
\end{align}
where  $\psi(\tau)  = \dm{\psi(\tau)}$ is the density matrix of the ITE state.%
\footnote{This equality was pointed out by D. Gosset to one of the authors of Ref.~\cite{SonDBQITE}, a paper on which this section is based.}
In terms of the density matrix $\psi(\tau)$, we get
\begin{align}\label{eq: DBF DE}
	\partial_{\tau} \psi(\tau) = \left[ [\psi(\tau),H],\psi(\tau)\right].
\end{align}
This equation is exactly in the form of a well-studied Brockett's DBF~\cite{Brockett1991DBF}. 
Given Eq.~\eqref{eq: DBF DE}, these results from DBF theory now apply to QITE and in essence signify its local optimality. 

The difficulty of implementing ITE on quantum computers lies in the non-unitarity of the propagator in Eq.~\eqref{eq: QITE} and its state-dependence in Eq.~\eqref{eq: GML}.
The challenge of circuit synthesis for QITE is to find a unitary $Q_{\tau}$ such that
\begin{align}\label{eq: QITE synthesis}
	\ket{\psi(\tau)} \approx Q_{\tau}\psivec{0}.
\end{align}
This transition from non-unitary propagation in the ITE of Eq.~\eqref{eq: QITE} to unitary implementation in Eq.~\eqref{eq: QITE synthesis} is termed QITE. 

Two approximations are made to synthesise the QITE unitary $Q_{\tau}$.
First, the continuous evolution of Eq.~\eqref{eq: GML DBF} is discretised. 
Note that for short durations $s$, 
\begin{align}\label{eq: QITE DBF approximation}
	\ket{\psi(s)}\approx e^{s[\psi(0),H]}\psivec{0}.
\end{align}
It is not straightforward to rigorously quantify the approximation in Eq.~\eqref{eq: QITE DBF approximation}, see Proposition~1 of ~\cite{Gluza2024DBI}.
However, the exponential converging property (Eq.~\eqref{eq: ITE fidelity convergence}) of ITE potentially provides a robustness to the evolution. 
Indeed, our results summarised in Table~\ref{table: DBQITE summary} justify the linearisation of Eq.~\eqref{eq: QITE DBF approximation}. 

As a result, we obtain a recursion starting from an initial state $\psivec{0}$ and following
\begin{align}\label{eq: DBI state main}
	\ket{\psi_{k+1}} = e^{s_{k}[\psi_{k}, H]} \ket{\psi_{k}}.
\end{align}
Here, we denote the time step size in the $(k+1)$th step as $s_{k}$ and 
$\psi_{k}=\dm{\psi_{k}}$. 
Rigorous results from Refs.~\cite{Moore1994DiscreteDBI, HelmkeMoore1994Book} apply to Eq.~\eqref{eq: DBI state main} and prove convergence to the ground state as $k\rightarrow\infty$. 

Eq.~\eqref{eq: DBI state main} is exactly in the form of quantum recursion with the recursion unitary $e^{s_{k}[\psi_{k}, H]}$, which can be written as the exponential $e^{-if(\psi_{k})s_{k}}$ of a linear Hermitian-preserving map $f(\psi_{k}) = i[\psi_{k},H]$.
To implement such a unitary using Proposition~\ref{proposition: higher order linear}, the exact form of the function $f(\cdot)$ must be known, which necessitates full knowledge of the Hamiltonian $H$.
This presents a challenge because our goal is to find the ground state of $H$, meaning we lack complete knowledge of $H$ itself and instead have only black-box access to its evolution operator $e^{-iHt}$.

One way to overcome this is to treat $i[\psi_{k},H]$ as a function of a new variable $G = \psi_{k}\otimes \1_{2} + \1_{1}\otimes H$, i.e.
\begin{align}
	i[\psi_{k},H] = \frac{i}{(\Tr[\1])^{2}}[\Tr_{2}[G],\Tr_{1}[G]] \eqcolon g(G).
\end{align}
Then, $e^{-ig(G)}$ can be compiled from Proposition~\ref{proposition: higher order linear} with the black-box evolution $e^{-itG} = e^{-it\psi_{k}}\otimes e^{-itH}$. 
The latter evolution is merely a tensor product of two black-box evolutions $e^{-it\psi_{k}}$ and $e^{-itH}$ that are available to us. 
Therefore, Eq.~\eqref{eq: DBI state main} can in principle be synthesised with the general procedure we outlined in the beginning of this chapter. 

However, this specific construction of unitaries involving the commutator is a well-studied topic in Lie group theory, specifically concerning product formulae for commutators (see Proposition~\ref{proposition: product formula} in Appendix~\ref{chapter: Lie groups}), which provides us more efficient pathways for the approximation.
Using the simplest product formula, we obtain the second approximation
\begin{align}\label{eq: GC approximation for DBF}
	e^{s_{k}[\psi_{k}, H]} = e^{i\sqrt{s_{k}}H}e^{i\sqrt{s_{k}}\psi_{k}}e^{-i\sqrt{s_{k}}H}e^{-i\sqrt{s_{k}}\psi_{k}} + O(s_{k}^{\frac{3}{2}}),
\end{align}
for any $\psi_{k}$ and $H$.
As $s_{k}$ is already set to be small for the first approximation, choosing the cheapest formula for the second approximation can be justified. 
Finally we notice that the rightmost unitary in the product formula of Eq.~\eqref{eq: GC approximation for DBF} is not necessary for our recursion because
\begin{align}
	e^{i\sqrt{s_{k}}H}e^{i\sqrt{s_{k}}\psi_{k}}e^{-i\sqrt{s_{k}}H}e^{-i\sqrt{s_{k}}\psi_{k}}\ket{\psi_{k}} = e^{-i\sqrt{s_{k}}}e^{i\sqrt{s_{k}}H}e^{i\sqrt{s_{k}}\psi_{k}}e^{-i\sqrt{s_{k}}H}\ket{\psi_{k}}.
\end{align}

Therefore, we arrive at the main recursion of interest
\begin{align}\label{eq: DBQITE  main equation}
	\ket{\psi_{k+1}} = e^{i\sqrt{s_{k}}H}e^{i\sqrt{s_{k}}\psi_{k}}e^{-i\sqrt{s_{k}}H}\ket{\psi_{k}} \eqcolon \Upsirec{\psi_{k}}\ket{\psi_{k}},
\end{align}
with one memory-call $e^{i\sqrt{s_{k}}\psi_{k}}$ and two constant unitaries $e^{\pm i\sqrt{s_{k}}H}$, i.e. $L = 1$ in Eq.~\eqref{eq: L memory calls}.
We refer to the evolution by the quantum recursion Eq.~\eqref{eq: DBQITE  main equation} as \emph{double-bracket quantum imaginary-time evolution (DB-QITE)}.

DB-QITE can be implemented via unfolding as described in the beginning of this chapter.
Since its memory-call $e^{i\sqrt{s_{k}}\psi_{k}}$ is already in a form to which Eq.~\eqref{eq: covariance of unitary def} is readily applicable (i.e. $M = 1$ in Eq.~\eqref{eq: unfolding exponential circuit depth growth}), the $(k+1)$th recursion step $\Upsirec{\psi_{k}}$ is written as
\begin{align}
	\Upsirec{\psi_{k}} = e^{i\sqrt{s_{k}}H}\Upsirec{\psi_{k-1}}e^{i\sqrt{s_{k}}\psi_{k-1}}  (\Upsirec{\psi_{k-1}})^{\dagger} e^{-i\sqrt{s_{k}}H},
\end{align}
where the original memory-call $e^{i\sqrt{s_{k}}\psi_{k}}$ is replaced by $(2LM+1) = 3$ queries to $e^{i\sqrt{s_{k-1}}\psi_{k-1}}$.
Following Eq.~\eqref{eq: unfolding exponential circuit depth growth}, the total circuit depth of $N$ recursion steps would scale as $O(3^{N})$.

\subsection{Performance guarantees}\label{subsection: DBQITE performance guarantees}

Despite the exponential depth growth, we argue that DB-QITE is a useful toolkit for ground state preparation. 
To do that, we prove two main theorems showing that DB-QITE closely follows the trajectory of the ideal continuous-time ITE. 
The proofs of both theorems are in Section~\ref{subsection: DBQITE theorem proofs}.

\begin{theorem}[fluctuation-refrigeration relation; original result]\label{theorem: fluctuation refrigeration relation} 
	The average energy $E_{k}\coloneq\bra{\psi_{k}}H\ket{\psi_{k}}$ of the states $\ket{\psi_{k}}$ following Eq.~\eqref{eq: DBQITE  main equation} is bounded as
	\begin{align}\label{eq: fluctuation refrigeration in theorem}
		E_{k+1} \leq  E_{k} - 2s_{k} V_{k} + O (s_{k}^{2}),
	\end{align}
	where $V_{k} \coloneq \bra{\psi_{k}}H^{2}\ket{\psi_{k}}-E_{k}^{2}$ is the variance of the energy in state  $\ket{\psi_{k}}$. 
	In particular, if the step size is chosen such that 
	\begin{align}
		s_{k}  \leq \frac{2V_{k}}{5 (1-|\braket{\lambda_{0}}{\psi_{k+1}}|^{2})\|H\|_{\infty}^{4}}
	\end{align}
	then $E_{k+1}\leq E_{k} - s_{k}V_{k}$.
\end{theorem}

This shows that in every step of DB-QITE  cooling rate matches the cooling rate of continuous ITE up to $O(s_{k}^{2})$.
Moreover, we give sufficient conditions on $s_{k}$ such that the higher order terms $O (s_{k}^{2})$ do not outweigh the first order cooling.

The energy reduction of DB-QITE quantified in Eq.~\eqref{eq: fluctuation refrigeration in theorem} is warranted by non-negativity of the energy variances $V_{k} >0$.
However, Theorem~\ref{theorem: fluctuation refrigeration relation} does not yet exclude the possibility of converging to an excited energy eigenstate. 
To have a guarantee of the ground state convergence, we establish the second main theorem. 

\begin{theorem}[ground state fidelity increase guarantee; original result]\label{theorem: fidelity convergence DBQITE}
	Suppose that the Hamiltonian $H$ has a unique ground state $\ket{\lambda_{0}}$, spectral gap $\Delta$ and spectral radius $\| H\|_{\infty} \geq 1$.
	Let $\ket{\psi_{0}}$ be a state with non-zero ground state fidelity $|\braket{\psi_{0}}{\lambda_{0}}|^{2} \eqcolon F_{0}>0$.
	Then the recursion $\ket{\psi_{k+1}} = \Upsirec{\psi_{k}}\ket{\psi_{k}}$ with the quantum recursion unitary $\Upsirec{\psi_{k}}$ defined in Eq.~\eqref{eq: DBQITE  main equation} and the duration set to be
	\begin{align}\label{eq: step duration DBQITE}
		s_{k} = s = \frac{\Delta}{12 \| H\|_{\infty}^{3}},
	\end{align}
	for all $k$, the ground state fidelity follows 
	\begin{align}\label{eq: fidelity recursion} 
		F_{k+1} \coloneq |\braket{\lambda_{0}}{\psi_{k+1}}|^{2}  \geq F_{k} \left(1 + \frac{(1-F_{k}) \Delta^{2}}{12 \| H\|_{\infty}^{3}}\right) \geq 1- q^{k},
	\end{align}
	where $q = 1- s  F_{0} \Delta$.
\end{theorem}

This result shows that DB-QITE systematically synthesises circuits that improve on previous ones to prepare a better approximation to the ground state. 
In particular, the first step is guaranteed to increase the fidelity with the ground state by $F_{0} (1-F_{0}) \Delta^{2}/ 12 \| H\|_{\infty}^{3}$ where $F_{0}$ is the fidelity of the initial guess state $\psivec{0}$ and $\Delta$ is the spectral gap.
Moreover, subsequent iterations are guaranteed to further increase the fidelity to the ground state. 
Thus we see that repeated iterations of DB-QITE inherently avoid converging to excited states, but rather transitions through many states with small energy fluctuations $V_{k}$ to converge to the ground state. 
Hence DB-QITE provides a means of systematically preparing states with increased fidelity with the ground state.

Another important point to note is that the convergence to the ground state is exponential.
The infidelity after $k$ steps of the recursion is upper bounded by $q^{k}$, which allows us to apply an alternative implementation of this algorithm that avoids the exponential circuit depth growth; we expound on this in Chapter~\ref{chapter: QDP}.
Apart from this possibility, the exponential convergence of DB-QITE hints that the algorithm can be practical even with the exponential scaling of the circuit depth.

\begin{corollary}[original result] \label{corollary: DBQITE depth scaling}
	For $l$ qubits, DB-QITE amplifies initial fidelity $F_{0}$ to desired fidelity $F_\mathrm{th}$ in circuit depth
	\begin{equation}\label{eq: depth scaling}
		O\left(l\left(\frac{1-F_{0}}{1-F_\mathrm{th}}\right)^{2/(s  F_{0}\Delta)}\right) , 
	\end{equation}
	where  $s = \frac{\Delta}{12 \| H\|_{\infty}^{3}}$ as defined in Eq.~\eqref{eq: step duration DBQITE}.
\end{corollary}
\begin{proof}
	We find $k$ such that
	\begin{align}
		\epsilon_{k}\coloneq 1-F_{k} \leq q^{k} \epsilon_0 \leq \epsilon_\mathrm{th} \coloneq 1 -F_\mathrm{th},
	\end{align}
	to be 
	\begin{align}
		k \geq \frac{\log( \epsilon_\mathrm{th}/ \epsilon_0 )}{\log(q)} .
	\end{align}
	Finally, we insert the lower bound to the query complexity estimate
	\begin{align}
		O(3^k) 
		= O\left(( \epsilon_\mathrm{th}/ \epsilon_0 )^{\log(3)/ \log(q)}\right) &= O\left(( \epsilon_\mathrm{th}/ \epsilon_0 )^{\log(3)/ \log(1-sF_{0}\Delta)}\right)\nonumber\\
		&= O\left(( \epsilon_0/\epsilon_\mathrm{th}  )^{\log(3)/(sF_{0}\Delta)}\right)\ ,\label{eq: circuit depth for eth}
	\end{align}
	where we use the identity $a^{\log_x(b)} = b ^{\log_x(a)}$ in the first line and $\log(1-x) \approx -x$ in the last line.
	We bound the depth of each subroutine query as $O(l)$ (i.e. dominated by reflections as opposed to Hamiltonian simulations which can be done in $O(1)$ time) so the overall depth is as stated.
\end{proof}

This corollary implies that the depth scales exponentially with the inverse of the spectral gap $\Delta$ and with the inverse of the initial ground-state fidelity $F_{0}$.
The base of the exponential scaling is the ratio of the initial and final infidelity.
The last factor in the exponent is the inverse dependence on the step duration given in Eq.~\eqref{eq: step duration DBQITE}.
For local Hamiltonians this step duration is polynomially decreasing in the number of qubits $l$ implying that the runtime grows exponentially with $l^{3}$.
Thus, the  DB-QITE scheduling involved in the rigorous analysis allows for only a small number of steps $k$ so that circuit depths are modest.

There are strong indications that this runtime analysis is unnecessarily pessimistic.
In order to prove the Theorem we needed to use bounds that facilitated multiplicative rather than additive relations between the infidelities of consecutive DB-QITE states.
This is rather intricate, and the bounds are likely not tight.
More specifically, Eq.~\eqref{eq: step duration DBQITE} arises from taking a coarse lower bound $E_{k}\geq \Delta (1-F_{k})$ which relates the energy to the fidelity $F_{k}$.
This lower bound is saturated if the $k$th DB-QITE state $\psivec{k}$ is supported only on the the ground-state $\ket{\lambda_0}$ and the first excited state $\ket{\lambda_{1}}$.
For such superposition of the two unknown lowest eigenstates, our algorithm effectively runs with the Hamiltonian $H_\text{eff} = -\Delta\dm{\lambda_{0}}$.
Then the entire algorithm has the structure of the Grover search~\cite{Grover96} (reflection around the target state $\ket{\lambda_{0}}$ and reflection around the initial state $\psivec{0}$), which is guaranteed to have a much more efficient scaling. 

In turn, when the bound is not saturated then knowledge of $E_{k}$ would allow to choose a longer step duration and hence gain a larger energy decrease~\cite{Gluza2024DBI,Xiaoyue2024DBQA}.%
\footnote{Note that the estimation of $E_{k}$ can be done with a basic Hamiltonian measurement of the initial state $\psivec{0}$ and a light classical computation tracking the evolution from $E_{0}$ to $E_{k}$.} 
For an unconstrained step duration $s_{k}$ an intermediate relation in the proof states
\begin{align}
	\label{additive fidelity increase}
	F_{k+1} = F_{k} + 2 E_{k} s_{k} + O(s_{k}^{2}) .
\end{align}
The magnitude of the higher-order terms determines the maximal step duration $s_{k}$. 
Rather than taking worst case estimates as we had to do in the proof, the higher-order terms can be estimated from simple measurements of the energy $E_{k}$ as a function of $s_{k}$; see Ref.~\cite{Xiaoyue2024DBQA} for numerical examples.
In the proof, we upper bound the higher-order terms by the norm of the Hamiltonian but this is very likely a big over-estimate.
Indeed, those upper bounds are saturated when $\ket{\psi_{k}}$ is a superposition of the lowest and highest eigestates, which is highly unlikely.
Indeed, there is strong numerical evidence that in practice rather long steps can be used~\cite{Gluza2024DBI,Xiaoyue2024DBQA,SonDBINumerics,Thomson2024Flow, motta2020determining}.
This makes it plausible that DB-QITE can be scheduled to gain much more fidelity in every step than is guaranteed by the worst-case lower bound~\eqref{eq: fidelity recursion}, which implies a much shorter circuit depth in those cases. 

\subsection{Proofs of main theorems}\label{subsection: DBQITE theorem proofs}

We first present two technical lemmas that we use in the proofs.

\begin{lemma}[original result]\label{lemma: more compact form of DBQA states}
	For any pure state $\psi_{k} = \dm{\psi_{k}}$, we have the identity
	\begin{align}\label{eq: more compact form of DBQA states}
		e^{i\sqrt{s_{k}} H}e^{i\sqrt{s_{k}}\psi_{k}}e^{-i\sqrt{s_{k}} H}\ket{\psi_{k}} = \left(1 - (1-e^{i\sqrt{s_{k}}}) \phi(-\sqrt{s_{k}}) e^{i\sqrt{s_{k}} H}\right) \ket{\psi_{k}},
	\end{align}
	where we define the characteristic function as
	\begin{align}\label{eq: DBQA characteristic function}
		\phi(t)\coloneq \bra{\psi_{k}}e^{it H}\ket{\psi_{k}}.
	\end{align}
\end{lemma}
\begin{proof}
	From direct calculation, we obtain the following identity
	\begin{align}
		e^{i\sqrt{s_{k}}\psi_{k}} = \1 - (1-e^{i\sqrt{s_{k}}})\psi_{k} ,
	\end{align}
	for a pure state $\psi_{k}$.
	Therefore, the DB-QITE recursion can be simplified to
	\begin{align}
		\ket{\psi_{k+1}} &=  e^{i\sqrt{s_{k}} H} \left( \1 - (1-e^{i\sqrt{s_{k}}})\psi_{k}\right)e^{-i\sqrt{s_{k}} H}\ket{\psi_{k}} \nonumber\\
		&= \ket{\psi_{k}}-(1-e^{i\sqrt{s_{k}}}) e^{i\sqrt{s_{k}} H} \dm{\psi_{k}} e^{-i\sqrt{s_{k}} H}\ket{\psi_{k}}  \nonumber\\
		&=\left(1 - (1-e^{i\sqrt{s_{k}}}) \phi(-\sqrt{s_{k}}) e^{i\sqrt{s_{k}} H}\right) \ket{\psi_{k}}, 
	\end{align}
	where we use the density matrix representation in the second line and the definition of the characteristic function in the last line.
\end{proof}

\begin{lemma}[original result]\label{lemma: characteristic function bound}
	The $n$th derivative of the characteristic function $\phi(t)$ in Eq.~\eqref{eq: DBQA characteristic function} can be upper-bounded by
	\begin{align}\label{eq: phi simpler bound}
		| \phi^{(n)}(\xi) |  \leq \| H^{n}\|_{\infty} ,
	\end{align}
	where $\xi \in [0,t]$.
	Moreover, suppose one knows the ground state infidelity $\epsilon_{k} = 1- |\braket{\psi_{k}}{0}|^{2}$ at $k$th DB-QITE iteration, then we can obtain a tighter bound for $| \phi^{(n)}(\xi) | $, i.e.
	\begin{align}\label{eq: phi tighter bound}
		| \phi^{(n)}(\xi) | \leq \epsilon_{k} \| H^{n}\|_{\infty} .
	\end{align}
\end{lemma}

\begin{proof}
	Directly evaluating the $n$th order derivative of $\phi(t)$ gives
	\begin{align}
		\phi^{(n)}(\xi)= i^{n} \bra{\psi_{k}} e^{i\xi H} H^{n}\ket{\psi_{k}} .
	\end{align}
	As the operator norm is equal to the largest eigenvalue, we obtain the bound
	\begin{align}
		| \phi^{(n)}(\xi) | = | \bra{\psi_{k}} e^{i\xi H} H^{n}\ket{\psi_{k}}|  \leq \|e^{i\xi H} H^{n}\|_{\infty} = \| H^{n}\|_{\infty},
	\end{align}
	where we use the unitary invariant property of the operator norm in the last equality and neglect the factor $i^{n}$ as it is of norm 1.
	Thus, the first part of this lemma has been proven.
	
	Next, to prove the second part, we denote $\Pi_{0}, \Pi_{\perp}$ as the ground state projector and its complement (i.e. $\Pi_0+ \Pi_\perp=\1$), then $1 - \epsilon_{k} = \bra{\psi_{k}}\Pi_0\ket{\psi_{k}}$ and $\epsilon_{k} = \bra{\psi_{k}}\Pi_{\perp}\ket{\psi_{k}}$. 
	Therefore, we obtain
	\begin{align}
		| \phi^{(n)}(\xi) | &= | \bra{\psi_{k}} e^{i\xi H} H^{n}\ket{\psi_{k}}| \leq |\bra{\psi_{k}}e^{i\xi H}H^{n}\Pi_{0}\ket{\psi_{k}}|+|\bra{\psi_{k}}e^{i\xi H}H^{n}\Pi_{\perp}\ket{\psi_{k}}|\nonumber\\
		&= |\bra{\psi_{k}}e^{i\xi H}H^{n}\Pi_{\perp}\ket{\psi_{k}} | \leq \sum_{j=1}^{d-1} \lambda_{j}^{n} | \braket{\psi_{k}}{\lambda_{j}} |^{2}
		\leq \epsilon_{k} \| H^{n}\|_{\infty}, 
	\end{align}    
	where in the last inequality, we use the fact that $\lambda_{d-1} = \|H\|_{\infty}$ is the largest energy eigenvalue. 
\end{proof}

Now we are ready to prove the first theorem. 
\begin{proof}[Proof of Theorem~\ref{theorem: fluctuation refrigeration relation}]
	We start from Lemma~\ref{lemma: more compact form of DBQA states} to write 
	\begin{align}
		\ket{\psi_{k+1}} = \left(1 - (1-e^{i\sqrt{s_{k}}}) \phi(-\sqrt{s_{k}}) e^{i\sqrt{s_{k}} H}\right) \ket{\psi_{k}},
	\end{align}
	where we employ the same notation for the characteristic function defined in Eq.~\eqref{eq: DBQA characteristic function}.
	To simplify calculational notation, we drop the square root and index from step sizes, i.e. $\sqrt{s_{k}}\rightarrow t$ for the moment. Moreover, we define $c = (1-e^{it}) \phi(-t)$ and hence  we have
	\begin{align}
		\ket{\psi_{k+1}} = (1-ce^{it H})\ket{\psi_{k}}.
	\end{align}
	
	To derive the cooling rate, we calculate
	\begin{align}
		E_{k+1} = \bra{\psi_{k+1}} H \ket{\psi_{k+1}} &= \bra{\psi_{k}}\left[1-c^{*}e^{-itH}\right] H \left[1-ce^{itH}\right]\ket{\psi_{k}}\nonumber\\
		& = E_{k} + |c|^{2} E_{k} - 2\mathrm{Re} \left( \bra{\psi_{k}} ce^{itH} H\ket{\psi_{k}}\right),\label{eq: cooling DBQA}
	\end{align}
	where we have already made use of the fact that $H$ commutes with $e^{-itH}$. 
	To achieve Eq.~\eqref{eq: fluctuation refrigeration in theorem}, we need to derive upper bounds from Eq.~\eqref{eq: cooling DBQA}, which we do for the individual terms:
	\begin{enumerate}
		\item The second term on the R.H.S. of Eq.~\eqref{eq: cooling DBQA} can be upper bounded by the fact that 
		\begin{align}\label{eq:firstbound}
			|c|^{2} = |(1-e^{it})\phi(-t)|^{2} \leq |(1-e^{it})|^{2}\cdot |\phi(-t)|^{2} \leq t^{2},
		\end{align}
		since $(1-e^{it})(1-e^{-it}) = 2(1-\cos(t))\leq t^{2}$, and $|\phi(-t)|\leq \|e^{-it H}\|_{\infty} \leq 1$. 
		\item The third term in Eq.~\eqref{eq: cooling DBQA} can be rewritten as
		\begin{align}
			f(t) \coloneqq -2\mathrm{Re} \left( c \bra{\psi_{k}} e^{it H}  H\ket{\psi_{k}}\right) &= -2\mathrm{Re}  \left( (1-e^{it}) \bra{\psi_{k}} e^{-it H} \ket{\psi_{k}}  \bra{\psi_{k}} e^{it H}  H\ket{\psi_{k}}\right) \nonumber\\
			&=-2\mathrm{Im}\left[(1-e^{it})\phi(-t)\phi^{(1)}(t)\right].
		\end{align}
		Our grand goal is to upper bound this term. 
		At this point, let us note that $f(t)$ is an even function with $f(0)=0$. 
		We may then omit odd derivatives of it, and write the following Taylor expansion,
		\begin{align}
			f(t) = \frac{f^{(2)}(0)}{2}t^{2} + \frac{f^{(4)}(\xi)}{24}t^{4}. 
		\end{align}
		To access the higher derivatives of $f(t)$, we start by defining 
		\begin{align}\label{eq:h_fn}
			h(t) \coloneqq \phi(-t)\phi^{(1)}(t),
		\end{align}
		and write derivatives of $f(t)$ as
		\begin{align}
			f^{(2)}(t) &= -2\mathrm{Im}\left[e^{it}h(t) - 2ie^{it}h^{(1)}(t) + (1-e^{it})h^{(2)}(t)\right],\label{eq:f2_eval}\\
			f^{(4)}(t) &= -2\mathrm{Im}\left[-e^{it}h(t) + 4ie^{it}h^{(1)}(t) + 6e^{it}h^{(2)}(t)  - 4ie^{it}h^{(3)}(t) + (1-e^{it})h^{(4)}(t)\right].\label{eq:f4_eval}
		\end{align}
		
		{\underline{i) Evaluating $f^{(2)}(0)$ in Eq.~\eqref{eq:f2_eval}}}\\[2pt]
		The first term is easy; we have explicitly derived $f^{(2)}(t)$ in Eq.~\eqref{eq:f2_eval}, and we need the expression for $h(t)$ as detailed in Eq.~\eqref{eq:h_fn} and \eqref{eq: DBQA characteristic function}. In particular, one can verify that $h(0) = iE_{k}$ and $h^{(1)}(0) = -V_{k}$. Therefore, we have 
		\begin{align}\label{eq:f2}
			f^{(2)}(0) = -2\mathrm{Im} \left[h(0)-2ih^{(1)}(0)\right] = -2E_{k}-4V_{k}.
		\end{align}
		At this point, it is good to note that by combining Eqs.~\eqref{eq:firstbound} and Eq.~\eqref{eq:f2} into the equation for cooling, i.e.  Eq.~\eqref{eq: cooling DBQA}, 
		\begin{align}
			E_{k+1}\leq E_{k} -2 V_{k}t^{2} + \frac{f^{(4)}(\xi)}{24}t^{4} = E_{k} - V_{k}t^{2} - V_{k}t^{2} + \frac{f^{(4)}(\xi)}{24}t^{4} ,
		\end{align}
		which will give us the desired fluctuation-refrigeration relation $E_{k+1}\leq E_{k} - V_{k}t^{2}$ if we can choose $t$ such that 
		\begin{equation}\label{eq:frr_cond}
			- V_{k}t^{2} + \frac{f^{(4)}(\xi)}{24}t^{4} \leq 0 \qquad \implies \qquad t^{2} \leq \frac{24V_{k}}{f^{(4)}(\xi)}.
		\end{equation}
		In other words, our next step is to formulate an upper bound $f^{(4)}(\xi)\leq X $, such that by choosing $t^{2} \leq \frac{24V_{k}}{X}$, we complete the proof.\\
		
		{\underline{ii) Upper bounding $f^{(4)}(0)$ in Eq.~\eqref{eq:f4_eval}}}\\[2pt]
		We proceed to carefully bound each individual term contained in $f^{(4)}(t)$, since
		\begin{align}
			f^{(4)}(t) &= -2\mathrm{Im}\left[-e^{it}h(t) + 4ie^{it}h^{(1)}(t) + 6e^{it}h^{(2)}(t)  - 4ie^{it}h^{(3)}(t) + (1-e^{it})h^{(4)}(t)\right] \nonumber\\
			& \leq 2 \left[ |h(t)|+ 4|h^{(1)}(t)| + 6|h^{(2)}(t)|+4|h^{(3)}(t)| + |(1-e^{-it}) h^{(4)}| \right],\label{eq:f4bound}
		\end{align}
		where note that most $e^{it}$ terms are omitted since its norm is bounded by 1.
		Recall that $h(t)$ is defined in Eq.~\eqref{eq:h_fn} in terms of derivatives of $\phi(t)$, and so $h^{(n)}(t)$ needs to be explicitly derived as functions containing derivatives of $\phi(t)$ via chain rule.
		To bound the term $h^{(n)}(t)$, we first recall the bound of $ | \phi^{(n)}(\xi) | $  from Eq.~\eqref{eq: phi simpler bound} in Lemma.~\ref{lemma: characteristic function bound}, i.e. 
		\begin{align}
			| \phi^{(n)}(\xi) |  \leq \| H^{n}\|_{\infty} ,
		\end{align}
		where $\xi \in [0,t]$.
		
		Furthermore, let us assume that $\|  H\|_{\infty}\geq 1$ so that $\|H^{n}\|_{\infty}\geq \| H^{n-1}\|_{\infty}$. 
		We summarise the bounds below:
		\begin{align}
			| h(t) | &\leq E_{k} \leq \epsilon_{k}\| H\|_{\infty} \leq \epsilon_{k}\| H^{4}\|_{\infty} ,\label{eq:firstterm}\\
			| h^{(1)}(t) | &\leq V_{k} \leq \epsilon_{k}\| H^{2}\|_{\infty} \leq \epsilon_{k}\| H^{4}\|_{\infty} ,\\
			| h^{(2)}(t) | &\leq | \bra{\psi_{k}} H^{3}\ket{\psi_{k}} - E_{k}\bra{\psi_{k}} H^{2}\ket{\psi_{k}}| \leq  \epsilon_{k}\| H^{3}\|_{\infty}\leq  \epsilon_{k}\| H^{4}\|_{\infty},\\
			| h^{(3)}(t) | &\leq | \bra{\psi_{k}}H^{4}\ket{\psi_{k}} - 4\bra{\psi_{k}}H^{3}\ket{\psi_{k}}E_{k} + 3\bra{\psi_{k}}H^{2}\ket{\psi_{k}}^{2}| \leq 4\epsilon_{k} \| H^{4}\|_{\infty}.
		\end{align}
		The final term requires an additional assumption that $t\|  H\|_{\infty}\leq 1$. 
		With this,
		\begin{align}
			| (1-e^{it})h^{(4)}(t)| &\leq t| -3\bra{\psi_{k}}H^{4}\ket{\psi_{k}}E_{k} +2\bra{\psi_{k}}H^{3}\ket{\psi_{k}}\bra{\psi_{k}}H^{2}\ket{\psi_{k}}  +\bra{\psi_{k}}H^{5}\ket{\psi_{k}} |. \nonumber\\
			&\leq t\bra{\psi_{k}}H^{5}\ket{\psi_{k}} +2t\max\{\bra{\psi_{k}}H^{4}\ket{\psi_{k}}E_{k},\bra{\psi_{k}}H^{3}\ket{\psi_{k}}\bra{\psi_{k}}H^{2}\ket{\psi_{k}}  \} \nonumber\\
			&\leq 3t\epsilon_{k}\| H^{5}\|_{\infty} \leq 3\epsilon_{k}\| H^{4}\|_{\infty}.\label{eq:lastterm}
		\end{align}
		Putting Eqs.~\eqref{eq:firstterm}-\eqref{eq:lastterm} back into Eq.~\eqref{eq:f4bound}, we obtain 
		\begin{align}\label{eq:boundingf4}
			f^{(4)}(\xi)\leq 60\epsilon_{k}\|  H\|_{\infty}^{4}.
		\end{align}   
	\end{enumerate}
	Plugging Eq.~\eqref{eq:boundingf4} into the choice of $t$ as detailed after Eq.~\eqref{eq:frr_cond} concludes the proof of the theorem (recall that $t= \sqrt{s_{k}}$).
\end{proof}

We proceed to prove the second theorem.

\begin{proof}[Proof of Theorem~\ref{theorem: fidelity convergence DBQITE}]
	From RHS of Eq.~\eqref{eq: more compact form of DBQA states}, the DB-QITE recursion is given by
	\begin{align}\label{eq: GCI simplified}
		\ket{\psi_{k+1}} = \left(1 - (1-e^{i\sqrt{s_{k}}}) \phi(-\sqrt{s_{k}}) e^{i\sqrt{s_{k}} H}\right) \ket{\psi_{k}} ,
	\end{align}
	where we use the same notation again for the characteristic function as in Eq.~\eqref{eq: DBQA characteristic function}
	For all subsequent calculations in this proof, we define $t=\sqrt{s_{k}}$ to simplify the notation.
	Our ultimate goal is to show that the fidelity between the ground state and the $k$th DBQA state ($F_{k}$) can be lower-bounded by
	\begin{align}
		F_{k} \geq 1- q^{k} ,
	\end{align}
	where $q$ is a real parameter such that $0 <q < 1$.
	To do so, we define the ground state infidelity $\epsilon_{k}\coloneqq 1-F_{k}$ and we will derive a recursive inequality relating $\epsilon_{k}$ and $\epsilon_{k+1}$.
	
	First, let us look at the overlap $\braket{\lambda_{0}}{\psi_{k}}$.
	Using Eq.~\eqref{eq: GCI simplified}, the overlap at $(k+1)$th and $k$th DBQA recursion is related by
	\begin{align}
		\braket{\lambda_{0}}{\psi_{k+1}} &=\braket{\lambda_{0}}{\psi_{k}} - (1-e^{it}) \phi(-t)  \bra{\lambda_{0}} e^{it H} \ket{\psi_{k}}  \nonumber\\
		&= \left(1 - (1-e^{it})\phi(-t) \right) \braket{\lambda_{0}} {\psi_{k}},\label{eq: DBQA overlap}
	\end{align}
	where we use the assumption $\lambda_0=0$ in the last line.
	Then, the ground state fidelities $F_{k+1} $ and $F_{k}$ are related by
	\begin{align}
		F_{k+1} = | \braket{\lambda_{0}}{\psi_{k+1}}|^{2} = \left|g(t) \right|^{2} F_{k} ,
	\end{align}
	where we define 
	\begin{align}
		g(t)\coloneq1 - (1-e^{it})\phi(-t).
	\end{align}
	Similarly, the ground state infidelities $\epsilon_{k}$ and $\epsilon_{k+1}$ are related by
	\begin{align}\label{eq: epsilon and p(t) relation}
		\epsilon_{k+1} = 1- |g(t) |^{2} (1-\epsilon_{k}) = \epsilon_{k}-p(t)(1-\epsilon_{k}).
	\end{align}
	where we also define 
	\begin{align}
		p(t) \coloneq | g(t) |^{2}-1
	\end{align} 
	for simplicity.
	
	Now, the remaining task is to show that $p(t)(1-\epsilon_{k})  \geq c \epsilon_{k}$ for some $c>0$. 
	We start by applying Taylor's theorem, i.e.
	\begin{align}\label{eq: f in Taylor series}
		p(t)= p(0)+tp^{(1)}(0)+\frac{t^{2}}{2}p^{(2)}(0) + \frac{t^{3}}{6}p^{(3)}(0) + \frac{t^{4}}{24}p^{(4)}(\xi),
	\end{align}
	where $\xi\in[0,t]$.
	Notice that $p(t)$ is an even function with $p(0)=0$ as 
	\begin{align}
		p(-t) = | g(-t) |^{2} - 1 =| g(t)^{*} |^{2} - 1 = | g(t) |^{2} - 1= p(t).
	\end{align}
	Hence, $p^{(2n+1)}(0) = 0$ for any non-negative integer $n$.
	Therefore, the Taylor series reduces to
	\begin{align}\label{eq: simpler taylor series k}
		p(t)= \frac{t^{2}}{2}p^{(2)}(0)  + \frac{t^{4}}{24}p^{(4)}(\xi) .
	\end{align} 
	\begin{enumerate}
		\item Directly evaluating $p^{(2)}(t)$ yields
		\begin{align}
			p^{(2)}(t) =2\mathrm{Re}[g^{*}(t)g^{(2)}(t)]+2 | g^{(1)}(t) |^{2} .
		\end{align}
		To determine the expression of the term $p^{(2)}(0)$, we explicitly compute the derivatives of $g(t)$ up to second order.
		The results are given by
		\begin{align}
			g^{(1)}(t) &= ie^{it}\phi(-t) + (1-e^{it})\phi^{(1)}(-t),\label{eq: app g1}\\
			g^{(2)}(t) &= -e^{it}\phi(-t) - 2ie^{it}\phi^{(1)}(-t) - (1-e^{it})\phi^{(2)}(-t).\label{eq: app g2}
		\end{align}
		Thus, we obtain
		\begin{align}
			p^{(2)}(0) &= 2\mathrm{Re}[g^{*}(0)g^{(2)}(0)]+2 | g^{(1)}(0) |^{2}  = 4E_{k}. \label{eq: k^{2} term}
		\end{align}
		where we use the relation $g^{(1)}(0)=i$, $\phi(0)=1$, and $\phi^{(1)}(0)= i E_{k}$.
		
		\item Here, we will derive an lower bound for 
		the term $p^{(4)}(\xi)$.
		
		First, recall that $p(t)=g(t) g^{*}(t) -1$ and hence its $n$th order derivative is given by
		\begin{align}\label{eq: k^n combinatorics factors}
			p^{(n)}(t) &=  \sum_{r=0}^n \binom{n}{r}g^{(r)}(t) g^{*(n-r)}(t) ,
		\end{align}
		where we introduce the factor $ \binom{n}{r} $ to account for combinatorial degeneracy.
		In particular, for $n=4$, we have
		\begin{align}\label{eq: k^{4}}
			p^{(4)}(t)= g(t) g^{*(4)}(t) + 4 g^{(1)}(t) g^{*(3)}(t)+ 6 g^{(2)}(t) g^{*(2)}(t) + 4 g^{(3)}(t) g^{*(1)}(t)+ g^{(4)}(t) g^{*}(t) .
		\end{align}
		To obtain the lower bound for $  p^{(4)}(\xi)$, we first derive an upper bound for $ |g^{(r)}(\xi)|$ with arbitrary non-negative integer $r$.   
		
		\begin{enumerate}
			\item  First, we determine the upper bound for $|g(\xi)|$. 
			
			Recall that $ g(t)=1 - (1-e^{it})\phi(-t)$ and we define $a(t)= -(1-e^{it})$ for notational simplicity.
			For the factor $a(\xi)$, it gives
			\begin{align}\label{eq: 1-exp bound}
				|a(\xi)|= |1-e^{i\xi}| = \sqrt{(1-\cos \xi)^{2} + \sin^{2} \xi} = 2  | \sin \frac{\xi}{2} | \leq \xi , 
			\end{align}
			where we use the relation $| \sin \frac{\xi}{2}| \leq \frac{\xi}{2}$ in the last inequality. 
			Thus, we have
			\begin{align} \label{eq: g(t) <2}
				|g(\xi)| \leq 1+ |a(\xi)|  | \phi(-\xi) | \leq 1 + \xi \leq 2 , 
			\end{align}
			where we use the bound $ | \phi(-\xi) | \leq 1$. 
			Note that $\xi \leq t \leq 1 $ by assumption.

			\item  Next, we compute the upper bound for $| g^{(r)}(\xi)|$.
			For $r$th order derivatives of $g(t)$, we have
			\begin{align}
				| g^{(r)}(\xi) | = \left | \sum_{m=0}^{r} \binom{r}{m}a^{(m)}(\xi) \phi^{(r-m)}(-\xi) \right | 
				\leq  \sum_{m=0}^{r} \binom{r}{m} \left|a^{(m)} (\xi)\right|  \left |\phi^{(r-m)}(-\xi)\right | .
			\end{align}
			We then split the sum into $m=0$ case and $m \neq 0$ cases to evaluate the upper bound, i.e. 
			it becomes
			\begin{align}
				| g^{(r)}(\xi) | &\leq  |a (\xi)| |\phi^{(r)}(-\xi)| + \sum_{m=1}^{r} \binom{r}{m} |a^ {(m)} (\xi)|  |\phi^{(r-m)}(-\xi)| \nonumber\\
				&\leq \xi \epsilon_{k} \|  H \|^{r} +\sum_{m=1}^{r} \binom{r}{m}   \epsilon_{k} \|  H \|^{r-m} .
			\end{align}
			For the first term, we use the bound $ |a (\xi)| \leq \xi$ (Eq.~\eqref{eq: 1-exp bound}) and Eq.~\eqref{eq: phi tighter bound}.
			Similarly, for the second term, we use the bound $|a^{(m)} (\xi)| \leq 1$ and Eq.~\eqref{eq: phi tighter bound}.
			Furthermore, we assumed that $\|  H \|_{\infty} \geq 1$ which leads to $\| H^{r} \|_{\infty}\geq \| H^{r-1} \|_{\infty}$.
			Thus, it simplifies to
			\begin{align}
				| g^{(r)}(\xi) | \leq \epsilon_{k} \| H \|_{\infty}^{r-1} +\sum_{m=1}^{r} \binom{r}{m}  \epsilon_{k} \| H \|_{\infty}^{r-1} &= \epsilon_{k} \|  H \|_{\infty}^{r-1} \left(1+\sum_{m=1}^{r} \binom{r}{m} \right) \nonumber\\
				&= 2^{r} \epsilon_{k} \|  H \|_{\infty}^{r-1}  . \label{eq: g^{r} bound}    
				\end{align}
			where we assume that $\xi \| H \|_{\infty} \leq 1$ in the first inequality and we employ binomial identity in the last inequality.
		\end{enumerate}
		
		Finally, observe that Eq.~\eqref{eq: k^{4}} can be upper-bounded by
		\begin{align}
			| p^{(4)}(\xi)| &\leq 2 |g(\xi)| \cdot | g^{(4)}(\xi)| + 8 | g^{(1)}(\xi)|\cdot| g^{(3)}(\xi)|  + 6 | g^{(2)}(\xi)|^{2} ,
		\end{align}
		from the equality $ | g^{(r)} (\xi)| =  | g^{*(r)} (\xi)|$.
		Using Eq.~\eqref{eq: g(t) <2} and Eq.~\eqref{eq: g^{r} bound}, it becomes
		\begin{align}
			| p^{(4)}(\xi)|
			&\leq 64 \epsilon_{k} \| H\|_{\infty}^{3}+   128 \epsilon_{k}^{2} \| H\|_{\infty}^{2}+96 \epsilon_{k}^{2} \| H\|_{\infty}^{2} \\
			&\leq 64 \epsilon_{k} \| H\|_{\infty}^{3}+   128 \epsilon_{k} \| H\|_{\infty}^{3}+96 \epsilon_{k} \| H\|_{\infty}^{3} = 288 \epsilon_{k} \| H\|_{\infty}^{3} ,
		\end{align}
		where we use the bounds $\epsilon_{k}^{2} \leq \epsilon_{k}$ and  $\| H\|_{\infty}^{2}  \leq \| H\|_{\infty}^{3} $ in the last line.
		Finally, the lower bound for $p^{(4)}(\xi)$ is given by
		\begin{align}\label{lower bound for k^{4}}
			| p^{(4)}(\xi)|\leq 288  \epsilon_{k} \| H\|_{\infty}^{3} \implies p^{(4)}(\xi) \geq - 288  \epsilon_{k} \| H\|^{3} .
		\end{align}
	\end{enumerate}
	
	Combining Eq.~\eqref{eq: simpler taylor series k}, Eq.~\eqref{eq: k^{2} term} and Eq.~\eqref{lower bound for k^{4}} yields
	\begin{align}
		p(t) & \geq 2t^{2} E_{k} - 12 t^{4} \epsilon_{k} \| H\|_{\infty}^{3}. 
	\end{align}
	We use another inequality $E_{k} \geq \Delta \epsilon_{k}$ to show that $p(t) \geq 2t^{2} \Delta \epsilon_{k} - 12 t^{4} \epsilon_{k} \| H\|_{\infty}^{3}$.
	This inequality is derived from 
	\begin{align}
		E_{k} = \bra{\psi_{k}} \sum_{j=0}^{d-1}\lambda_{j}\ket{\lambda_{j}}\braket{\lambda_{j}}{\psi_{k}} =  \sum_{j=1}^{d-1}\lambda_{j} |\braket{\psi_{k}}{\lambda_{k}}|^{2} \geq \sum_{j=1}^{d-1}\lambda_{1} |\braket{\psi_{k}}{\lambda_{k}}|^{2} = \Delta \epsilon_{k}.
	\end{align}
	
	By setting $t^{2} = s = \frac{\Delta}{12 \| H\|_{\infty}^{3}}$ as a constant for all $k$th DB-QITE recursions, $p(t)$ can be lower-bounded as
	\begin{align}
		p(t) \geq 2\epsilon_{k} \left(\frac{\Delta^{2}}{12 \| H\|_{\infty}^{3}}-\frac{6\Delta^{2}}{144 \| H\|_{\infty}^{3}}\right) = \frac{\Delta^{2} \epsilon_{k}}{12 \| H\|_{\infty}^{3}} .
	\end{align}
	Substituting it into Eq.~\eqref{eq: epsilon and p(t) relation} yields
	\begin{align}
		\epsilon_{k+1} \leq \epsilon_{k} -  (1-\epsilon_{k})\frac{\Delta^{2} \epsilon_{k}}{12 \| H\|_{\infty}^{3}} =\left(1-\frac{\Delta^{2} F_{k}}{12 \| H\|_{\infty}^{3}}\right) \epsilon_{k} \leq \left(1-\frac{\Delta^{2} F_{0}}{12 \| H\|_{\infty}^{3}}\right) \epsilon_{k} = q \epsilon_{k} ,
	\end{align}
	where $q =1- s  F_{0} \Delta$ as promised.
	Ultimately, the ground state fidelity is given by
	\begin{align}
		\label{infidelity relation app}
		\epsilon_{k} \leq q^{k} \epsilon_0 & \implies  F_{k} \geq 1- q^{k}F_{0} \geq 1-q^{k},
	\end{align}
	where we used that $\epsilon_0= 1- F_{0} \leq 1$.
\end{proof}

\section{Concluding Remarks}\label{section: concluding remarks dbqite}

Quantum recursion, as defined in Eqs.~\eqref{eq: Upsirec def}--\eqref{eq: quantum recursion def}, is a natural extension of the recursive steps in classical computing; it is perhaps the most general form, when the computation is unitary and the recursion depends only on the immediately previous result. 
Nevertheless, relatively little effort has been dedicated to analysing quantum recursions as a distinct class of quantum algorithm, except for a few case studies~\cite{Grover2005FP, Yoder2014Grover}.
We suspect that a primary reason for this is the lack of effective implementation strategies for such state-dependent unitaries.

In this section, we have introduced a systematic way of implementing quantum recursions using techniques in Section~\ref{section: higher order transformation} and the identity Eq.~\eqref{eq: covariance of unitary def}.
We have also explicated the implementation with a specific example of quantum imaginary-time evolution in Section~\ref{section: DBQITE}.
Although it suffers from the exponential circuit growth, common to unfolding implementation as in Eq.~\eqref{eq: unfolding exponential circuit depth growth}, the performance guarantees suggest that the total circuit depth required for modestly good performance might not be unduly severe; see, e.g. Eq.~\eqref{eq: circuit depth for eth}.
These findings provide strong motivation for further studies into quantum recursions. 

Another motivation for quantum recursions comes from how we derive the recursion unitary Eq.~\eqref{eq: DBI state main} from the non-linear Schrödinger equation Eq.~\eqref{eq: GML DBF}. 
The core idea is to discretise the continuous non-linear differential equation, akin to the Euler method used for numerically solving differential equations with classical computers. 
Similarly, any non-linear Schrödinger equation can be recast as quantum recursions, albeit with errors arising from discretisation. 
This observation vastly expands the potential use-cases of quantum recursions.
However, there is a caveat: for each non-linear Schrödinger equation, the implementation error (from discretisation of the differential equation and from approximations needed for unfolding implementation) must be carefully managed.%
\footnote{This is the same for classical numerical algorithms; see e.g. Ref.~\cite{Moore1994DiscreteDBI} for the example of classical double-bracket algorithm.}

The performance guarantees presented in Section~\ref{section: DBQITE} achieve this for the imaginary-time evolution, leveraging the exponential convergence of the algorithm. 
Since the trajectory following the differential equation is exponentially converging, some deviations resulting from implementation errors do not significantly affect performance. 
Hence, studying exponentially converging non-linear Schrödinger equations from the perspective of quantum recursions is an especially promising avenue for future research. 

We provide an even stronger motivation for such endeavours in the next chapter, by presenting a technique that can resolve the exponential circuit depth problem by employing (exponentially many) auxiliary systems as memory.
\chapter{Quantum dynamic programming}\label{chapter: QDP}

In this chapter, we develop an approach to quantum recursions that is complementary to unfolding in Chapter~\ref{chapter: quantum recursions}.
In particular, while unfolding performs un-computation and re-computation to address memory calls, our approach makes use of auxiliary systems instead.
The main result is the exponential circuit depth reduction that can be achieved for a certain class of recursions, thanks to these auxiliaries.
However, the number of auxiliary systems must be large for this circuit depth reduction, making our result a form of trade-off between circuit depth and width.
Intuitively, we use auxiliary systems as quantum memories.
They evolve with the system of interest, while storing the operations applied to them, and later instruct the computation using the stored information.

Specifically, our technique is a quantum adaptation of memoisation~\cite{Michie1968memoization}, a type of dynamic programming~\cite{Bellman1952_DP, Bellman_dynamic}.
Memoisation utilises a small amount of memory to yield vastly shorter computation times by avoiding re-computations.
We outline how it is done in classical computing with the example of the Fibonacci sequence.

\begin{example}[memoisation for Fibonacci sequence evaluation]\label{example: Fibonacci}
	Fibonacci sequence is defined by the recurrence relation 
	\begin{align}\label{eq: Fibonacci recurrence}
		F(k) = F(k-1)+F(k-2), \quad k\geq 2,
	\end{align}
	and its base cases $F(0) = 0$ and $F(1) = 1$. 
	It is possible to define an algorithm directly in this recursive manner.
	In this case $F(\cdot)$ is a function that calls itself twice: $F(k)$ would call $F(k-1)$ and $F(k-2)$, and each of those two would call ($F(k-2)$, $F(k-3)$) and ($F(k-3)$, $F(k-4)$), respectively.
	We can already see the obvious redundancy of this algorithm that re-calculate the same entry, e.g. $F(k-2)$, multiple times. 
	Indeed, the time complexity of this algorithm can be evaluated.
	If the time complexity for $F(k)$ is $T_{k}$, it must be larger than the sum of $T_{k-1}$ and $T_{k-2}$.
	Setting the base complexities $T_{0} = T_{1} = 1$, we get the recurrence relation for $T_{k}$, identical to Eq.~\eqref{eq: Fibonacci recurrence}, with base cases corresponding to $T_{0} = F(1)$ and $T_{1} = F(2)$.
	Hence, the complexity
	\begin{align}
		T_{N} \geq F(N+1) = O\left(\left(\frac{1+\sqrt{5}}{2}\right)^{N}\right),
	\end{align}
	exponential to $N$.
	
	However, a better algorithm exists.
	Suppose that there exists a ``memo pad'' that records the results $F(k)$ for each $k$ after evaluation.
	Furthermore, when the function $F(k)$ is called, instead of immediately starting the recursive function calls, we look up the memo pad and retrieve the result there if it already exists.
	This way, $F(k)$ for each $k$ is calculated only once, making the time complexity of the algorithm for $F(N)$ linear, i.e. $T_{N} = O(N)$.
	In other words, an exponential speedup is achieved with the cost of memory of the size $O(N)$.
\end{example}

The redundancy observed in the memoryless implementation without memoisation is reminiscent of the unfolding implementation in Section~\ref{section: quantum recursions definition}.
In the unfolding implementation, the memory-call $e^{it\rho_{k}}$ needed for each recursion step $k$ is synthesised using Eq.~\eqref{eq: covariance of unitary def}, i.e. $e^{it\rho_{k}} = \Upsirec{\rho_{k-1}}e^{it\rho_{k-1}} (\Upsirec{\rho_{k-1}})^{\dagger}$, every time it is called.
This repetition is the source of exponential circuit depth growth for the unfolding implementation, just like the exponential time complexity of the memoryless implementation in Example~\ref{example: Fibonacci}.

Hence, we expect a solution for quantum recursions analogous to classical recursive algorithms.
However, the memoisation technique as in Example~\ref{example: Fibonacci} cannot be used straightforwardly because of quintessentially quantum challenges. 
Memory utilisation in memoisation consists of two steps: writing down the $k$th result after evaluating it and reading it out from the memory when needed. 
In quantum recursions, the $k$th result of the recursion is a quantum state $\rho_{k}$.

This result can be stored in two ways: classically or quantumly. 
The classical way is what we call the tomography-based learning-and-compiling method. 
Learning a quantum state is a challenge because of the uncertainty principle of quantum mechanics. 
In this approach, the classical description of $\rho_{k}$ (i.e. its density matrix) is extracted via tomography~\cite{Fano1957Tomography, QuantumEstimationBook, Haah2016_tomography}.
This process requires a large number of copies of $\rho_{k}$, scaling polynomially with the dimension of the system ($2^{n}$ for an n-qubit system), as detailed in Eq.~\eqref{eq: tomography scaling}.

It is infeasible to go through this scaling for each step $k$, and hence we must consider storing the result state quantumly, i.e. as a memory state $\rho_{k}$.
Now, the remaining problem is how to read out this quantum memory to synthesise the quantum recursion unitary $\Upsirec{\rho_{k}}$.

Counter-intuitively, our solution uses the memories without reading them out.
Specifically, we use the techniques introduced in Section~\ref{section: state instructed circuits}.
Note that those techniques---density matrix exponentiation (DME) or Hermitian-preserving map exponentiation (HME)---also require multiple copies of the instruction state $\rho_{k}$ to approximate the recursion unitary to good precision. 
Nevertheless, the scaling is typically better: for example, DME requires $M = \frac{1}{\epsilon}$ copies without the explicit dependence (Eq.~\eqref{eq: DME sample complexity}) on the system dimension, as opposed to Eq.~\eqref{eq: tomography scaling}. 

In this sense, our approach belongs to a lineage of studies on circuits instructed by quantum states~\cite{Lloyd2014quantum, Marvian2016_emulator, Pichler2016DME, Kimmel2017DME_OP, Kjaergaard2022DME, Wei2023hermpreserving, Patel2023WML1, Patel2023WML2, Rodriguezgrasa2023cloningDME, Go2024DME, Schoute2024QProgrammableReflections}, where desired operations are implemented by injecting quantum instruction states that encode the operation, rather than by compiling circuits based on classical information.
However, we use this technique not only with given (static) quantum states, but also with quantum memories that are dynamically evolving. 
We call this alternative algorithmic paradigm for quantum recursions \emph{quantum dynamic programming (QDP)}.

The main breakthrough of this work is the following observation: instead of unfolding the recursion (i.e. synthesising $\Upsirec{\rho_{k}}$ with exponentially many queries to $\Upsirec{\rho_{0}}$), we can evolve memory states from $\rho_{0}$ to $\rho_{k}$ in parallel and directly implement $\Upsirec{\rho_{k}}$ by injecting these memory states. 
In essence, QDP ``folds up'' the unfolded recursion step $\Upsirec{\rho_{k}}$ into the memory state as it evolves to $\rho_{k}$.
By doing so, we achieve an exponential circuit depth reduction compared to the unfolding approach, as formalised in Theorems~\ref{theorem: qdp pure recursion} and~\ref{theorem: qdp mixed recursion}, on a par with the classic example of computing Fibonacci numbers.
In the rest of this chapter, we introduce QDP in more detail and prove the exponential circuit depth reduction for two separate cases: pure state recursions and mixed state recursions. 

My original results in this section include Theorems~\ref{theorem: qdp pure recursion}, \ref{theorem: qdp mixed recursion} and Lemma~\ref{lemma: mixedness reduction subroutine}.

\section{Pure state recursions}\label{section: pure state recursions}

In this section, we consider quantum recursions with pure quantum states evolving as
\begin{align}
	\psivec{k} = \Upsirec{\psi_{k-1}}\psivec{k-1},
\end{align}
with the notation for its density matrix $\psi_{k} = \dm{\psi_{k}}$.
As in the unfolding implementation introduced in Section~\ref{section: quantum recursions definition}, we approximate the recursion unitary $\Upsirec{\psi}$ by 
\begin{align}\label{eq: dynamic unitary 0}
	\Upsirec{\{f_{i}\}_{i},\psi}=  V_{L} e^{i f_{L}(\psi)}  V_{L-1}  \ldots  V_{1} e^{i f_{1}(\psi)}  V_{0},
\end{align}
where $\{f_{i}\}_{i}$ is a collection of Hermitian-preserving (linear or polynomial) maps $ f_{i}$, while $ V_{i}$ are static unitaries independent of the instruction $\ket{\psi}$.
The quantum recursion unitary Eq.~\eqref{eq: dynamic unitary 0} contains $L$ \emph{memory-calls} of the form $e^{i f(\psi)}$.
Recall that DB-QITE Eq.~\eqref{eq: DBQITE  main equation} introduced in Section~\ref{section: DBQITE} is one example of a pure state recursion with one memory-call $e^{i\sqrt{s_{k}}\psi_{k}}$ and two constant unitaries $e^{\pm i\sqrt{s_{k}}H}$.

To resolve the exponential depth of unfolded circuits discussed in Chapter~\ref{chapter: quantum recursions}, QDP invokes dynamically evolving memory states to directly instruct memory-calls. 
That is, instead of synthesising $e^{if_{i}(\psi_{k})}$ using multiple queries to $\Upsirec{\psi_{k-1}}$ as in unfolding implementation (Proposition~\ref{proposition: higher order linear} followed by Eq.~\eqref{eq: covariance of unitary def}), multiple copies of $\psivec{k}$ and HME are used to implement $e^{if_{i}(\psi_{k})}$ directly. 
To reiterate the scheme, we first define Eq.~\eqref{eq: HME def} as a \emph{memory-usage query} 
\begin{align}\label{eq: memory-usage query}
	\hE{\tau}{f, \rho}(\sigma_{S}) \coloneq \Tr_{\bar{S}}\left[e^{-i\tau F}\left(\rho_{\bar{S}}\otimes\sigma_{S}\right)e^{i\tau F} \right].
\end{align}
Here, $F\coloneq d(C_{f})^{\transp_{\bar{S}}}$, where $C_{f}$ is the Choi matrix of $f$ as defined in Definition~\ref{theorem: Choi theorem}, $d$ is the dimension of $S$, and the partial transpose $\transp_{\bar{S}}$ acts only on the system $\bar{S}$ that has the same size as the system $S$.
We call $S$ as the \emph{working register}, upon which the operation acts, and $\bar{S}$ as the \emph{memory register}, which contains the instruction state $\rho$.
Each memory-usage query invokes an error that scales quadratically with the duration $\tau$.
By setting $\tau = 1/M$ and the same memory-usage queries (Eq.~\eqref{eq: memory-usage query}) $M$ times, we obtain the channel $(\hE{1/M}{f,\rho})^{M}$, which approximates the memory-call unitary $e^{if(\rho)}$ with a total error $ O(\tau^{2}/M)$.
Therefore, the error in the memory-call approximation can be kept arbitrarily small by increasing $M$.

To be explicit, we approximate the recursion unitary Eq.~\eqref{eq: dynamic unitary 0} by intialising the recursion with $\sigma_{0} = \psi_{0}$ and define the QDP iteration as
\begin{align}\label{eq: iterated memory usage}
	\sigma_{k+1} = 	 V_{L} \left(\hE{1/M}{f_{L},\sigma_{k}}\right)^{M}V_{L-1}  \ldots  V_{1} \left(\hE{1/M}{f_{L},\sigma_{k}}\right)^{M}  V_{0},
\end{align}
i.e. each memory-call is approximated by $M$ memory-usage queries, totalling $ML$ memory-usage queries for a step. 
In general, memory states $\sigma_{k\neq0}$ are no longer pure because the channel $\hE{1/M}{ f_{i},\sigma_{k}}$ is not exactly unitary. 
However, we show in Theorem~\ref{theorem: qdp pure recursion} that $\sigma_{N}$ can be made arbitrarily close to the desired state $\psivec{N}$. 

To prepare one copy of $\sigma_{1}$, we make $ML$ memory-usage queries with the memory state $\sigma_{0}$.
This requires $(ML+1)$ root state copies ($ML$ in the memory register and one in the working register).
Likewise, preparing $\sigma_{2}$ requires $(ML+1)^{2}$ copies of $\sigma_{0}$; thus, the iteration of Eq.~\eqref{eq: iterated memory usage} consumes $(ML+1)^{k}$ copies of the root state for preparing one copy of $\sigma_{k}$.
Meanwhile, the circuit depth remains linear: multiple instruction states can be prepared in parallel, resulting in a maximum depth of $kML$ memory-usage queries.
Compare this to the unfolding scenario in Section~\ref{section: quantum recursions definition}, where a memory-call to $\psivec{k}$ is executed by making $(2ML+1)^{k}$ memory-calls to the root state (see Eq.~\eqref{eq: unfolding exponential circuit depth growth}).
In contrast, QDP folds all root state calls into the memory state $\sigma_{k}$, allowing a memory-call to be implemented with a fixed-depth circuit.

QDP inherently involves approximation errors at each step, which depend on $M$, the number of memory-usage queries per memory-call.
Setting $M =  O( 1/\epsilon)$ in Eq.~\eqref{eq: iterated memory usage} ensures that the QDP prepared state $\sigma_{1}$ approximates the exact result $\psivec{1}$ within an error $\|\sigma_1-\psi_{1}\|_{1} \leq O(\epsilon) $, in terms of the Schatten 1-norm $\|\cdot\|_{1}$ defined in Definition~\ref{definition: Schatten p-norms}.
However, subsequent steps may amplify this error, as $\sigma_{1}\mapsto\sigma_{2}$ becomes instructed by the approximate state $\sigma_1$, not the exact state $\psivec{1}$.
Indeed, the triangle inequality gives $\|\sigma_{k+1}-\psi_{k+1}\|_{1} \leq O((ML+1)\|\sigma_{k}-\psi_{k}\|_{1})$, which in principle allows the exponential error accumulation 
\begin{align}
	\|\sigma_{N} - \psi_{N}\|_{1} = O((ML+1)^{N}),
\end{align}
indicating the failure of QDP. 

To address this, we identify sufficient conditions to prevent such destructive error propagation. 
It turns out that if the first few steps are sufficiently accurate, then later steps benefit from stabilization properties of typical quantum recursions.
In particular, recursions with fixed-points (e.g. via the Polyak-Łojasiewicz inequality in gradient descent iterations~\cite{Karimi2016Convergence} or asymptotic stability in time-discrete dynamical systems~\cite{Moore1994DiscreteDBI}) typically exhibit exponential convergence. 
For quantum recursions, this implies $\|\psi_{N}-\psi_{\infty}\|_{1} \leq \alpha^N\|\psi_{0}-\psi_{\infty}\|_{1}$, for some $\alpha<1$.
However, this relies on the assumption we call \emph{fast spectral convergence}, whose formal definition is stated in Definition~\ref{definition: fast spectral convergence}.
Informally, fast spectral convergence can be understood as: 
\begin{enumerate}[label=(\roman*)]
	\item There exists a stable fixed-point $\psivec{\infty}$ as $N\rightarrow \infty$.
	\item The fixed-point is unique for any initial state $\psivec{0}$.
	\item The fixed-point is sufficiently strongly attracting. 
\end{enumerate}
Without (iii), recursions become unstable, suggesting limited physical relevance, as practically achievable protocols must withstand small perturbations.
In such cases, quantum computation would require infinite precision and resources for a successful operation.

It is worth noting that unfolding is also subject to similar stability constraints: if $\Upsirec{\psi_{0}}$ is not compiled exactly, subsequent recursion unitaries $\Upsirec{\psi_{k}}$, which rely on exponentially many applications of $\Upsirec{\psi_{0}}$, become exponentially unstable. 
Thus, whether using QDP or unfolding, quantum recursions must exhibit fast spectral convergence; otherwise, the computational task is ill-conditioned. When this condition is met, QDP delivers an exponential depth reduction compared to unfolding:

\begin{theorem}[Exponential circuit depth reduction; original result]\label{theorem: qdp pure recursion}
	Suppose a quantum recursion starting from $\psivec{0}$ satisfies fast spectral convergence. 
	Then QDP combined with a mixedness reduction protocol yields a final state $\sigma_{N}$ satisfying
	\begin{align}
		\|\sigma_N - \psi_{N}\|_{1} \leq \epsilon,
	\end{align}
	for any final error $0<\epsilon<\frac{2}{3}$ and with any success probability $1 - p_\mathrm{th}$, using a circuit of depth $ O(N\epsilon^{-1})$ and width $\epsilon^{-N}e^{ O(N)}\log(p_\mathrm{th}^{-1})$.
	Here, $\psi_{N} = \dm{\psi_{N}}$ is the exact solution to the recursion. 
\end{theorem}

We prove this theorem in Section~\ref{section: proofs of QDP theorems} by addressing two error types. 
Unitary errors are efficiently suppressed by fast spectral convergence; on the other hand, non-unitary errors destabilise convergence and require explicit mitigation. 
To suppress the latter, we employ an additional protocol (Proposition~\ref{proposition: swap test mixedness reduction}) designed to mitigate these non-unitary errors for pure states.
This protocol is applied following each recursion step implemented through Eq.~\eqref{eq: iterated memory usage}.
The proof of Theorem~\ref{theorem: qdp pure recursion} demonstrates this protocol incurs only a slightly faster exponential growth in circuit width.
Moreover, this additional mixedness reduction protocol is probabilistic, rendering QDP for pure states probabilistic. 
However, it is possible to achieve an arbitrarily large success probability with overhead $\log(p_\mathrm{th}^{-1})$ where $p_\mathrm{th}$ is the upper bound for the probability of failure.

\section{Mixed state recursions}\label{section: mixed state recursions}

Similarly to the pure state case in Section~\ref{section: pure state recursions}, we define the mixed state recursion starting from a mixed state $\rho_{0}$ as
\begin{align}\label{eq: q recursion mixed}
	\rho_{k}\mapsto\rho_{k+1} = \Upsirecch{\rho_{k}} (\rho_{k}) ,
\end{align}
where $\Upsirecch{\rho_{k}}$ is the unitary channel corresponding to the recursion unitary $\Upsirec{\rho_{k}}$ that depends on $\rho_{k}$.
Furthermore, the fast spectral convergence conditions in Section~\ref{section: pure state recursions} need to be slightly refined to account for the mixed state cases. 
Specifically, we update the uniqueness of the fixed-point (condition (ii)) to be
\begin{itemize}
	\item[(ii')] The fixed-point is \emph{spectrally unique}, i.e. if the initial states $\rho_{0}$ and $\rho'_{0}$ have the same spectrum, they have the same fixed-point $\rho_{\infty} = \rho'_{\infty}$. 
\end{itemize}
For pure states, conditions (ii) and (ii') are equivalent; the formal statement is in Definition~\ref{definition: fast spectral convergence}.

The exponential circuit depth reduction established in Theorem~\ref{theorem: qdp pure recursion} can also be achieved for mixed state recursions, albeit with lower efficiency. 
\begin{theorem}[Mixed state quantum dynamic programming; original result]\label{theorem: qdp mixed recursion}
	Suppose the quantum recursion satisfies fast spectral convergence.
	For any $\epsilon>0$, implementing $N$ recursion steps with each memory-call replaced by $ O(N\epsilon^{-1})$ memory-usage queries, yields a final state $\sigma_{N}$, such that 
	\begin{align}
		\|\sigma_N - \rho_{N}\|_1 \leq \epsilon,
	\end{align}
	using a circuit of depth $ O(N^{2}\epsilon^{-1})$.
	Here, $\rho_N$ is the exact solution to the recursion. 
\end{theorem}

This theorem is proved in Section~\ref{section: proofs of QDP theorems}.
Note that we do not have an explicit exponential scaling of the circuit width in this case unlike Theorem~\ref{theorem: qdp pure recursion}. 
This is due to the lack of an additional protocol, corresponding to the mixedness reduction for the pure state case, that can preserve the spectrum of the density matrices.
Hence, the non-unitary error from memory-usage queries must be suppressed by requiring each memory-usage query to be very close to a unitary channel. 
Unfortunately, the recently developed channel purification protocol~\cite{Liu2024ChannelPurification}, which is a quantum channel version of the state mixedness reduction protocols, also does not improve the scaling, as the number of copies needed for the channel purification is comparable to that of improving each memory-usage query. 
Hence, using our construction, the circuit width scales doubly exponentially in the worst case, which is highly impractical. 

\section{Examples}
We emphasise the potential impact of QDP by presenting examples of quantum recursions that can benefit from QDP.

\begin{example}[nested fixed-point Grover search~\cite{Yoder2014Grover}]
	The nested fixed-point Grover search~\cite{Yoder2014Grover}, is a recursive version of the Grover search algorithm. 
	The Grover search algorithm aims to transform a state $\psivec{0}$ into the one close to the \emph{unknown} target state $\ket{\tau}$.
	This transformation is achieved by 
	\begin{align}
		\psivec{0} \mapsto \psivec{1} = \left(\prod_{i = 1}^{L}e^{-i\alpha_{i}\psi_{0}}e^{-i\beta_{i}\tau}\right)\psivec{0},
	\end{align}
	with density matrices $\psi_{0} = \dm{\psi_{0}}$, $\tau = \dm{\tau}$, and suitable angles $\{\alpha_{i},\beta_{i}\}_{i}$.
	This transformation achieves 
	\begin{align}\label{eq: Grover performance non-recursive}
		\frac{1}{2}\| \psi_{1} - \tau \|_{1} \simeq e^{-\lvert \braket{\psi_{0}}{\tau}\rvert (2L+1)},
	\end{align}
	for large $L$, which implies $L = O(\lvert \braket{\psi_{0}}{\tau}\rvert^{-1})$ achieving the quadratic advantage over the classical serach scaling as $\lvert \braket{\psi_{0}}{\tau}\rvert^{-2}$.
	
	A recursive variant of this algorithm starts from the idea that, instead of setting a large $L$ for one round of the Grover search, we can choose small $L$ to obtain $\psivec{1}$ and re-initiate the Grover search with the new initial state $\psivec{1}$ and the reflection $e^{-i\alpha_{i}\psi_{1}}$.
	Although $\psivec{1}$ is not very close to $\tau$ due to small $L$, the distance to $\tau$ is decreased compared to $\psivec{0}$, and the new round of the Grover search would transform it into a state that is even closer to $\tau$.
	An iteration of this process is called nested fixed-point Grover search, and it can be written as a quantum recursion 
	\begin{align}
		\psivec{k+1} = \left(\prod_{i = 1}^{L}e^{-i\alpha_{i}\psi_{k}}e^{-i\beta_{i}\tau}\right)\psivec{k} \eqcolon \Upsirec{\psi_{k}}\psivec{k}.
	\end{align}
	The performance of this algorithm is known to be equivalent to its non-recursive counterpart~\cite{Yoder2014Grover}, that is, the circuit depth required for a desired final distance to $\tau$ for non-recursive and recursive Grover search algorithms are almost equivalent.
	For the recursive version, the distance to the target final state $\ket{\tau}$ evolves as  
	\begin{align}\label{eq: Grover performance recursive}
		\frac{1}{2}\| \psi_{N} - \tau \|_{1} \simeq e^{-\lvert \braket{\psi_{0}}{\tau}\rvert (2L+1)^{N}},
	\end{align}
	with $(2L+1)$ in the non-recursive version (Eq.~\eqref{eq: Grover performance non-recursive}) replaced by $(2L+1)^{N}$.
	Nevertheless, when the recursion is implemented through unfolding as in Section~\ref{section: quantum recursions definition}, the total circuit depth scales exponentially as $ (2L+1)^{N}$, annulling the improvement. 
	This unfolding implementation and its analysis have been studied in Ref.~\cite{Yoder2014Grover}, albeit with small differences in details. 
	
	Since Eq.~\eqref{eq: Grover performance recursive} exhibits fast spectral convergence, the QDP implementation we developed in this chapter can be applied; this reduces the circuit depth to $ O(N)$ but with the requirement of exponential copies of the initial state $\psivec{0}$.
	Unfortunately, it is known that~\cite{Kimmel2017DME_OP} there is no quantum advantage in sample-based Grover search algorithms, which include the QDP implementation. 
	Nevertheless, this example serves as a valuable demonstration of QDP’s applicability for a historically significant quantum algorithm. 
\end{example}

The second example is the double-bracket quantum imaginary-time evolution (DB-QITE) algorithm introduced in Chapter~\ref{chapter: quantum recursions}.
There are two different ways to apply QDP for two different setups of implementing DB-QITE. 

\begin{example}[DB-QITE (Section~\ref{section: DBQITE})]
	The recursion unitary is defined in Eq.~\eqref{eq: DBQITE  main equation}, and Theorem~\ref{theorem: fidelity convergence DBQITE} guarantees the fast spectral convergence of this recursion. 
	Hence, it is possible to apply QDP to this algorithm by replacing the memory-call $e^{i\sqrt{s_{k}}\psi_{k}}$ in the recursion unitary with the DME memory-usage query Eq.~\eqref{eq: memory-usage query} with $f = \id$.
	Since the recursion states $\psivec{k}$ are pure, Theorem~\ref{theorem: qdp pure recursion} can be applied, which implements $N$ steps of DB-QITE recursion with circuit depth linear to $N$ and width exponential to $N$.
\end{example}  

Interestingly, QDP can directly implement another recursion, which could not have been implemented via unfolding. 
Specifically, QDP can be applied to the QITE recursion Eq.~\eqref{eq: DBI state main}, which we call QITE DBI. 
QITE DBI, which yields DB-QITE following a product formula approximation, exhibits a superior convergence rate to the ground state in terms of recursion steps when compared to DB-QITE (see Table II of Ref.~\cite{SonDBQITE}).
However, its performance concerning circuit depth or width requires separate evaluation and warrants a dedicated study. 
This example is presented primarily to showcase QDP's ability to unlock new quantum recursions for implementation.

\begin{example}[QITE DBI]
	The recursion Eq.~\eqref{eq: DBI state main} is known as QITE DBI in Ref.~\cite{SonDBQITE}.
	Furthermore a convergence theorem similar to Theorem~\ref{theorem: fidelity convergence DBQITE} has been established in the same paper. 
	Importantly, both algorithms converge to the ground state with the fidelity $1 - q^{k}$ after $k$ steps, and QITE DBI typically has a larger lower bound for $q$ than DB QITE. 
	
	To apply QDP, we consider the sample-based Hamiltonian simulation setup~\cite{Kimmel2017DME_OP}.
	In other words, instead of having access to evolutions $e^{-itH}$, multiple copies of the state $\Theta\propto H$ that encodes the Hamiltonian are given. 
	Multiple copies of $\Theta$ can indeed emulate the evolution $e^{-itH}$ with DME.
	We also prepare copies of some pure state $\psivec{0}$ having non-zero overlap with the ground state. 
	Using $\Theta$ and $\psivec{0}$ copies, recursions of the form
	\begin{align}
		\psivec{k} \mapsto \psivec{k+1} = e^{s[\dm{\psi_{k}},H]}\psivec{k},
	\end{align}
	with some duration $s$ can be implemented oblivious to both $H$ and $\psivec{k}$ with Proposition~\ref{proposition: PFE}.
\end{example}
	
Finally, we introduce a novel algorithm that is natively dynamic.	
\begin{example}[oblivious Schmidt decomposition]
	For pure bipartite states, the Schmidt decomposition (Definition~\ref{definition: Schmidt decomposition}) fully characterises their entanglement.
	Thus we may envision a protocol that unitarily transforms any unknown $\psivec{0} \in \mH_{AB}$ into a form where its Schmidt basis aligns with the computational basis: $V_{A} \otimes V_{B}\psivec{0} = \sum_{k=1}^{D} \sqrt{\lambda_{k}} \ket{k}\otimes \ket{k}$ with $\{\lambda_{k}\}_{k}$ denoting the Schmidt spectrum.
	
	Traditionally, prior knowledge of $\psivec{0}$ is required to construct $V_{A}$ and $V_{B}$. 
	However, QDP enables this task while completely circumventing such a need. 
	In this section, we demonstrate that this task is achievable by diagonalizing the reduced state of a given pure state $ \psi $ via an adaptation of the double-bracket iteration~\cite{Gluza2024DBI}, similar to Eq.~\eqref{eq: QITE DBF approximation}.   
	
	We first describe how double-bracket iterations on the reduced system can be implemented with memory-usage queries using the entire bi-partite state. 
	Suppose that multiple copies of the instruction state $\psivec{k}_{AB}$ are given. 
	Let $D$ be a non-degenerate diagonal operator on subsystem $A$.
	The recursion unitary is defined as
	\begin{align}\label{eq:stateDBI_main}
		\Upsirec{f,\psi_{k}} = e^{s\left[D,\Tr_B[\psi_{k}]\right]}\otimes \1_B\ .
	\end{align}
	This is a single memory-call type recursion with $f(\rho_{AB})=-is[D,\Tr_{B}[\rho_{AB}]]\otimes\1_{B}$ that is Hermitian-preserving; hence, with copies of $\psivec{k}$, the recursion step $\psivec{k} \mapsto \psivec{k+1}= \Upsirec{\psi_{k}}\psivec{k}$ can be approximated using HME.
	The fast spectral convergence of such a unitary is shown in Refs.~\cite{HelmkeMoore1994Book, Smith_Thesis}.
	When $s$ is small, Theorem~\ref{theorem: qdp pure recursion} applies and $\{\psivec{k}\}$ exponentially converges to the fixed-point of the recursion Eq.~\eqref{eq:stateDBI_main}.
	This fixed-point is the state that commutes with the matrix $D$, i.e. a state diagonal to the computational basis.
	In other words, QDP effectively applies $V_{A}\otimes\1_{B}\psivec{0}$.
	A similar procedure implements $V_{B}$, completing the oblivious Schmidt decomposition.
	
	Let us discuss the implications of oblivious Schmidt decomposition in the context of quantum information processing.
	The replica method~\cite{Horodecki2002method} extracts classical information $\lambda_{k}$ but requires exponentially many swap operations.
	Similarly, we expect that oblivious Schmidt decomposition may also require a long runtime to converge.
	However, it not only enables sampling from the Schmidt spectrum but also provides the bipartite quantum state $\sum_{k} \sqrt{\lambda_{k}}\ket{k}\otimes\ket{k}$ \emph{coherently} in the computational basis.
	This is useful, e.g. for entanglement distillation of an \emph{unknown} state $\psi$, contrasting with standard settings~\cite{Bennett1996concentrating} that require the knowledge about the initial state for compiling local operations.
	While oblivious Schmidt decomposition via QDP may not attain the optimal asymptotic rates as derived from Ref.~\cite{Matsumoto2007universal}, it provides a constructive approach to oblivious entanglement distillation with explicit circuit implementations.
\end{example}

\section{Concluding remarks: Implications of the depth-width trade-off}\label{section: QDP implications}

Theorems~\ref{theorem: qdp pure recursion} and~\ref{theorem: qdp mixed recursion} establish the circuit depth-width trade-off.
At its core, this trade-off is valuable because it exponentially reduces computational time, which can turn a practically impossible task into a feasible one.
Beyond this immediate advantage, we focus on how QDP facilitates implementation. 

One simple measure is the quantum circuit size, defined as the product of the maximum circuit depth and width~\cite{Yoder2016universal, Takagi2017Error}. 
We illustrate how this measure benefits from the trade-off.
Suppose that in each recursion step, $L$ memory-calls to $\rho$ are made, each approximated by $M$ black-box access to the evolution of the form $e^{-it\rho}$. 
The unfolding circuit size therefore scales as $(2ML+1)^{N}$ for $N$ recursion steps as described in Eq.~\eqref{eq: unfolding exponential circuit depth growth}.
Meanwhile, memory-usage queries directly implement each memory-call, but require $M'$ memory-usage queries to manage implementation error, i.e. $M'L$ memory-usage queries per recursion step. 
In this scenario, the QDP circuit width scales as $(M'L+1)^{N}$, which dominates the total circuit size for large $N$.
Whenever $L$ is large and $M'<2M$, QDP approximately achieves a $(2M/M')^{N}$-fold improvement in circuit size compared to unfolding.
For instance, in the most error-tolerant case of $M = M' = 1$, QDP significantly reduces the circuit size.

An interesting comparison arises from studies on quantum circuit complexity, which has been studied in contexts ranging from quantum circuit compilation to black holes~\cite{Nielsen2006Geometry, Dowling2006Geometry, Brown2018SecondLaw, Brandao2021Complexity, Haferkamp2022Complexity}. 
Quantum circuit complexity of a state is defined as the minimum number of 2-local gates required to prepare that state from a fiducial state such as $\ket{0}$. 
An implicit assumption in this definition is that the system is closed, without auxiliary systems. 
Ref.~\cite{Du2024Embedded} explores the consequences of relaxing this assumption, investigating the depth-width trade-off when auxiliaries are employed. 
It demonstrates that using auxiliary systems enables the preparation of more complex states with shallower circuits, while the overall circuit size remains comparable (up to constant factors). 
Intriguingly, this advantage is gained by using the auxiliaries to effectively ``fold up'' the complexity within them, which is then transferred using the gate teleportation scheme~\cite{Gottesman1999}. 
This mechanism of using auxiliary systems to store and later deploy complexity closely parallels the way quantum memories are employed in QDP.

Beyond circuit size, QDP offers additional practical advantages: its local modularity is particularly well-suited for distributed quantum computing~\cite{Wehner2018_QInternet, Cacciapuoti2019Distributed, Davarzani2020Distributed}. 
By localizing and decoupling circuits~\cite{Wang2024Decoupling}, QDP allows copies of $\psivec{k}$ to be prepared independently before being injected into recursion steps $\Upsirec{\psi_{k}}$.
As the recursion progresses, the circuit width decreases exponentially as most copies are consumed as memory and traced out during memory-usage queries.
This enables greater flexibility in parallelization, since processors with shorter coherence times can be allocated for preparing $\psivec{k}$ with smaller $k$, optimizing the use of available quantum devices.

The flexibility of QDP is reinforced by a hybrid strategy.
Realistically, quantum devices can neither execute exponentially many sequential gates nor operate on exponentially many qubits simultaneously. 
A practical approach is to initiate QDP after several rounds of unfolding, distributing the exponential factor between circuit depth and width to prevent either from becoming prohibitive.
For example, one could begin with $N_{1}$ unfolding steps, using a circuit of depth $e^{ O(N_{1})}$, nearing the device’s depth limit. 
These unfolding steps are performed in parallel, maximizing circuit width to produce $M$ copies of intermediate states $\psivec{N_{1}}$.
For many quantum recursions, this starting strategy also conveniently steers the system into a regime where the recursion starts to exhibit fast spectral convergence. 
This allows  QDP to subsequently take over, executing an additional $ N_{2} $ recursions to attain the final state $ \sigma_{N_{1}+N_{2}} $. 
This hybrid strategy fully utilises the device capacity, which is otherwise limited to producing either $ \psivec{N_{1}} $ without QDP or $ \sigma_{N_{2}} $ without unfolding. 

Currently, we are unaware of quantum algorithms where QDP ensures a rigorous computational advantage over classical methods. 
These may appear in settings robust against imperfect unitary implementations, similar to those present in diagonalizing double-bracket iterations~\cite{HelmkeMoore1994Book,Smith_Thesis,Gluza2024DBI}.
Furthermore, QDP allows us to add oblivious Schmidt decomposition to the quantum algorithmic toolkit. 
We hope that once it will be feasible to experimentally implement memory-usage queries with high fidelity, oblivious Schmidt decomposition and QDP in general will facilitate practical state preparations that will advance our knowledge of quantum properties in materials, e.g. magnets or superconductors.

\section{Proofs of exponential circuit depth reduction}\label{section: proofs of QDP theorems}

We begin by formally defining fast spectral convergence for pure and mixed state recursions.  

\begin{definition}[Fast spectral convergence]\label{definition: fast spectral convergence}
	Consider a quantum recursion unitary channel $\Upsirecch{\rho}$ that defines a fixed-point iteration
	\begin{align}\label{eq: fp iteration}
		\rho\mapsto\Upsirecch{\rho}(\rho).
	\end{align} 
	This recursion defines sequences $\{\rho_{0}\}_{k=0}^{\infty}$ for each initial state $\rho_{0}$.
	
	The recursion has the fast spectral convergence property if it satisfies the following conditions.
	\begin{enumerate}
		\item For any initial state $\rho_{0}$, the sequence must converge to a fixed-point $\tau$, i.e. the trace distance $ \delta_{k} \coloneq \frac{1}{2}\|\tau - \rho_{k} \|_{1} < \delta_{k-1}$ and $\displaystyle\lim_{k\rightarrow\infty} \delta_{k} = 0$. 
		\item The fixed-point $\tau$ must be the same for any two initial states $\rho_{0}$ and $\tilde{\rho_{0}}$ (with the exception of states chosen from a measure zero set), with $\spec(\rho_{0}) = \spec(\tilde{\rho}_{0})$.
		\item The convergence must be fast, i.e.  	
		for any iso-spectral states $\rho$ and $\sigma$, 
		\begin{align}
			\| \Upsirecch{\rho}(\rho) - \Upsirecch{\sigma}(\sigma) \| \leq r \|\rho - \sigma\| 
		\end{align}
		for some $r<1$.
	\end{enumerate}
\end{definition}

Spectral convergence implies that the QDP implementation will also approach the same fixed-point, when the non-unitary error is not too large. 
Note that Theorem~\ref{theorem: fidelity convergence DBQITE} guarantees the fast spectral convergence for DB-QITE algorithm. 

We first prove Theorem~\ref{theorem: qdp mixed recursion} valid for mixed state recursions.
\begin{proof}[Proof of Theorem~\ref{theorem: qdp mixed recursion}]
	From Proposition~\ref{proposition: HME}, we are able to locally accurately implement the unitary channel $\Upsirecch{\rho}$ by a \emph{non-unitary} channel $\Upsirecapp{\rho}$, such that 
	\begin{align}\label{eq: locally accurate implementation}
		\frac{1}{2}\left\|\Upsirecch{\rho} - \Upsirecapp{\rho}\right\|_{\Tr} \leq \eta,
	\end{align}
	for any $\eta>0$, by making $ O(\eta^{-1})$ memory-usage queries each consuming a copy of $\rho$.
	The circuit depth for this implementation is also $ O(\eta^{-1})$. 
	
	Suppose that the sequence of states $\{\rho'_{k}\}_{k}$ is obtained from such emulation starting from $\rho'_{0} = \rho^{}_{0}$ and thus recursively defined 
	\begin{align}\label{eq:recursion_without_purification}
		\rho'_{k} = \Upsirecapp{\rho'_{k-1}}\left(\rho'_{k-1}\right)\ .
	\end{align}
	This sequence might deviate from the desired sequence $\{\rho_{k}\}_{k}$ very quickly and, in general, $\spec\{\rho'_{k}\} \neq \spec\{\tau\}$.
	
	The quantity of interest is the distance
	\begin{align}
		\delta_{k} \coloneq \frac{1}{2}\| \rho_{k} - \rho'_{k} \|_{1},
	\end{align}
	and we want to upper bound $\delta_{N}$ with an arbitrarily small $\epsilon>0$.
	For some state $\tilde{\rho}_{k}$, the triangle inequality gives
	\begin{align}
		\delta_{k} \leq \frac{1}{2}\| \rho_{k} - \tilde{\rho}_{k} \|_{1} + \frac{1}{2}\| \tilde{\rho}_{k} - \rho'_{k} \|_{1} \eqcolon \mu_{k} + \varepsilon_{k},
	\end{align}
	where we define $\mu_{k} \coloneq \frac{1}{2}\| \rho_{k} - \tilde{\rho}_{k} \|_{1} $ and $\varepsilon_{k} = \frac{1}{2}\| \tilde{\rho}_{k} - \rho'_{k} \|_{1} $.
	
	Now we set $\{\tilde{\rho}_{k}\}_{k}$ to be a sequence of states defined by the exact unitary recursion with an erroneous instruction state $\tilde{\rho}_{k} = \Upsirecch{\rho'_{k-1}}(\tilde{\rho}_{k-1})$ starting from $\tilde{\rho}_{0} = \rho_{0}$. 
	Hence, $\spec\{\tilde{\rho}_{k}\} = \spec\{\tau\}$ for any $k$, and $\varepsilon_{0} = 0$.
	We first analyse how $\varepsilon_{k}$ scales.
	Observe that
	\begin{align}\label{eq: varepsilon scaling}
		\varepsilon_{k+1} = \frac{1}{2}\left \| \Upsirecch{\rho'_{k}}(\tilde{\rho}_{k}) - \rho'_{k+1} \right \|_{1} &\leq \frac{1}{2}\left \| \Upsirecch{\rho'_{k}}(\tilde{\rho}_{k}) - \Upsirecch{\rho'_{k}}(\rho'_{k})  \right \|_{1} +\frac{1}{2} \left \| \Upsirecch{\rho'_{k}}(\rho'_{k}) - \Upsirecapp{\rho'_{k}}(\rho'_{k}) \right \|_{1}\nonumber\\
		&\leq \varepsilon_{k} + \eta,
	\end{align}
	from the triangle inequality, the unitary invariance of the trace norm, and Eq.~\eqref{eq: locally accurate implementation}.
	Therefore, $\varepsilon_{k}$ scales linearly i.e.
	\begin{align}
		\varepsilon_{k}\leq k\eta.
	\end{align} 
	
	Now we analyse how $\mu_{k}$ evolves.
	Observe that 
	\begin{align}
		\mu_{k+1} = \frac{1}{2}\left\| \Upsirecch{\rho_{k}}(\rho_{k}) - \Upsirecch{\rho'_{k}}(\tilde{\rho}_{k}) \right\|_{1} &\leq \frac{1}{2}\left\| \Upsirecch{\rho_{k}}(\rho_{k}) - \Upsirecch{\tilde{\rho}_{k}}(\tilde{\rho}_{k}) \right\|_{1}  + \frac{1}{2}\left\| \Upsirecch{\tilde{\rho}_{k}}(\tilde{\rho}_{k})  - \Upsirecch{\rho'_{k}}(\tilde{\rho}_{k}) \right\|_{1}\nonumber\\ 
		&\leq r\mu_{k} + \frac{1}{2}\left\| \Upsirecch{\tilde{\rho}_{k}}(\tilde{\rho}_{k})  - \Upsirecch{\rho'_{k}}(\tilde{\rho}_{k})\right\|_{1},\label{eq: mu k evolve}
	\end{align}
	with some $r<1$ from the fast spectral convergence condition in Definition~\ref{definition: fast spectral convergence}.
	
	Next, recall that $\left \| e^{if(\rho)} - e^{if(\sigma)}\right \| \leq C'\| \rho - \sigma\|$
	for some constant $C'$, from the mean value theorem for operators.
	This implies 
	\begin{align}\label{eq:MVT}
		\left \| \Upsirecch{\rho}- \Upsirecch{\sigma}\right\|_{\Tr} \leq C\| \rho - \sigma\|_{1}
	\end{align}
	for some constant $C$.
	Hence, Eq.~\eqref{eq: mu k evolve} can be further bounded as
	\begin{align}
		\mu_{k+1} \leq r\mu_{k} + C\varepsilon_{k} \leq r\mu_{k} + Ck\eta. 
	\end{align}
	This recursive relation gives
	\begin{align}
		\mu_{N} \leq r^{N}\mu_{0} + \sum_{k = 0}^{N-1}r^{N-k-1}Ck\eta = O(N\eta).
	\end{align}
	
	Therefore, we obtain the final error
	\begin{align}
		\delta_{N} \leq \mu_{N} + \varepsilon_{N} \leq O(N\eta).
	\end{align}
	The desired upper bound $\delta_{N}\leq\epsilon$ for any $\epsilon>0$ can be obtained by setting $\eta = O(\epsilon N^{-1})$.
	
	Since the circuit depth for each recursion step scales as $\eta^{-1}$, regardless of the step number $k$, the total circuit depth scales as $N\eta^{-1} = O(N^{2}\epsilon^{-1})$, as stated in the theorem.
\end{proof}

Before we prove the theorem for pure state recursions, we introduce a useful lemma using the mixedness reduction protocol in Section~\ref{section: mixedness reduction}.
\begin{lemma}[QDP mixedness reduction subroutine; original result]\label{lemma: mixedness reduction subroutine}
	Let $ \rho $ be a density matrix with the largest eigenvalue $\lambda_{1} = 1-x$ for some mixedness parameter $x \in [0,\frac{1}{3}]$, corresponding to the eigenvector $\ket{v_{1}}$. 
	Given any maximum tolerable failure probability $q_\mathrm{th}\geq 0$, one can choose any parameter $\mathsf{g}\in\mathbb{R}$ and prepare $M$ copies of $ \rho' $, such that:
	\begin{enumerate}
		\item $\rho'$ has the largest eigenvalue $\lambda'_{1} \geq 1-\frac{x}{\mathsf{g}}$, corresponding to the same eigenvector $\ket{v_{1}}$. 
		\item a number of $ R =  O(\log(\mathsf{g})) $ mixedness reduction rounds in Proposition~\ref{proposition: swap test mixedness reduction} is used,
		\item a total of $M (\frac{2}{c})^{R}$ copies of $ \rho $ is consumed, 
		with $ c\in(0,1) $ satisfying
		\begin{align}\label{eq: c for mixedness reduction}
			c = 1 - x - M^{-\frac{1}{2}}\sqrt{\log\left(\frac{R}{q_\mathrm{th}}\right)}
		\end{align}
		\item the success probability of the entire subroutine is $q_\mathrm{succ} \geq 1 - q_\mathrm{th}$.
	\end{enumerate}
	Note that $M$ must be sufficiently large to guarantee $c>0$.
\end{lemma}
\begin{proof}
	To prove this lemma, we analyze $R$ sequential rounds of the mixedness reduction protocol, with initial and final states $\rho$ and $\rho'$. 
	Let us denote the intermediate states generated via the mixedness reduction protocols as $\lbrace \chi_j\rbrace_{j=0}^R$, where $\chi_0 = \rho$ and $\chi_R = \rho'$; and let $\lbrace y_j \rbrace_{j=0}^R$ be the mixedness parameter (i.e. $1$ minus the largest eigenvalue) for each $\chi_{j}$ with $y_{0} = x$ and $y_{R} \leq \frac{x}{\mathsf{g}}$. 
	Furthermore, let $M_j$ be the number of copies of an intermediate state $\chi_j$ generated; hence, we obtain $M_{R} = M$ copies of the desired state $\rho'$ at the end, and aim to show that $M_{0} = M\times (\frac{2}{c})^{R}$ suffices. 
	
	From Eq.~\eqref{eq: x ratio upperbound} in Proposition~\ref{proposition: swap test mixedness reduction}, the mixedness parameter after $j$ rounds of the mixedness reduction protocol becomes $y_{j}\leq \frac{1+y_{j-1}}{2-2y_{j-1}+y_{j-1}^{2}}y_{j-1}$. 
	The final reduction $ \frac{x'}{x} = \frac{y_{R}}{y_{0}} \leq \mathsf{g}^{-1} $ is achievable if $R$ is sufficiently big to satisfy
	\begin{align}
		\frac{y_{R}}{y_{0}} = \prod_{j=1}^{R}\frac{y_{j}}{y_{j-1}} \leq \prod_{j=1}^{R} \frac{1+y_{j-1}}{2-2y_{j-1}+y_{j-1}^{2}} \leq \left(\frac{1+x}{2-2x+x^{2}}\right)^{R} \leq \mathsf{g}^{-1}.
	\end{align}
	The second last inequality follows from two facts: i) the factor $\frac{1+y}{2-2y+y^{2}}$ monotonically increases as $y$ increases ii) $y_{j}\leq y_{0}$ for all $j$.
	Since $\log(\frac{1+x}{2-2x+x^{2}})>0$ by the assumption $x\leq \frac{1}{3}$, the number of rounds $R =   O(\log(\mathsf{g}))$ is sufficient as claimed. 
	
	The parameter $c$ in the statement of the lemma can be interpreted as the survival rate after each round of the mixedness reduction protocol.
	More specifically, the $j$th round of the protocol is successful if at least $ \frac{c}{2}M_{j-1} $ output states $ \chi_{j} $ is prepared from $ M_{j-1} $ copies of $ \chi_{j-1} $.
	If successful, we discard all the surplus output copies and set $ M_{j} = \frac{c}{2}M_{j-1} $.
	If not, we declare that the whole subroutine has failed.
	Hence, if the entire subroutine succeeds, the final number of copies $ M = M_{R} $ is related to the initial number of copies as $M_{0} = (2/c)^{R} M$. 
	
	Finally, we estimate the success probability $ q_\mathrm{succ} $ given $ R, M, c, x $ and require it to be lower bounded with $ (1 - q_\mathrm{th}) $.
	The failure probability of each round will be bounded using Hoeffding's inequality. 
	From Proposition~\ref{proposition: swap test mixedness reduction}, the success probability of preparing one copy of $ \chi_{j} $ from a pair $ \chi_{j-1}^{\otimes 2} $ has a lower bound $p(y_{j-1}) \geq 1-y_{j-1}$.
	At each round, $ \frac{M_{j-1}}{2} $ independent trials of this mixedness reduction protocol are conducted.
	The probability of more than $ c\frac{M_{j-1}}{2} $ attempts succeeds, i.e. the success probability of a round, is
	\begin{align}
		q_{r}^{(j)} = 1 - F\left(\left\lceil\frac{c M_{j-1}}{2}-1\right\rceil;\frac{M_{j-1}}{2},p(y_{j-1})\right)\geq   1 - F\left(\frac{c M_{j-1}}{2};\frac{M_{j-1}}{2},1-y_{j-1}\right),
	\end{align}
	where $ F(k;n,q) $ is the cumulative binomial distribution function defined as
	\begin{align}
		F(c n;n,p) = \sum_{i=0}^{\lfloor c n\rfloor}\binom{n}{i}p^{i}(1-p)^{n-i}.
	\end{align}
	The expected number of success for $n$ trials of a process with success probability $p$ is $pn$, and $F(c n;n,p)$ denotes the probability of having less than $cn$ successes. 
	Hence, the Hoeffding's inequality gives $F(c n;n,p) \leq e^{-2n(p-c)^{2}}$, and consequently sets the bound
	\begin{align}
		q_{r}^{(j)} \geq 1 - e^{-M_{j}\left(1-y_{j-1} - c\right)^{2}}.
	\end{align}
	The success probability of an entire subroutine can then be bounded as
	\begin{align}
		q_\mathrm{succ} &= \prod_{j = 1}^{R}q_{r}^{(j)} \geq \prod_{j=1}^{R}\left(1 - e^{-M_{j}\left(1-y_{j-1} - c\right)^{2}}\right) >1 - \sum_{j=1}^{R}e^{-M_{j}\left(1-y_{j-1} - c\right)^{2}},
	\end{align}
	where the last inequality uses $ \prod_{j}(1-q_{j})>1-\sum_{j}q_{j} $ that holds when $ q_{j}\in(0,1) $ for all $ j $.
	Moreover, by recalling $ x\geq y_{j-1} $ and $ M\leq M_{j} $ for all $ j $, the final bound
	\begin{align}\label{eq: q success bound c}
		q_\mathrm{succ}> 1 - Re^{-M\left(1-x- c\right)^{2}} 
	\end{align}
	is obtained. 
	The prescribed failure threshold $ q_\mathrm{th} $ holds when 
	$e^{-M\left(1-x - c\right)^{2}} = \frac{q_\mathrm{th}}{R}$ ,
	or equivalently, when $M$ is sufficiently large and Eq.~\eqref{eq: c for mixedness reduction} is true. 
\end{proof}
During the earlier recursions, $M$ is always sufficiently large to guarantee that $c$ is well separated from $0$. 
For later recursions, in particular for the last recursion where only $M = 1$ is needed, Eq.~\eqref{eq: c for mixedness reduction} can become close to zero, or even negative. 
To guarantee positive and non-vanishing $c$, we allow some redundancy of the final state $\rho_{N}$ and prepare $M>1$ copies of it.  

Next we prove the theorem for pure state recursions using Lemma~\ref{lemma: mixedness reduction subroutine}.
\begin{proof}[Proof of Theorem~\ref{theorem: qdp pure recursion}]
	We again set $\Upsirecapp{\rho}$ to satisfy Eq.~\eqref{eq: locally accurate implementation}, but now we define the sequence $\{\rho'_{k}\}_{k}$ to have an initial state $\rho'_{0} = \psi_{0}$ and follow the recursion relation 
	\begin{align}
		\rho'_{k} = \mathrm{MR}\circ\Upsirecapp{\rho'_{k-1}}(\rho'_{k-1}),
	\end{align}
	instead of Eq.~\eqref{eq:recursion_without_purification}.
	$\mathrm{MR}$ denotes the $R$ rounds of the mixedness reduction subroutine as in Lemma~\ref{lemma: mixedness reduction subroutine}. 
	Suppose that $R$ is sufficiently large to guarantee that $\rho'_{k}$ has the largest eigenvalue $1-x$ with a mixedness parameter $x<\nu$ for some small $\nu$ and the corresponding eigenvector $\ket{\tilde{\psi}_{k}}$.
	Then we can write
	\begin{align}\label{eq: delta into mu and varepsilon in pure state recursion}
		\delta_{k} \coloneq \frac{1}{2}\| \psi_{k} - \rho'_{k} \|_{1} \leq \frac{1}{2}\| \psi_{k} - \tilde{\psi}_{k} \|_{1} +\frac{1}{2}\| \tilde{\psi}_{k} - \rho'_{k} \|_{1} \eqcolon \mu_{k} + \varepsilon_{k},
	\end{align}
	as in the proof for Theorem~\ref{theorem: qdp mixed recursion}.
	
	The first term, $\mu_{k}$, can be analysed using the inequality
	\begin{align}
		\mu_{k+1} = \frac{1}{2}\left\| \Upsirecch{\psi_{k}}(\psi_{k}) - \tilde{\psi}_{k+1} \right\|_{1} &\leq  \frac{1}{2}\left\| \Upsirecch{\psi_{k}}(\psi_{k}) - \Upsirecch{\tilde{\psi}_{k}}(\tilde{\psi}_{k}) \right\|_{1} +  \frac{1}{2}\left\|  \Upsirecch{\tilde{\psi}_{k}}(\tilde{\psi}_{k}) - \tilde{\psi}_{k+1} \right\|_{1} \nonumber\\
		&\leq r\mu_{k} +  \frac{1}{2}\left\|  \Upsirecch{\tilde{\psi}_{k}}(\tilde{\psi}_{k}) - \tilde{\psi}_{k+1} \right\|_{1}, \label{eq: mu k evolve in pure recursion proof}
	\end{align}
	with some $r<1$ using the fast spectral convergence in Definition~\ref{definition: fast spectral convergence}.
	The second term of Eq.~\eqref{eq: mu k evolve in pure recursion proof} can be further bounded by the triangle inequality 
	\begin{align}\label{eq:psi_tilde_evolve_triangle}
		\frac{1}{2}\left\| \tilde{\psi}_{k+1} - \Upsirecch{\tilde{\psi}_{k}}(\tilde{\psi}_{k}) \right\|_{1} \leq \frac{1}{2}\left\| \tilde{\psi}_{k+1} - \Upsirecapp{\rho'_{k}}(\rho'_{k}) \right\|_{1} + \frac{1}{2}\left\| \Upsirecapp{\rho'_{k}}(\rho'_{k}) - \Upsirecch{\tilde{\psi}_{k}}(\tilde{\psi}_{k})\right\|_{1}. 
	\end{align}
	Note that mixedness reduction subroutines do not change the eigenvector corresponding to the largest eigenvalue of the state; hence $\tilde{\psi}_{k+1}$ is the eigenvector of $\Upsirecapp{\rho'_{k}}(\rho'_{k})$ corresponding to the largest eigenvalue of it. 
	The first term of the bound Eq.~\eqref{eq:psi_tilde_evolve_triangle} is then the mixedness parameter of $\Upsirecapp{\rho'_{k}}(\rho'_{k})$.
	Since $\rho'_{k}$ has the mixedness parameter smaller than $\nu$ and $\Upsirecapp{\rho'_{k}}$ is only $\eta$ different from the unitary operator $\Upsirecch{\rho'_{k}}$, we obtain $\frac{1}{2}\| \tilde{\psi}_{k+1} - \Upsirecapp{\rho'_{k}}(\rho'_{k}) \|_{1} \leq \nu + \eta$.
	The second term of the bound Eq.~\eqref{eq:psi_tilde_evolve_triangle} can be further expanded as
	\begin{align}
		&\frac{1}{2}\left\| \Upsirecapp{\rho'_{k}}(\rho'_{k}) - \Upsirecch{\tilde{\psi}_{k}}(\tilde{\psi}_{k})\right\|_{1} \nonumber\\ 
		&\leq \frac{1}{2}\left\| \Upsirecapp{\rho'_{k}}(\rho'_{k}) - \Upsirecapp{\rho'_{k}}(\tilde{\psi}_{k}) \right\|_{1} + \frac{1}{2}\left\| \Upsirecapp{\rho'_{k}}(\tilde{\psi}_{k}) - \Upsirecch{\rho'_{k}}(\tilde{\psi}_{k}) \right\|_{1} + \frac{1}{2}\left\| \Upsirecch{\rho'_{k}}(\tilde{\psi}_{k}) - \Upsirecch{\tilde{\psi}_{k}}(\tilde{\psi}_{k}) \right\|_{1} \nonumber \\
		&\leq \frac{1}{2}\left\| \rho'_{k} - \tilde{\psi}_{k} \right\|_{1} + \frac{1}{2}\left\| \Upsirecapp{\rho'_{k}} - \Upsirecch{\rho'_{k}} \right\|_{\Tr} + \frac{1}{2}\left\| \Upsirecch{\rho'_{k}} - \Upsirecch{\tilde{\psi}_{k}} \right\|_{\Tr} \leq C\nu + \eta,
	\end{align}
	for some constant $C$.
	The second and third inequalities follow from data processing inequality, definition of the channel distance, and Eqs.~\eqref{eq: locally accurate implementation} and~\eqref{eq:MVT}.
	Combining everything, we arrive at the recursive bound
	\begin{align}
		\mu_{k+1} \leq r\mu_{k} + (C+1)\nu + 2\eta,
	\end{align}
	which leads to the explicit upper bound
	\begin{align}
		\mu_{N} \leq r^{N}\mu_{0} + \sum_{k = 0}^{N-1}r^{N-k-1}((C+1)\nu + 2\eta) = O(\nu+\eta).
	\end{align}
	
	The other error term $\varepsilon_{k}$ in Eq.~\eqref{eq: delta into mu and varepsilon in pure state recursion} is the distance between $\rho'_{k}$ and its principal component eigenvector $\tilde{\psi}_{k}$, which is bounded by $\nu$ by the applications of the mixedness reduction protocol. 
	Hence, we obtain the desired bound
	\begin{align}
		\delta_{N} \leq O(\nu) + O(\eta) \leq \epsilon,
	\end{align}
	by setting $\nu = O(\epsilon)$ and $\eta = O(\epsilon)$.
	
	Finally, we derive the number of initial state copies needed for the algorithm. 
	Let us denote $\mathtt{I}_{k}$ and $\mathtt{O}_{k}$ to be the number of $\rho'_{k-1}$ and $\Upsirecapp{\rho'_{k-1}}(\rho'_{k-1})$ copies before and after $k$th recursion step. 
	We require $\mathtt{I}_{N+1}\geq1$ and $\mathtt{I}_{1}$ will be the number of $\psivec{0}$ copies we need.
	As the implementation of $\Upsirecapp{\rho'_{k-1}}$ requires $O(\eta^{-1}) = O(\epsilon^{-1})$ copies, 
	\begin{align}
		\mathtt{I}_{k} = O(\epsilon^{-1})\mathtt{O}_{k}.
	\end{align}
	
	$\mathtt{I}_{k+1}$ and $\mathtt{O}_{k}$ are related by the number of mixedness reduction rounds $R$ and survival rate $c$ as
	\begin{align}\label{eq:On_in_QDPpureThm}
		\mathtt{O}_{k} = \left(\frac{2}{c}\right)^{R}\mathtt{I}_{k+1}.
	\end{align}
	$R = O(1)$, while $c$ can be determined by the success probability $q_\mathrm{succ}$ of $R$ round mixedness reduction subroutines using Eq.~\eqref{eq: c for mixedness reduction} with $M = \mathtt{I}_{k+1}$. 
	Assume that we impose the success probability $q_\mathrm{succ}$ to be higher than $1 - q_\mathrm{th}$ for some threshold value $ q_\mathrm{th} $. 
	The probability of all $N$ mixedness reduction protocols to be successful is
	\begin{align}\label{eq:pth_in_QDPpureThm}
		p_\mathrm{succ} = q_\mathrm{succ}^{N} \geq \left(1 - q_\mathrm{th}\right)^{N} \geq 1 - Nq_\mathrm{th},
	\end{align}
	where the second inequality follows from the Taylor expansion. 
	Thus setting $ q_\mathrm{th}= \frac{p_\mathrm{th}}{N} $ ensures that $ p_\mathrm{succ} \geq 1 - p_\mathrm{th} $ as required. 
	
	Now we fix the value of $c$.
	Recalling Eq.~\eqref{eq: c for mixedness reduction} and $\nu < \frac{\epsilon}{2} < \frac{1}{3}$, we have
	\begin{align}\label{eq:fix_c}
		\mathtt{I}_{N+1}^{-\frac{1}{2}}\sqrt{\log(\frac{R}{q_\mathrm{th}})} < \frac{1}{6} \quad \Rightarrow\quad c = \frac{1}{2},
	\end{align}
	or $\mathtt{I}_{N+1} = O(\log(p_\mathrm{th}^{-1}N))$ implies $c = \frac{1}{2}$.
	Combining everything, 
	\begin{align}
		\mathtt{I}_{k} = \epsilon^{-1}e^{O(1)}\mathtt{I}_{k+1},
	\end{align}
	and therefore
	\begin{align}
		\mathtt{I}_{1} = \epsilon^{-N}e^{O(N)}\mathtt{I}_{N+1} = \epsilon^{-N}e^{O(N)}O(\log(p_\mathrm{th})),
	\end{align}
	which concludes the proof. 
\end{proof}

\addpart{Outlook}
\chapter{Conclusions and open problems}\label{chapter: conclusions}

\section{Compositions in resource theories}

In this thesis, we have explored how auxiliary systems are utilised for quantum information processing.
The journey began within resource theories, where everything is more ordered (pun intended, although it would be more precise to say `preordered').
The auxiliary systems of choice are catalysts; they provide surprisingly strong advantages, given the strict recovery condition required for them.
We have uncovered secrets behind catalytic advantages by employing an appropriate set of allowed operations, leveraging the clear-cut structure of resource theories.

Firstly, we restricted the operations to be decomposable. 
Because each operation in the decomposition is implemented with a fresh environment, this decomposability is comparable to imposing memory restrictions  or certain degrees of Markovianity; cf. Definition~\ref{definition: Markovian quantum channel}.
Such decomposability, or memory restriction, has provided two methods by which to inspect catalytic advantages. 
In Section~\ref{section: unravelling CETO}, decomposability is directly utilised to unravel the inner working of catalysis. 
Such access to the process is rare in resource theories, since the main objects of interest are most often the initial and final state pairs, with little focus on the intermediate evolution. 
From snapshots of the process, obtained after each smaller operation in the decomposition, we observed that catalysts assist the system's evolution by functioning as a temporary storage for the resource.

In Section~\ref{section: a hierarchy collapses}, further evidence of the catalyst memory-effect has been obtained by using catalysts to bridge memory-unrestricted operations and memory-restricted ones. 
To be specific, we compared several classes of operations modelling thermodynamic processes, where the sole distinction between them lay in their varying degrees of memory restriction. 
If a particular factor nullifies the operational gaps between these classes, this nullification is attributed to the memory effect. 
Theorems~\ref{theorem: GCETO is TO} and~\ref{theorem: CETO is CTO} demonstrate exactly this.

To examine the second reason behind the power of catalysis, we assumed our catalytic operations to be input-state agnostic. 
This assumption takes into account that real-world operations always involve noise in the state preparation. 
Surprisingly, it turns out that most catalyses are extremely sensitive to state preparation noise; i.e. a small change in the input system state can break the catalyticity of the operation. 
Notably, we show in Theorem~\ref{theorem: mincomp no catchan} that in any resource theory with a certain composition structure, all catalyses are sensitive to the input state. 
This implies that the ability to prepare the initial state precisely, according to the catalyst state, is an essential requirement for catalysis.

Finally, we restricted our theories to those robust to input state noise. 
Theorem~\ref{theorem: catchan and broadcasting} reveals a connection between two seemingly disparate phenomena: robust catalysis occurs only when resource broadcasting is possible. 
Resource broadcasting is a better-studied phenomenon, wherein a resource can be transferred from one system to another without altering the original one. 
Hence, the connection we have established implies another reason for catalytic advantage: namely, the catalyst acting as a seed for broadcasting. 
Furthermore, in Theorem~\ref{theorem: broadcasting possible}, we have found a concrete construction and a necessary and sufficient condition for resource broadcasting applicable to various classes of resource theories, thereby paving the way towards more systematic studies on utilising catalysts for broadcasting.

These first steps into the origins of catalytic advantage have opened up several interesting avenues for future research. 
For example, we have observed that catalytic assistance can render different sets of thermodynamic free operations equivalent. 
A similar question can be posed from the opposite direction: if a common restriction is imposed, can distinct sets of free operations become equivalent once more? 
Should we succeed in finding the common restriction that renders operational and axiomatic classes the same, it would provide valuable insight into the underlying reasons for the separation between them. 
This separation is apparent, yet not fully understood, in some of the most significant and well-studied theories, such as those of entanglement and magic. 
In particular, the gap between thermal operations (TO, Definition~\ref{definition: TO}) and Gibbs-preserving covariant operations (GPC, Definition~\ref{definition: GPC})---two different sets of free operations in resource theory of athermality, where the former is defined operationally and the latter axiomatically---is currently poorly understood; the sole insight to date stems from the existence of a single specific example illustrating this gap~\cite{Ding21_EnTO}. 

One immediate restriction for consideration is Markovianity. 
Markovian versions of TO and GPC, namely Markovian TO (MTO, Definition~\ref{definition: MTO}) and Markovian GPC (MGPC)~\cite{Lostaglio2022MTO1, Korzekwa2022MTO2}, have been proposed and studied independently, yet it remains unclear whether these two operations are equivalent. 
Since both must be describable by Lindblad master equations, the question reduces to determining whether Lindbladians that are Gibbs-preserving and covariant can also be described by a Markovian heat bath, a problem we expect to be more tractable.%
\footnote{Some preliminary studies have been taken in Ref.~\cite{vomEnde2023Markovian}.} 

A related restriction is Gaussianity. 
Interestingly, Gaussian TO is known to be Markovian~\cite{Serafini2020GTO}, as is Gaussian GPC. 
An advantage of considering Gaussianity is that one can leverage the equivalence between Gaussian completely positive trace-preserving maps on density matrices and Gaussian operations on the covariance matrices of bosonic modes.
If either Markovianity or Gaussianity is shown to be the deciding factor in this distinction, a deeper understanding of the missing axiom needed to fully characterise operational thermodynamic operations could be attained.

Another direction for future studies was hinted at in our analysis of robust catalysis and resource broadcasting. 
In that work, we classified general resource theories in terms of their composition rules, i.e. how the set of composite free states is constructed given the sets for subsystems. 
It was found that the choice of composition rule significantly alters the behaviour of these theories. 
Surprisingly, such rules have not been studied extensively, beyond instances where a specific rule is imposed as an assumption~\cite{Marvian2014GlobalvLocal, Pinske2024Censorship, Fang2025GAEP}. 
We expect that dedicated research into the impact of compositions would illuminate the field of resource theory at large and, in particular, the effects of auxiliary systems.

A key extension would be to study the composition of \emph{free operations} instead of the composition of free states. 
This focus is motivated by the fact that resource theories are defined by free operations, not free states. 
In our study of free state composition, free operations must first be defined in terms of the free states, e.g. by selecting CRNG operations as free. 
Such an approach, however, cannot capture theories defined more operationally, such as local operations and classical communication (LOCC) or TO. 
Formalising the composition of free operations would enable this framework to encompass any properly defined resource theory.

One interesting example is the interplay between locality and symmetry~\cite{Marvian2022_locality, Marvian2024Abelian}.
Locality constraints the number of subsystems upon which an operation can act; for example, each $2$-local operation acts, at most, on two subsystems. 
Symmetry can be represented as a covariance of the operation with respect to all actions of a group describing said symmetry. 
For general unitary operations, without symmetry governing the dynamics, any $k$-local operations for $k>1$ achieve universality, i.e. any global unitary operation can be expressed as a concatenation of $k$-local unitaries.
Surprisingly, when local and symmetric operations are composed, universality is lost.
To be specific, if the system is $n$-partite, there are symmetric unitary operations that cannot be decomposed into a sequence of $(n-1)$-local symmetric unitary operations. 
Hence, the composition rule, determining the locality of the operations, dictates the final achievable operations. 

We are particularly interested in hybridising different resource theories through composition.
This problem presents both practical and fundamental motivations. 
Practically, one can imagine a situation where two distinct physical platforms interact.
Each platform is governed by a resource theory suited to addressing its platform-specific difficulties. 
Hence, to model such hybrid platforms, it is essential to understand how different resource theories are composed together. 
On the fundamental side, the hybridisation of resource theories indicate that we can study the interconversion of disparate resources.
Determining the ultimate rate of converting coherence to entanglement, or athermality to magic, for instance, would provide deeper insight into the nature of each resource.

\section{Compositions in quantum computing}

The second part of the thesis has made a departure from the clear-cut world of resource theories and undertook studies in quantum computing. 
In particular, we focused on quantum recursions, which are a generalisation of classical recursions ubiquitous in classical algorithms. 
Quantum recursions apply unitary operators that depend on the previous quantum states, a process which introduces non-linearity to the evolution. 
This non-linearity enables myriads of exotic evolutions that were not accessible due to the inherent linearity of conventional unitary evolutions (which are independent of the quantum state).
One such example is the imaginary-time evolution mapping $\ket{\psi} \mapsto \frac{e^{-\tau H}}{\| e^{-\tau H} \ket{\psi}\|_{2}}\ket{\psi}$, which, though clearly non-linear, qualifies as a unitary evolution. 
In Chapter~\ref{chapter: quantum recursions}, we presented a way to implement this evolution using quantum recursions. 
Our implementation, while approximate, inherits the essential properties of the imaginary-time evolution: namely exponential convergence to the ground state (Theorem~\ref{theorem: fidelity convergence DBQITE}) and an average energy decrease proportional to the variance of the energy (Theorem~\ref{theorem: fluctuation refrigeration relation}). 

To overcome the inherent linearity of standard unitary evolution, quantum recursions require special methods for implementation. 
A standard protocol is to unfold the recursion unitary, whereby operations dependent on the previous state are emulated by reapplying and then reversing the unitary operations used to prepare that state. 
This repeated process of application and reversal introduces redundancies, giving rise to exponential growth in circuit depth with the number of recursion steps. 
However, in Chapter~\ref{chapter: quantum recursions}, it was suggested that despite this exponential growth, the imaginary-time evolution algorithm can nonetheless be of practical utility, thanks to its rapid convergence properties.

In Chapter~\ref{chapter: QDP}, an alternative method for solving quantum recursions was presented.
The solution is quantum dynamic programming, which is a generalisation of classical dynamic programming. 
The crux of this technique is to introduce many quantum systems that function as auxiliary memory states, evolving in tandem with the main system of interest until they are consumed to instruct subsequent recursion steps. 
As a result, the exponential growth in circuit depth can be mitigated, albeit at the cost of an exponential increase in circuit width, when the recursion satisfies certain convergence conditions. 
This trade-off between depth and width (or time and space) can be adjusted to match the specifications of available quantum hardware.

Our results on quantum dynamic programming provided initial evidence that an exponential reduction in circuit depth for quantum recursions is possible. 
However, this is a proof-of-principle result focused on the sufficiency of our technique, without extensive effort dedicated to probing its optimality. 
Hence, a natural subsequent research question concerns the fundamental limits of the space-time trade-off.

In fact, this question can be posed in more abstract settings beyond quantum recursions. 
Quantum circuit complexity measures the minimum number of $2$-local unitary operations needed for the implementation of a given global unitary operation or a given state transformation. 
Existing scattered observations indicate that auxiliary systems can be utilised to induce space-time trade-offs~\cite{Du2024Embedded}, or even employed catalytically to expand the set of exactly compilable operations~\cite{Amy2023CatalyticEmbedding}.
We aim to systematically study the role of auxiliary systems in circuit complexity and related topics. 
Initially, a comparison between single-shot transformation complexity and its catalytic or multi-copy counterparts would be interesting. 
For catalytic transformation, the baseline is that achievable via gate teleportation~\cite{Gottesman1999}, utilising approximately the same number of $2$-local gates. 
Similarly, for multi-copy transformations, a lower bound can be achieved by parallel, independent applications of the same operations, which achieve an $n$-copy transformation with $n$-fold complexity. 
It is probable, however, that more effective alternative pathways exist for achieving the desired transformations by exploiting the interplay between different subsystems in a more intricate manner. 
Such methods could offer more efficient means of compiling desired operations by embedding the primary system within a larger composite one.

A variant of this problem involves examining the geometric picture of quantum states and their evolutions.
Nielsen's complexity~\cite{Nielsen2005Geometry, Nielsen2006Geometry, Dowling2006Geometry, Nielsen2006Control} concerns the minimum distance between two states, defined by a Riemannian metric that takes into account the locality of operations. 
This notion of complexity aligns more closely with the continuous evolution of a state, as opposed to the discrete evolutions typically assumed in gate-based quantum computation. 
It also naturally relates to speed limits~\cite{Deffner2017SpeedLimit}, which quantify the minimum time required for a state to evolve into another. 
This framework also lends itself to investigating catalyst-assisted or multi-copy complexity. 
In comparison to gate-based complexity, this geometric approach benefits from more developed technical tools. 
Notably, partial answers for ancilla-assisted transformations~\cite{Dowling2006Geometry} and multi-copy transformations~\cite{Gyhm2022Battery} already exist, yet further studies are required to obtain conclusive answers.

\section{Concluding remarks}

Hopefully, this thesis would work as a helpful guide for navigating the infinitely larger spaces accessible with auxiliary systems.

In this thesis, we have investigated the mechanisms behind the catalytic use of auxiliaries, applications to quantum computing that enable space-time trade-offs, and promising avenues for future research involving auxiliary systems.

Stepping back, I wish to conclude by recalling why I have been fascinated by auxiliary systems since the beginning of my PhD. 
Initially, I was drawn to the puzzling nature of catalysis in resource theories. 
The catalytic activation of an previously impossible task into a feasible one seemed to be a 'free lunch'. 
In the course of the research that culminated in this thesis, I came to realise that catalytic advantages, and auxiliary system utilisation in general, are akin to exploring new dimensions---quite literally, by expanding the dimensions of the system itself. 
As I stated in the introduction of this thesis, the unique joy of being a theorist is tied to the freedom of choosing what to consider. 
Auxiliary systems subsequently stood out as a versatile tool for expanding the horizons of theorists.

It is my hope that this thesis will serve as a helpful guide for navigating this new horizon, one so vastly expanded thanks to our quantum comrades.

\appendix 

\addpart{Appendix}

\chapter{State transformation conditions and improvements for ETO characterisation}\label{chapter: ETO cone characterisation}

\section{Previous results}

We summarise known characterisations for state transformations under thermodynamic free operations considered in Chapter~\ref{chapter: CETO}.
First, Gibbs-preserving operations (GP, Definition~\ref{definition: GP}) are defined as a channel $\mE$ with the constraint $\mE(\gamma) = \gamma$. 
Therefore, the conditions for state transformations under GP (i.e. the question of whether a GP map $\mE$ exists such that $\mE(\rho) = \rho'$) can be framed as a special case of the quantum dichotomy problem~\cite{Alberti1980dichotomy}, which asks whether a CPTP map $\mE$ exists that satisfies $\mE(\rho_{i}) = \rho'_{i}$ for given pairs $(\rho_{1},\rho'_{1})$ and $(\rho_{2},\rho'_{2})$.
When the system S is two-dimensional, this question can be answered by evaluating three inequalities~\cite{Heinosaari2012GP}.
In general, one can construct a (potentially infinite) family of inequalities that completely characterise the state transformation~\cite{Gour2017Affine}.

For Gibbs-preserving and covariant operations (GPC, Definition~\ref{definition: GPC}), a similar family of inequalities can be constructed for general asymmetric input states~\cite{Gour2018Qmaj}. 
For symmetric states, Proposition~\ref{proposition: GPC TO equiv GStochastic} implies that the conditions for GPC are equivalent to those for thermal operations (TO, Definition~\ref{definition: TO}).

For TO, there is currently no complete characterisation for state transformations among general asymmetric states.
Moreover, there is a GPC state transformation that cannot be performed with a TO~\cite{Ding21_EnTO}, indicating the inadequacy of the construction in Ref.~\cite{Gour2018Qmaj} for TO.
On the other hand, when either the initial or the final state is known to be symmetric, TO state transformations are completely characterised by thermomajorisation relations, which we explain below.

The covariance of TO implies that we can safely use population vectors (Definition~\ref{definition: propulation vectors}) and Gibbs-stochastic matrices (Definition~\ref{definition: Gibbs stochastic}) in lieu of density matrices and channels. 
We then define a probability–probability plot, similar to a Lorenz curve, for each population vector.
\begin{definition}[thermomajorisation curve]\label{definition: thermomajorisation curve}
	For a state $\pstate\in\probspace_{d}$, the thermomajorisation curve $\mL_{\pstate}: [0,1] \to [0,1] $ is a piecewise-linear function that interpolates between the coordinates $\{(0,0)\}$ and elbow points $\{(\sum_{k=1}^{l}\gbs_{\pi_{\pstate}(k)}, \sum_{k=1}^{l}p_{\pi_{\pstate}(k)})\}_{l=1}^{d}$.
\end{definition}
Note that $\mL_{\pstate}$ is a \emph{concave} function by definition of the $\beta$-ordering (Definition~\ref{definition: beta-ordering}).

From thermomajorisation curves, a binary relation between two population vectors is defined. 

\begin{definition}[thermomajorisation relation]\label{definition: thermomajorisation}
	For two population vectors $\pstate, \qstate\in\probspace_{d}$, thermomajorisation relation $\pstate\dsucc\qstate$ if $\mL_{\pstate}(x)\geq\mL_{\qstate}(x)$ for all $x\in[0,1]$.
\end{definition}

The thermomajorisation relation defines a preorder, analogous to $\to$ relation for state transformations. 
The reflexivity ($\pstate\dsucc\pstate$) and the transitivity ($\pstate\dsucc\qstate$ and $\qstate\dsucc\rstate$ together implies $\pstate\dsucc\rstate$) can easily be proven.
Indeed, for TO and GPC, thermomajorisation relation defines the preorder for population vectors given by $\to$. 

\begin{proposition}[Ref.~\cite{Ruch78_mixing}, Theorem~2]\label{proposition: dmajorisation and GStochastic}
	For two population vectors $\pstate,\qstate\in\probspace_{d}$, there exists a Gibbs-stochastic matrix $E$, such that $E\pstate = \qstate$ if and only if $\pstate \dsucc \qstate$.
\end{proposition}

Proofs for this proposition can also be found in Refs.~\cite{Joe1990, Horodecki13_fundamental, Shiraishi2020_construction}.
Combined with Proposition~\ref{proposition: GPC TO equiv GStochastic}, we arrive at the desired corollary. 

\begin{corollary}\label{corollary: dmajorisation and TO GPC}
	Suppose that the set of free operations is either TO or GPC and that $\rho_{S},\rho'_{S}\in\SYM{S}$.
	Then, $\rho_{S}\to\rho'_{S}$ if and only if $\pstate\dsucc\qstate$ for population vectors $\pstate$ and $\qstate$ corresponding to $\rho_{S}$ and $\rho'_{S}$, respectively. 
\end{corollary}

Now we move on to the construction of the set of reachable states $\setXTO(\pstate)$ (Definition~\ref{definiton: set of reachable states}) via some operation X starting from a state $\pstate$.
Earlier works have shown that the set $\setTO(\pstate)$ characterised by thermomajorisation can be constructed efficiently. 
This is summarised in the theorem below. 

\begin{theorem}[\cite{Lostaglio_18_ETO}, Lemma~12 and \cite{Mazurek19_channels}, Theorem~2]\label{theorem: TO cone} 
	For any given $ \pstate $, the set of reachable states $\setTO(\pstate)$ is a convex combination of $ d! $ unique extreme points that correspond to distinct $\beta$-orders and are tightly thermomajorised by $ \pstate $.
\end{theorem}

\begin{figure}[t]
	\centering
	\includegraphics[width=0.44\linewidth]{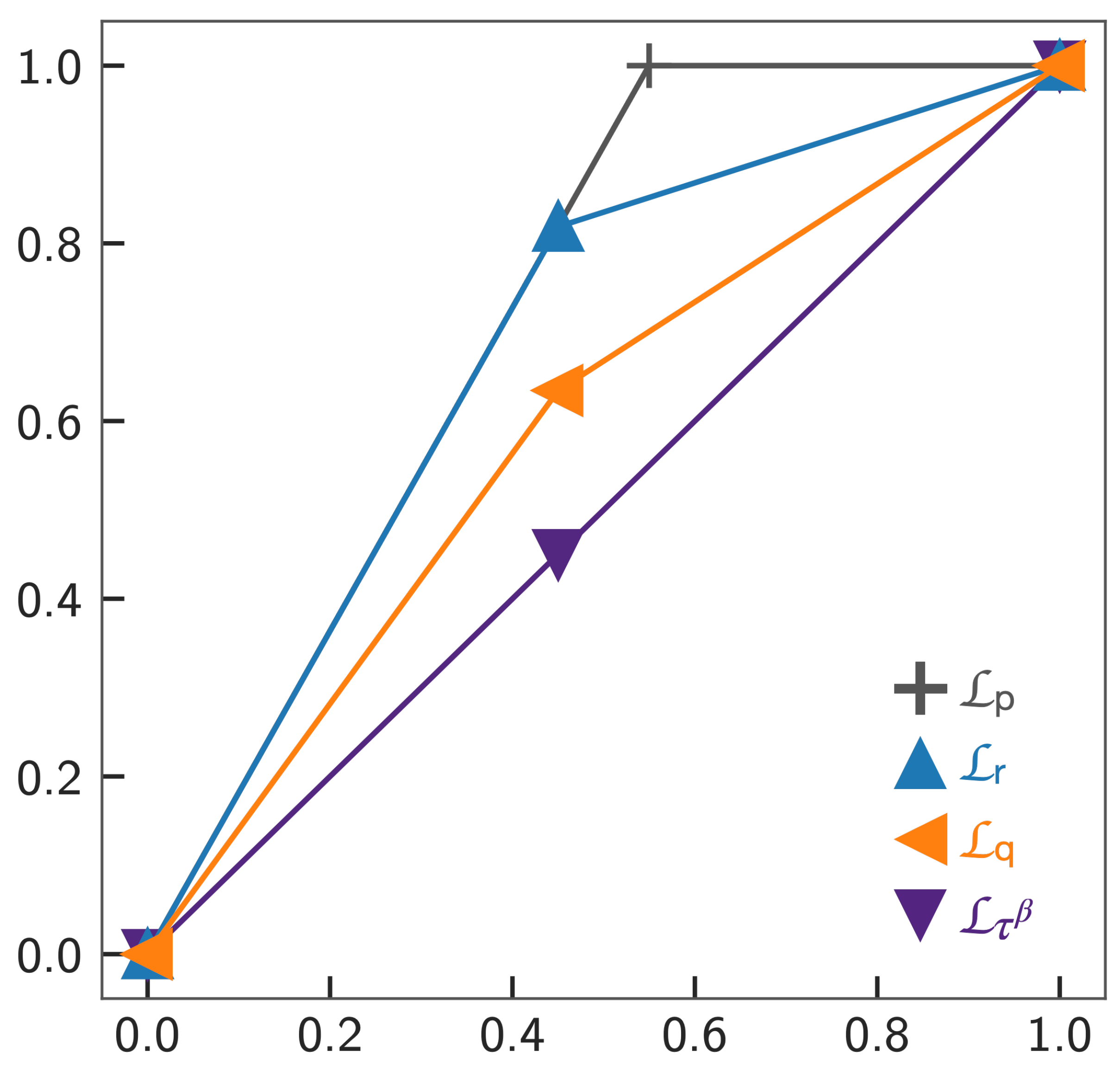}
	\caption{Thermomajorisation curves of $ \pstate,\rstate,\qstate $, and $ \gbs $, where $ \pstate\succ_{\beta}\rstate\succ_{\beta}\qstate\succ_{\beta}\gbs $. In particular, $ \rstate $ is tightly thermomajorised by $ \pstate $, implying that $ \rstate $ thermomajorises any $ \qstate\in\setETO(\pstate) $ with the same order. Here we plotted one such state $ \qstate = (\rstate+\gbs)/2 $. Temperature and energy levels are set to be $E_{1} = 0$, $ \beta E_{2} = 0.2$.
	Figure adapted from Figure~1 of Ref.~\cite{Son2024_CETO}.
	}
	\label{figure: thermomaj example}
\end{figure}

\begin{kaobox}[frametitle = An example of tight thermomajorisation]
	Given a qubit Hamiltonian $\mathrm{H} = (E_{1},E_{2})$ with its corresponding Gibbs state $\gbs = (\gbs_{1} , \gbs_{2})$ of some fixed temperature $1/\beta$, consider the pure ground state $ \pstate = (1,0) $, which has a $\beta$-order $\pi_{\pstate} = (1,2)$ and a simple thermomajorisation curve 
	\begin{align}
		\mL_\pstate(a) = \begin{cases}
			a(\gbs_{1})^{-1}, &\text{for } a\leq \gbs_{1},\\
			1, &\text{for }  \gbs_{1}<a\leq 1.
		\end{cases}
	\end{align}
	A state $\qstate = (q_{1},q_{2})$ has $\pi_{\qstate} = (2,1)$ if $q_{2}>\gbs_{2}$, which leads to a thermomajorisation curve
	\begin{align}
		\mL_\qstate(a) = \begin{cases}
			aq_{2}(\gbs_{2})^{-1}, &\text{for } a\leq \gbs_{2},\\
			1 - \frac{(1-a)q_{1}}{1-\gbs_{2}}, &\text{for }  \gbs_{2}<a\leq 1.
		\end{cases}
	\end{align}
	
	Then $\pstate\succ_\beta\qstate$ if $q_{2}\leq \exp[\beta(E_{1}-E_{2})]:= \Delta_{12}$. Furthermore, when $r_{2} = \Delta_{12}$, $\rstate = (r_{1},r_{2})$ is \emph{tightly thermomajorised} by $\pstate$, i.e. the elbows of $\mL_\rstate$ coincide with the curve $\mL_\qstate$. See Fig.~\ref{figure: thermomaj example} for an example of tight-thermomajorisation.
\end{kaobox}

With Theorem~\ref{theorem: TO cone}, determining $\setTO(\pstate)$ is computationally inexpensive. 
This theorem follows from the fact that if a state is tightly thermomajorised by $ \pstate $, then it thermomajorises any other state in $ \setETO(\pstate) $ that has the same $ \beta $-order $\pi^{(m)}$. 
The importance of tightly thermomajorised states can also be seen from another perspective:

\begin{lemma}[Theorem~12 of~\cite{Perry18_PRX_Crude}]\label{lemma:same_border}
	If two states $\pstate$ and $\qstate$ have the same $\beta$-order $\pi_{\pstate} = \pi_{\qstate}$ and $\pstate\succ_{\beta}\qstate$, then $\qstate$ can be obtained from $\pstate$ by a sequence of two-level partial level thermalisations (PLT).
\end{lemma}

\begin{figure}
	\centering
	\includegraphics[width = \linewidth]{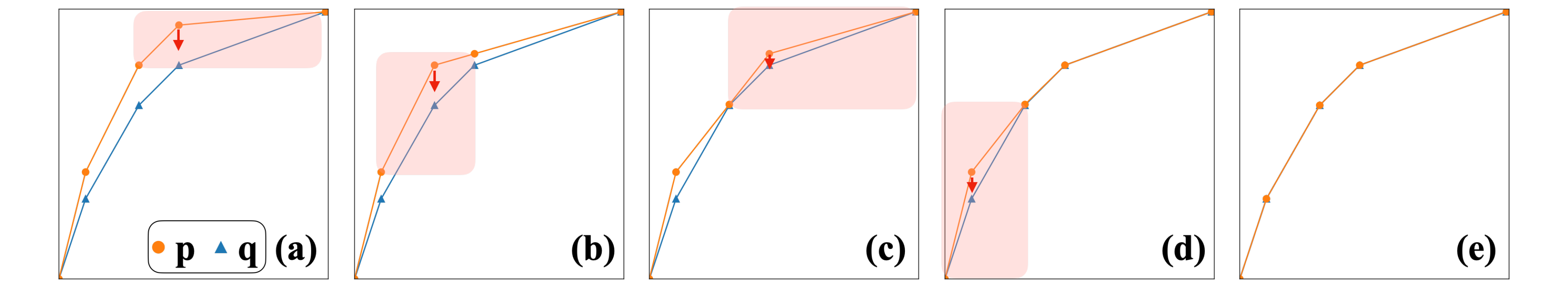}
	\caption{Lemma~\ref{lemma:same_border} can be intuitively understood as follows: Partial level thermalisation between levels $\pi_{\pstate}(i)$ and $\pi_{\pstate}(i+1)$ does not alter any elbow points of $\mL_{\pstate}$ except for the $i$th one. 
		The $i$th elbow point can move downward in the $y$-direction as long as the entire curve remains concave, while the $x$-coordinate is fixed. 
		Panels (a)--(d) illustrate thermomajorisation curves of such a transformation in four steps. At each step, two levels of $\pstate$ (marked by pink shades) undergo partial thermalisation.
		Figure adapted from Figure~2 of Ref.~\cite{Son2024_CETO}.
		}
	\label{figure: PLT}
\end{figure}

PLT~\cite{Perry18_PRX_Crude} between two levels can be conceptualised as a process where these two levels are coupled to a Markovian heat bath, bringing their population ratio closer to that of a thermal state; hence, any PLT is both MTO and ETO. 
Importantly, PLT preserves the $\beta$-order of the state; see Figure~\ref{figure: PLT}.
This leads to the following Corollary.

\begin{corollary}\label{corollary: TO-ETO extreme points}
	If a state $\qstate \in \setETO (\pstate)$ is tightly thermomajorised by $\pstate$, then it is a unique extreme point of $\setETO (\pstate)$ among states of the same $\beta$-order.
\end{corollary}

Since elementary thermal operations (ETO) and Markovian thermal operations (MTO) form a strict subset of TO, even for symmetric state transformations, the thermomajorisation relation (Definition~\ref{definition: thermomajorisation}) gives a necessary but not sufficient condition for $\pstate \to \qstate$. 
Hence, we need to find another way to characterise the set of reachable states $\setETO(\pstate)$ and $\setMTO(\pstate)$.
The only known method is to explicitly construct the set of extreme points for each set $\extr[\setETO(\pstate)]$ or $\extr[\setMTO(\pstate)]$ by exhaustive search over finite, but large number of swap sequences. 
Another helpful fact is that $\setMTO(\pstate)\subsetneq\setETO(\pstate)$ for any $\pstate\in\probspace_{d}$~\cite{Lostaglio2022MTO1}.

Now we define a class of T- or $\beta$-swaps that typically minimise dissipation/change in athermality.
These are thus strong candidates for the swaps needed to perform extremal state transformations under MTO/ETO.

\begin{definition}[neighbouring levels and neighbouring swaps]\label{definition: neighbouring swaps}
	Two levels $i$ and $j$ are neighbouring for a vector $\pstate$, if there exists a $\beta$-order $\pi_{\pstate}$, such that $\pi_{\pstate}^{-1}(i) = \pi_{\pstate}^{-1}(j) \pm1$, i.e. they are two consecutive levels in the $\beta$-order of the input state. 
	Neighbouring T- or $\beta$-swaps are the ones that act on two neighbouring levels.
\end{definition}

Indeed, extremal MTO state transformations are characterised by sequences of such neighbouring swaps. 

\begin{proposition}[Ref.~\cite{Lostaglio2022MTO1}, Theorem~4 and Corollary~9]\label{proposition: mto extr set}
	All elements of $\extr[\setMTO(\pstate)]$ of a state $\pstate$ are given by some sequence of neighbouring T-swaps without repetition.
	Furthermore if $\pi_{\qstate} = \pi_{\pstate}$ and $\pstate\dsucc\qstate$, then $\qstate\in\setMTO(\pstate)$.
\end{proposition}
Hence, deciding whether $\qstate\in\setMTO(\pstate)$ is much more difficult than TO or GPC, but Proposition~\ref{proposition: mto extr set} guarantees that this decision problem can be solved in finite time.

\section{Our improvements}

Similarly to MTO, the only known way to determine whether $\pstate \toeto \qstate$, is to construct the full set $\setETO(\pstate)$ by finding all the extreme points of this convex polytope, and check whether $\qstate\in\setETO(\pstate)$.
The extreme points $\extr[\setETO(\pstate)]$ are found via checking over a set of swap sequences applied to $\pstate$.
However, there is an additional complexity for ETO, because non-neighbouring swaps are also used for some of the extremal transformations. 

Ref.~\cite{Lostaglio_18_ETO} provided a systematic way of finding all extreme points of $\setETO(\pstate)$ for an arbitrary dimension $d$, which involves an exhaustive search among all possible $\beta$-swap sequences with a bounded length $\ell_\mathrm{max}$. 
In other words, this procedure identifies an upper bound on the number of extreme points 
\begin{align}\label{eq: number of extr points}
	\#~ \text{extreme~points}~ \leq {d \choose 2}^{\ell_\mathrm{max}}.
\end{align}
In Ref.~\cite{Lostaglio_18_ETO}, $ \ell_\mathrm{max} \leq d! $ is shown, which means that Eq.~\eqref{eq: number of extr points} grows super-exponentially with the dimension of the system.
This presents a serious roadblock to both understanding and determining the possibility of state transitions via ETO, a reason why ETO, despite its strong physical motivation, has not been extensively studied.

We show that the $\setETO(\pstate)$ simplifies drastically when $ \pstate $ is known to have a particular $ \beta $-ordering, i.e. when it is \emph{monotonic in energy},
\begin{align}\label{eq: mono order}
	\pi_{\pstate} = (1,\cdots,d) ~ \text{or} ~ \pi_{\pstate} = (d,\cdots,1). 
\end{align}

\begin{theorem}\label{theorem: nice order} 
	If $\pi_{\pstate} $ is monotonic in energy, extreme points of $\setETO(\pstate)$ are achieved if and only if the corresponding $\beta$-swap series that produce them are
	\begin{enumerate}
		\item always neighbouring, 
		\item containing no repetition of each swap.
	\end{enumerate} Furthermore, when $ \vec\beta_{1}\pstate, \vec\beta_{2}\pstate \in \extr[\setETO(\pstate)] $ and $ \pi_{\vec\beta_{1}\pstate} = \pi_{\vec\beta_{2}\pstate}$ for such $ \pstate $, the two series are identical ($ \vec\beta_{1} = \vec\beta_{2} $). 
\end{theorem}

This is a reiteration of Theorem~\ref{theorem: nice order_main} in Chapter~\ref{chapter: CETO}.
Several important simplifications follow from the above theorem, whenever $\pi_{\pstate}$ is monotonic in energy. 
From the no-repetition condition, $\ell_\mathrm{max} = d(d-1)/2$ is obtained. 
The equivalence of $\beta$-swaps outputting the same target state $\beta$-order also guarantees the uniqueness of extreme points of $ \setETO(\pstate) $ at each order, setting the maximum number of extreme points to be $d!$. 
More importantly, given the target $\beta$-order, one can immediately identify a corresponding extreme point without the need of searching over all possible series, since we developed an explicit algorithm to evaluate this extreme point, which we call the standard formation (see Def.~\ref{definition: standard formation} for details). 

The rest of this chapter is a route to a proof of Theorem~\ref{theorem: nice order}.
Nevertheless, each technical lemma or remark provides insights on the construction of $\setETO$.
We have omitted the results in Ref.~\cite{Son2024_CETO} that are not directly used for Theorem~\ref{theorem: nice order}.
These include the full characterisation of the simplest non-trivial case of $d=3$ or the tightening of the upper bound on $\ell_\mathrm{max}$ analytically by the factor of $d-3$ for any dimension $d$, see Ref.~\cite{Son2024_CETO}.

\subsection{Useful technical tools for constructing $\setETO$}

Before we prove Theorem~\ref{theorem: nice order}, we establish several technical tools that are useful to understand the extreme points of $\setETO$.

Given a system characterised by Hamiltonian $ H $, we denote its thermal state as $\gbs$, and describe the initial state with respect to its energy population vector $\pstate$. 
Under a $\beta$-swap $\beta^{(k,l)}$ as defined in Eq.~\eqref{eq: beta swap def}, 
\begin{align}\label{eq:betaswap}
	\begin{pmatrix}
		q_k\\q_l
	\end{pmatrix} &\equiv \beta^{(k,l)}\begin{pmatrix}
		p_{k}\\p_{l}
	\end{pmatrix}
	= \begin{pmatrix}
		(1-\Delta_{kl})p_{k}+p_{l} \\ \Delta_{kl}p_{k}
	\end{pmatrix},\ \text{and }	q_m = p_m\ \forall m\neq k,l, 
\end{align}
when $E_k\leq E_l$.
Then the element-wise ratios, $ \slope(\qstate) $ as defined in Eq.~\eqref{eq: slopes}, are transformed accordingly
\begin{align}
	\slope(\qstate)_{k} &= (1-\Delta_{kl})\slope(\pstate)_{k}+ \Delta_{kl}\slope(\pstate)_{l}\label{eq:slope_after_swap1},\\
	\slope(\qstate)_{l} &= \slope(\pstate)_{k},\label{eq:slope_after_swap2}\\
	\slope(\qstate)_{m} &= \slope(\pstate)_{m},\ \forall m\neq k,l\label{eq:slope_after_swap3},
\end{align}
where 
\begin{align}
	\begin{cases}
		\slope(\pstate)_k\geq\slope(\qstate)_k\geq\slope(\pstate)_l, \quad \text{if}\quad \slope(\pstate)_k\geq\slope(\pstate)_l,\\
		\slope(\pstate)_k\leq\slope(\qstate)_k\leq\slope(\pstate)_l,\quad \text{if}\quad \slope(\pstate)_k\leq\slope(\pstate)_l.
	\end{cases}\label{eq:slope_after_swap4}
\end{align}
Furthermore, equalities for the above equations hold only under the following circumstances:
\begin{align}
	\slope(\pstate)_k=\slope(\pstate)_l\qquad &\implies \qquad 	\slope(\pstate)_k = \slope(\qstate)_k = \slope(\pstate)_l, \\
	E_k=E_l \qquad&\implies\qquad  \slope(\qstate)_k = \slope(\pstate)_l.
\end{align}  
Naturally, the $ \beta $-swap operations also alter $ \beta $-orderings of states.
Let us denote the initial $\beta$-order as $\pi_{\pstate} = (\pi_{1},\cdots,\pi_d)$. 
If $k = \pi_i$ and $l = \pi_{i \pm 1}$ for some $i$, that is if $\beta^{(k,l)}$ is a neighbouring swap for a state $\pstate$, then $\pi_{\qstate}$ can be easily determined: 
\begin{align}\label{eq:order_after_swap1}
	\pi_{\qstate}= S_{i,i \pm 1} (\pi_{\pstate}),
\end{align}
where $ S_{i,j} $ is a swap between $ i $'th and $ j $'th elements.

Below are sundry remarks on $\beta$-swaps that are utilised in proofs of lemmas and theorems. 
These results hold as equalities in the channel level and do not depend on the states these channels are acting on. 

\begin{remark}\label{remark: no_repetition}
	The $ \beta $-swap series $\beta^{(k,l)}\beta^{(k,l)}=M_{\lambda}^{(k,l)}$ for some $\lambda\neq 0,1$, and thus always produces a non-extreme point except for in the trivial case where $E_k = E_l$. In that trivial case, two repeated swaps always result in identity, and $ \lambda=0 $.
\end{remark}

%

\begin{remark}\label{remark: commute}
	When $k,l,m,n$ are all distinct, $ \beta $-swaps commute, i.e.
	\begin{align}
		\beta^{(k,l)}\beta^{(m,n)} = \beta^{(m,n)}\beta^{(k,l)}.
	\end{align}
\end{remark}

\begin{remark}\label{remark: swap_series_eq}
	If $E_{k}\leq E_{l}\leq E_{m}$, 
	\begin{align}
		\beta^{(k,l)}\beta^{(k,m)}\beta^{(l,m)} = \beta^{(l,m)}\beta^{(k,m)}\beta^{(k,l)}.
	\end{align}
\end{remark}
The equality is obtained through direct calculations,
\begin{align}
	\beta^{(k,l)}\beta^{(k,m)}\beta^{(l,m)} = \beta^{(l,m)}\beta^{((k,m)}\beta^{(k,l)}
	= \begin{pmatrix}
		(1-\Delta_{kl})(1-\Delta_{km}) & 1-\Delta_{km} & 1\\
		\Delta_{kl}(1-\Delta_{km}) & \Delta_{km} & 0\\
		\Delta_{km} & 0 & 0
	\end{pmatrix},
\end{align}
where we omit identities acting on irrelevant levels when writing ETO maps (and do so consistently in the rest of the appendix for notational brevity).

Next we want to prove that a $ \beta $-swap involving two levels $ E_j,E_k $ produces an extreme point only if its sole effect on the $ \beta $-ordering on the final state is a swap of $ j $ and $ k $. 
This technical result is later used in establishing Lemma~\ref{lemma: only one nn}.
\begin{lemma}\label{lemma: order swap}
	For any $\pstate\in\probspace_{d}$ with $\pi_{\pstate}(k) = \pi_{k}$, the state $\beta^{( \pi_{i}, \pi_{i+c})}\pstate$ can be an extreme point of $\setETO(\pstate)$ only if $\pi_{\beta^{( \pi_i, \pi_{i+c})}\pstate} = S_{i,i+c}(\pi_{\pstate})$, where $ S_{i,j} $ is the operation that swaps the $ i $th element with the $ j $th element.
\end{lemma}

\begin{proof}
	We start by proving the lemma for the case of $\pstate\in\probspace_{3}$.
	Denote the initial $\beta$-order as $\pi_{\pstate} = (\pi_{1},\pi_{2},\pi_{3})$. 
	Consider the following three cases:
	\begin{enumerate}
		\item Suppose $E_{\pi_{1}} < E_{\pi_{3}}$ and $\pi_{\qstate} = (\pi_{3},\pi_{1},\pi_{2})$, where $\qstate = \beta^{(\pi_{1},\pi_{3})}\pstate$. 
		The same final ordering is obtained after two neighbouring swaps, i.e. 
		\begin{align} 
			\qstate'= \beta^{(\pi_{1},\pi_{3})}\beta^{(\pi_{2},\pi_{3})}\pstate, \qquad \pi(\qstate') =  \pi(\qstate).
		\end{align}
		Note that $q_{\pi_{3}} = q'_{\pi_{3}}$~(this is seen from Eq.~\eqref{eq:betaswap}). 
		Therefore, the thermomajorisation curves $\mL_{\qstate},\mL_{\qstate'}$ are identical up to the first elbow
		\begin{align}
			\mL_\qstate(x) = \mL_{\qstate'}(x), \qquad x\in[0,\gbs_{\pi_{3}}].
		\end{align}
		The third elbow also coincides as $\mL_\qstate(1) = \mL_{\qstate'}(1) = 1$. Finally, the second elbow points of $\qstate$ and $\qstate'$ curves are given by 
		\begin{align}\label{eq:lemma6_proof_eq1}
			\mL_\qstate(1-\gbs_{\pi_{2}}) = 1 - p_{\pi_{2}} < 1 - (\beta^{(\pi_{2},\pi_{3})}\pstate)_{\pi_{2}}= \mL_{\qstate'}(1-\gbs_{\pi_{2}}), 
		\end{align} 
		from $q'_{\pi_{2}} = (\beta^{(\pi_{2},\pi_{3})}\pstate)_{\pi_{2}}<p_{\pi_{2}}$. Therefore, $\qstate'$ strictly thermomajorises $\qstate$ and the non-extremity of $\qstate$ then follows from Lemma~\ref{lemma:same_border}. 
		\item For $E_{\pi_{3}} > E_{\pi_{1}}$, comparison between $\qstate$ with $\pi_{\qstate} = (\pi_{2},\pi_{3},\pi_{1})$ and $\qstate' = \beta^{(\pi_{3},\pi_{1})}\beta^{(\pi_{1},\pi_{2})}\pstate$ gives the same result.
		\item If $E_{\pi_{3}} = E_{\pi_{1}}$, we always get $ \pi_{\qstate} = (\pi_{3},\pi_{2},\pi_{1}) $.
	\end{enumerate}
	Note that from Eqs.~\eqref{eq:slope_after_swap1}--\eqref{eq:slope_after_swap4}, cases 1 and 2 cover all possible ways of obtaining $\pi_{\beta^{\pi_{1},\pi_{3}}\pstate} \neq S_{1,3}(\pi_{\pstate})$.
	
	For the general case of $\pstate\in\probspace_d$, if $\pi_{\qstate =\beta^{(\pi_i,\pi_{i+c})}\pstate} \neq S_{i,i+c}(\pi_{\pstate})$, then the equivalent of $ \qstate' $ above can be chosen as follows:
	
	\begin{enumerate}
		\item If $E_{\pi_i}<E_{\pi_{i+c}}$ and $ \pi_{i}\neq \pi_{\qstate}(i+c) $, then $ \qstate' = \beta^{(\pi_i,\pi_{i+c})}\beta^{(\pi_{i+c-1},\pi_{i+c})}\pstate \succ_\beta \qstate$.
		Although $ \pi_{\qstate}\neq\pi_{\qstate'} $ in general, $ \qstate $ can always be obtained from $ \qstate' $ by partial level thermalisation between $ \pi_{i} $ and $ \pi_{i+c-1} $. To see this, notice that $ q_{\pi_{k}} = q'_{\pi_{k}}$,  $\forall k\neq i,i+c-1 $, i.e. 
		\begin{align}\label{eq:lemma6_proof_eq2}
			q_{\pi_{i}} + q_{\pi_{i+c-1}} = q'_{\pi_{i}} + q'_{\pi_{i+c-1}}.
		\end{align}  
		Using the same argument to Eq.~\eqref{eq:lemma6_proof_eq1}, $q'_{\pi_{i}}>q_{\pi_{i}} $, and thus
		\begin{align}
			\slope(\qstate')_{\pi_{i}} &> \slope(\qstate)_{\pi_{i}} > \slope(\qstate)_{\pi_{i+c-1}} > \slope(\qstate')_{\pi_{i+c-1}}.
		\end{align}
		Combining with Eq.~\eqref{eq:lemma6_proof_eq2}, $ \qstate $ is obtained $ \qstate' $ by partial thermalisation between levels $ i $ and $ i+c-1 $. 
		
		\item If $E_{\pi_i}>E_{\pi_{i+c}}$ and $ \pi_{i+c}\neq\pi_{\qstate}(i) $, then $ \qstate' = \beta^{(\pi_{i+c},\pi_{i})}\beta^{(\pi_{i},\pi_{i+1})}\pstate \succ_\beta \qstate$. Likewise, $ q_{\pi_{k}} = q'_{\pi_{k}}$,  $\forall k\neq i+1,i+c $, and $ q_{\pi_{i+1}} + q_{\pi_{i+c}} = q'_{\pi_{i+1}} + q'_{\pi_{i+c}}$. With 
		\begin{align}
			\slope(\qstate')_{\pi_{i+1}} &> \slope(\qstate)_{\pi_{i+1}} > \slope(\qstate)_{\pi_{i+c}} > \slope(\qstate')_{\pi_{i+c}},
		\end{align}
		$ \qstate $ is obtained $ \qstate' $ by partial thermalisation between levels $ i+1 $ and $ i+c $.
		
		\item If $ E_{\pi_{i}} = E_{\pi_{i+c}} $, we always get $ \pi_{i} = \pi_{\qstate}(i+c) $.
	\end{enumerate}
	Therefore, the state $ \qstate $ with order $\pi_{\qstate} \neq S_{i,i+c}(\pi_{\pstate})$ is always non-extremal in $\setETO(\pstate)$.
\end{proof}

We now show that it is impossible to have a $ \beta $-order that has no vertex of $ \setETO $. 
Lemma~\ref{lemma:ext_point_for_each_order} hints the lower bound scaling of the number of extreme points in worst cases. 
\begin{lemma}\label{lemma:ext_point_for_each_order}
	For any state $ \pstate\in\probspace_d $, the reachable state set $ \setETO(\pstate) $ has at least one extreme point $ \rstate $ having $ \pi_{\rstate} = \psi $ for any ordering $ \psi $. 
\end{lemma}

\begin{proof}
	To start, we construct a series of sets $ S_d\subset S_{d-1}\subset\cdots\subset S_0 $ defined as follows:
	\begin{itemize}
		\item $ S_0 = \{\qstate\vert \qstate\in\setETO(\pstate)\ \text{and}\ \pi_{\qstate} = \psi\} $, 
		\item $ S_j = \{\qstate\vert\qstate\in S_{j-1}\ \text{and}\ q_{\psi_j}\geq q^\prime_{\psi_j}, \forall \qstate^\prime\in S_{j-1}\} $ for $ 1\leq j\leq d $.
	\end{itemize}
	In other words, $ S_{1} $ is the set of states in $ \setETO(\pstate) $ with a specific $ \beta $-order $ \psi $, and with maximal $ \psi_{1} $ population $ r_{\psi_{1}} $. Likewise, $ S_{2} $ is a subset of $ S_{1} $, having \textit{additionally} the maximal $ \psi_{2} $ population, and so on. Note that $ S_d $ always has a single element for each fixed choice of $ \psi $, which we denote as $ \rstate $.%
	\footnote{It should be noted at this point that $ \rstate $ may not be the only extreme point that has $ \beta $-ordering $ \psi $---there could be states $ \rstate' $ of the same $ \beta$-order, where the first elbow is lower that of $ \rstate $, and the second elbow higher. Such states are not, however, contained in $ S_d $.}
	
	Suppose there is no extreme point of $ \setETO(\pstate) $ corresponding to $ \psi $. Then $ \rstate $ can be written as a strict convex combination of extreme states $ \estate^{(i)} $, i.e.
	\begin{align} 
		\rstate = \sum_i 	p_i	\estate^{(i)} , \qquad p_i \in (0,1).
	\end{align}
	Starting from $ j=1 $, check the following:
	\begin{enumerate}
		\item For $ j > 1 $, we have $ e^{(i)}_{\psi_k} = r_{\psi_{k}} $, $ \forall k< j $ from the last iteration. For $ j=1 $, we do not need any condition yet. 
		
		\item If $ e^{(i)}_{\psi_j}>r_{\psi_j} $ for some $ i $, a state $ \estate^\prime = \vec\beta\estate^{(i)} $ with $ \pi_{\estate^\prime} = \psi $ and $ e^\prime_{\psi_k} = e^{(i)}_{\psi_k} $, $ \forall k\leq j $ can be found. Let us show how to do this:
		\begin{itemize}
			\item If $ \slope(\estate^{(i)})_{\psi_{j-1}} = \slope(\rstate)_{\psi_{j-1}} \geq \slope(\estate^{(i)})_{\psi_{j}} >\slope(\rstate)_{\psi_{j}}$, we can simply thermalise all the levels $ \psi_{k} $, $ \forall k>j $ of $ \estate^{(i)} $ to have the same slope, which is smaller than $ \slope(\estate^{(i)})_{\psi_{j}} $. Then we obtain the desired state $ \estate^{\prime} $, since levels with degenerate slopes -- all $ \psi_{k>j} $ in this case -- can be permuted within themselves in the $ \beta $-order. 
			\item If $ \slope(\estate^{(i)})_{\psi_{j}}> \slope(\rstate)_{\psi_{j-1}} >\slope(\rstate)_{\psi_{j}}$%
			\footnote{Requiring $ \slope(\rstate)_{\psi_{j-1}} >\slope(\rstate)_{\psi_{j}} $ is always possible by putting all the levels having the same slope $ \slope(\rstate)_{\psi_{j}} $ to come after $ \psi_{j} $ in the order $ \psi $.}, 
			we can first reduce $ e^{(i)}_{\psi_{j}} $ by partially thermalising with populations of levels $ \psi_{k>j} $ until $ \slope(\rstate)_{\psi_{j-1}} \geq e^{\prime}_{\psi_{j}} > \slope(\rstate)_{\psi_{j}} $. Then, as in the previous case, thermalising all the levels $ \psi_{k>j} $ will give $ \estate^{\prime} $.
		\end{itemize}
		
		However, such $ \estate^{\prime} $ satisfies $ \estate^\prime\in S_{j-1} $ and $ e^\prime_{\psi_j}>r_{\psi_j} $, which contradicts the assumption that $ \rstate \in S_j $.
		\item If $ e^{(i)}_{\psi_j}\leq r_{\psi_j} $ for all $ i $, from convexity of the combination, $ e^{(i)}_{\psi_j}=r_{\psi_j} $ for all $ i $. 
		Proceed to $ j\rightarrow j+1 $.
	\end{enumerate}
	If $ e^{(i)}_{\psi_j} = r_{\psi_j} $ for all $ i $ and $ j $, $ \estate^{(i)} = \rstate $, which contradicts the assumption that $ \rstate $ is not extremal.
\end{proof}

The two remaining results in this subsection are established specifically for $\pstate\in\probspace_{3}$.
However, they will be used in the proof of Theorem~\ref{theorem: nice order} and thus we present them here. 

\begin{remark}\label{remark: two swap TO ext}
	For $\pstate\in\probspace^3$, 
	\begin{enumerate}
		\item the states $\beta^{(2,3)}\beta^{(1,2)}\pstate$ and  $\beta^{(2,3)}\beta^{(1,3)}\pstate $  are extreme points of $\setTO(\pstate)$, if $\pi_{\pstate} = (2,1,3)$ or $(3,1,2)$;
		\item the state $\beta^{(1,3)}\beta^{(2,3)}\pstate$ is an extreme point of $\setTO(\pstate)$, if $\pi_{\pstate} = (1,2,3)$ or $(3,2,1)$; and
		\item the state $\beta^{(1,2)}\beta^{(2,3)}\pstate$ is an extreme point of $\setTO(\pstate)$, if $\pi_{\pstate} = (1,3,2)$ or $(2,3,1)$.
	\end{enumerate}  
\end{remark}

\begin{proof}
	Direct calculation gives
	\begin{align}
		\beta^{(2,3)}\beta^{(1,2)} = \begin{pmatrix} 1-\Delta_{12} & 1 & 0\\ \Delta_{12} - \Delta_{13} & 0 & 1\\ \Delta_{13} & 0 & 0\end{pmatrix}, \quad
		\beta^{(2,3)}\beta^{(1,3)} = \begin{pmatrix} 1-\Delta_{13} & 0 & 1\\ \Delta_{13} & 1-\Delta_{23} & 0\\ 0 & \Delta_{23} & 0\end{pmatrix},\label{eq:ext_biplanar_matrices}\\
		\beta^{(1,2)}\beta^{(2,3)} = \begin{pmatrix} 1-\Delta_{12} & 1-\Delta_{23} & 1\\ \Delta_{12} & 0 & 0\\ 0 & \Delta_{23} & 0\end{pmatrix}, \quad
		\beta^{(1,3)}\beta^{(2,3)} = \begin{pmatrix} 1-\Delta_{13} & \Delta_{23} & 0\\ 0 & 1-\Delta_{23} & 1\\ \Delta_{13} & 0 & 0\end{pmatrix}.\nonumber
	\end{align}
	By using the algorithm in Definition~6 of \cite{Mazurek19_channels}, one can verify that the above channels are \emph{biplanar extreme points} of the set of thermal processes. 
	According to Theorem~4 in \cite{Mazurek19_channels}, such biplanar extremal channels generate extreme points of $\setTO(\pstate)$, when the initial state corresponds to a particular $ \beta $-ordering that can be found in the process of decomposing the graph structure of the channel matrix. 
	Performing this procedure according to \cite{Mazurek19_channels} reveals that for $ \beta^{(2,3)}\beta^{(1,2)}  $ and $ \beta^{(2,3)}\beta^{(1,3)} $, the relevant input state $ \beta $-order is given by $ (2,1,3) $ and $ (3,1,2) $; similarly for statements 2 \& 3 in remark.
\end{proof}

\begin{lemma}\label{lemma: only one nn}
	Given $\pstate\in\probspace_{3}$ with $\pi_{\pstate} = (\pi_{1},\pi_{2},\pi_{3})$, $\vec{\beta}\pstate$ is extremal for $\setETO(\pstate)$ only if i) $\vec{\beta}$ is always neighbouring when applied to $\pstate$ or ii) $\vec{\beta} = \beta^{(\pi_{1},\pi_{3})}$.
\end{lemma}
\begin{proof}
	To prove this, we need to show that i) a neighbouring swap following a non-neighbouring swap produces non-extreme point and ii) a non-neighbouring swap following a neighbouring one also yields a non-extreme point. Since only extreme points are of our interest, using Lemma~\ref{lemma: order swap}, we can safely assume that $ \pi_{\beta^{(\pi_{i},\pi_{j})}\pstate} = S_{i,j}(\pi) $ for any $ \pi = \pi_{\pstate} $ and $ i,j $.
	We tackle each problem by further dividing cases.

	\emph{Case i-(a):} $\pi = (1,2,3)$ or $(3,2,1)$ experiencing a neighbouring swap followed by a non-neighbouring swap. 
	Swap $ \beta^{(2,3)}\beta^{(1,2)} $ produces final states with order $ (3,1,2) $ or $ (2,1,3) $. 
	From Remark~\ref{remark: two swap TO ext}, these orders have a unique extreme point produced from $ \beta^{(1,3)}\beta^{(2,3)} $. 
	Another swap gives 
	\begin{align}
		\rstate = \beta^{(1,2)}\beta^{(2,3)}\pstate = \begin{pmatrix}
			(1-\Delta_{12})p_{1}+(1-\Delta_{23})p_{2} + p_{3} \\ \Delta_{12}p_{1}\\ \Delta_{23}p_{2}
		\end{pmatrix},
	\end{align} 
	whereas two consecutive neighbouring swaps give
	\begin{align}
		\qstate = \beta^{(1,3)}\beta^{(1,2)}\pstate = \begin{pmatrix}
			(1-\Delta_{12})(1-\Delta_{13})p_{1}+(1-\Delta_{13})p_{2} + p_{3} \\ \Delta_{12}p_{1}\\ \Delta_{13}(1-\Delta_{12})p_{1}+\Delta_{13}p_{2}
		\end{pmatrix},
	\end{align}
	with order $ (2,3,1) $ or $ (1,3,2) $ depending on the initial order. 
	Note that $ r_{2} = q_{2} $ and 
	\begin{align}
		q_{3} - r_{3} = (1-\Delta_{12})\Delta_{23}(\Delta_{12}p_{1} - p_{2}),
	\end{align}
	i.e. $ q_{3} \geq r_{3} $ if initial $ \pi = (1,2,3) $, and $ q_{3} \leq r_{3} $ if initial $ \pi = (3,2,1) $. Either way, $ \qstate\succ_\beta\rstate $ while $ \pi_{\qstate} = \pi_{\rstate} $ and thus $ \rstate $ cannot be extremal from Lemma~\ref{lemma:same_border}.
	
	\emph{Case i-(b):} $\pi = (1,2,3)$ or $(3,2,1)$ experiencing a non-neighbouring swap followed by a neighbouring swap. We can in fact prove that non-neighbouring swap already always produces non-extreme points. Compare two states 
	\begin{eqnarray}
		\rstate = \beta^{(1,3)}\pstate &=& \begin{pmatrix}
			(1-\Delta_{13})p_{1}+p_{3}\\ p_{2}\\ \Delta_{13}p_{1}
		\end{pmatrix},\label{eq:non-neighbouring-non-extremal1}\\
		\qstate = \beta^{(2,3)}\beta^{(1,3)}\beta^{(1,2)}\pstate
		&=& \begin{pmatrix}
			(1-\Delta_{12})(1-\Delta_{13})p_{1} + (1-\Delta_{13})p_{2}+p_{3}\\ \Delta_{12}(1-\Delta_{13})p_{1} + \Delta_{13}p_{2}\\ \Delta_{13}p_{1}
		\end{pmatrix},\label{eq:non-neighbouring-non-extremal2}
	\end{eqnarray}
	with $\pi_{\rstate} = \pi_{\qstate} = (3,2,1)$ or $ (1,2,3) $ depending on the initial $ \pi $. 
	Then we may observe the following:
	\begin{align}
		r_{3} &=q_{3}, \label{eq:non-neighbouring-non-extremal3}\\ 
		r_{2} - q_{2} &= (1-\Delta_{13})(p_{2}-\Delta_{12}p_{1})\leq 0,  \ \text{for}\ \pi = (1,2,3), \label{eq:non-neighbouring-non-extremal4}\\
		r_{2} - q_{2} &= (1-\Delta_{13})(p_{2}-\Delta_{12}p_{1})\geq 0,  \ \text{for}\ \pi = (3,2,1), \label{eq:non-neighbouring-non-extremal5}
	\end{align}
	i.e. $\qstate \succ_{\beta} \rstate$ and thus $\rstate$ is not extremal for $\setETO(\pstate)$.

	\emph{Case ii-(a):} $\pi = (1,3,2)$ or $(2,3,1)$ experiencing a neighbouring swap followed by a non-neighbouring swap. 
	The two possible neighbouring swaps for these initial orders are $\beta^{(2,3)}$ and $\beta^{(1,3)}$. 
	First, consider the neighbouring swap $\beta^{(2,3)}$: this modifies the order into $(1,2,3)$ or $(3,2,1)$ and \emph{Case i-(b)} forbids a non-neighbouring swap to come next. 
	The other neighbouring and non-neighbouring swap pair $\beta^{(2,3)}\beta^{(1,3)}$ gives output orders $(2,1,3)$ or $(3,1,2)$. 
	But Remark~\ref{remark: two swap TO ext} states that $\beta^{(1,2)}\beta^{(2,3)}\pstate$ is a unique extreme point for that output order. 
	
	\emph{Case ii-(b):} $\pi = (1,3,2)$ or $(2,3,1)$ experiencing a non-neighbouring swap followed by a neighbouring swap. 
	After a non-neighbouring swap, $\pi_{\beta^{(1,2)}\pstate} = (2,3,1)$ or $(1,3,2)$. If a following neighbouring swap is $\beta^{(1,3)}$, the resulting orders are $(2,1,3)$ or $(3,1,2)$, which again cannot be extremal from the Remark~\ref{remark: two swap TO ext}. The remaining possibility is to apply $\beta^{(2,3)}$, which results in 
	\begin{align}
		\rstate &= \beta^{(2,3)}\beta^{(1,2)}\pstate=\begin{pmatrix}
			(1-\Delta_{12})p_{1}+p_{2} \\ (\Delta_{12}-\Delta_{13})p_{1}+p_{3}\\ \Delta_{13}p_{1}
		\end{pmatrix}.
	\end{align}
	Compare this with a state having the same $\beta$-order $ \pi_{\rstate} = \pi_{\qstate} =  (3,2,1) $ or $ (1,2,3) $,
	\begin{align}
		\qstate = \beta^{(1,2)}\beta^{(1,3)}\pstate\label{eq:rmk6_klkm}
		=\begin{pmatrix}
			(1-\Delta_{12})(1-\Delta_{13})p_{1} + p_{2} + (1-\Delta_{12})p_{3} \\ \Delta_{12}(1-\Delta_{13})p_{1} + \Delta_{12}p_{3}\\ \Delta_{13}p_{1}
		\end{pmatrix}.
	\end{align}
	Then 
	\begin{align}
		q_{1} - r_{1} &= -\Delta_{13}(1-\Delta_{12})p_{1}+(1-\Delta_{12})p_{3}\leq0,\label{eq:ineq_rmk6_{3}}
	\end{align}
	for initial $\pi = (2,1,3)$ and becomes positive for $(3,1,2)$. Plus, $ q_{3} = r_{3} $, which in turn gives $\qstate\succ_{\beta}\rstate$; thus, $\rstate$ is not extremal for $\setETO(\pstate)$.
	
	\emph{Case iii-(a):} $\pi = (2,1,3)$ or $(3,1,2)$ experiencing a neighbouring swap followed by a non-neighbouring swap. 
	If the first neighbouring swap is $\beta^{(1,2)}$, the output $ \beta $-order becomes $ (1,2,3) $ or $ (3,2,1) $, which does not allow non-neighbouring swap to follow as stated in \emph{Case i-(b)}. 
	The other series, $\beta^{(1,2)}\beta^{(1,3)}$ outputs orders $(1,3,2)$ or $(2,3,1)$, but these orders have unique extreme points for $\setETO$ given by Remark~\ref{remark: two swap TO ext}.
	
	\emph{Case iii-(b):} $\pi = (2,1,3)$ or $(3,1,2)$ experiencing a non-neighbouring swap followed by a neighbouring swap. 
	Two candidate series are $\beta^{(1,3)}\beta^{(2,3)}$ and $\beta^{(1,2)}\beta^{(2,3)}$, which respectively produces orders $(1,3,2)$ and $(3,2,1)$ when applied to an initial state with $\pi = (2,1,3)$; $(2,3,1)$ and $(1,2,3)$ when applied to $\pi = (3,1,2)$. 
	All output states obtained here have different unique extreme points for $\setETO$ given in Remark~\ref{remark: two swap TO ext}, and the states generated by the considered swaps are therefore non-extremal.
	
	We exhausted all possible cases and none of the swaps can create an extreme point of $\setETO(\pstate)$. 
\end{proof}

\subsection{Proof of Theorem~\ref{theorem: nice order}}

The technical proof of Theorem~\ref{theorem: nice order} can be sketched in the following steps:
\begin{itemize}
	\item Firstly, we introduce a specific series of $\beta$-swaps that transforms an initial $\beta$-ordering to a target ordering (Definition~\ref{definition: standard formation}). 
	\item This structure, which we refer to as the \emph{standard formation},	is then shown to be equivalent to any $\beta$-swap series, where i) all swaps are \emph{neighbouring} to initial states of the form Eq.~\eqref{eq: mono order} and ii) each swap is applied \emph{at most once} (Lemma~\ref{lemma: standard form}). 
	\item Finally, to prove Theorem~\ref{theorem: nice order}, we show that whenever the initial $ \beta $-ordering is monotonic in energy, then a transformation that is not according to a standard formation always leads to a non-extreme state. 
	This Lemma allows us to conclude that the number of extreme points for such a $ \setETO(\pstate) $ is at most $ d! $, similar to that of $ \setTO (\pstate)$.
\end{itemize}

\begin{definition}[Standard formation]\label{definition: standard formation}
	Given a tuple of $d$-dimensional $ \beta $-orderings $ (\pi,\pi') $, a standard formation is a $\beta$-swap series $\vec{\beta}_\mathrm{sf}$ that transforms an initial state $ \pstate $ with an order $\pi$ into some final state $ \pstate' $ having an order $\pi^\prime$,  
	with the construction below:
	\begin{enumerate}
		\item Set an initial index of $ j=1 $, and identify $ m $ such that $\pi^\prime_j = \pi_{m}$. If $m=j$, define $\vec\beta^{(j)}$ as an identity. Otherwise, since $ \pi_{1},\cdots,\pi_{m-1} $ are already occupied by  $ \pi^{\prime}_{1},\cdots,\pi^{\prime}_{m-1} $, we get $ m>j $. Then, define a swap-series $\vec\beta^{(j)} = \beta^{(\pi_{j},\pi_{m})}\beta^{(\pi_{j+1},\pi_{m})}\cdots\beta^{(\pi_{m-1},\pi_{m})}$. Note that these swaps are always neighbouring when applied to a state initially having the order $\pi$, due to Eq.~\eqref{eq:order_after_swap1}. After $ j=1 $ round, this swap series will take the initial ordering $ \pi $ to the new ordering $ (\pi_{1}',\pi_{1},\cdots,\pi_{m-1},\pi_{m+1},\cdots,\pi_d) $ if $ \vec{\beta}^{(1)}\neq\1 $.
		\item Iterate the above step for $ j=2,\cdots,d-1 $, defining $\lbrace \vec\beta^{(j)}\rbrace_{j=1}^{d-1}$. 
	\end{enumerate} 
	The standard formation series is then simply the concatenation
	\begin{align}
		\vec{\beta}_\mathrm{sf} = \vec\beta^{(d-1)} \cdots \vec\beta^{(2)} \vec\beta^{(1)}.
	\end{align}
	By construction, there is no repetition of a swap in this series and they are all neighbouring when applied to an initial state with the order $\pi$. 
\end{definition}

\begin{kaobox}[frametitle = An example of the standard formation]
	For the orderings $ \pi = 1234 $ and $ \pi'=4231 $, the standard formation is $ \vec{\beta}_\mathrm{sf} = \vec{\beta}^{(3)}\vec{\beta}^{(2)}\vec{\beta}^{(1)} $, where
	\begin{align} 
		\vec{\beta}^{(1)} &=\beta^{(1,4)}\beta^{(2,4)}\beta^{(3,4)},\qquad
		\vec{\beta}^{(2)} =\beta^{(1,2)},\qquad
		\vec{\beta}^{(3)} =\beta^{(1,3)}. 
	\end{align}
	The intermediate $ \beta $-orderings given by this process are
	\begin{align} 
		\pi = 1234 \quad \xrightarrow[\vec{\beta}^{(1)}]{}~ \quad 4123 \quad \xrightarrow[\vec{\beta}^{(2)}]{}~ \quad 4213 \quad
		\xrightarrow[\vec{\beta}^{(3)}]{}~\quad \pi ' = 4231.
	\end{align}
\end{kaobox}

This formation has a nice property, namely all swaps in each block $\vec{\beta}^{(j)}$ acts on level $\pi^\prime_j$ and the ones in $\vec{\beta}^{(k)}$ for any $k>j$ does not act on level $\pi^\prime_j$. In the lemma below, we illustrate how certain classes of swap series, which turns out to be the ones producing extreme states, can always rearranged into a standard formulation.

\begin{lemma}\label{lemma: standard form}
	Given an initial state $ \pstate $ with ordering $ \pi $ monotonic in energy, denote a $\beta$-swap series 
	\begin{align} 
		\vec\beta = \prod_{i} \beta_i,
	\end{align} 
	and $ \pi' $ to be the final $ \beta $-ordering of the state $\pstate' = \vec{\beta}\pstate $. If $ \vec{\beta} $ is such that: 
	\begin{enumerate}
		\item each $ \beta_i $ is a distinct swap, and
		\item when applied to $ \pstate $, is always a neighbouring swap,
	\end{enumerate}
	then $ \vec{\beta} $ can always be expressed in the form of a standard formation for $ (\pi,\pi') $.
\end{lemma}

\begin{proof}
	We prove this by induction. 
	Suppose the above lemma is true for $\pstate\in\probspace_{d-1}$. The first goal is to prove this for initial ordering $\pi =  (1,2,\cdots,d)$ and $ \vec{\beta} $ that satisfies the conditions in the statement of the lemma. 
	By identifying the first and the last swaps acting on the level $d$, one can decompose $ \vec{\beta} $ into 
	\begin{align} 
		\vec{\beta} = \vec{\beta}^{(\text{Post})}\vec{\beta}^{(d-\text{rel})}\vec{\beta}^{(\text{Pre})}.
	\end{align}
	Here, $\vec{\beta}^{(\text{Pre})}$ are the swaps coming before the first swap acting on the level $d$, and $\vec{\beta}^{(\text{Post})}$ are the ones after the last swap acting on the level $d$. 
	
	\emph{Case i:} $d = \pi^\prime_{m}$, where $ m\neq 1 $ is the position of level $ d $ in the $ \beta $-ordering of the final state $ \pstate' $. 
	To make a rearrangement of swaps, we first remark a few points using sets $A = \{i\vert\slope_i>\slope_d\}$ and $B = \{i\vert\slope_i<\slope_d\}$. 
	We will update these sets after each swap. 
	In the beginning, there is no element in $B$ and all the other levels except $ d $ are in the set $A$. 
	Next, we note the following:
	\begin{enumerate}
		\item Swapping $i\in A$ and $j\in B$ is not allowed, since they are non-neighbouring.
		
		\item Any $i\in A$ can move to $B$ only when $ \beta^{(i,d)} $ is implemented.
		
		\item Since initially $ B = \emptyset $, any given level either stays in $A$ at all times, or it moves to $B$ at some point and remains so thereafter. 
		This comes from the restriction that in order for $ i $ to move between $ A $ and $ B $, the swap $  \beta^{(i,d)}$ must be used.
		
		\item If $k,l \in A$ when $\beta^{(k,l)}$ is applied, $\beta^{(k,l)}$ precedes $\beta^{(k,d)}$ and $\beta^{(l,d)}$ when they exist in $\vec\beta$.
	\end{enumerate}
	The sets $A$ and $B$ after the whole transformation is determined by the target $ \beta $-ordering $ \pi' $, where we denote them as $A_f = \{\pi^\prime_i\vert i<m\}$ and $B_f = \{\pi^\prime_i\vert i>m\}$. 
	
	Given the constraints above, we know that $ \vec{\beta} $ describes a special process. See Fig.~\ref{fig:lemma8_illustration}, for instance, for a visualisation of this operation. $A$ and $B$ are separated by level $d$ (point 1 above). 
	Starting from $B = \emptyset$, some elements $i\in A$ are transferred to $B$ whenever $ \beta^{(i,d)} $ is implemented. 
	Once this happens, $ i $ cannot go back to $A$, since it would require the repetition of $  \beta^{(i,d)} $ to do so. 
	Visually, this is understood by saying that the bar representing level $d$ in Fig.~\ref{fig:lemma8_illustration} is penetrable from the left only. 
	At the end, $A = A_f$ and $B=B_f$, where elements of $A_f$ never passed through level $d$ (point 3 above). 
	Lastly, if $k,l\in A$, then they have not experienced a swap with level $d$ yet, explaining the point 4 above.  
	
	\begin{figure}[h!]\centering
		\includegraphics[width = 0.88\columnwidth]{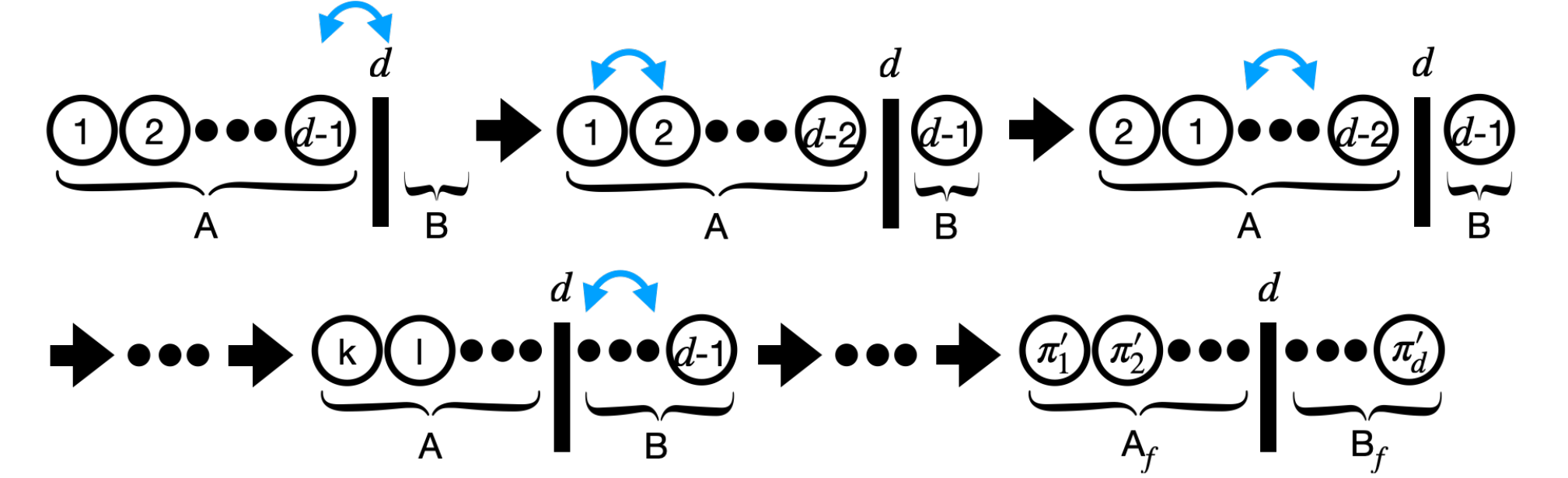}
		\caption{
			Illustration of changes in $\beta$-order starting from $\pi = (1,2,\cdots,d)$ undergoing a swap series $\vec\beta$, where it is assumed that $ \vec\beta $ is always neighbouring and allows no repetition of a particular swap. 
			Level $d$ is depicted as a bar, dividing the levels into sets $A$ and $B$. 
			For an initial state whose $ \beta $-order is monotonically increasing in energy, $ A = \lbrace 1, \cdots, d-1\rbrace $ while $ B = \emptyset$. 
			Three different types of swaps are possible: i) swapping between level $d$ and its neighbouring element which is in $A$ (1st swap above), ii) swapping among neighbouring levels in $A$ (2nd, 3rd steps above), and iii) swaps among levels in $B$ (4th explicit step above). 
			When the type ii) and iii) swaps occur, $A$ and $B$ remain the same. 
			On the other hand, type i) swaps move one element of $A$ into $B$. 
			The process of elements going from $B$ to $A$ is forbidden due to the no-repetition constraint. 
			At the end of applying $\vec\beta$, $A$ and $B$ becomes $A_f$ and $B_f$. 
			The elements of $A_f$ never experience a swap with level $d$, while the elements of $ B_f $ experienced exactly once swapping with level $ d $.
			Figure adapted from Figure~D1 of Ref.~\cite{Son2024_CETO}.
			}
		\label{fig:lemma8_illustration}
	\end{figure}
	
	The next step we want to show is that w.l.o.g., a rearrangement, where $ \vec{\beta}^\mathrm{(Pre)} $ contains all $ \beta$-swaps that are part of $ \vec{\beta} $, acting on elements $ i\in A_f $, but not involving level $ d $, is possible.
	To do so, we identify all swaps $\beta^{(k,l)}$ in $\vec{\beta}^{(\text{Post})}\vec{\beta}^{(d-\text{rel})}$  such that $k,l\in A$ at the time of swap. 
	From the rightmost one, make a decomposition
	\begin{align} 
		\vec{\beta}^{(\text{Post})}\vec{\beta}^{(d-\text{rel})} = \vec{\beta}^{(a)}\beta^{(k,l)}\vec{\beta}^{(b)}.
	\end{align}
	Notice that $\vec{\beta}^{(b)}$ only contains swaps among $B\cup\{d\}$, which includes neither $k$ nor $l$. From Remark \ref{remark: commute}, we then have that $ \beta^{(k,l)} $ and $  \vec{\beta}^{(b)}$ commute, i.e. $\beta^{(k,l)}\vec{\beta}^{(b)} = \vec{\beta}^{(b)}\beta^{(k,l)}$. Therefore, w.l.o.g., 
	\begin{align} 
		\vec{\beta}^{(\text{Post})}\vec{\beta}^{(d-\text{rel})} \vec{\beta}^{(\text{Pre})} = \vec{\beta}^{(a)}\beta^{(k,l)}\vec{\beta}^{(b)} \vec{\beta}^{(\text{Pre})} = \vec{\beta}^{(a)}\vec{\beta}^{(b)} \beta^{(k,l)}\vec{\beta}^{(\text{Pre})}
	\end{align}$ \beta^{(k,l)} $ can be integrated into $ \beta^{(\text{Pre})} $ to update $ \beta^{(\text{Pre})} \rightarrow \beta^{(k,l)}\beta^{(\text{Pre})} $ and $ \vec{\beta}^{(\text{Post})}\vec{\beta}^{(d-\text{rel})} \rightarrow \vec{\beta}^{(a)}\vec{\beta}^{(b)} $.

	If $k\in A_f$, the existence of $\beta^{(k,l)}$ in $\vec\beta$ indicates $l\in A$ when the swap is applied. Hence, by repeating this until the end, all swaps $\beta^{(k,l)}$ with $k\in A_f$ are merged into $\vec{\beta}^{(\text{Pre})}$. 
	Since $\vec{\beta}^{(\text{Pre})}$ does not contain swaps acting on level $d$, it acts on at most $d-1$ levels and can be reordered in the standard formation by the assumption that the lemma holds for states in $ \probspace_{d-1} $. 
	Until now, there is no swap acting on level $ d $ and thus $d = \pi_{\vec{\beta}^{(\text{Pre})}\pstate}(d)$. 
	Then levels in $ B_{f} $, which should be swapped with $ d $ in $ \vec{\beta}^{(d-\text{rel})} $, occupy later $d-m$ slots in the $ \beta $-order: $ \pi_{\vec{\beta}^{(\text{Pre})}\pstate}(m), \pi_{\vec{\beta}^{(\text{Pre})}\pstate}(m+1),\cdots,\pi_{\vec{\beta}^{(\text{Pre})}\pstate}(d-1)$. 
	By construction of the standard formation, $\vec{\beta}^{(\text{Pre})}$ then can be decomposed into $\vec{\beta}^{(\text{Pre})} = \vec{\beta}^{(B_f)}\vec{\beta}^{(A_f)}$ with $\vec{\beta}^{(B_f)}$ swapping only between $ B_{f} $ elements.
	
	Finally, $\vec{\beta}^{(\text{Post})}\vec{\beta}^{(d-\text{rel})}\vec{\beta}^{(B_f)}$ consists of swaps among the levels in $B_f\cup \{d\}$ ($d-m+1< d$ elements). 
	Again, by assumption this swap can be rearranged as a standard formation. 
	Concatenating standardised series $\vec{\beta}^{(\text{Post})}\vec{\beta}^{(d-\text{rel})}\vec{\beta}^{(B_f)}$ and $ \vec{\beta}^{(A_f)} $, we obtain the standard formation for the entire series. \\
	
	\emph{Case ii:} $d = \pi^\prime_{1}$. The only difference here is that level $d$ swaps with every other level and $A_f$ is an empty group. 
	Again, we locate $\beta^{(k,l)}$ such that  $\vec{\beta}^{(\text{Post})}\vec{\beta}^{(d-\text{rel})} = \vec{\beta}^{(a)}\beta^{(k,l)}\vec{\beta}^{(b)}$ from the rightmost swap. 
	If $k,l\in A$ after $\vec{\beta}^{(b)}$, move $\beta^{(k,l)}$ to be included in $\vec{\beta}^{(\text{Pre})}$ as before. 
	In addition, repeat this process for $ \vec{\beta}^{(d-\text{rel})} $ but starting from the leftmost swap $\beta^{(k,l)}$ with $k,l\neq d$ in $\vec{\beta}^{(d-\text{rel})} = \vec{\beta}^{(a)}\beta^{(k,l)}\vec{\beta}^{(b)}$. 
	Since we already moved all $k,l\in A$ swaps to go before $ \vec{\beta}^{(d-\text{rel})} $, at the point of swap $ \beta^{(k,l)} $ all $ k,l\in B $, which leads to the equality $\vec{\beta}^{(a)}\beta^{(k,l)} = \beta^{(k,l)}\vec{\beta}^{(a)}$ and enables $ \vec{\beta}^{(\text{Post})} \rightarrow \vec{\beta}^{(\text{Post})}\vec{\beta}^{(k,l)} $.
	After merging all such $\beta^{(k,l)}$ into $\vec{\beta}^{(\text{Pre})}$ or $\vec{\beta}^{(\text{Post})}$, we get
	\begin{align} 
		\vec{\beta}^{(d-\text{rel})} =\beta^{(\delta_{1},d)}\beta^{(\delta_{2},d)}\cdots\beta^{(\delta_{d-1},d)},
	\end{align} 
	where $\delta = (\delta_{1},\delta_{2},\cdots,\delta_{d-1},d)$ is the $\beta$-order after $\vec{\beta}^{(\text{Pre})}$. 
	Both $\vec{\beta}^{(\text{Pre})}$ and $\vec{\beta}^{(\text{Post})}$ act at most $d-1$ levels, and can be modified into the standard formation. 
	Now to put the entire series into the standard formation, $\delta$ need to be rearranged. 
	We do this starting from $j=1$.
	\begin{enumerate}
		\item Find $m$ such that $\delta_m = j$. If $m = j$, proceed to the last step. 
		If not, previous iterations guarantee that $ \delta_{k}=k $, $ \forall k<j $, which leads to $ m>j $ and $ \delta_{m-1}>j $.
		Defining $L = \{\delta_n\vert n<m-1\}$ and $R = \{\delta_n\vert n>m\}$, we get 
		\begin{align} 
			\vec{\beta}^{(d-\text{rel})} = \vec{\beta}^{(dL)}\beta^{(\delta_{m-1},d)}\beta^{(j,d)}\vec{\beta}^{(dR)},
		\end{align}
		where $ \vec{\beta}^{dL(R)} $ denotes the series swapping $ d $ and elements of $ L(R) $.
		Moreover, from the standardisation $\vec{\beta}^{(\text{Pre})} = \vec{\beta}^{(R)}\beta^{(j,\delta_{m-1})}\vec{\beta}^{(L)}$, where $\vec{\beta}^{(R)}$ does not act on levels $j$ and $\delta_{m-1}$ and we can rearrange it into  $\vec{\beta}^{(dR)}\vec{\beta}^{(\text{Pre})} = \beta^{(j,\delta_{m-1})}\vec{\beta}^{(dR)}\vec{\beta}^{(R)}\vec{\beta}^{(L)}$. 
		From Remark~\ref{remark: swap_series_eq}, $\beta^{(\delta_{m-1},d)}\beta^{(j,d)}\beta^{(j,\delta_{m-1})} = \beta^{(j,\delta_{m-1})}\beta^{(j,d)}\beta^{(\delta_{m-1},d)}$ since $ d>\delta_{m-1}>j $. 
		This procedure updates $\delta \rightarrow  (\cdots,\delta_{m-2},j,\delta_{m-1},\delta_{m+1},\cdots )$.
		
		\item Repeat the first step until $\delta_{j} = j$.
		
		\item Repeat the first and the second step with $ j\rightarrow j+1 $.
	\end{enumerate}
	At the end, one gets $ \delta = (1,2,\cdots,d-1) $ and $ \vec{\beta}^{(\text{Pre})}  = \1 $. 
	By standardizing $ \vec{\beta}^{(\text{Post})} $ and concatenating with $ \vec{\beta}^{(d-\text{rel})} $, the standard formation is obtained. 
	
	For $d=2$, the Lemma is trivially true. 
	Thus by induction, the lemma is proved for $\pi = (1,2,\cdots,d)$. Following the same logic, this can also be proven for $\pi = (d,d-1,\cdots,1)$.
\end{proof}

Now we state the main proof.

\begin{proof}[Proof of Theorem~\ref{theorem: nice order}] 
	We prove the theorem for the case $\pi_{\pstate} = (1,2,\cdots,d)$ and argue that the proof also holds for $\pi_{\pstate} = (d,d-1,\cdots,1)$.
	
	We first prove the only if statement of the theorem, as follows:
	\begin{enumerate}
		\item We show that the repetition of any particular $ \beta $-swap always leads to a non-extreme state if the series is all-neighbouring. 
		This is done by contradiction: suppose that $\vec\beta = \beta^{(k,l)}\vec{\beta}^\prime$, w.l.o.g. assuming $k<l$. 
		Furthermore, assume $\vec{\beta}^\prime$ to be a series satisfying the only if part of the statement but contains $\beta^{(k,l)}$, causing $\beta^{(k,l)}$ to occur twice in  $\vec\beta$. 
		From Lemma~\ref{lemma: standard form}, $\vec{\beta}^\prime$ can be written in a standard formation, which reads
		\begin{align}
			\vec\beta = \beta^{(k,l)}\vec{\beta}^\prime = \beta^{(k,l)}\vec{\beta}^{(\text{Post})}\beta^{(k,l)}\vec{\beta}^{(\text{Pre})}.
		\end{align}   
		Notice that since $\vec\beta$ is an all-neighbouring swap, this implies that after $\vec\beta^\prime$, levels $k$ and $l$ should be neighbouring with $\slope(\vec\beta^\prime\pstate)_k\leq\slope(\vec\beta^\prime\pstate)_l$, which then implies  
		\begin{align}
			\vec{\beta}^{(\text{Post})} = \vec{\beta}^{(\text{Irrel})} \left(\prod_{m_i\in M}\beta^{(m_i,k)}\right)\left(\prod_{m_i\in M}\beta^{(m_i,l)}\right),
		\end{align} 
		for some set of levels $M \subset \{m\vert m<k<l\}$, 
		by construction of the standard formation. Here, $\vec{\beta}^{(\text{Irrel})}$ is a series that acts on neither $k$ nor $l$. Using Remarks~\ref{remark: commute} and~\ref{remark: swap_series_eq},
		\begin{align}
			\vec\beta &= \vec{\beta}^{(\text{Irrel})}\beta^{(k,l)}\prod_{m_{i}\in M}\left(\beta^{(m_i,k)}\beta^{(m_i,l)}\right)\beta^{(k,l)}\vec{\beta}^{(\text{Pre})}\nonumber\\ 
			&= \vec{\beta}^{(\text{Irrel})}   \left(\beta^{(k,l)}\right)^2\prod_{m_{i}\in M}\left(\beta^{(m_i,l)}\beta^{(m_i,k)}\right)  \vec{\beta}^{(\text{Pre})},\label{eq:rearrange_using_comm}
		\end{align}
		which always generates a non-extreme point from Remark~\ref{remark: no_repetition}.
		\item Now we show that non-neighbouring swaps are also not allowed. 
		\begin{enumerate}
			\item \textit{For $ d=3 $}: from Eqs.~\eqref{eq:non-neighbouring-non-extremal1}--\eqref{eq:non-neighbouring-non-extremal5}, $ \beta^{(1,3)} $ yields non-extreme point of $ \setETO(\pstate) $ with $ \pi_{\pstate} = (1,2,3) $ or $ (3,2,1) $. Also, Lemma~\ref{lemma: only one nn} forbids any other occasions having a non-neighbouring swap. 
			
			\item \textit{For $ d>3 $:} suppose that $\beta^{(k,m)}$ is the only non-neighbouring swap in the series $ \vec{\beta} $, i.e. 
			\begin{align} 
				\vec{\beta} = \beta^{(k,m)}\vec{\beta}^{( \text{NS})} 
			\end{align} and  \begin{align} 
				\exists l\quad \text{s.t.}\quad \slope(\vec{\beta}^{(\text{NS})}\pstate)_k>\slope(\vec{\beta}^{(\text{NS})}\pstate)_l>\slope(\vec{\beta}^{(\text{NS})}\pstate)_m,
			\end{align} after all-neighbouring series $ \vec{\beta}^{(\text{NS})} $. From the first part of the proof, $ \vec{\beta}^{(\text{NS})} $ also cannot include any repetition. Using Lemma~\ref{lemma: standard form}, we rearrange $ \vec{\beta}^{(\text{NS})} $ into the standard formation. 
			\begin{enumerate}
				\item  $ \exists l $ satisfying $k<l<m$ or $k>l>m$: firstly, note that the procedures from Eqs.~\eqref{eq:non-neighbouring-non-extremal1}--\eqref{eq:non-neighbouring-non-extremal5} can be generalised for higher dimensions. If $\slope(\rstate)_k>\slope(\rstate)_l>\slope(\rstate)_m$ and $k<l<m$ for some $\rstate$, 
				\begin{align}
					(\beta^{(k,m)}\rstate)_{i} &= (\beta^{(l,m)}\beta^{(k,m)}\beta^{(k,l)}\rstate)_{i},\quad \forall i\neq k,l,\\
					(\beta^{(k,m)}\rstate)_{k} + (\beta^{(k,m)}\rstate)_{l} &= (\beta^{(l,m)}\beta^{(k,m)}\beta^{(k,l)}\rstate)_{k} + (\beta^{(l,m)}\beta^{(k,m)}\beta^{(k,l)}\rstate)_{l},
				\end{align}
				and
				\begin{align}
					\slope(\beta^{(l,m)}\beta^{(k,m)}\beta^{(k,l)}\rstate)_{l} \geq \slope(\beta^{(k,m)}\rstate)_{l} \geq  \slope(\beta^{(k,m)}\rstate)_{k} \geq \slope(\beta^{(l,m)}\beta^{(k,m)}\beta^{(k,l)}\rstate)_{k}, 
				\end{align}
				which implies that $\beta^{(k,m)}\rstate$ can be obtained from $\beta^{(l,m)}\beta^{(k,m)}\beta^{(k,l)}\rstate$ via partial thermalisation of levels $ k $ and $ l $ (cf. $ d $-dimensional case for Lemma~\ref{lemma: order swap}), and thus not extremal in $ \setETO(\rstate) $. 
				Similarly, the result also holds when $ k>l>m $. By putting $ \rstate = \vec{\beta}^{(\text{NS})}\pstate $, the state $ \vec{\beta}\pstate $ is not extremal in $ \setETO(\vec{\beta}^{(\text{NS})}\pstate) $ and thus not extremal in $ \setETO(\pstate) $.
				
				\item $ \exists l $ satisfying $m<l$: $l$ initially has a smaller slope than $m$ in $ \pstate $ and thus swapped with $ m $ during $ \vec{\beta}^{(\text{NS})} $. 
				Denote the last such $ l $ swapped with $ m $ as $ l_{0} $. 
				\begin{enumerate}
					\item If $ l_{0} $ and $ m $ are neighbouring in $ \vec{\beta}^{(\text{NS})}\pstate $, standard formation indicates
					\begin{align} 
						\vec{\beta}^{(\text{NS})} = \vec{\beta}^{(\text{Irrel})}\left(\prod_{n_{i}\in N}\beta^{(n_{i},m)}\right)\left(\prod_{n_{i}\in N}\beta^{(n_{i},l_{0})}\right)\beta^{(m,l_{0})}\vec{\beta}^{(\text{Pre})},
					\end{align}
					for some set of levels $ N\subset\{n\vert n<m\} $. Here, $ \vec{\beta}^{(\text{Irrel})} $ acts on neither $ m $ nor $ l_{0} $.
					Then as in Eq.~\eqref{eq:rearrange_using_comm}, 
					\begin{align}
						\beta^{(k,m)}\vec{\beta}^{(\text{NS})} &= \beta^{(k,m)}\beta^{(m,l_{0})}\vec{\beta}^{(\text{Irrel})}\left(\prod_{n_{i}\in N}\beta^{(n_{i},l_{0})}\right)\left(\prod_{n_{i}\in N}\beta^{(n_{i},m)}\right)\vec{\beta}^{(\text{Pre})}\nonumber\\
						&= \beta^{(k,m)}\beta^{(m,l_{0})}\vec{\beta}^{(\text{Pre-}2)}.\label{eq:rearrange_using_comm2}
					\end{align}
					$ \pi_{\vec{\beta}^{(\text{Pre-}2)}\pstate} = (\cdots,k,\cdots,m,l_{0},\cdots) $ and thus $ \beta^{(k,m)}\beta^{(m,l_{0})} $ produces a non-extreme state from Lemma~\ref{lemma: only one nn}.
					\item If $ l_{0} $ and $ m $ are not neighbouring and there is no $ l $ satisfying $k<l<m$ or $k>l>m$, the levels between $ l_{0} $ and $ m $ are $ \{j_{i}\} $ such that $ j_{i}<k,m $. By the construction of the standard formation and the resulting order  
					\begin{align} 
						\slope(\vec{\beta}^{(\text{NS})} \pstate)_{k}>\slope(\vec{\beta}^{(\text{NS})} \pstate)_{l_{0}}\geq\slope(\vec{\beta}^{(\text{NS})} \pstate)_{j_{i}}>\slope(\vec{\beta}^{(\text{NS})} \pstate)_{m},
					\end{align}
					we know that for all $ j_{i} $: i) $ \beta^{(j_{i},l_{0})} $ and $ \beta^{(j_{i},k)} $ exist in $ \vec{\beta}^{(\text{NS})} $, ii) $ \beta^{(j_{i},k)} $ proceeds $ \beta^{(j_{i},l_{0})} $, and iii) $ \beta^{(j_{i},m)} $ does not exist in $ \vec{\beta}^{(\text{NS})}$. The last condition also implies that after $ \beta^{(j_{i},l_{0})} $, swaps acting on $ j_{i} $ are only acting within the set $ \{j_{i}\} $, which we will denote as $ \vec{\beta}^{(\{j_{i}\})} $.
					Then 
					\begin{align}
						\beta^{(k,m)}\vec{\beta}^{(\text{NS})} &= \beta^{(k,m)}\vec{\beta}^{(\text{Irrel})}\vec{\beta}^{(\{j_{i}\})}\left(\prod_{i}\beta^{(j_{i},l_{0})}\right)\beta^{(l_{0},m)}\vec{\beta}^{(\text{Pre})}\nonumber\\
						&=  \vec{\beta}^{(\{j_{i}\})}\left(\prod_{i}\beta^{(j_{i},l_{0})}\right)\beta^{(k,m)}\vec{\beta}^{(\text{Irrel})}\beta^{(l_{0},m)}\vec{\beta}^{(\text{Pre})},
					\end{align}
					since $ \beta^{(k,m)}\vec{\beta}^{(\text{Irrel})} $ does not act on levels $ l_{0} $ and all $ j_{i} $. Finally, $ l_{0} $ is again neighbouring to $ m $ in $ \pi_{\vec{\beta}^{(\text{Irrel})}\beta^{(l_{0},m)}\vec{\beta}^{(\text{Pre})}\pstate} $ and we can use the argument from case ii.A to prove this state is non-extremal.
				\end{enumerate}
				\item All $ l<k,m $: denote the last such $ l $ swapped with $ k $ as $ l_{0} $ and they are neighbouring from the structure of the standard formation. Similar to Eq.~\eqref{eq:rearrange_using_comm2}, we get $ \beta^{(k,m)}\beta^{(k,l_{0})} $ part, which produces a non-extreme state by Lemma~\ref{lemma: only one nn}.
			\end{enumerate}
			As a result, non-neighbouring swaps are completely ruled out from the candidate of extreme point producing $\beta$-swaps when starting from monotonic order states. 
		\end{enumerate}
	\end{enumerate}
	
	The sufficient condition of the theorem can be shown by recalling two properties: i) there exists at least one extreme point of $\setETO$ for each $\beta$-order (Lemma~\ref{lemma:ext_point_for_each_order}) and ii) $\beta$-swap series satisfying the conditions of the lemma are all equivalent if the pair ($\pi,\pi^\prime$) is identical (Lemma~\ref{lemma: standard form}), which makes them the only candidate for an extreme point with order $ \pi^{\prime} $. 
	
	Lastly, we note that the proof holds for $\pi = (d,d-1,\cdots,1)$ initial states since all we have used are Remark~\ref{remark: swap_series_eq}, Lemma~\ref{lemma: only one nn}, and Lemma~\ref{lemma: standard form}, which hold even when the energy ordering is inverted. 
\end{proof}

\chapter{Lie groups and Lie algebras}\label{chapter: Lie groups}

In this chapter, we introduce some techniques in Lie groups and Lie algebras that is necessary for deriving our results in Chapter~\ref{chapter: CETO}.
\begin{definition}[matrix Lie group~\cite{Book_Liegroups}]
	Let $\mathrm{GL}(n)$ be the set of all invertible $n \times n$ matrices with complex number entries.
	A set of $n \times n$ matrices $G$ is a matrix Lie group if it is a closed subgroup of $\mathrm{GL}(n)$.
\end{definition}
Here, the binary operation used to define the group is the matrix multiplication. 
Hence, $G$ always includes an identity matrix $\1$ and the inverse matrix $M^{-1}$ for any $M\in G$.
$G$ is also closed under the matrix multiplication.
An example includes $\mathrm{U}(n)$ the set of all $n \times n$ unitary matrices.

Now we show that a subset of $\mathrm{U}(n)$ defined by the commutation relation with respect to some $n\times n$ Hermitian matrix $H_{0}$. 
\begin{align}\label{eq: Lie group def}
	G = \{U \,\vert\, U\in \mathrm{U}(n), [U,H_{0}] = 0\},
\end{align}
is also a compact and connected Lie group~\cite{Ende2022_bath}.

If $H_{0}$ is set to be the Hamiltonian $H_{S}\otimes\1_{R}+\1_{S}\otimes H_{R} $, this group $G$ becomes the Lie group corresponding to energy-preserving unitaries given the system and bath Hamiltonian, which is the unitary dilation of thermal operations in Chapter~\ref{chapter: CETO}.
Let us denote the dimensions of the system and bath Hilbert spaces as $ d_{S} $ and $ d_{R} $.

First, we show that it is a Lie group by showing that it is a closed subgroup of $\mathrm{U}(n)$.
We verify that $G$ is a subgroup of $\mathrm{U}(n)$ by checking the conditions:
\begin{enumerate}
	\item $U_{1}U_{2} \in G$ whenever $U_{1},U_{2} \in G$ because 
	\begin{align}
		[U_{1}U_{2},H_{0}] = U_{1}[U_{2},H_{0}] + [U_{1},H_{0}]U_{2} = 0.
	\end{align}
	
	\item $\1 \in G$, since $\1$ commutes with any matrix.
	
	\item $U^{\dagger}\in G$ whenever $U \in G$ because 
	\begin{align}
		[U^{\dagger},H_{0}] = - \left([U,H_{0}]\right)^{\dagger} = 0.
	\end{align}
\end{enumerate}
It is also a closed subgroup of $\mathrm{U}(n)$: if a sequence $\{U_{m}\}_{m}\subset G$ converges to $U$, from $[U_{m},H_{0}] = 0$ we also have $[U,H_{0}] = 0$.

Since the unitary group $\mathrm{U}(n)$ is compact, all its closed subgroups are also compact. 
Finally, we use the fact that any unitary matrix can be written as $U = \sum_{i}e^{i\theta_{i}}\dm{i}$ for an orthonormal basis $\{\ket{i}\}_{i}$ and if $[U,H_{0}] = 0$, we can choose this basis to be an eigenbasis of $H_{0}$.
By choosing the path $U(t) = \sum_{i}e^{i\theta_{i}t}\dm{i}$ connecting $U = U(1)$ and $\1 = U(0)$, we show that $G$ is path-connected and thus connected~\cite{Book_Liegroups}.

We also define Lie algebras corresponding to matrix Lie groups. 
\begin{definition}[Lie algebra of a group~\cite{Book_Liegroups}]
	For a matrix Lie group $G$, the corresponding Lie algebra is defined as 
	\begin{align}
		\mathfrak{g} = \{K \,|\, e^{tK}\in G,\ \forall t\in\mbR\}.
	\end{align}
\end{definition}
Hence, the Lie algebra corresponding to our group Eq.~\eqref{eq: Lie group def} can be written as
\begin{align}\label{eq: Lie algebra def}
	\mathfrak{g} = \{-iH \,\vert\, H\in\mL_{SR},\ H^{\dagger} = H,\ [H_{0},H] = 0 \},
\end{align} 
the set of interaction Hamiltonians (times $ -i $) that commute with $ H_{0} $. 
Finally, we define the set $\mK = \{K_{1},K_{2},\cdots,K_{L}\}$ that generates the Lie algebra $\mathfrak{g}$; that is, $\mathfrak{g}$ is the real linear span of a set consisting of all elements of $\mK$ and their (repeated) commutators. 
The following lemma then holds for $ G $ and $ \mathfrak{g} $.

\begin{lemma}[Appendix D, Lemma 1 of~\cite{Book_Control}]\label{lemma: Lie controllability} 
	If a set $ \{K_{1},K_{2},\cdots,K_{L}\} $ generates a Lie algebra $ \mathfrak{g} $, any element $ U $ of the corresponding connected Lie group $ G $ can be expressed as
	\begin{align}\label{eq: finite decomposition}
		U = \prod_{n=1}^{S}\exp(K_{i_{n}}t_{n}),
	\end{align}
	where $ S\in\mathbb{N} $ is a finite number, $ i_{n}\in\{1,\cdots,L\} $, and $ t_{n}\in\mathbb{R} $.
\end{lemma}

Lemma~\ref{lemma: Lie controllability} is a stronger result than the usual results showing that $e^{A+B}$ or $e^{[A,B]}$ can be written as a product of \emph{infinite} sequence of $e^{At}$ and $e^{Bt}$ for some $t$.
However, Lemma~\ref{lemma: Lie controllability} lacks the explicit construction, i.e. the existence of Eq.~\eqref{eq: finite decomposition} is known, but not its exact form. 
Therefore, the potentially infinite decomposition of $e^{A+B}$ and $e^{[A,B]}$ are often used as a universal tool for compiling desired unitaries. 
\begin{proposition}[Lie product formula~\cite{Book_Liegroups}]\label{proposition: Lie product formula}
	For $n \times n$-dimensional matrices $A$ and $B$,
	\begin{align}
		\left(e^{\frac{A}{m}}e^{\frac{B}{m}}\right)^{m} = e^{A+B+O(\frac{1}{m})}.
	\end{align}
	This implies that 
	\begin{align}
		\lim_{m\to\infty}\left(e^{\frac{A}{m}}e^{\frac{B}{m}}\right)^{m} = e^{A+B}.
	\end{align}
\end{proposition}
This result can be extended to certain unbounded operators, in which case it is called as the Trotter product formula~\cite{Trotter59_Trotter}.
Higher-order formulae with better error scaling to the number of $e^{tA}$ and $e^{tB}$ queries also exist~\cite{Hatano2005}.

Similarly, the exponential of a commutator can be decomposed into the exponential of each matrix. 
\begin{proposition}[product formula for commutators~\cite{Book_Control}]\label{proposition: product formula}
	For two matrices $A$ and $B$, 
	\begin{align}
		e^{-tA}e^{-tB}e^{tA}e^{tB} = e^{t^{2}[A,B] + O(t^{3})}.
	\end{align}
	This implies that 
	\begin{align}
		\left(e^{-\frac{1}{\sqrt{m}}A}e^{-\frac{1}{\sqrt{m}}B}e^{\frac{1}{\sqrt{m}}A}e^{\frac{1}{\sqrt{m}}B}\right)^{m} = e^{[A,B] + O(m^{-\frac{3}{2}})}.
	\end{align}
\end{proposition}
Again, higher-order formulae can be useful in practice~\cite{Chen2022ProductFormulae}.

}

\backmatter 
\setchapterstyle{plain} 


\printbibliography[heading=bibintoc, title=Bibliography] 

\end{document}